\documentclass[11pt,a4paper,twoside]{article}
\newif\ifdraft \global\drafttrue
\def\production{\global\draftfalse}
\production
\usepackage[T1]{fontenc}
\usepackage[latin1]{inputenc}
\usepackage{a4wide}
\usepackage{times}
\usepackage{graphicx}
\usepackage{theorem}
\usepackage[colorlinks=true,hyperindex=true]{hyperref}
\usepackage{color}
\usepackage{fancyhdr}
\usepackage{latexsym}
\usepackage{amsmath}
\usepackage{bbm}
\usepackage{amssymb}
\usepackage{amsfonts}
\usepackage{enumerate}
\usepackage{cancel}
\usepackage{caption}
\usepackage{makeidx}
\ifdraft
\fi
\numberwithin{equation}{section}

\def\init{\setcounter{equation}{0}}

\newcounter{smallarabics}

\newcounter{smallroman}

\newcommand{\ben}{\begin{enumerate}[{\rm (1)}]}
\newcommand{\een}{\end{enumerate}}

\newtheorem{theoreme}{Theorem}[section]
\newtheorem{proposition}[theoreme]{Proposition}
\newtheorem{lemma}[theoreme]{Lemma}
\newtheorem{corollary}[theoreme]{Corollary}
{\theorembodyfont{\upshape}
\newtheorem{definition}[theoreme]{Definition}
\newtheorem{remark}[theoreme]{Remark}

}

\def\bep{\begin{proposition}}
\def\eep{\end{proposition}}
\def\bel{\begin{lemma}}
\def\eel{\end{lemma}}
\def\bet{\begin{theoreme}}
\def\eet{\end{theoreme}}
\def\bed{\begin{definition}}
\def\eed{\end{definition}}
\def\bec{\begin{corollary}}
\def\eec{\end{corollary}}
\def\ber{\begin{remark}}
\def\eer{\end{remark}}

\def\rr{{\mathbb R}}
\def\zz{{\mathbb Z}}
\def\cc{{\mathbb C}}
\def\nn{{\mathbb N}}
\def\Z{{\mathbb Z}}
\def\EE{\mathbb{E}}
\def\PP{\mathbb{P}}
\def\QQ{\mathbb{Q}}
\def\WW{\mathbb{W}}

\def\textsl{{}}

\def\Im{{\rm Im}\,}
\def\Re{{\rm Re}\,}

\def\c0inf{C_0^\infty}

\def\cH{{\cal  H}}

\def\cR{{\cal R}}
\def\cA{{\cal A}}

\def\dist{{\rm dist}}

\def\i{{\rm i}}
\def\bx{\boldsymbol{x}}

\newcommand{\beq}{\begin{equation}}
\newcommand{\eeq}{\end{equation}}
\newcommand{\bear}[1]{\begin{array}{#1}}
\newcommand{\ear}{\end{array}}

\newcommand{\bal}{\begin{align*}}
\newcommand{\eal}{\end{align*}}
\newcommand*\centermathcell[1]{\omit\hfil$\displaystyle#1$\hfil\ignorespaces}

\def\sp{{\hat e}}

\newcommand{\e}{\mathrm{e}}
\renewcommand{\i}{\mathrm{i}}

\renewcommand{\d}{\mathrm{d}}

\def\qed{$\Box$\medskip}
\def\cP{{\cal P}}
\def\cJ{{\cal J}}\def\cC{{\cal C}}

\def\cG{{\cal G}}
\def\cI{{\rm I}}

\def\cM{{\cal M}}
\def\cA{{\cal A}}
\def\cQ{{\cal Q}}
\def\cK{{\cal K}}
\def\cL{{\cal L}}
\def\R{{\rm R}}

\def\bar{\overline}

\def \p{ \partial}
\def\X{{\cal X}}
\def\12{{\frac12}}

\def\dd{{\bf D}}

\def\e{{\rm e}}
\def\cD{{\cal D}}

\def\d{{\rm d}}
\def\Ran{{\rm Ran}\,}

\def\cH{{\cal H}}

\def\Ker{{\rm Ker}\,}
\def\Dom{{\rm Dom}\,}

\def\sp{{\rm sp}}
\def\ep{{\rm ep}}

\def\cS{{\cal S}}
\def\cR{{\cal R}}

\def\R{{r}}

\def\C{{\rm C}}

\def\Ent{{\rm Ent}}
\def\Ep{{\rm Ep}}
\def\cE{{\cal E}}
\def\cW{{\cal W}}
\def\cX{{\cal X}}

\def\cV{{\cal V}}
\def\cI{{\cal I}}
\def\fh{\mathfrak{h}}

\def\fM{\mathfrak{M}}
\def\fH{\mathfrak{H}}
\def\fS{\mathfrak{S}}

\def\fI{\mathfrak{I}}
\def\fX{\mathfrak{X}}

\def\fm{\mathfrak{m}}
\def\fC{\mathfrak{C}}
\def\fR{\mathfrak{R}}

\def\tr{{\rm tr}}

\def\cO{{\cal O}}
\newcommand{\ds}{\displaystyle}

\makeindex

\begin{document}
\def\today{}
\title{Entropic fluctuations in thermally driven harmonic networks}
\author{V. Jak\v{s}i\'c$^{1}$,  C-A. Pillet$^{2}$, A. Shirikyan$^{1,3}$
\\ \\ 
$^1$Department of Mathematics and Statistics, 
McGill University, \\
805 Sherbrooke Street West, 
Montreal,  QC,  H3A 2K6, Canada
\\ \\
$^2$Aix-Marseille Universit\'e, CNRS, CPT, UMR 7332, Case 907, 13288 Marseille, France\\
Universit\'e de Toulon, CNRS, CPT, UMR 7332, 83957 La Garde, France\\
FRUMAM
\\ \\
$^3$Department of Mathematics, University of Cergy--Pontoise\\
CNRS UMR 8088, 2 avenue Adolphe Chauvin\\
95302 Cergy--Pontoise, France
}
\maketitle
\centerline {\em Dedicated to David Ruelle and Yakov Sinai}
\centerline{\em on the occasion of their 80th birthday}

\begin{quote}
{\small
{\bf Abstract.} We consider a general network of harmonic oscillators driven
out of thermal equilibrium by coupling to several heat reservoirs at different
temperatures. The action of the reservoirs is implemented by Langevin
forces. Assuming the existence and uniqueness of the steady state of the 
resulting process, we construct a canonical entropy production functional  
$S^t$ which satisfies the Gallavotti--Cohen fluctuation theorem. More 
precisely, we prove that there exists $\kappa_c>\frac12$ such that the 
cumulant generating function of $S^t$ has a large-time limit $e(\alpha)$ which 
is finite on a closed interval $[\12-\kappa_c,\12+\kappa_c]$, infinite on its
complement and satisfies the Gallavotti--Cohen symmetry $e(1-\alpha)=e(\alpha)$ 
for all $\alpha\in\rr$. Moreover, we show that $e(\alpha)$ is essentially 
smooth, i.e., that $e'(\alpha)\to\mp\infty$ as $\alpha\to\tfrac12\mp\kappa_c$. 
It follows from the G\"artner-Ellis theorem that $S^t$ satisfies a global 
large deviation principle with a rate function $I(s)$ obeying the 
Gallavotti--Cohen fluctuation relation $I(-s)-I(s)=s$ for all $s\in\rr$. We 
also consider perturbations of $S^t$ by quadratic boundary terms and prove 
that they satisfy extended fluctuation relations, i.e., a global large 
deviation principle with a rate function that typically differs from $I(s)$
outside a finite interval. This applies to various physically relevant 
functionals and, in particular, to the heat dissipation rate of the network. 
Our approach relies on the properties of the maximal solution of a 
one-parameter family of algebraic matrix Riccati equations. It turns out that 
the limiting cumulant generating functions of $S^t$ and its perturbations can 
be computed in terms of spectral data of a Hamiltonian matrix depending on the 
harmonic potential of the network and the parameters of the Langevin 
reservoirs. This approach is well adapted to both analytical and numerical 
investigations.
}
\end{quote}
\thispagestyle{empty}
\newpage

\tableofcontents

\section{Introduction} 
\label{SEC-Intro}

Boundary driven mechanical systems are paradigmatic in nonequilibrium 
statistical mechanics. Existence and uniqueness of nonequilibrium steady 
states have been extensively studied for a variety of such systems: 
harmonic~\cite{LS1} and anharmonic~\cite{BK} crystals, 1-dimensional chains 
of anharmonic oscillators~\cite{EPR1,EPR2,EH1,EH2,RT1,Ca,BL},
rotors~\cite{CEP,CE2} and other
Hamiltonian systems~\cite{EY,LY1,LY2,CE}. More general Hamiltonian networks
have been considered in~\cite{EZ,MNV,CE1}. In this paper, we shall study stochastically driven networks of 
harmonic oscillators which are the simplest models in the last category. 
The questions of existence and uniqueness of the steady state is well
understood in such systems. Estimates of the rate of relaxation to the 
steady state are also available~\cite{RT2,AE}. The focus of this work is on 
the concept of entropy production and its fluctuations,
although our approach can be extended to cover  the fluctuations 
of energy/entropy fluxes between individual heat reservoirs and the network. 
The universal {\sl fluctuation relations} satisfied by the entropy production 
rate (or phase-space contraction rate) in transient~\cite{ECM,ES} and 
stationary~\cite{GC1,GC2} processes have 
been one of the central issues in the recent developments of nonequilibrium
statistical mechanics. Various approaches to these relations have been
proposed in the literature and we refer the reader 
to~\cite{RM,Se,LS2,Ma,CFG,CG,JPR,JOPP} for reviews and detailed 
discussions. The interested reader should also consult~\cite{RT3}, where 
fluctuation relations are derived for boundary driven anharmonic chains, 
and~\cite{JPS} for a discussion of these topics in the framework of Gaussian 
dynamical systems.  For theoretical and experimental works dealing 
specifically with mechanically driven harmonic systems we refer the reader to~\cite{JGC,JGDPC,KSD}. 

In this paper we follow the scheme  advocated in~\cite{JPR,JOPP} and 
fully elaborated in~\cite{JNPPS}. The details are as follows. 

Consider a probability space $(\Omega,\cP,\PP)$ 
equipped with a measurable involution $\Theta:\Omega\to\Omega$. Suppose that 
the measures $\PP$ and $\widetilde{\PP}=\PP\circ\Theta$ are equivalent.
We define the {\sl canonical entropic functional} of the quadruple 
$(\Omega,\cP,\PP,\Theta)$ by
\beq
S(\omega)=\log\frac{\d\PP}{\d\widetilde{\PP}}(\omega),
\label{EQ-Sdef}
\eeq
and denote by $P$ the law of this random variable under $\PP$. Since
\beq
S\circ\Theta(\omega)
=\log\frac{\d\PP\circ\Theta}{\d\widetilde{\PP}\circ\Theta}(\omega)
=\log\frac{\d\widetilde{\PP}}{\d\PP}(\omega)
=-S(\omega),
\label{EQ-Ssymmetry}
\eeq
the support of $P$ is symmetric w.r.t.\;the origin. It reduces to $\{0\}$
whenever $\widetilde{\PP}=\PP$. In the opposite case the symmetry $\Theta$
is broken and the well known fact that the relative entropy of $\PP$ 
w.r.t.\;$\widetilde{\PP}$, given by\index{$\Ent$}
$$
\Ent(\PP|\widetilde{\PP})=-\int_\Omega S(\omega)\PP(\d\omega)
=-\int_{\rr}sP(\d s)
$$
is strictly negative (it vanishes iff $\PP=\widetilde{\PP}$) shows that the 
law $P$ favors positive values of $S$. To obtain a more quantitative statement 
of this fact, it is useful to consider R\'enyi's relative $\alpha$-entropy\index{$\Ent_\alpha$}\index{R\'enyi entropy}
$$
\Ent_\alpha(\PP|\widetilde{\PP})
=\log\int_\Omega\e^{\alpha S(\omega)}\widetilde{\PP}(\d\omega).
$$
Note that $\Ent_0(\PP|\widetilde{\PP})=\Ent_1(\PP|\widetilde{\PP})=0$, and
since the function $\rr\ni\alpha\mapsto\Ent_\alpha(\PP|\widetilde{\PP})$ is 
convex by H\"older's inequality, one has
$\Ent_\alpha(\PP|\widetilde{\PP})\le0$ for $\alpha\in[0,1]$. 
It is straightforward to check that $\Ent_\alpha(\PP|\widetilde{\PP})$ is a real-analytic 
function of $\alpha$ on some open interval containing $]0,1[$ and infinite
on the (possibly empty) complement of its closure. In particular, it is
strictly convex on its analyticity interval.

From the definition of $\widetilde{\PP}$ and Relation~\eqref{EQ-Ssymmetry} 
we deduce
\beq
\Ent_\alpha(\PP|\widetilde{\PP})
=\log\int_\Omega\e^{\alpha S\circ\Theta(\omega)}\PP(\d\omega)
=\log\int_\Omega\e^{-\alpha S(\omega)}\PP(\d\omega)
=\log\int_\Omega\e^{-\alpha S(\omega)}\frac{\d\PP}{\d\widetilde{\PP}}(\omega)
\widetilde{\PP}(\d\omega),
\label{EQ-RenyiOne}
\eeq
and the definition of $S$ yields
$$
\log\int_\Omega\e^{-\alpha S(\omega)}\frac{\d\PP}{\d\widetilde{\PP}}(\omega)
\widetilde{\PP}(\d\omega)
=\log\int_\Omega\e^{(1-\alpha) S(\omega)}\widetilde{\PP}(\d\omega)
=\Ent_{1-\alpha}(\PP|\widetilde{\PP}).
$$
It follows that R\'enyi's entropy satisfies the symmetry relation
\beq
\Ent_{1-\alpha}(\PP|\widetilde{\PP})=\Ent_{\alpha}(\PP|\widetilde{\PP}),
\label{EQ-Mother}
\eeq
which, in applications to dynamical systems, will turn into the so-called
Gallavotti--Cohen symmetry.
The second equality in Eq.~\eqref{EQ-RenyiOne} allows us to express
R\'enyi's entropy in terms of the law $P$ as
$$
\Ent_\alpha(\PP|\widetilde{\PP})=e(\alpha)
=\log\int_\rr\e^{-\alpha s}P(\d s).
$$
Note that, up to the sign of $\alpha$, $e(\alpha)$ is the the cumulant 
generating function of the random variable $S$. Denoting by $\widetilde{P}$ the 
law of $-S$ under $\PP$, the symmetry~\eqref{EQ-Mother} leads to
$$
\int_\rr\e^{\alpha s}\widetilde{P}(\d s)
=\int_\rr\e^{-\alpha s}P(\d s)
=\int_\rr\e^{-(1-\alpha)s}P(\d s)
=\int_\rr\e^{\alpha s}\e^{-s}P(\d s),
$$
from which we obtain
\beq
\frac{\d\widetilde{P}}{\d P}(s)=\e^{-s}
\label{EQ-GenericFT}
\eeq
on the common support of $P$ and $\widetilde{P}$. Thus, negative values of 
$S$ are exponentially suppressed by the universal weight $\e^{-s}$. In 
the physics literature such an identity is called a {\sl fluctuation relation} 
or a {\sl fluctuation theorem} for the quantity described by $S$. Most often 
$S$ is a measure of the power injected in a system or of the rate at which it 
dissipates heat in some thermostat. The equivalent symmetry of the cumulant
generating function $e(\alpha)$ of $S$ which follows from the symmetry~\eqref{EQ-Mother} of R\'enyi's entropy
\beq
e(1-\alpha)=e(\alpha)
\label{EQ-GCsym0}
\eeq
is referred to as the {\sl Gallavotti--Cohen symmetry}\index{Gallavotti--Cohen symmetry}. The name {\sl symmetry 
function}\index{symmetry function} is sometimes given to
$$
\mathfrak{s}(s)=\log\frac{\d P}{\d\widetilde{P}}(s).
$$
In terms of this function, the fluctuation relation is expressed as 
$$
\frak{s}(s)=s.
$$
The above-mentioned fact that
$$
0=\Ent_1(\PP|\widetilde{\PP})=\log\int_\rr\e^{-s}P(\d s),
$$
rewritten as
\beq
\int_\rr\e^{-s}P(\d s)=1,
\label{EQ-JarzynskiIdentity}
\eeq
constitute the associated {\sl Jarzynski identity}\index{Jarzynski identity} and the strict negativity
of relative entropy
\beq
0<-\Ent(\PP|\widetilde{\PP})=\int sP(\d s),
\label{EQ-JarzynskiInequalityt}
\eeq
becomes {\sl Jarzynski's inequality.}\index{Jarzynski's inequality}

In all known applications of the above scheme to nonequilibrium statistical 
mechanics, the space $(\Omega,\cP,\PP)$ describes the space-time statistics
of the physical system under consideration over some finite time interval
$[0,t]$ (in the 
following, we shall denote by a superscript or a subscript  the dependence of various objects
on the length $t$ of the considered time interval). The involution $\Theta^t$ is related to time-reversal and the canonical entropic functional 
$S^t$  to entropy production or phase space contraction. The 
fluctuation relation~\eqref{EQ-GenericFT} as a fingerprint of time-reversal 
symmetry breaking and the strict inequality in~\eqref{EQ-JarzynskiInequalityt} 
is a  signature of nonequilibrium.

\begin{figure}
\centering
\includegraphics[scale=0.5]{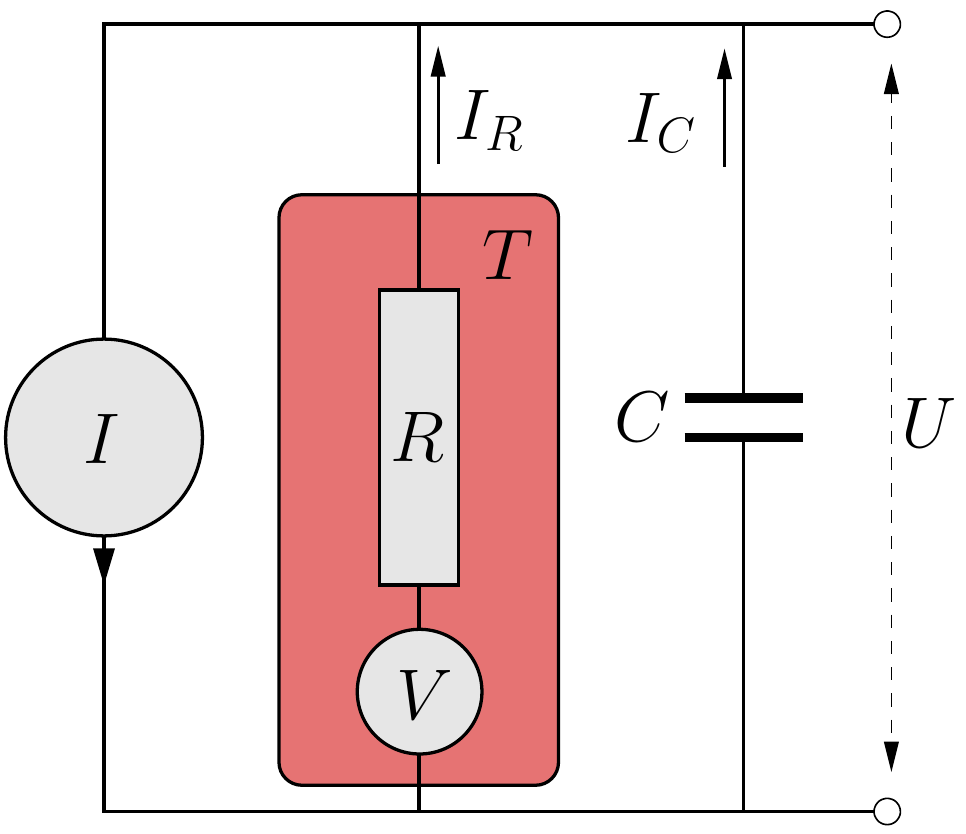}
\caption{A parallel RC circuit is fed with a constant current $I$. The 
resistor $R$ is in contact with a heat bath at temperature $T$. The 
Johnson--Nyquist thermal noise in this resistor generates a fluctuating 
electromotive force~$V$ which contributes to the potential difference 
$U=RI_R+V$ driving the capacitor~$C$.}
\label{FigCircuit}
\end{figure}

The practical implementation of our scheme to nonequilibrium statistical 
mechanics requires 4 distinct steps which will structure our treatment
of thermally driven harmonic networks. In order to clearly formulate the
purpose of each of these steps, we illustrate the procedure at hand on a
very simple model of electrical RC-circuit described in 
Figure~\ref{FigCircuit}. We shall not provide detailed proofs of our claims 
in this example since they all reduce to elementary calculations. We refer the 
reader to~\cite{ZCC} for a detailed physical analysis and to~\cite{GC} for 
experimental verification of the fluctuation relations for this system. 

\bigskip
\noindent{\bf Step 1: Construction of the canonical entropic functional}

\medskip\noindent
The internal energy of the circuit of Figure~\ref{FigCircuit} is stored in
the electric field within the capacitor and is given by
\beq
E=\frac{z^2}{2C},
\label{EQ-CircEnergy}
\eeq
where $z$ denotes the charge on the  plate of the capacitor and~$C$ is the capacitance. The equation of motion for~$z$ is
$$
\dot z_t=I-\frac{z_t}{RC}+\frac{V_t}R,
$$
where $I$ is the constant current fed into the circuit and $V_t$ the
electromotive force (emf) generated by the Johnson--Nyquist thermal noise within the 
resistor $R$. Integrating the equation of motion gives
\beq
z_t=\e^{-t/RC}z_0+(1-\e^{-t/RC})RCI+\frac1R\int_0^t\e^{-(t-s)/RC}V_s\d s.
\label{EQ-qtForm}
\eeq
To simplify our discussion (and to avoid stochastic integrals and the 
technicalities related to time-reversal), we shall assume that $V_t$ has the
form
$$
\frac{V_t}R=\sum_{k=1}^\infty \xi_k\delta(t-k\tau),
$$
where $\tau\ll\tau_0=RC$ and $\xi_k$ denotes a sequence of i.i.d.\;centered 
Gaussian random variables with variance $\sigma^2$. Sampling the charge at 
times $n\tau+0$ yields a sequence $z_0,z_1,z_2,\ldots$ satisfying the recursion relation
$$
z_{k+1}=\eta z_k+(1-\eta)\bar z+\xi_{k+1},
$$
where $\bar z=I\tau_0$ and $\eta=\e^{-\tau/\tau_0}$. According 
to~\eqref{EQ-qtForm}, the charge between two successive kicks is given by
\beq
z_{k\tau+s}=\e^{-s/\tau_0}z_k+(1-\e^{-s/\tau_0})\bar z,\quad
s\in]0,\tau[.
\label{EQ-qtForm2}
\eeq
Assuming $z_0$ to be independent of~$\{\xi_k\}$, the sequence 
$z_0,z_1,z_2\ldots$ is a Markov chain with transition kernel
\beq
p(z'|z)=\frac1{\sqrt{2\pi\sigma^2}}
\e^{-(z'-\eta z-(1-\eta)\bar z)^2/2\sigma^2}.
\label{EQ-TransForm}
\eeq
One easily checks that the unique invariant measure for this chain has the pdf
\beq
p_{\rm st}(z)=\frac1{\sqrt{2\pi\sigma^2/(1-\eta^2)}}\,
\e^{-(z-\bar z)^2(1-\eta^2)/2\sigma^2}.
\label{EQ-NessForm0}
\eeq
In the case $I=0$ (no external forcing), according to the
zero${}^{\rm th}$ law of thermodynamics, the system should relax 
to its thermal equilibrium at the temperature $T$ of the heat bath. Thus, in this case
the invariant measure should be the equilibrium Gibbs state of the circuit at 
temperature $T$ which, by~\eqref{EQ-CircEnergy}, has the pdf
$$
p_{\rm eq}(z)=\frac1{\sqrt{2\pi k_BTC}}\,
\e^{-z^2/2k_BTC},
$$
$k_B$ denoting Boltzmann's constant. This requirement fixes the value of variance of $\xi_k$'s and 
$$
\sigma^2=k_BTC(1-\eta^2).
$$

One can show (see Section~8 in~\cite{Bi}) that, in the limit $\tau\to0$, the covariance of the 
fluctuating emf $V_t$ converges to
$$
\langle V_sV_t\rangle=2k_BTR\delta(s-t),
$$
in accordance with the Johnson-Nyquist formula (\cite{Ny}, see 
also~\cite[Section~IX.2]{vK}). For 
$I\not=0$, Eq.~\eqref{EQ-NessForm0} describes a nonequilibrium steady state 
(NESS) of the system. In the following, we shall consider the stationary 
Markov chain started with the invariant measure and denote by
$\langle\,\cdot\,\rangle_{\rm st}$ the corresponding expectation.

The pdf of a finite segment $Z_n=(z_0,\ldots,z_n)\in\rr^{n+1}$ of the 
stationary process is given by
\beq
p_n(Z_n)=p(z_n|z_{n-1})\cdots p(z_1|z_0)p_{\rm st}(z_0),
\label{EQ-pweight}
\eeq
which is the Gaussian measure on $\rr^{n+1}$ with mean and covariance
$$
\langle z_k\rangle_{\rm st}=\bar z,\qquad
\langle z_kz_j\rangle_{\rm st}
-\langle z_k\rangle_{\rm st}\langle z_j\rangle_{\rm st}
=k_BTC\e^{-|k-j|\tau/\tau_0}.
$$
We chose the involution $\Theta:\rr^{n+1}\to\rr^{n+1}$ to be the composition 
of charge conjugation $z\mapsto-z$ with time-reversal of the Markov chain,
$$
\Theta:(z_0,\ldots,z_n)\mapsto(-z_n,\ldots,-z_0).
$$
The time-reversed process is the Markov chain which assigns the 
weight~\eqref{EQ-pweight} to the reversed segment $\Theta(Z_n)$. Thus, the 
transition kernel $\tilde p(z'|z)$ and invariant measure
$\tilde p_{\rm st}(z)$ of the time-reversed process must satisfy
\beq
\tilde p(-z_0|-z_1)\cdots\tilde p(-z_{n-1}|-z_n)
\tilde p_{\rm st}(-z_n)
=p(z_n|z_{n-1})\cdots p(z_1|z_0)p_{\rm st}(z_0)
\label{EQ-TRrecursive}
\eeq
for all $n\ge1$ and $Z_n\in\rr^{n+1}$. For $n=1$, this equation becomes 
\beq
\tilde p(-z_0|-z_1)\tilde p_{\rm st}(z_1)=p(z_1|z_0)p_{\rm st}(z_0).
\label{EQ-DetailedBalance}
\eeq
Integrating both sides  over $z_1$ gives
$$
\tilde p_{\rm st}(-z_0)=p_{\rm st}(z_0),
$$
from which we further deduce
$$
\tilde p(-z_0|-z_1)
=p(z_1|z_0)\frac{p_{\rm st}(z_0)}{p_{\rm st}(z_1)}.
$$
One then easily checks that~\eqref{EQ-TRrecursive} is indeed satisfied
for all $n\ge1$. Note that in the case $I=0$ one has 
$$
p(-z'|-z)=p(z'|z),\qquad
p_{\rm st}(z)=p_{\rm st}(-z),
$$
and it follows that $\tilde p(z'|z)=p(z'|z)$,
Eq.~\eqref{EQ-DetailedBalance} turning into the detailed balance condition.
In this case, the time-reversed process coincides with the direct one: in 
thermal equilibrium, the time-reversal symmetry holds.
However, in the nonequilibrium case $I\not=0$, time-reversal invariance is 
broken and $\tilde p_{\rm st}(z)\not=p_{\rm st}(z)$.

We are now ready to describe  the canonical entropic functional. 
Applying our general scheme to the marginal $\PP^{n\tau}$ of the finite
segment~$Z_n$ (which has the pdf~$p_n$), we can write~\eqref{EQ-Sdef} as
\begin{align*}
S^{n\tau}
=\log\frac{\d\PP^{n\tau}}{\d\widetilde{\PP}^{n\tau}}(Z_n)
=\log\frac{p_n(Z_n)}{p_n(\Theta(Z_n))}
&=\log\frac%
{p(z_n|z_{n-1})\cdots p(z_1|z_0)p_{\rm st}(z_0)}%
{p(-z_0|-z_1)\cdots p(-z_{n-1}|-z_n)p_{\rm st}(-z_n)}\\
&=\sum_{k=0}^{n-1}\log\frac{p(z_{k+1}|z_k)}{p(-z_k|-z_{k+1})}
+\log\frac{p_{\rm st}(z_0)}{p_{\rm st}(-z_n)}.
\end{align*}
Eqs.~\eqref{EQ-TransForm} and~\eqref{EQ-NessForm0} yield
\begin{align*}
\log\frac{p(z'|z)}{p(-z|-z')}
&=-\frac1{k_BTC}\left(\frac{z^{\prime 2}}2-\frac{z^2}2
-\frac{1-\eta}{1+\eta}(z+z')\bar z\right),\\[4pt]
\log\frac{p_{\rm st}(z_0)}{p_{\rm st}(-z_n)}
&=\frac1{k_BTC}\left(\frac{z_n^2}2-\frac{z_0^2}2
+(z_0+z_n)\bar z\right),
\end{align*}
from which we deduce 
$$
S^{n\tau}=\frac1{k_BT}
\left[\frac{1-\eta}{1+\eta}\sum_{k=0}^{n-1}z_k
+\frac{z_n+\eta z_0}{1+\eta}
\right]\frac{2\bar z}C.
$$

\bigskip
\noindent{\bf Step 2: Deriving a large deviation principle}

\medskip\noindent
From a more mathematical point of view, as stressed by 
Gallavotti--Cohen~\cite{GC1,GC2}, the interesting question is whether the
entropic functional~$S^t$ satisfies a large deviation principle in the limit 
$t\to\infty$. More precisely, is it possible to control the large 
fluctuations of~$S^t$ by a rate function $\rr\ni s\mapsto I(s)$ such that
$$
\PP\left[\frac1t S^t\in \cS\right]
\approx\e^{-t\inf_{s\in\cS}I(s)},
$$
as $t\to\infty$ for any open set $\cS\subset\rr$\,? Moreover, does this
rate function satisfy the relation
\beq
I(-s)=I(s)+s,
\label{EQ-FTLDP}
\eeq
which is the limiting form of~\eqref{EQ-GenericFT}, {\sl for all $s\in\rr$}\,? 
Finally, can one relate this rate function to the large-time asymptotics of 
R\'enyi's entropy via a Legendre transformation
$$
I(s)=\sup_{\alpha\in\rr}
\left(\alpha s-e(-\alpha)\right),\qquad
e(\alpha)=\limsup_{t\to\infty}
\frac1t\Ent_\alpha(\PP^t|\widetilde{\PP}^t),
$$
as suggested by the theory of large deviations? To illustrate these points,
we return to our simple example.

\medskip
For this very particular system,
the fluctuation relation~\eqref{EQ-GenericFT} essentially fixes
the law of the random variable $S^{n\tau}$. Indeed, since $S^{n\tau}$ is 
Gaussian under the law of the stationary process (as a linear combination of 
Gaussian random variables~$\xi_k$), its pdf~$P^{n\tau}$ is completely  
determined by the mean~$\bar s_n$ and variance~$\sigma_n^2$ of $S^{n\tau}$. 
A simple calculation based on~\eqref{EQ-GenericFT} shows that
$\sigma_n^2=2\bar s_n$, whence it follows that
\begin{equation} \label{pdf_n}
P^{n\tau}(s)=\frac1{\sqrt{4\pi\bar s_n}}\e^{-(s-\bar s_n)^2/4\bar s_n^2},
\end{equation}
where we set
$$
\bar s_n=\langle S^{n\tau}\rangle_{\rm st}
=\frac1{k_BT}\left(\frac{1-\eta}{1+\eta}n+1\right)\frac{2\bar z^2}C.
$$
We conclude that
\beq
e_{n\tau}(\alpha)=\Ent_\alpha(\PP^{n\tau}|\widetilde{\PP}^{n\tau})
=\log\int\e^{-\alpha s}P^{n\tau}(s)\d s
=-\alpha(1-\alpha)\bar s_n,
\label{EQ-RotorCumulant}
\eeq
and hence
$$
e(\alpha)=
\lim_{n\to\infty}\frac1{n\tau}e_{n\tau}(\alpha)
=-\alpha(1-\alpha)\bar s,\qquad
\bar s=\frac1{k_BT}\frac{1-\eta}{1+\eta}\frac{2\bar z^2}{C\tau}.
$$
A direct calculation using~\eqref{pdf_n} implies that, for any open set $\cS\subset\rr$, 
$$
\PP^{n\tau}\left[\frac{S^{n\tau}}{n\tau}\in\cS\right]
\approx\e^{-n\tau\inf_{s\in\cS}I(s)}\quad\mbox{as $n\to\infty$},
$$
where the rate function
$$
I(s)=\sup_\alpha(\alpha s-e(-\alpha))=\frac{(s-\bar s)^2}{4\bar s}
$$
satisfies the fluctuation relation~\eqref{EQ-FTLDP}. The large-time
symmetry function for $S^{n\tau}$ is
$$
\frak{s}(s)=I(-s)-I(s)=s.
$$

\bigskip
\noindent{\bf Step 3: Relating the canonical entropic functional   to a relevant dynamical or 
thermodynamical quantity}

\medskip\noindent
Denoting by~$U_t=z_t/C$ the voltage and using~\eqref{EQ-qtForm}, the work performed on the system 
by the external current~$I$ in the period $]k\tau,(k+1)\tau[$ is equal to 
$$
\delta W_k
=\int_0^\tau U_t I\d t=\int_0^\tau \frac{z_t}C I\d t
=(1-\eta)\frac{\bar z z_k}C-(1-\tau/\tau_0-\eta)\frac{\bar z^2}C.
$$
Thus, we can rewrite
$$
\frac{S^{n\tau}}{n\tau}=\frac1{k_BT}\left[\frac2{1+\eta}
(w_{n}-\bar w)+2\frac{1-\eta}{1+\eta}\frac{\tau_0}{\tau}\bar w
+\frac1n\frac{2\bar z}{C\tau}
\frac{z_n+\eta z_0}{1+\eta}
\right],
$$
where 
$$
w_n=\frac{W^{n\tau}}{n\tau},\qquad W^{n\tau}=\sum_{k=0}^{n-1}\delta W_k,
\qquad
\bar w=\langle w_n\rangle_{\rm st}
=\frac{\bar z^2}{C\tau_0}=RI^2.
$$
$W^{n\tau}$ is the work performed by the external current during the period
$[0,n\tau]$. Accordingly, $w_n$ is the average injected power and 
$\bar w$  is its expected stationary value. It follows from the 
first law of thermodynamics that the heat dissipated by the resistor $R$ in the
thermostat during the interval $[0,n\tau+0[$ is given by
$$
Q^{n\tau}=-\left(\frac{z_n^2}{2C}-\frac{z_0^2}{2C}\right)+W^{n\tau},
$$
and so we may also write
$$
\frac{S^{n\tau}}{n\tau}
=\frac1{k_BT}\left[\frac2{1+\eta}
(q_n-\bar q)+2\frac{1-\eta}{1+\eta}\frac{\tau_0}{\tau}\bar q
+\frac1n\left(\frac{2\bar z}{C\tau}
\frac{z_n+\eta z_0}{1+\eta}
+\frac{z_n^2-z_0^2}{C\tau(1+\eta)}\right)
\right],
$$
where
$$
q_n=\frac{Q^{n\tau}}{n\tau},\qquad
\bar q=\langle q_n\rangle_{\rm st}=\bar w,
$$
denote the average dissipated power and its expected stationary value.

Thus, up to a multiplicative and additive constant and a ``small'' (i.e., 
formally $\cO(n^{-1})$) correction, $S^{n\tau}/n\tau$ is the time averaged 
power injected in the system by the external forcing and the time averaged
power dissipated into the heat reservoir during the time period $[0,n\tau+0[$.

\bigskip
\noindent{\bf Step 4: Deriving a large deviation principle for physically 
relevant quantities}

\medskip\noindent
The problem encountered here stems from the fact that the relation between~$S^t$ and a physically relevant quantity (denoted by~$\fS^t$) typically involves some ``boundary terms'', which depend on 
the state of the system at the initial time $0$ and final time $t$. In cases 
where these boundary terms are uniformly bounded as $t\to\infty$, one finds 
that $\fS^t$ satisfies the same large deviation principle as $S^t$. This is
what happens, for example, in strongly chaotic dynamical systems over
a compact phase space (e.g., under the Gallavotti--Cohen chaotic hypothesis); 
we refer the reader to~\cite[Section 10]{JPR} for a discussion of this case.
However, unbounded boundary terms can compete with the tails of the law of~$S^t$, which may lead to complications, as our example shows.

\medskip
Given the Gaussian nature of $w_n$, it is an easy exercise to show 
that the entropic functional  directly related to work and defined by
$$
\frac{\fS_{\rm w}^{n\tau}}{n\tau}
=\frac{S^{n\tau}}{n\tau}-\frac1n\frac1{k_BT}
\frac{2\bar z}{C\tau}\frac{z_n+\eta z_0}{1+\eta}
=\frac1{k_BT}\left[\frac2{1+\eta}
(w_n-\bar w)+2\frac{1-\eta}{1+\eta}\frac{\tau_0}{\tau}\bar w\right],
$$
has a cumulant generating function which satisfies
$$
\lim_{n\to\infty}\frac1{n\tau}
\log\langle\e^{-\alpha\fS_{\rm w}^{n\tau}}\rangle_{\rm st}=e(\alpha),
$$
for all $\alpha\in\rr$. It follows 
that $\fS_{\rm w}^{n\tau}$ satisfies the very same large deviation estimates
as $S^{n\tau}$. However, note that unlike 
function~\eqref{EQ-RotorCumulant},
the finite-time cumulant generating function 
$\log\langle\e^{-\alpha\fS_{\rm w}^{n\tau}}\rangle_{\rm st}$ 
does not satisfy the Gallavotti--Cohen symmetry~\eqref{EQ-GCsym0}. Only in 
the large time limit do we recover this symmetry. A simple change of variable 
allows us to write down the cumulant generating function of the work 
$W^{n\tau}$,
$$
e_{\rm work}(\alpha)=\lim_{n\to\infty}\frac1{n\tau}\log\langle
\e^{-\alpha W^{n\tau}/k_BT}\rangle_{\rm st}=-\alpha
\left(1-\alpha\frac{1-\eta^2}{2\tau/\tau_0}\right)
\frac{\bar w}{k_BT}.
$$
We conclude that the work $W^{n\tau}$ satisfies the large deviations estimate
$$
\PP^{n\tau}\left[\frac1{n\tau}\frac{W^{n\tau}}{k_BT}\in\cW\right]
\approx\e^{-n\tau\inf_{w\in\cW} I_{\rm work}(w)}
$$
for all open sets $\cW\subset\rr$ with the rate function
$$
I_{\rm work}(w)=\frac14\left(w-\frac{\bar w}{k_BT}\right)^2\frac{k_BT}{\bar w}
\frac{2\tau/\tau_0}{1-\eta^2}.
$$
The symmetry function for work is thus
$$
\frak{s}_{\rm work}(w)
=I_{\rm work}(-w)-I_{\rm work}(w)=\frac{2\tau/\tau_0}{1-\eta^2}w.
$$
Note that, as the kick period $\tau$ approaches zero, we recover the universal
fluctuation relation~\eqref{EQ-FTLDP}, i.e., $\frak{s}_{\rm work}(w)=w$.

\medskip
Consider now the entropic functional
\begin{align}
\frac{\fS_{\rm h}^{n\tau}}{n\tau}
&=\frac{S^{n\tau}}{n\tau}-\frac1n\frac1{k_BT}\left(
\frac{2\bar z}{C\tau}\frac{z_n+\eta z_0}{1+\eta}
+\frac{z_n^2-z_0^2}{C\tau(1+\eta)}\right)\label{EQ-BadBT}\\
&=\frac1{k_BT}\left[\frac2{1+\eta}
(q_n-\bar q)+2\frac{1-\eta}{1+\eta}\frac{\tau_0}{\tau}\bar q\right],\nonumber
\end{align}
related to the dissipated heat. The explicit evaluation of a Gaussian integral 
shows that its cumulant generating function is given by
$$
\frac1{n\tau}\log\langle\e^{-\alpha\fS_{\rm h}^{n\tau}}\rangle_{\rm st}
=\left\{
\begin{array}{ll}
\ds e(\alpha)-\frac1{2n\tau}\left[\log\left(1-\frac{\alpha^2}{\alpha_n^2}\right)
+\frac{a_n\alpha+b_n}{\alpha_n^2-\alpha^2}\alpha^3
\right]&\mbox{if }|\alpha|<\alpha_n;\\[10pt]
+\infty&\mbox{otherwise};
\end{array}
\right.
$$
where $a_n$ and $b_n$ are bounded (in fact converging) sequences and
$$
\alpha_n=\frac12\frac{1+\eta}{(1-\eta^{2n})^{\frac12}}.
$$
The divergence of the cumulant generating function for $|\alpha|\ge\alpha_n$ 
is of course due to the competition between the tail of the Gaussian law~$p_n$ and 
the quadratic terms in $\fS^{n\tau}_{\rm h}$.

\begin{figure}
\centering
\includegraphics[scale=0.7]{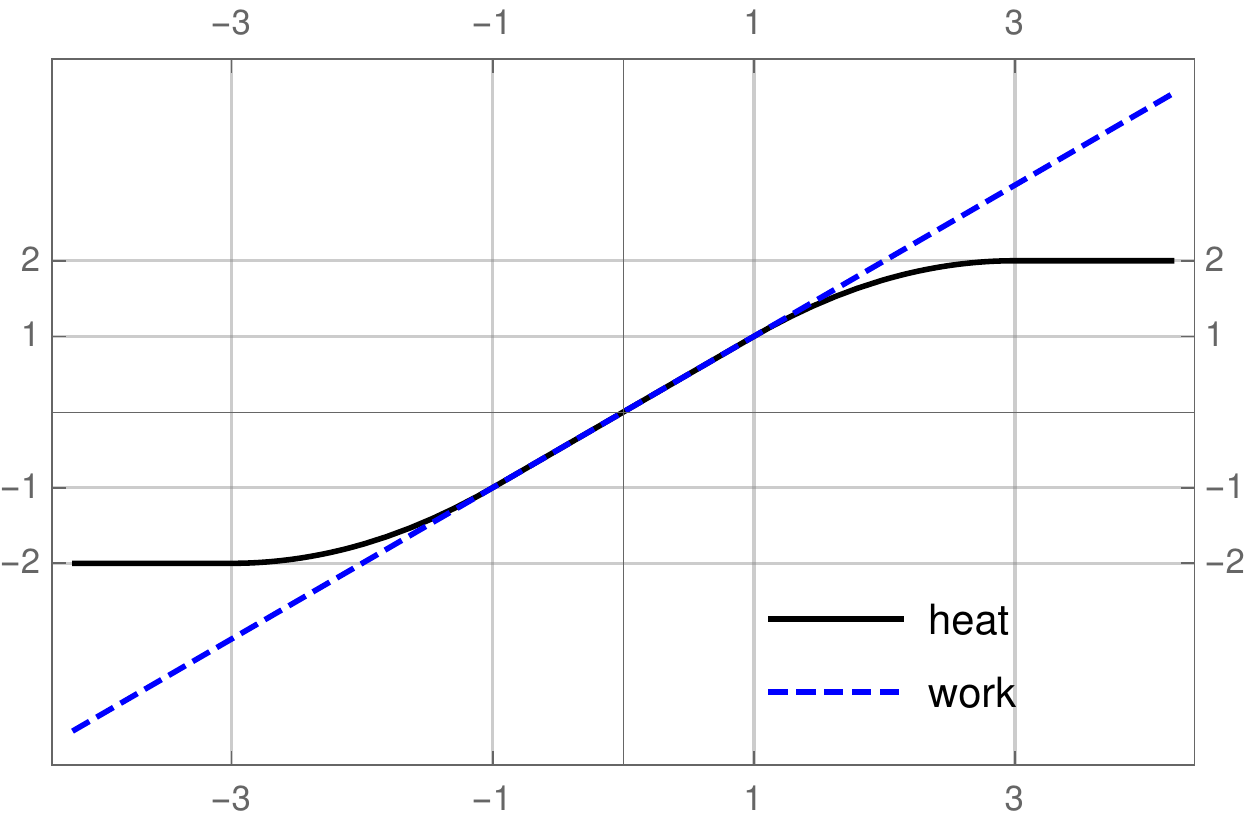}
\caption{The symmetry functions (i.e., twice the odd part of the rate 
function) of work and heat for the circuit of
Figure~\ref{FigCircuit} in the limit $\tau\to0$ 
(the unit on the abscissa is $RI^2/k_BT$).}
\label{FigCircuitSymmetry}
\end{figure}

Note that the sequence $\alpha_n$ is monotone decreasing to its limit
$$
\alpha_{\rm c}=\frac{1+\eta}{2},
$$
and it follows that
$$
\lim_{n\to\infty}\frac1{n\tau}
\log\langle\e^{-\alpha\fS_{\rm h}^{n\tau}}\rangle_{\rm st}
=\left\{
\begin{array}{ll}
e(\alpha)&\mbox{if }|\alpha|<\alpha_{\rm c};\\[4pt]
+\infty&\mbox{if }|\alpha|>\alpha_{\rm c}.
\end{array}
\right.
$$
The unboundedness of the boundary terms involving $z_0^2$ and $z_n^2$
in~\eqref{EQ-BadBT} leads to a breakdown of the Gallavotti--Cohen symmetry for 
$|\alpha-\frac12|>|\alpha_{\rm c}-\frac12|$. More dramatically, the limiting
cumulant generating function is not steep, i.e., its derivative fails
to diverge as $\alpha$ approaches $\pm\alpha_{\rm c}$. Under such circumstances, 
 the derivation
of a global large deviation principle  for nonlinear dynamical systems is a difficult problem which remains
largely open and deserves further investigations. For linear systems,
however, as shown in~\cite{JPS}, it is sometimes possible to exploit the 
Gaussian nature of the process to achieve this goal. Indeed, following
the strategy developped in Section~\ref{SEC-LDP}, one can show that 
$\fS_{\rm h}^{n\tau}$ satisfies a large deviation principle with rate 
function
$$
I_{\rm h}(s)=\sup_{|\alpha|<\alpha_{\rm c}}(\alpha s-e(-\alpha))
=\left\{
\begin{array}{ll}
I(s_-)+(s-s_-)I'(s_-)&\mbox{for }s<s_-;\\[4pt]
I(s)&\mbox{for }s\in[s_-,s_+];\\[4pt]
I(s_+)+(s-s_+)I'(s_+)&\mbox{for }s>s_+;
\end{array}
\right.
$$
where
$$
s_-=-e'(\alpha_c)=-\eta\bar s,\qquad
s_+=-e'(-\alpha_c)=(2+\eta)\bar s.
$$
Performing a simple change of variable, we conclude that the cumulant
generating function of the heat $Q^{n\tau}$ satisfies
$$
e_{\rm heat}(\alpha)=\lim_{n\to\infty}\frac1{n\tau}\log\langle
\e^{-\alpha Q^{n\tau}/k_BT}\rangle_{\rm st}=
\left\{
\begin{array}{cc}
e_{\rm work}(\alpha)&\mbox{for } |\alpha|<1;\\[4pt]
+\infty&\mbox{for } |\alpha|>1.
\end{array}
\right.
$$
The corresponding large deviations estimate reads
$$
\PP^{n\tau}\left[\frac1{n\tau}\frac{Q^{n\tau}}{k_BT}\in\cQ\right]
\approx \e^{-n\tau\inf_{q\in\cQ}I_{\rm heat}(q)}
$$
for all open sets $\cQ\subset\rr$ with the rate function
$$
I_{\rm heat}(q)=\sup_{|\alpha|<1}(\alpha q-e_{\rm heat}(-\alpha))
=I_{\rm h}\left(\frac2{1+\eta}q+2\frac{1-\tau/\tau_0-\eta}{1+\eta}
\frac{\tau_0}{\tau}\frac{\bar q}{k_BT}\right),
$$
which satisfies what is called in the physics literature an extended
fluctuation relation~\cite{Fa1,Fa2,CvZ1,CvZ2,Vi,HRS,HS,HR,NE} with the symmetry 
function
$$
\frak{s}_{\rm heat}(q)=I_{\rm heat}(-q)-I_{\rm heat}(q)
=\left\{
\begin{array}{ll}
\ds2q\frac{q_++q_-}{q_+-q_-}&\mbox{for }0\le q\le -q_-;\\[14pt]
\ds-\frac{q^2-2qq_++q_-^2}{q_+-q_-}&
\mbox{for }-q_-<q\le q_+;\\[18pt]
q_++q_-&\mbox{for }q>q_+;
\end{array}
\right.
$$
where
$$
q_-=-\frac{RI^2}{k_BT}\left(\frac{1-\eta^2}{\tau/\tau_0}-1\right),\qquad
q_+=\frac{RI^2}{k_BT}\left(\frac{1-\eta^2}{\tau/\tau_0}+1\right).
$$
Thus, the linear behavior persists for small fluctuations $|q|\le|q_-|$, but 
saturates to the constant values $\mp(q_++q_-)$ for $|q|>q_+$, the 
crossover between these two regimes being described by a parabolic
interpolation. Note also that, as the kick period $\tau$ approaches zero, 
$q_\mp\to(1\mp2)\bar q/k_BT$. In this limit the symmetry function 
$\frak{s}_{\rm heat}(q)$ agrees with the conclusions of~\cite{ZCC}
(see Figure~\ref{FigCircuitSymmetry}). \qed

\bigskip
As this example shows, the main problem in understanding the 
mathematical status and physical implications of fluctuation relations in 
oscillator networks and other boundary driven Hamiltonian systems stems from 
the lack of compactness of phase space and its consequence: the unboundedness 
of the observable describing the energy transfers between the system and the 
reservoirs (i.e., the last term in the right-hand side of 
Eq.~\eqref{EQ-BadBT}). We will show that one can achieve  complete 
control of these boundary terms by an appropriate change of drift (a Girsanov 
transformation) in the Langevin equation describing the dynamics of harmonic 
networks. This change is parametrized by the maximal solution of a 
one-parameter family of algebraic Riccati equation naturally associated to
deformations of the Markov semigroup of the system. For a network of $N$ 
oscillators,  our approach reduces the calculation of the limiting cumulant 
generating function of the canonical functional $S^t$ and its perturbations by 
quadratic boundary terms to the determination of some spectral data of the
$4N\times4N$ Hamiltonian matrix of the above-mentioned Riccati equations.
Combining this asymptotic information with  Gaussian estimates of the 
finite time cumulant generating functions, we are able to derive a global 
large deviation principle for arbitrary quadratic boundary perturbations of
$S^t$. We stress that our scheme is completely constructive and well suited to
numerical calculations.

The remaining parts of this paper are organized as follows.
In Section~\ref{SEC-Model} we introduce a general class of harmonic 
networks and the stochastic processes describing their nonequilibrium
dynamics. Section~\ref{SEC-Setup} contains our main results. There, we consider more general  framework 
and study the large time asymptotics of the entropic functional $S^t$ 
canonically associated to stochastic differential equations with linear drift 
satisfying some structural constraints (fluctuation--dissipation relations). 
We prove a global large deviation
principle for this functional and show, in particular, that it satisfies the 
Gallavotti--Cohen fluctuation theorem. We then consider perturbations
of $S^t$ by quadratic boundary terms and show that they also satisfy
a global large deviation principle. This applies, in particular, to the
heat released by the system in the reservoirs. We turn back to harmonic 
networks in Section~\ref{SEC-Examples} where we apply our results to specific
examples. Finally, Section~\ref{SEC-Proofs} collects the proofs of 
our results.

\bigskip
\noindent{\bf Acknowledgements.}
This research was supported by the CNRS collaboration grant RESSPDE. The
authors gratefully acknowledge the support of NSERC and ANR (grants 09-
BLAN-0098 and ANR 2011 BS01 015 01). The work of C.-A.P. has been carried out in the framework of the Labex Archim\`ede
(ANR-11-LABX-0033) and of the A*MIDEX project (ANR-11-IDEX-0001-02),
funded by the ``Investissements d'Avenir'' French Government programme
managed by the French National Research Agency (ANR). The research of AS was carried out within the MME-DII Center of Excellence and supported by the RSF grant 14-49-00079. 

\section{The model}
\label{SEC-Model}

We consider a collection of one-dimensional harmonic oscillators indexed 
by a finite set $\cI$. The configuration space $\rr^\cI$ is endowed with
its Euclidean structure and the phase space $\Xi=\rr^\cI\oplus\rr^\cI$ is 
equipped with its canonical symplectic $2$-form $\d p\wedge\d q$. The 
Hamiltonian is given by
\beq
\Xi\ni(p,q)\mapsto h(p,q)=\tfrac12|p|^2+\tfrac12|\omega q|^2,
\label{EQ-The_Hamiltonian}
\eeq
where $|\cdot|$ is the Euclidean norm and $\omega:  \rr^\cI \rightarrow 
\rr^\cI$\index{$\omega$} is
a non-singular linear map. Time-reversal of the Hamiltonian flow 
of  $h$ is implemented by the anti-symplectic involution of $\Xi$ given by\index{$\theta$} 
\beq
\theta:(p,q)\mapsto(-p,q).
\label{EQ-thetaDef}
\eeq

We consider the stochastic perturbation of the Hamiltonian flow
of $h$ obtained by coupling a non-empty subset of the oscillators, indexed by 
$\partial\cI\subset\cI$, to Langevin heat reservoirs. The reservoir coupled 
to the $i^\mathrm{th}$ oscillator is characterized by two parameters: its
temperature $\vartheta_i>0$ and its relaxation rate $\gamma_i>0$. We encode 
these parameters in two linear maps: a bijection 
$\vartheta:\rr^{\partial\cI}\to\rr^{\partial\cI}$ and an injection 
$\iota:\rr^{\partial\cI}\to\rr^\cI=\rr^{\partial\cI}\oplus\rr^{\cI\setminus\partial\cI}$ defined by\index{$\vartheta$}\index{$\iota$}
$$
\vartheta:
(u_i)_{i\in\partial\cI}\mapsto(\vartheta_iu_i)_{i\in\partial\cI},\qquad
\iota:
(u_i)_{i\in\partial\cI}\mapsto(\sqrt{2\gamma_i}u_i)_{i\in\partial\cI}\oplus0.
$$
The external force acting on the $i^\mathrm{th}$ oscillator has the usual
Langevin form
\beq
f_i(p,q)=(2\gamma_i\vartheta_i)^\12\dot w_i-\gamma_ip_i,
\label{EQ-The_Force}
\eeq
where the $\dot w_i$ are independent white noises.

In mathematically more precise terms, we shall deal with the dynamics 
described by the following system of stochastic differential equations
\beq
\d q(t)=p(t)\d t,\quad
\d p(t)=-\left(\tfrac12\iota\iota^\ast p(t)+\omega^\ast\omega q(t)\right)\d t
+\iota\vartheta^\12\d w(t),
\label{EQ-First_SDE}
\eeq
where ${}^\ast$ denotes conjugation w.r.t.\;the Euclidean inner products and 
$w$ is a standard $\rr^{\partial\cI}$-valued Wiener process over the canonical 
probability space $(W,\cW,\WW)$. We denote by $\{\cW_t\}_{t\ge0}$ the 
associated natural filtration.

To the Hamiltonian~\eqref{EQ-The_Hamiltonian} we associate the graph
$\cG=(\cI,\cE)$ with vertex set $\cI$ and edges
$$
\cE=\{\{i,j\}\subset\cI\,|\,(\omega^\ast\omega)_{ij}\not=0\}.
$$
To avoid trivialities, we shall always assume that $\cG$ is connected.

As explained in the introduction, we shall construct the canonical entropic
functional of the process $(p(t),q(t))$ and relate it to the heat released
by the network into the thermal reservoir. We end this section with a calculation of the latter quantity.

Applying It\^o's formula to the Hamiltonian $h$ we obtain  the expression  
$$
\d h(p(t),q(t))
=\sum_{i\in\partial\cI}\gamma_i\left(\vartheta_i-p_i(t)^2\right)\d t
+(2\gamma_i\vartheta_i)^\12p_i(t)\d w_i(t)
$$
which describes   the change in energy of the system.
The $i^\mathrm{th}$ term on the right-hand side of this identity is the 
work performed on the network by the $i^\mathrm{th}$ Langevin 
force~\eqref{EQ-The_Force}. Since these Langevin forces describe 
the action of heat reservoirs, we shall identify
\beq
\delta Q_i(t)=\gamma_i\left(\vartheta_i-p_i(t)^2\right)\d t
+(2\gamma_i\vartheta_i)^\12p_i(t)\d w_i(t)
\label{EQ-dWiDef}
\eeq
with the heat injected in the network by the $i^\mathrm{th}$ reservoir. A direct  application of the fundamental thermodynamic relation between heat 
and entropy leads to consider $\d S_i(t)=-\vartheta_i^{-1}\delta Q_i(t)$ as 
the entropy dissipated into the $i^\mathrm{th}$ reservoir. Accordingly, the 
total entropy dissipated in the reservoirs during the time interval $[0,t]$ is 
given by the  functional\index{$\fS^t$}
\beq
\fS^t=-\sum_{i\in\partial\cI}\int_0^t\frac{\delta Q_i(s)}{\vartheta_i}
=\sum_{i\in\partial\cI}\int_0^t\left(
-(2\gamma_i\vartheta_i^{-1})^\12p_i(s)\d w_i(s)
-\gamma_i(1-\vartheta_i^{-1}p_i(s)^2)\d s\right).
\label{EQ-SrealDef}
\eeq
For a lack of better name, we shall call the
physical quantity described by this functional  the {\sl thermodynamic entropy}\index{thermodynamic entropy} (TDE), in order to distinguish it  from various information theoretic entropies 
that will be introduced latter.

\section{Abstract setup and main results}
\label{SEC-Setup}

It turns out that a large part of the analysis of the 
process~\eqref{EQ-First_SDE} and its entropic functionals
is independent of the  details of the model and relies only on its  few 
structural properties. In this section we recast the harmonic networks 
in a more abstract framework, retaining only the  structural properties of the original 
system which are necessary for our analysis.

\noindent{\bf Notations and conventions.} 
Let $E$ and $F$ be real or complex Hilbert
spaces. $L(E,F)$ denotes the set of (continuous) linear operators $A:E\to F$ 
and $L(E)=L(E,E)$. For $A\in L(E,F)$, $A^\ast\in L(F,E)$ denotes the adjoint
of $A$, $\|A\|$ its operator norm, $\Ran A\subset F$ its range and 
$\Ker A\subset E$ its kernel. We denote the spectrum of $A\in L(E)$ by 
$\sp(A)$. $A$ is non-negative (resp.\;positive), written $A\ge0$ 
(resp.\;$A>0$), if it is self-adjoint and $\sp(A)\in[0,\infty[$
(resp.\;$\sp(A)\subset]0,\infty[$). We write $A\ge B$ whenever
$A-B\in L(E)$ is non-negative. The relation $\ge$ defines a partial order on 
$L(E)$. The controllable subspace of a pair $(A,Q)\in L(E)\times L(F,E)$
is the smallest $A$-invariant subspace of $E$ containing $\Ran Q$. We denote 
it by $\cC(A,Q)$. If $\cC(A,Q)=E$, then $(A,Q)$ is said to be controllable. 
We denote by $\cc_\mp$ the open left/right half-plane. $A\in L(E)$ is said
to be stable/anti-stable whenever $\sp(A)\subset\cc_\mp$.

\bigskip
We start by rewriting the equation of motion~\eqref{EQ-First_SDE} in a more 
compact form. Setting\index{$A$}\index{$Q$}
\beq
x=\left[
\begin{array}{c}
p\\
\omega q
\end{array}
\right],\qquad
A=
\left[
\begin{array}{cc}
-\12\iota\iota^\ast&-\omega^\ast\\
\omega&0
\end{array}
\right],\qquad
Q=\left[
\begin{array}{c}
\iota \\
0
\end{array}
\right]\vartheta^\12,
\label{EQ-AQdef}
\eeq
Eq.~\eqref{EQ-First_SDE} takes the form 
\beq
\d x(t)=Ax(t)\d t+Q\d w(t),
\label{EQ-The_SDE}
\eeq
and functional~\eqref{EQ-SrealDef} becomes
\beq
\fS^t
=-\int_0^t\vartheta^{-1}Q^\ast x(s)\cdot\d w(s)
+\tfrac12\int_0^t|\vartheta^{-1}Q^\ast x(s)|^2\d s
-\tfrac12 t\,\tr(Q\vartheta^{-1}Q^\ast).
\label{EQ-SAbstractDef}
\eeq

Note that the vector field~$Ax$ splits into a conservative (Hamiltonian) part~$\Omega x$ and a dissipative part~$-\Gamma x$ defined by\index{$\Omega$}\index{$\Gamma$}
\begin{align}
\Omega&=\tfrac12(A-A^\ast)=\left[
\begin{array}{cc}
0&-\omega^\ast\\
\omega&0
\end{array}
\right],
\label{EQ-OmegaDef}\\
\Gamma&=-\tfrac12(A+A^\ast)=\tfrac12 Q\vartheta^{-1}Q^\ast.
\label{EQ-OmegaGammadef}
\end{align}
These operators satisfy the relations
\beq
\Omega^\ast=\theta\Omega\theta=-\Omega,\qquad
\Gamma^\ast=\theta\Gamma\theta=\Gamma.
\label{EQ-OmegaGammaCommute}
\eeq
The solution of the  Cauchy problem associated to~\eqref{EQ-The_SDE} with 
initial condition $x(0)=x_0$ can be written explicitly as
\beq
x(t)=\e^{tA}x_0+\int_0^t\e^{(t-s)A}Q\d w(s).
\label{EQ-The_Process}
\eeq
This relation defines a family of $\Xi$-valued Markov processes indexed 
by the initial condition $x_0\in\Xi$. This family is completely characterized
by the data
\beq
(A,Q,\vartheta,\theta)\in 
L(\Xi)\times L(\partial\,\Xi,\Xi)
\times L(\partial\,\Xi)\times L(\Xi),
\label{EQ-Data}
\eeq
where $\Xi$ and $\partial\,\Xi$ are finite-dimensional Euclidean 
vector spaces and $(A,Q,\vartheta,\theta)$ is subject to the 
following structural constraints\index{structural constraints}:
\beq
\begin{split}
\centermathcell{\Ker(A-A^\ast)\cap\Ker Q^\ast=\{0\},\quad
A+A^\ast=-Q\vartheta^{-1}Q^\ast,\quad
\vartheta>0,\quad
Q^\ast Q>0,}\\[6pt]
\centermathcell{\theta=\theta^\ast=\theta^{-1},\quad
\theta Q=\pm Q,\quad
\theta A\theta=A^\ast,\quad [\vartheta,Q^\ast Q]=0.}
\end{split}
\label{EQ-Structure}
\eeq
In the remaining parts of Section~\ref{SEC-Setup}, we shall consider the 
family of processes~\eqref{EQ-The_Process}, which are strong solutions of SDE~\eqref{EQ-The_SDE}, associated with the data~\eqref{EQ-Data} satisfying~\eqref{EQ-Structure}.

\ber\label{REM-qM} The concrete models of the previous section fit into
the abstract setup defined by~\eqref{EQ-The_SDE}, \eqref{EQ-Data}, and~\eqref{EQ-Structure} with $\Ker(A-A^\ast)=\{0\}$ and $\theta Q=-Q$. 
We have weakened the first condition
and included the case $\theta Q=+Q$ in~\eqref{EQ-Structure} in order to 
encompass the quasi-Markovian models introduced in~\cite{EPR1,EPR2}. There, 
the Langevin reservoirs are not directly coupled to the network, but to
additional degrees of freedom described by dynamical variables 
$r\in\rr^\cJ$, where $\cJ$ is a finite set. The augmented phase space of the 
network is $\Xi=\rr^\cJ\oplus\rr^\cI\oplus\rr^\cI$, and
$\partial\,\Xi=\rr^\cJ$. The equations of motion take the 
form~\eqref{EQ-The_SDE} with
$$
x=\left[
\begin{array}{c}
r\\p\\\omega q
\end{array}
\right],\qquad
A=\left[
\begin{array}{ccc}
-\12\iota\iota^\ast&-\Lambda^\ast&0\\
\Lambda&0&-\omega^\ast\\
0&\omega&0
\end{array}
\right],\qquad
Q=\left[
\begin{array}{c}
\iota\\0\\0
\end{array}
\right]\vartheta^{\12},
$$
where $\iota:\rr^\cJ\to\rr^\cJ$ is bijective and $\Lambda:\rr^\cJ\to\rr^\cI$ 
injective. The time reversal map in this case is given by
$$
\theta=\left[
\begin{array}{ccc}
I&0&0\\
0&-I&0\\
0&0&I
\end{array}
\right].
$$
Writing the system internal energy as $H(x)=\frac12|p|^2+\frac12|\omega q|^2+\frac12|r|^2$,
the calculation of the previous section yields the following formula 
for the total entropy dissipated into the reservoirs
\beq
\fS^t+
\tfrac12|\vartheta^{-\12}r(t)|^2-\tfrac12|\vartheta^{-\12}r(0)|^2,
\label{EQ-QMTDE}
\eeq
where $\fS^t$ is given by~\eqref{EQ-SAbstractDef}.
\eer

\bigskip
Let $\cP(\Xi)$ be the set of Borel probability measures on $\Xi$ and
denote by $P^t(x,\,\cdot\,)\in\cP(\Xi)$ the transition kernel of the 
process~\eqref{EQ-The_Process}. For bounded or non-negative measurable 
functions $f$ on $\Xi$ and $\nu\in\cP(\Xi)$ we write
$$
\nu(f)=\int f(x)\nu(\d x),\qquad
f_t=P^tf=\int P^t(\,\cdot\,,\d y)f(y),\qquad
\nu_t=\nu P^t=\int \nu(\d y)P^t(y,\,\cdot\,),
$$
so that $\nu(f_t)=\nu_t(f)$. A measure $\nu$ is invariant if $\nu_t=\nu$ for 
all $t\ge0$. We denote the actions of time-reversal by\index{$\Theta$}
$$
\widetilde{f}=\Theta f=f\circ\theta,\qquad
\widetilde{\nu}=\nu\Theta=\nu\circ\theta,
$$
so that $\nu(\widetilde{f})=\widetilde{\nu}(f)$. A measure $\nu$ is
time-reversal invariant if $\widetilde{\nu}=\nu$.
The generator $L$ of the Markov semigroup $P^t$ acts on smooth 
functions as
\beq
L=\tfrac12\nabla\cdot B\nabla+Ax\cdot\nabla,
\label{EQ-LDef}
\eeq
where\index{$B$}
\beq
B=QQ^\ast.
\label{EQ-BDef}
\eeq
We further denote by $\PP_{x_0}$ the induced probability measure on 
the path space $C(\rr^+,\Xi)$ and by $\EE_{x_0}$ the associated expectation.
Considering~$x_0$ as a random variable, independent of the driving Wiener 
process~$w$ and distributed according to $\nu\in\cP(\Xi)$, we denote by~$\PP_\nu$ and~$\EE_\nu$ the induced path space measure and expectation.
In the language of statistical mechanics, functions $f$ on $\Xi$ are the
observables of the system, $\nu$ is its initial state, and the flow 
$t\mapsto\nu_t$ describes its time evolution. Invariant measures
thus correspond to steady states of the system.

The following result is well 
known (see Chapter~6 in the book~\cite{DPZ} and the papers~\cite{EZ,MNV}). For the reader convenience, we 
provide a sketch of its proof in Section~\ref{SSECT_Proof_of_THM_InvMeas}.

\bet\label{THM-InvMeas}
\ben
\item 
Under the above hypotheses, the operator
$$
M:=\int_0^\infty e^{sA}Be^{sA^*}\d s
$$
is well defined and non-negative, and its restriction to $\Ran M$ satisfies the inequality 
\beq
\vartheta_{\rm min}=\min\sp(\vartheta)
\le M\big|_{\Ran M}\le \max\sp(\vartheta)=\vartheta_{\rm max}.
\label{EQ-Mbounds}
\eeq
Moreover, the centred Gaussian measure~$\mu$ with covariance~$M$ is invariant for the Markov processes associated with~\eqref{EQ-The_SDE}.

\item 
The invariant measure~$\mu$ is unique iff the pair~$(A,Q)$ is controllable. In this case, the mixing property holds in the sense that, for any $f\in L^1(\Xi,\d \mu)$, we have
$$
\lim_{t\to+\infty}P^tf=\mu(f),
$$
where the convergence holds in $L^1(\Xi,\d \mu)$ and uniformly on compact subsets of~$\Xi$.

\item 
Let $x(t)$ be defined by relation~\eqref{EQ-The_Process}, in which the initial condition~$x_0$ is independent of~$w$ and is distributed as~$\mu$. Then $x(t)$ is a centred stationary Gaussian process. Moreover, its covariance operator defined by the relation $(\eta_1,K(t,s)\eta_2)=\EE_\mu\bigl\{(x(t),\eta_1)(x(s),\eta_2)\bigr\}$ has the form
\beq
K(t,s)= e^{(t-s)_+A}M e^{(t-s)_-A^\ast}.
\label{EQ-Mts}
\eeq
\een
\eet

\ber\label{REM-EquilibriumM}
In the harmonic network setting, if 
$\vartheta=\vartheta_0 I$ for some $\vartheta_0\in]0,\infty[$
(i.e., the reservoirs are in a joint thermal equilibrium at temperature 
$\vartheta_0$), then it follows 
from~\eqref{EQ-Mbounds} that $M=\vartheta$, which means that $\mu$ is the Gibbs state 
at temperature $\vartheta_0$ induced by the Hamiltonian $h$.
\eer

\bigskip
In the sequel, we shall assume without further notice that 
process~\eqref{EQ-The_Process} has a unique invariant measure~$\mu$, 
i.e., that the following hypothesis holds:

\begin{quote}
\label{HYP-C}
{\bf Assumption (C)} {\sl The pair $(A,Q)$ is controllable.}
\end{quote}

\ber
To make contact with~\cite{MNV}, note that in terms
of Stratonovich integral the TDE functional~\eqref{EQ-SAbstractDef}
is given by
$$
\fS^t=
-\int_0^t\vartheta^{-1}Q^\ast x(s)\circ\d w(s)
+\tfrac12\int_0^t|\vartheta^{-1}Q^\ast x(s)|^2\d s.
$$
This identity is a standard result of stochastic calculus (see, e.g., Section II.7 
in~\cite{Pr}) and is used as a definition of the entropy current 
in~\cite{MNV}. 
\eer

\subsection{Entropies and entropy production}
\label{SSEC-Entropies}

In this section we introduce information theoretic quantities which play
an important role in our approach to fluctuation relations. We briefly 
discuss their basic properties and in particular their relations with
the TDE $\fS^t$.

Let $\nu_1$ and $\nu_2$ be two probability measures on the same measurable
space. If $\nu_1$ is absolutely continuous w.r.t. $\nu_2$,  the relative entropy of the pair $(\nu_1, \nu_2)$ is defined 
by \index{$\Ent$}
$$
\Ent(\nu_1|\nu_2)=-\int\log\left(\frac{\d\nu_1}{\d\nu_2}\right)\d\nu_1.
$$
We recall that $\Ent(\nu_1|\nu_2)\in[-\infty,0]$, with $\Ent(\nu_1|\nu_2)=0$ 
iff $\nu_1=\nu_2$ (see, e.g., \cite{OP}). 

Suppose that  $\nu_1$ and $\nu_2$ are mutually absolutely continuous. 
For $\alpha\in\rr$, the R\'enyi 
\cite{Re} relative $\alpha$-entropy of  the pair $(\nu_1, \nu_2)$ is\index{$\Ent_\alpha$}
$$
\Ent_\alpha(\nu_1|\nu_2)=\log\int\left(\frac{\d\nu_1}{\d\nu_2}\right)^\alpha
\d\nu_2.
$$
The function $\rr\ni\alpha\mapsto\Ent_\alpha(\nu_1|\nu_2)\in]-\infty,\infty]$ 
is convex. It is non-positive on $[0,1]$, vanishes for $\alpha\in\{0,1\}$, and 
is non-negative on $\rr\setminus]0,1[$. It is real analytic on 
$]0,1[$ and vanishes identically on this interval  iff $\nu_1=\nu_2$. Finally, 
\beq
\Ent_{1-\alpha}(\nu_1|\nu_2)=\Ent_\alpha(\nu_2|\nu_1)
\label{EQ-RenyiSymmetry}
\eeq
for all $\alpha\in\rr$.

Let $\nu\in\cP(\Xi)$ be such that $\nu(|x|^2)<\infty$ (recall that in our
abstract framework the Hamiltonian is $h(x)=\12|x|^2$). 
The Gibbs--Shannon entropy of $\nu_t=\nu P^t$ is defined by \index{$S_{\rm GS}$}
\beq
S_{\rm GS}(\nu_t)=-\int\log\left(\frac{\d\nu_t}{\d x}\right)\nu_t(\d x).
\label{EQ-G_S_Form}
\eeq
The Gibbs--Shannon entropy is finite for all $t>0$ (see Lemma~\ref{LEM-LStar} (1) below)
and is a measure of the internal entropy of the system at time $t$.

To formulate our next result (see 
Section~\ref{SSECT_Proof_of_PROP-Strict_Positivity} for its proof) we define
$$
\cP_{+}(\Xi)=\left\{\nu\in\cP(\Xi)\,\bigg|\,
\int\e^{\12 m|x-a|^2}\nu(\d x)<\infty
\text{ for some }m>0\text{ and }a\in\Xi\right\}.
$$
Note that any Gaussian
measure on $\Xi$ belongs to $\cP_+(\Xi)$.

\bep\label{PROP-Strict_Positivity} Let a non-negative operator 
$\beta\in L(\Xi)$ be such that\index{$\beta$}\footnote{An operator~$\beta$ satisfying~\eqref{EQ-betadef} always exists. For instance, one can define~$\beta$ by the relations $\beta x=Q\vartheta^{-1}y$ if $x=Qy$ for some~$y\in \p\Xi$ and $\beta x=x$ if $x\bot \Ran Q$.}
\beq
\beta Q=Q\vartheta^{-1},\qquad \theta\beta\theta=\beta.
\label{EQ-betadef}
\eeq
Define the quadratic form\index{$\sigma_\beta$}
\beq
\sigma_\beta(x)=\tfrac12 x\cdot\Sigma_\beta x,\qquad
\Sigma_\beta=[\Omega,\beta],
\label{EQ-sigmadef}
\eeq
and a reference measure $\mu_\beta$ on $\Xi$ by\index{$\mu_\beta$}
\beq
\frac{\d\mu_\beta}{\d x}(x)=\e^{-\12|\beta^\12 x|^2}.
\label{EQ-muR}
\eeq
Then the following assertions hold. 
\ben
\item $\mu_\beta\Theta=\mu_\beta$ and 
$\Theta\sigma_\beta=-\sigma_\beta$.
\item Let $L^\beta$ denote the formal adjoint of the Markov 
generator~\eqref{EQ-LDef} w.r.t.\;the inner product of the Hilbert space 
$L^2(\Xi,\mu_\beta)$. Then 
\beq
\Theta L^\beta\Theta=L+\sigma_\beta.
\label{EQ-GenDBC}
\eeq
\item The TDE~\eqref{EQ-SAbstractDef} can be written as
\beq
\fS^t=-\int_0^t\sigma_\beta(x(s))\d s
+\log\frac{\d\mu_{\beta}}{\d x}(x(t))
-\log\frac{\d\mu_{\beta}}{\d x}(x(0)).
\label{EQ-SForm}
\eeq
\item Suppose that Assumption (C) holds. Then for any $\nu\in\cP_+(\Xi)$ the de Bruijn relation\index{de Bruijn relation}
\beq
\frac{\d\ }{\d t}\Ent(\nu_t|\mu)
=\tfrac12 \nu_t(|Q^\ast\nabla\log\frac{\d\nu_t}{\d\mu}|^2)
\label{EQ-DeBruijn}
\eeq
holds for $t$ large enough. In particular, $\Ent(\nu_t|\mu)$
is non-decreasing for large $t$.
\item Under the same assumptions 
\beq
\frac{\d\ }{\d t}\left(S_{\rm GS}(\nu_t)+\EE_{\nu}[\fS^t]\right)
=\tfrac12\nu_t(|Q^\ast\nabla\log\frac{\d\nu_t}{\d\mu_\beta}|^2)
\label{EQ-EntProd}
\eeq
holds for $t$ large enough.
\een
\eep

\ber
Part~(2) states that our system satisfies a 
{\sl generalized detailed balance condition}\index{generalized detailed balance} as defined in~\cite{EPR2} 
(see also \cite{BL}). 
\eer

\begin{figure}
\centering
\includegraphics[scale=0.5]{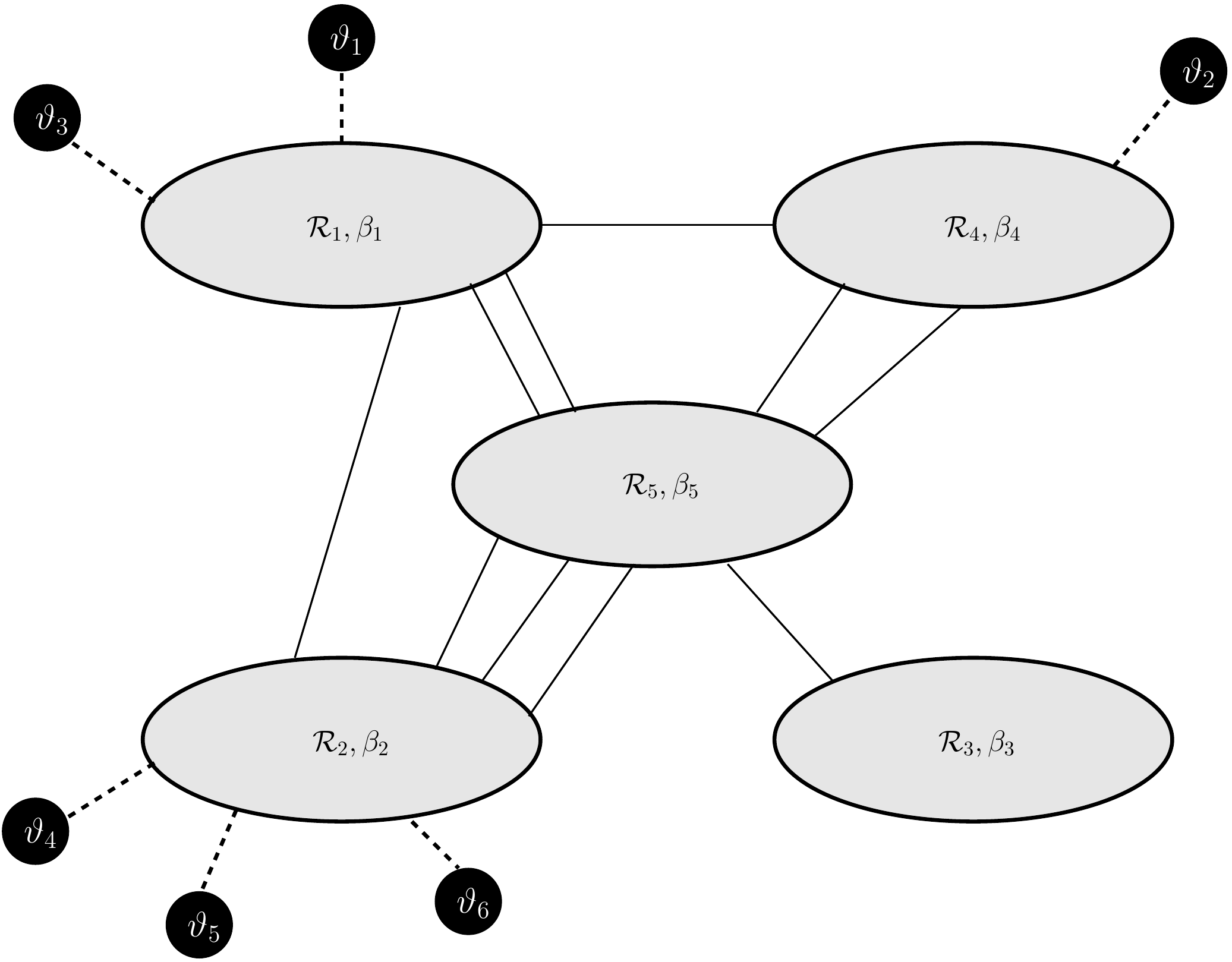}
\caption{A partition of the network. Black disks represents heat reservoirs. 
In this situation one has, $\beta_1^{-1}=\vartheta_1=\vartheta_3$,
$\beta_2^{-1}=\vartheta_4=\vartheta_5=\vartheta_6$, $\beta_4^{-1}=\vartheta_2$.
$\beta_3\ge0$ and $\beta_5\ge0$ arbitrary.}
\label{Fig1}
\end{figure}

\smallskip
Let us comment on the physical interpretation of 
Part~(3) in the harmonic network setting. 
Let $\cI=\cup_{k\in K}\cI_k$ be a partition of the network
and denote by $\pi_k$ the orthogonal projection on $\rr^\cI$ with range~$\rr^{\cI_k}$. Defining
$$
h_k(p,q)=\tfrac12|\pi_kp|^2+\tfrac12|\omega\pi_kq|^2,\qquad
v_{k,l}(q)=\tfrac12q\cdot(\pi_k\omega^\ast\omega\pi_l
+\pi_l\omega^\ast\omega\pi_k)q,
$$
for $k,l\in K$, we decompose the network into $|K|$ clusters
$\cR_k$ with internal energy $h_k$, interacting through the potentials 
$v_{k,l}$. Denote by
$$
\tilde h_k(p,q)=h_k(p,q)+\tfrac12\sum_{l\not=k}v_{k,l}(q)
$$
the total energy stored in $\cR_k$.
Assume that all the reservoirs attached to $\cR_k$, if any, are at the
same temperature, i.e.,
\beq
i\in\cI_k\cap\partial\cI\Rightarrow
\vartheta\iota^\ast\pi_k=\vartheta_i\iota^\ast\pi_k,
\label{EQ-PartCompat}
\eeq
and for $k\in K$ let $\beta_k\ge0$ be such that $\beta_k=\vartheta_i^{-1}$ 
whenever $i\in\cI_k\cap\partial\cI$
(see Figure~\ref{Fig1}). Defining the non-negative operator $\beta$ by
\beq
\tfrac12x\cdot\beta x=\sum_{k\in K}\beta_k\tilde h_k(p,q),
\label{EQ-betaGeneral}
\eeq
we observe that~\eqref{EQ-betadef} holds as a consequence of~\eqref{EQ-PartCompat} and the time-reversal invariance of $\tilde h_k$.
The corresponding reference measure $\mu_\beta$ is, up to irrelevant 
normalization, a local Gibbs measure where each 
cluster $\cR_k$ is in equilibrium at the inverse temperatures $\beta_k$.

It\^o's formula yields the local energy balance relation
\beq
\d\tilde h_k(x(t))
=-\tfrac14p(t)\cdot(\pi_k\omega^\ast\omega-\omega^\ast\omega\pi_k)q(t)\d t
+\sum_{i\in\cI_k\cap\partial\cI}\delta Q_i(t),
\label{EQ-LocalBalance}
\eeq
where $\delta Q_i(t)$ is given by~\eqref{EQ-dWiDef}.
The last term on the right-hand side of this identity is the total heat
injected into subsystem $\cR_k$ by the reservoirs attached to it. 
Thus, we can identify
$$
\mathfrak{j}_k(t)=\sum_{l\not=k}\mathfrak{j}_{k\to l}(t),\qquad
\mathfrak{j}_{k\to l}(t)
=\tfrac14p(t)\cdot(\pi_k\omega^\ast\omega\pi_l
-\pi_l\omega^\ast\omega\pi_k)q(t),
$$
with the total flux of energy flowing out of $\cR_k$ into its environment
which is composed of the other subsystems $\cR_{l\not=k}$.
Multiplying Eq.~\eqref{EQ-LocalBalance} with 
$\beta_k$, summing over $k$, integrating over $[0,t]$ and comparing the 
result with~\eqref{EQ-SrealDef} we obtain 
$$
\fS^t=-\sum_{k\in K}\beta_k\int_0^t\mathfrak{j}_k(t)\d t
+\log\frac{\d\mu_\beta}{\d x}(x(t))-\log\frac{\d\mu_\beta}{\d x}(x(0)).
$$
Comparison with~\eqref{EQ-SForm} yields
$$
\sigma_\beta(x(t))=\sum_{k\in K}\beta_k\mathfrak{j}_k(t)
=\tfrac12\sum_{k\not=l}(\beta_k-\beta_l)\mathfrak{j}_{k\to l}(t),
$$
which, according to the heat-entropy relation,
is the total inter-cluster entropy flux.
Two different ways of partitioning the system and assigning reference
local temperatures to each subsystems leads to total entropy dissipation
which only differs by a boundary term
$$
\sigma_\beta(x(t))-\sigma_{\beta'}(x(t))
=\sum_{k\in K}\beta_k \mathfrak{j}_k(t)
-\sum_{k\in K'}\beta'_k \mathfrak{j'}_k(t)
=\frac{\d}{\d t}\log\frac{\d\mu_\beta}{\d\mu_{\beta'}}(x(t)),
$$ 
provided the local inverse temperatures $\beta_k$, $\beta_k'$ are 
consistent with the temperatures of the reservoirs.

\bigskip
Eq.~\eqref{EQ-EntProd} can be read as an entropy 
balance equation. Its left-hand side is the sum of the rate of increase
of the internal Gibbs--Shannon entropy of the system and of the 
TDE flux leaving
the system. Thus, the quantity on the right-hand side of 
Eq.~\eqref{EQ-EntProd} can be interpreted as the total entropy
production rate of the process. Using Eqs.~\eqref{EQ-G_S_Form} and~\eqref{EQ-SForm}, we can rewrite Eq.~\eqref{EQ-EntProd} as
\beq
\frac{\d\ }{\d t}\EE_{\nu}[\Ep(\nu,t)]
=\nu_t(-\sigma_\beta)+\frac{\d\ }{\d t}\Ent(\nu_t|\mu_\beta)
=\tfrac12\nu_t(|Q^\ast\nabla\log\frac{\d\nu_t}{\d\mu_\beta}|^2),
\label{EQ-EpExp}
\eeq
where the entropy production functional $\Ep$ is defined by\index{$\Ep(\nu,t)$}
\beq
\begin{split}
\Ep(\nu,t)&=-\int_0^t\sigma_\beta(x(s))\d s-\log\frac{\d\nu_t}{\d\mu_\beta}(x(t))
+\log\frac{\d\nu}{\d\mu_\beta}(x(0))\\
&=\fS^t-\log\frac{\d\nu_t}{\d x}(x(t))
+\log\frac{\d\nu}{\d x}(x(0)).
\end{split}
\label{EQ-EpDef}
\eeq
In the physics literature, the quantity
$$
\varsigma_{\rm stoch}(t)=-\log\frac{\d\nu_t}{\d x}(x(t)),
$$
is sometimes called {\sl stochastic entropy}\index{stochastic entropy} (see, e.g., 
\cite[Section~2.4]{Se}).
In the case $\nu=\mu$, i.e., for the stationary process, stochastic
entropy does not contribute to the expectation of $\Ep(\mu,t)$, and 
Eq.~\eqref{EQ-EpDef} yields
\beq
\frac1t\EE_{\mu}[\Ep(\mu,t)]
=\frac1t\EE_{\mu}[\fS^t]
=-\mu(\sigma_\beta),
\label{EQ-musigma}
\eeq
so that~\eqref{EQ-EpExp} reduces to
\beq
-\mu(\sigma_\beta)
=\tfrac12\mu(|Q^\ast\nabla\log\frac{\d\mu}{\d\mu_\beta}|^2),
\label{EQ-EpPos}
\eeq
where the right-hand side is the steady state entropy production rate.
In the following, we set\index{$\ep$}
\beq
\ep=-\mu(\sigma_\beta).
\label{EQ-epDef}
\eeq
By~\eqref{EQ-musigma} this quantity is independent of the choice 
of $\beta\in L(\Xi)$ satisfying Conditions~\eqref{EQ-betadef}. The relation~\eqref{EQ-EpPos} 
shows that $\ep\ge0$. Computing the Gaussian integral
on the right-hand side of~\eqref{EQ-EpPos} yields
\beq
\ep=\tfrac12\tr(\vartheta^{-1}(MQ-Q\vartheta)^\ast 
M^{-1}(MQ-Q\vartheta)\vartheta^{-1})
=\tfrac12\|M^{-\frac12}(MQ-Q\vartheta)\vartheta^{-\frac12}\|_2^2,
\label{EQ-NESSEp}
\eeq
where $\|\cdot\|_2$ denotes the Hilbert-Schmidt norm. Thus, $\ep>0$
iff $MQ-Q\vartheta\not=0$. By Remark~\ref{REM-EquilibriumM}, 
the latter condition implies in particular that the eigenvalues of $\vartheta$ 
(i.e., the temperatures $\vartheta_i$) are not all equal.
Part~(2) of the next proposition provides a converse. For the proof see  
Section~\ref{SSECT_Proof_of_PROP-StrictPositivity2}.

\bep\label{PROP-StrictPositivity2}
\ben
\item
$$
\ep=0\Leftrightarrow MQ=Q\vartheta
\Leftrightarrow [\Omega,M]=0\Leftrightarrow \mu\Theta=\mu.
$$
In particular, the steady state entropy production rate vanishes
iff the steady state $\mu$ is time-reversal invariant and invariant 
under the (Hamiltonian) flow $\e^{t\Omega}$.
\item Let $\vartheta_1,\vartheta_2$ be two distinct eigenvalues of 
$\vartheta$ 
and denote by $\pi_1,\pi_2$ the corresponding spectral projections.
If $\cC(\Omega,Q\pi_1)\cap\cC(\Omega,Q\pi_2)\not=\{0\}$, then 
$\ep>0$.
\een
\eep

\ber
The time-reversal invariance $\mu\Theta=\mu$ of the steady state 
 is equivalent to $\theta M\theta=M$. For Markovian
harmonic networks, the latter condition is easily seen to imply
$$
\mu(p_i q_j)=0,\qquad (i,j\in\cI),
$$
i.e., the statistical independence of simultaneous positions and 
momenta. In the quasi-Markovian case, $\theta M\theta=M$ implies
$$
\mu(p_i q_j)=\mu(p_ir_k)=\mu(q_jr_k)=0,\qquad (i,j\in\cI, k\in\cJ).
$$
\eer

\subsection{Path space time-reversal}
\label{SSEC-Time_Reversal}

Given $\tau>0$, the space-time statistics of the process~\eqref{EQ-The_Process}
in the finite period $[0,\tau]$ is described by 
$(\fX^\tau,\cX^\tau,\PP_\nu^\tau)$, where $\PP_\nu^\tau$\index{$\PP_\nu^\tau$} is the measure
induced  by the initial law $\nu\in\cP(\Xi)$ on the path-space 
$\fX^\tau=C([0,\tau],\Xi)$ equipped with its Borel $\sigma $-algebra $\cX^\tau$.
Path space time-reversal is given by the involution
$$
\Theta^\tau:\bx=\{x(t)\}_{t\in[0,\tau]}\mapsto
\boldsymbol{\tilde x}=\{\theta x(\tau-t)\}_{t\in[0,\tau]}
$$
of $\fX^\tau$. The time reversed path space measure $\widetilde\PP_\nu^\tau$
is defined by\index{$\widetilde{\PP}_\nu^\tau$}
$$
\widetilde{\PP}_\nu^\tau=\PP_\nu^\tau\circ\Theta^\tau.
$$ 
Since
\beq
\widetilde{\EE}_\nu^\tau[f(x(0))]
=\EE_\nu^\tau[f(\theta x(\tau))]=\nu P^\tau\Theta(f),
\label{EQ-PathTRDef}
\eeq
$\widetilde{\PP}_\nu^\tau$ describes the statistics of the time reversed 
process $\boldsymbol{\tilde x}$ started with the law $\nu P^\tau\Theta$.
It is therefore natural to compare it with $\PP_{\nu P^\tau\Theta}^\tau$.
The following result (proved in 
Section~\ref{SSECT_Proof_of_PROP-Time_Reversal}) provides a connection between 
the functional $\Ep(\,\cdot\,,\tau)$ and time-reversal of the path space 
measure.

Set
$$
\cP^1_{\mathrm{loc}}(\Xi)=\left\{\zeta\in\cP(\Xi)\,\bigg|\,
\frac{\d\zeta}{\d x}+
\left|\nabla\frac{\d\zeta}{\d x}\right|\in L^2_{\rm loc}(\Xi,\d x)
\right\}.
$$
\bep\label{PROP-Time_Reversal}
For any $\tau>0$ and any $\nu\in\cP^1_{\mathrm{loc}}(\Xi)$,
$\widetilde{\PP}_\nu^\tau$ is absolutely continuous
w.r.t.\;$\PP_{\nu P^\tau\Theta}^\tau$ and
\beq
\log\frac{\d\widetilde{\PP}_\nu^\tau}{\d\PP_{\nu P^\tau\Theta}^\tau}
=\Ep(\nu,\tau)\circ\Theta^\tau
=-\fS^\tau
-\log\frac{\d\nu_\tau}{\d x}(\theta x(0))
+\log\frac{\d\nu}{\d x}(\theta x(\tau)).
\label{EQ-PTR}
\eeq
\eep

\ber
The above result is a mathematical formulation
of~\cite[Section~3.1]{MNV} in the framework of harmonic networks. 
Rewriting~\eqref{EQ-PTR} as
$$
\log\frac{\d\PP_\nu^\tau}{\d\PP_{\nu P^\tau\Theta}^\tau\circ\Theta^\tau}
=\Ep(\nu,\tau)
=\fS^\tau+\log\frac{\d\nu}{\d x}(x(0))-\log\frac{\d\nu_\tau}{\d x}(x(\tau)),
$$
we obtain Eq.~(3.12) of~\cite{MNV}.
Proposition~\ref{PROP-Time_Reversal} is a consequence of Girsanov formula,
the generalized detailed balance condition~\eqref{EQ-GenDBC},
and the fact that the time-reversed process $\boldsymbol{\tilde x}$
is again a diffusion. Apart from the last fact, which was proven
in~\cite{PH}, the main technical difficulty in its proof is to check the 
martingale property of the exponential of the right-hand side 
of~\eqref{EQ-PTR}.
\eer
\ber
It is an immediate consequence of
Eq.~\eqref{EQ-PtStar} below that
$\nu P^\tau\in\cP^1_{\mathrm{loc}}(\Xi)$
for any $\nu\in\cP(\Xi)$ and $\tau>0$.
\eer

\smallskip
Equipped with Eq.~\eqref{EQ-PTR} it is easy to transpose the
relative entropies formulas of the previous section to path space measures.
As a first application, let us compute the relative entropy 
of $\PP_{\eta\Theta}^\tau$ w.r.t.\;$\widetilde{\PP}_\nu^\tau$:
\begin{align*}
\Ent(\widetilde{\PP}_\nu^\tau|\PP_{\eta\Theta}^\tau)
&=\widetilde{\EE}_\nu^\tau\left[
-\log\frac{\d\widetilde{\PP}_\nu^\tau}{\d\PP_{\eta\Theta}^\tau}\right]
=\widetilde{\EE}_\nu^\tau\left[
-\log\frac{\d\widetilde{\PP}_\nu^\tau}{\d\PP_{\nu P^\tau\Theta}^\tau}
+\log\frac{\d\PP_{\eta\Theta}^\tau}{\d\PP_{\nu P^\tau\Theta}^\tau}\right]\\
&=\widetilde{\EE}_\nu^\tau\left[-\Ep(\nu,\tau)\circ\Theta^\tau
+\log\frac{\d\eta}{\d\nu_\tau}(\theta x(0))\right]
=-\EE_\nu^\tau\left[\Ep(\nu,\tau)\right]+\Ent(\nu_\tau|\eta).
\end{align*}
If $\nu\in\cP_+(\Xi)$ then~\eqref{EQ-EpExp} yields
$$
-\Ent(\widetilde{\PP}_\nu^\tau|\PP_{\nu P^\tau\Theta}^\tau)
=\EE_\nu^\tau\left[\Ep(\nu,\tau)\right]
=\tfrac12\int_0^\tau\nu_t(|Q^\ast\nabla\log\frac{\d\nu_t}{\d\mu_\beta}|^2)\d t,
$$
which, according to the previous section, is the entropy produced by 
the process during the period $[0,\tau]$. Setting $\nu=\mu$, 
we obtain
$$
-\Ent(\widetilde{\PP}_\mu^\tau|\PP_{\mu\Theta}^\tau)=\mathrm{ep}(\mu)\tau.
$$
Together with Proposition~\ref{PROP-StrictPositivity2}~(1), this relation proves

\bet\label{THM-ReversibleProcess}
The following statements are equivalent:
\ben
\item $\PP_{\mu}^\tau\circ\Theta^\tau=\PP_\mu^\tau$ { }for all
$\tau>0$, i.e., the stationary process~\eqref{EQ-The_Process} is reversible.
\item
$\PP_{\mu}^\tau\circ\Theta^\tau=\PP_\mu^\tau$ { }for some $\tau>0$.
\item $\ep=0$.
\een
\eet

\subsection{The canonical entropic functional}

We are now in position to deal with the first step in our scheme: the
construction of the canonical entropic functional $S^\tau$ associated to 
$(\fX^\tau,\cX^\tau,\PP_\mu^\tau,\Theta^\tau)$. By 
Proposition~\ref{PROP-Time_Reversal}, R\'enyi's relative $\alpha$-entropy per
unit time of  the pair ($\PP_\mu^\tau, \widetilde{\PP}_\mu^\tau)$, 
$$
\Ent_\alpha(\PP_\mu^\tau|\widetilde{\PP}_\mu^\tau)
=\log\EE_\mu\left[\e^{-\alpha S^\tau}\right],
$$
is the cumulant generating function of\index{$S^\tau$}
\beq
S^\tau=\log\frac{\d\PP_\mu^\tau}{\d\widetilde{\PP}_\mu^\tau}
=\fS^\tau
-\log\frac{\d\mu}{\d x}(\theta x(\tau))
+\log\frac{\d\mu}{\d x}(x(0)).
\label{EQ-TwistedEp}
\eeq
In the following, we shall set
\beq
e_\tau(\alpha)
=\frac1\tau\log\EE_\mu\left[\e^{-\alpha S^\tau}\right],
\label{EQ-etaudef}
\eeq
which, by construction, satisfies the Gallavotti--Cohen symmetry
$e_\tau(1-\alpha)=e_\tau(\alpha)$.

Before formulating our main result
on the large time asymptotics of $e_\tau(\alpha)$, we need several 
technical facts which will be proved in 
Section~\ref{SSECT_Proof_of_THM-ealpha}.

\bet\label{THM-ealpha} Suppose  that Assumption~(C) holds.
\ben
\item For $\beta\in L(\Xi)$ satisfying Conditions~\eqref{EQ-betadef}, the map\index{$E(\omega)$}
\beq
\rr\ni\omega\mapsto E(\omega)
=Q^\ast(A^\ast-\i\omega)^{-1}\Sigma_\beta(A+\i\omega)^{-1}Q
\label{EQ-EomegaDef}
\eeq
takes values in the self-adjoint operators on the complexification of 
$\partial\Xi$. As such, it is continuous and independent of the choice of 
$\beta$.
\item Set\index{$\varepsilon_+,\varepsilon_-$}\index{$\kappa_c$}
$$
\varepsilon_-=\min_{\omega\in\rr}\min\sp(E(\omega)),\qquad
\varepsilon_+=\max_{\omega\in\rr}\max\sp(E(\omega)),\qquad
\kappa_c=\frac1{\varepsilon_+}-\frac12.
$$
The following alternative holds: either $\kappa_c=\infty$ in which case
$E(\omega)=0$ for all $\omega\in\rr$, or $\frac12<\kappa_c<\infty$, 
$\varepsilon_-<0$, $0<\varepsilon_+<1$, and
$$
\frac1{\varepsilon_-}+\frac1{\varepsilon_+}=1.
$$
\item Set\index{$\fI_c$} 
$\fI_c=]\frac12-\kappa_c,\frac12+\kappa_c[\,
=\,]\frac1{\varepsilon_-},\frac1{\varepsilon_+}[$. The function
\beq
e(\alpha)=
-\int_{-\infty}^{\infty}\log\det\left(
I-\alpha E(\omega)\right)\frac{\d\omega}{4\pi}
\label{EQ-lambdaalphadef}
\eeq
is analytic on the cut plane 
$\fC_c=(\cc\setminus\rr)\cup\fI_c$. It is convex on the open interval $\fI_c$ 
and extends to a continuous function on the closed interval $\bar\fI_c$. 
It further satisfies 
\beq
e(1-\alpha)=e(\alpha)
\label{EQ-TheSymmetry2}
\eeq
for all $\alpha\in\mathfrak{C}_c$,
$$
\left\{\begin{array}{ll}
e(\alpha)\le0&\mbox{for }\alpha\in[0,1];\\[6pt]
e(\alpha)\ge0&\mbox{for }\alpha\in\bar\fI_c\setminus]0,1[;\\[4pt]
\end{array}
\right.
$$
and in particular $e(0)=e(1)=0$. Moreover 
$$
\ep=-e'(0)=e'(1),
$$
and either $\ep=0$, $\kappa_c=\infty$, and $e(\alpha)$ vanishes
identically, or $\ep>0$, $\kappa_c<\infty$, $e(\alpha)$ is
strictly convex on $\bar\fI_c$, and
\beq
\lim_{\alpha\downarrow\frac12-\kappa_c}e'(\alpha)=-\infty,
\qquad
\lim_{\alpha\uparrow\frac12+\kappa_c}e'(\alpha)=+\infty.
\label{EQ-spmDef}
\eeq
\item If $\ep>0$, then there exists a unique signed Borel 
measure $\varsigma$ on $\rr$, supported on $\rr\setminus\fI_c$, such that
$$
\int\frac{|\varsigma|(\d r)}{|r|}<\infty,
$$
and
$$
e(\alpha)=-\int\log\left(1-\frac\alpha r\right)\varsigma(\d r).
$$
\item For $\alpha\in\rr$ define
\beq
K_\alpha=\left[\begin{array}{cc}
-A_\alpha&QQ^\ast\\
C_\alpha&A_\alpha^\ast
\end{array}\right],
\label{EQ-KalphaDef}
\eeq
where
\beq
A_\alpha=(1-\alpha)A-\alpha A^\ast,\qquad
C_\alpha=\alpha(1-\alpha)Q\vartheta^{-2}Q^\ast.
\label{EQ-ACdef}
\eeq
For all $\omega\in\rr$ and $\alpha\in\rr$ one has
$$
\det(K_\alpha-\i\omega)=|\det(A+\i\omega)|^2\det(I-\alpha E(\omega)).
$$
Moreover, for $\alpha\in\fI_c$,
\beq
e(\alpha)
=\tfrac14\tr(Q\vartheta^{-1}Q^\ast)
-\tfrac14\sum_{\lambda\in\sp(K_\alpha)}|\Re\lambda|\, m_\lambda,
\label{EQ-ealphaSpecForm}
\eeq
where $m_\lambda$ denotes the algebraic multiplicity of 
$\lambda\in\sp(K_\alpha)$.
\een
\eet

\ber\label{REM-kappa0}
We shall prove, in Proposition~\ref{PROP-RicX}~(11), that
\beq
\kappa_c\ge
\kappa_0=\frac12
\frac{\vartheta_{\mathrm{max}}+\vartheta_{\mathrm{min}}} %
{\vartheta_{\mathrm{max}}-\vartheta_{\mathrm{min}}}.
\label{EQ-kappaclow}
\eeq
This lower bound is sharp, i.e., there are networks for which equality
holds (see Theorem~\ref{THM-Jacobi}~(3)).
\eer
\ber
It follows from~\eqref{EQ-kappaclow} that $\kappa_c=\infty$ for
harmonic networks at equilibrium, i.e., whenever
$\vartheta_{\rm min}=\vartheta_{\rm max}=\vartheta_0>0$. Up to the
controllability assumption of Proposition~\ref{PROP-StrictPositivity2}~(2), these are the only examples with $\kappa_c=\infty$
(see also Remark~\ref{REM-EquilibriumXalpha} and Section~\ref{SEC-Examples}).
\eer
\ber Remark 2 after Theorem 2.1 in \cite{JPS} applies to Part (4) of Theorem \ref{THM-ealpha}.
\eer

\smallskip
In the sequel it will be convenient to consider the following 
natural extension of the function $e(\alpha)$.
\bed\label{DEF-e} The function
$$
\rr\ni\alpha\mapsto e(\alpha)\in]-\infty,+\infty]
$$ 
is given by~\eqref{EQ-lambdaalphadef} for $\alpha\in\bar\fI_c$ and
$e(\alpha)=+\infty$ for $\alpha\in\rr\setminus\bar\fI_c$.
\eed

This definition makes $\rr\ni\alpha\mapsto e(\alpha)$ an {\sl essentially smooth closed 
proper convex function} (see~\cite{Ro}).

The main result of this section relates the spectrum of the matrix $K_\alpha$,
through the function $e(\alpha)$, to the large time asymptotics of the 
R\'enyi entropy~\eqref{EQ-etaudef} and the cumulant generating function of the canonical entropic functional 
$S^t$.

\bep\label{PROP-Renyi}
Under Assumption (C) and with Definition~\ref{DEF-e} one has 
\beq
\lim_{\tau\to\infty}e_\tau(\alpha)=e(\alpha),
\label{EQ-ealphaForm}
\eeq
for all $\alpha\in\rr$.
\eep
\noindent

\bigskip A closer look at the proof of Proposition~\ref{PROP-Renyi} 
in Section~\ref{SSECT_Proof_of_PROP-Renyi} gives more. For any $x\in\Xi$
and $\alpha\in\bar\fI_c$ 

$$
\lim_{\tau\to\infty}\EE_x\left[\e^{-\alpha S^\tau-\tau e(\alpha)}\right]
=f_\alpha(x)=c_\alpha\e^{-\frac12 x\cdot T_\alpha x},
$$
see~\cite[Section~20.1.5]{MT} and references therein.
The functions $\alpha\mapsto c_\alpha\in[0,\infty[$ and
$\alpha\mapsto T_\alpha\in L(\Xi)$ are real analytic on $\fI_c$,
continuous on $\bar\fI_c$, $c_\alpha>0$ for $\alpha\in\fI_c$, and
$T_\alpha>M^{-1}$ for $\alpha\in\bar\fI_c$.
Moreover, the convergence also holds in $L^1(\Xi,\d\mu)$ and is exponentially 
fast for $\alpha\in\fI_c$. For $\alpha\in\bar\fI_c$ and as $\tau\to\infty$,  one has
$$
e_\tau(\alpha)=e(\alpha)+\frac1\tau g_\tau(\alpha)
=e(\alpha)+\frac1\tau\left(\log\mu(f_\alpha)
+\mathcal{O}(\e^{-\epsilon(\alpha)\tau})\right),
$$
where $\epsilon(\alpha)>0$ for $\alpha\in\fI_c$.
However, $c_\alpha$ vanishes on $\partial\fI_c$ and hence the 
"prefactor" $g_\tau(\alpha)$ diverges as $\alpha\to\partial\fI_c$.
Nevertheless, \eqref{EQ-ealphaForm} holds because
$$
-\infty=\lim_{\tau\to\infty}\lim_{\alpha\to\partial\fI_c}
\frac1\tau g_\tau(\alpha)\not=
\lim_{\alpha\to\partial\fI_c}\lim_{\tau\to\infty}
\frac1\tau g_\tau(\alpha)=0.
$$
Like in our introductory example, the occurrence of singularities in the 
"prefactor" $g_\tau(\alpha)$ is related to the tail of the 
law of $S^t$. This phenomenon was observed by Cohen and van Zon in 
their study of the fluctuations of the work done on a dragged Brownian 
particle and its heat dissipation \cite{CvZ1} (see also~\cite{CvZ2,Vi} for 
more detailed analysis). In their model, which is closely related to ours, the 
cumulant generating function of the dissipated heat $e_\tau(\alpha)$ diverges 
for $\alpha^2\ge(1-\e^{-2\tau})^{-1}$ and hence 
$$
\lim_{\tau\to\infty}e_\tau(\alpha)=+\infty\quad\mbox{for $|\alpha|>1$}.
$$
This leads to a breakdown of the Gallavotti--Cohen
symmetry and to an extended fluctuation relation. We will come back to this 
point in the next section and see that this is a general feature of the 
TDE functional $\fS^t$ (see Eq.~\eqref{EQ-RawLimit} below).
Proposition~\ref{PROP-Renyi} and Theorem~\ref{THM-ealpha}~(3) show that the 
canonical entropic functional $S^t$ does not suffer from this defect: its
limiting cumulant generating function $e(\alpha)$ satisfies 
Gallavotti--Cohen symmetry for all $\alpha\in\rr$.

\subsection{Large deviations of the canonical entropic functional}
\label{SEC-LDP}

We now turn to Step 2 of our scheme.
We recall some fundamental results on the large deviations of
a family $(\xi_t)_{t\ge0}$ of real-valued random variables (the 
G\"artner-Ellis theorem, see, e.g., \cite[Theorem~V.6]{dH}).
We shall focus on the situations relevant for our discussion of entropic 
fluctuations. We refer the reader to \cite{dH,DZ} for more general  exposition.

By H\"older's inequality, the cumulant generating function
$$
\rr\ni\alpha\mapsto\Lambda_t(\alpha)
=\frac1t\log\EE[\e^{\alpha\xi_t}]\in]-\infty,\infty],
$$
is convex and vanishes at $\alpha=0$. It is finite on some
(possibly empty) open interval and takes the value $+\infty$ on the 
(possibly empty) interior of its complement.

\ber
The above definition follows the convention 
used in the mathematical literature on large deviations. Note, however, that
in the previous section we have adopted the convention of the physics
literature on entropic fluctuations where the cumulant generating function of 
an entropic functional~$\xi_t$ is defined by 
$\alpha\mapsto t^{-1}\log\EE[\e^{-\alpha\xi_t}]$. This clash of conventions is
the origin of various minus signs occurring in Theorems~\ref{THM-TwistedLDP}
and~\ref{THM-GaussianLDP} below.
\eer

\smallskip
The function
$$
\rr\ni\alpha\mapsto\Lambda(\alpha)=\limsup_{t\to\infty}\Lambda_t(\alpha)
=\lim_{t\to\infty}\sup_{s\ge t}\Lambda_s(\alpha)\in[-\infty,\infty]
$$
is convex and vanishes at $\alpha=0$. Let~$D$ be the interior of its
effective domain $\{\alpha\in\rr\,|\,\Lambda(\alpha)<\infty\}$, and assume 
that $0\in D$. Then $D$ is a non-empty open interval,
$\Lambda(\alpha)>-\infty$ for all $\alpha\in\rr$, and the function
$D\ni\alpha\mapsto\Lambda(\alpha)$ is convex and continuous. The Legendre transform
$$
\Lambda^\ast(x)=\sup_{\alpha\in\rr}(\alpha x-\Lambda(\alpha))
=\sup_{\alpha\in\bar D}(\alpha x-\Lambda(\alpha))
$$
is convex and lower semicontinuous, as supremum of a family of affine
functions. Moreover, $\Lambda(0)=0$ implies that $\Lambda^\ast$ is 
non-negative. The large deviation upper bound
\beq
\limsup_{t\to\infty}\frac1t\log\PP\left[\frac1t\xi_t\in C\right]
\le-\inf_{x\in C}\Lambda^\ast(x)
\label{EQ-LDP-upper}
\eeq
holds for all closed sets $C\subset\rr$.

Assume, in addition, that on some finite open interval 
$0\in D_0=]\alpha_-,\alpha_+[\subset D$ 
the function $D_0\ni\alpha\mapsto\Lambda(\alpha)$ is real analytic  and  not linear. Then $\Lambda$ is strictly convex and its derivative 
$\Lambda'$ is strictly increasing on $D_0$. We denote by $x_\mp$ the (possibly 
infinite) right/left limits of $\Lambda'(\alpha)$ at $\alpha=\alpha_\mp$. 
By convexity,
\beq
\Lambda(\alpha)\ge\Lambda(\alpha_0)+(\alpha-\alpha_0)\Lambda'(\alpha_0)
\label{EQ-supporting}
\eeq
for any $\alpha_0\in D_0$ and $\alpha\in\rr$, and
$$
\Lambda(\alpha_\pm)\ge\Lambda_\pm
=\lim_{D_0\ni\alpha\to\alpha_\pm}\Lambda(\alpha).
$$
Since $\Lambda^\ast$ is non-negative, it follows that 
$\Lambda^\ast(\Lambda'(0))=0$. One easily shows that~\eqref{EQ-supporting} 
also implies
$$
\Lambda^\ast(x)=\sup_{\alpha\in D_0}(\alpha x-\Lambda(\alpha))
$$
for $x\in E=]x_-,x_+[$. If the limit
$$
\lim_{t\to\infty}\Lambda_t(\alpha)
$$
exists for all $\alpha\in D_0$, then it coincides with $\Lambda(\alpha)$, and the large deviation lower bound
\beq
\liminf_{t\to\infty}\frac1t\log\PP\left[\frac1t\xi_t\in O\right]
\ge-\inf_{x\in O\cap E}\Lambda^\ast(x)
\label{EQ-LDP-lower}
\eeq
holds for all open sets $O\subset \rr$. Note that in cases where
$x_-=-\infty$ and $x_+=+\infty$ one has $E=\rr$ and convexity implies
$\Lambda(\alpha)=+\infty$ for $\alpha\in\rr\setminus[\alpha_-,\alpha_+]$.

We shall say that the family $(\xi_t)_{t\ge0}$ satisfies a local LDP on $E$
with rate function $\Lambda^\ast$ if~\eqref{EQ-LDP-upper} holds for all closed
sets $C\subset\rr$  and~\eqref{EQ-LDP-lower} holds for all open sets
$O\subset \rr$. If the latter holds with $E=\rr$, we say that this family 
satisfies a global LDP with rate function~$\Lambda^\ast$.

\begin{figure}
\centering
\includegraphics[scale=0.45]{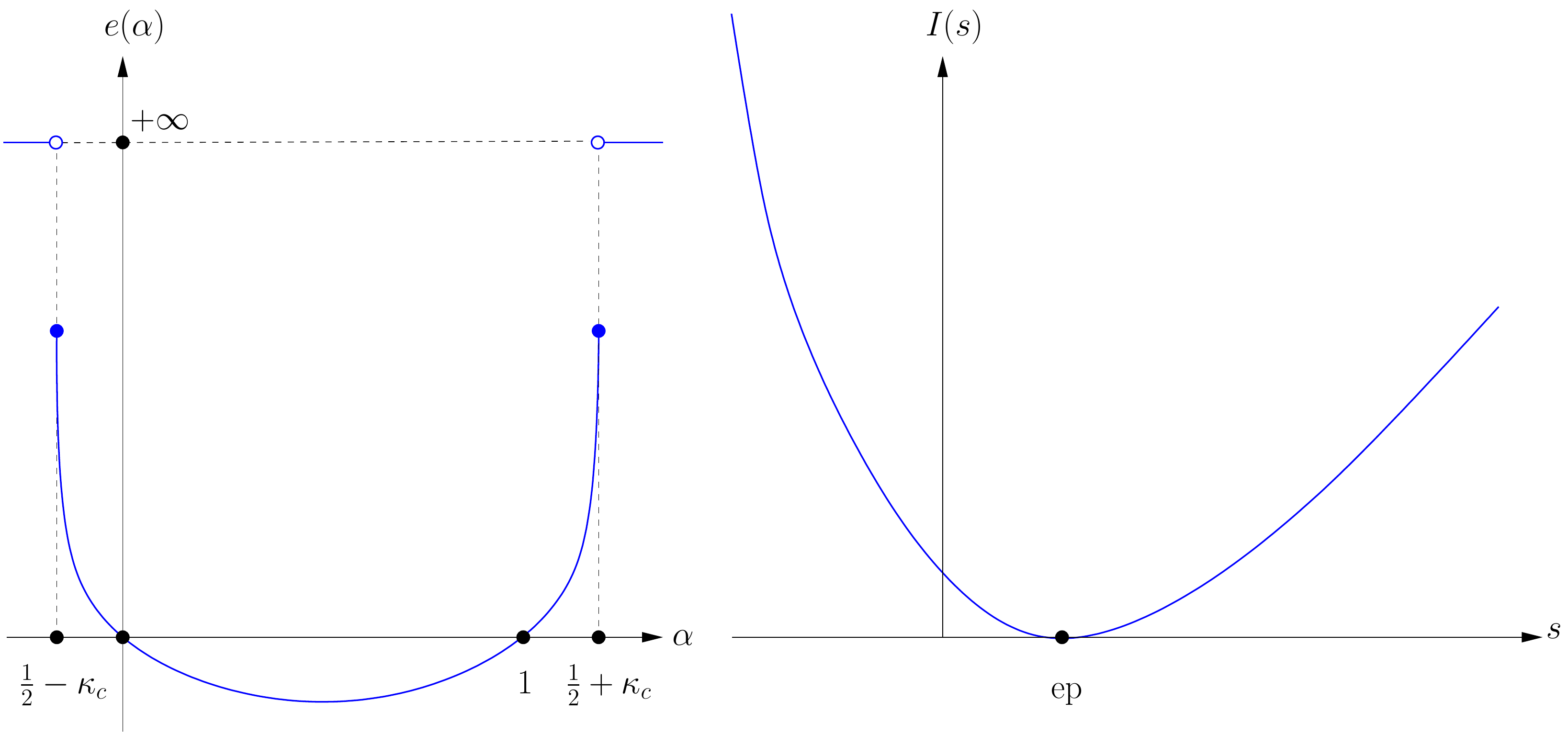}
\caption{The cumulant generating function $e(\alpha)$ and the rate
function $I(s)$ for the canonical entropic functional of a harmonic network 
satisfying Assumption~(C) and $\ep>0$. Notice the bias due to the symmetry 
$I(-s)=I(s)+s$.}
\label{Fig2}
\end{figure}

\bigskip
By the above discussion, Proposition~\ref{PROP-Renyi} and 
Theorem~\ref{THM-ealpha}~(3) immediately yield:

\bet\label{THM-TwistedLDP}
Suppose that Assumption~(C) holds. Then, 
under the law $\PP_\mu$, the family $(S^t)_{t\ge0}$ satisfies a global
LDP with rate function (see Figure~\ref{Fig2})
\beq
I(s)=\sup_{-\alpha\in\fI_c}(\alpha s-e(-\alpha)).
\label{EQ-TheRate}
\eeq
\eet

It follows from the Gallavotti--Cohen 
symmetry~\eqref{EQ-TheSymmetry2} that the 
function $\rr\ni s\mapsto I(s)+\12 s\in[0,\infty]$ is even, i.e., the universal
fluctuation relation
\beq
\frak{s}(s)=I(-s)-I(s)=s,
\label{EQ-UFT}
\eeq
holds for all $s\in\rr$.

\ber\label{REM-ellSymmetry}
If $\ep>0$, then the strict convexity and analyticity 
of the function $e(\alpha)$ stated in Theorem~\ref{THM-ealpha}~(3)
imply that the rate function $I(s)$ is itself real analytic and strictly 
convex. Denoting by $s\mapsto \ell(s)$ the inverse of the function 
$\alpha\mapsto-e'(-\alpha)$, we derive
$$
I(s)=s\ell(s)-e(-\ell(s)),\qquad I'(s)=\ell(s),
$$
and the Gallavotti--Cohen symmetry translates to $\ell(-s)+\ell(s)=-1$.
\eer
{}

\subsection{Intermezzo: A naive approach to the cumulant generating function of ${{\mathfrak{S}}^t}$}

Before dealing with perturbations of the functional $S^t$, we briefly digress
 from the main course of our scheme in order to better motivate 
what will follow. We shall try to compute the cumulant generating function
of the TDE functional $\fS^t$ by a simple Perron-Frobenius type argument.

By It\^o calculus, for any $f\in C^2(\Xi)$ one has
$$
\d(\e^{-\alpha\fS^t}f(x(t)))
=\e^{-\alpha\fS^t}\left[(L_\alpha f)(x(t))\d t
+\left(Q^\ast(\nabla f)(x(t))+\alpha\vartheta^{-1}Q^\ast x(t)f(x(t))\right)\cdot\d w(t)
\right],
$$
where  
$$
L_\alpha=\tfrac12\left(\nabla\cdot B\nabla+2A_\alpha x\cdot\nabla
-x\cdot C_\alpha x+\alpha\tr(Q\vartheta^{-1}Q^\ast)\right)
$$
is the  deformation of the Fokker-Planck 
operator~\eqref{EQ-LDef}, and   $A_\alpha$, $B$, $C_\alpha$ are given by~\eqref{EQ-BDef}, \eqref{EQ-ACdef}. Note that the structural relations~\eqref{EQ-Structure}
imply
\beq
\Theta L_\alpha\Theta=L^\ast_{1-\alpha},
\label{EQ-LalphaSym}
\eeq
where $L_\alpha^\ast$ denotes the formal adjoint of $L_\alpha$.
Assuming $L_\alpha$ to have a non-vanishing spectral 
gap, a naïve application of Girsanov formula leads to
\beq
\EE_\mu\left[\e^{-\alpha\fS^t}\right]=\mu(\e^{tL_\alpha}1)
=\e^{t\lambda_\alpha}\left(\mu(\Psi_\alpha)\int\Psi_{1-\alpha}(x)\d x
+o(1)\right),\quad (t\to\infty).
\label{EQ-PerronFrobenius}
\eeq
where $\Psi_\alpha$ is the properly normalized eigenfunction of $L_\alpha$ to 
its dominant eigenvalue $\lambda_\alpha$. It follows that
$$
\lim_{t\to\infty}\frac1t\log\EE_\mu\left[\e^{-\alpha\fS^t}\right]
=\lambda_\alpha,
$$
the Gallavotti--Cohen symmetry $\lambda_{1-\alpha}=\lambda_\alpha$ being a
direct consequence of~\eqref{EQ-LalphaSym}.

Given the form of $L_\alpha$, the Gaussian Ansatz
$$
\Psi_\alpha(x)=\e^{-\frac12 x\cdot X_\alpha x}\qquad
$$
is mandatory. Insertion into the eigenvalue equation 
$L_\alpha\Psi_\alpha=\lambda_\alpha\Psi_\alpha$ leads to the following 
equation for the real symmetric matrix $X_\alpha$,
\beq
X_\alpha BX_\alpha-X_\alpha A_\alpha-A_\alpha^\ast X_\alpha-C_\alpha=0,
\label{EQ-RicNaive}
\eeq
while the dominant eigenvalue is given by
\beq
\lambda_\alpha=\tfrac12\left(\alpha\tr(Q\vartheta^{-1}Q^\ast)
-\tr(BX_\alpha)\right).
\label{EQ-ealphaNaive}
\eeq
There are two difficulties with this naïve argument. The first one is that
it is far from obvious that Girsanov theorem applies here. The second one is
again related to the "prefactor" problem. In fact we shall see that
Eq.~\eqref{EQ-RicNaive} does not have positive definite solutions
for $\alpha\le0$, making the right-hand side of~\eqref{EQ-PerronFrobenius} 
infinite for $\alpha\ge1$. Nevertheless, the above calculation reveals
Eq.~\eqref{EQ-RicNaive} and~\eqref{EQ-ealphaNaive} which will play a central role in what follows.

\subsection{More entropic functionals}
\label{SEC-MoreEF}

In this section we deal with step 3 of our scheme. The main result,
Proposition~\ref{PROP-GaussianCocycles} below, concerns the large time
behavior of cumulant generating functions of the kind
$$
\rr\ni\alpha\mapsto g_t(\alpha)
=\frac1t\log\EE_\nu\left[\e^{-\alpha[S^t+\Phi(x(t))-\Psi(x(0))]}\right],
$$
where $\Phi$ and $\Psi$ are quadratic forms on the phase space $\Xi$,\index{$\Phi,\Psi$}\index{$F,G$}
\beq
\Phi(x)=\tfrac12 x\cdot Fx,\qquad \Psi(x)=\tfrac12 x\cdot Gx,
\label{EQ-PhiPsiDef}
\eeq
and the initial measure $\nu\in\cP(\Xi)$ is Gaussian. We then apply 
this result to some entropic functionals of physical interest:
\ben
\item The steady state TDE (recall Eq.~\eqref{EQ-TwistedEp}),\index{$\fS^t$}
\beq
\fS^t=S^t+\log\frac{\d\mu}{\d x}(\theta x(t))
-\log\frac{\d\mu}{\d x}(x(0)),
\label{EQ-Heat}
\eeq
with $\nu=\mu$.
\item The steady state TDE for quasi-Markovian 
networks~\eqref{EQ-QMTDE} which we can rewrite as\index{$\fS^t_{\rm qM}$}
\beq
\fS^t_{\rm qM}
=\fS^t+\tfrac12|\vartheta^{-1/2}\pi_Q x(t)|^2
-\tfrac12|\vartheta^{-1/2}\pi_Q x(0)|^2,
\label{EQ-qMTDE}
\eeq
where $\pi_Q$ denotes the orthogonal projection to
$\Ran Q=\partial\Xi$, with $\nu=\mu$.
\item Transient TDEs, i.e., the functionals $\fS^t$ and 
$\fS_{\rm qM}^t$, but in the 
transient process started with a Dirac measure $\nu=\delta_{x_0}$.
\item The steady state entropy production functional\index{$\Ep(\mu,t)$}
$$
\Ep(\mu,t)=S^t+\log\frac{\d\mu\Theta}{\d\mu}(x(t))
$$
with $\nu=\mu$.
\item The canonical entropic functional for the transient process, started 
with the non-degenerate Gaussian measure $\nu\in\cP(\Xi)$,\index{$S^t_\nu$}
$$
S^t_\nu=\log\frac{\d\PP_\nu^t}{\d\widetilde{\PP}_\nu^t}
=\log\frac{\d\PP_\mu^t}{\d\widetilde{\PP}_\mu^t}
+\log\frac{\d\PP_\nu^t}{\d\PP_\mu^t}
-\log\frac{\d\widetilde{\PP}_\nu^t}{\d\widetilde{\PP}_\mu^t}
=S^t-\log\frac{\d\nu}{\d\mu}(\theta x(t))+\log\frac{\d\nu}{\d\mu}(x(0)).
$$
\een

To formulate our general result, we need some facts about the matrix
equation~\eqref{EQ-RicNaive}.

Define a map $\cR_\alpha:L(\Xi)\to L(\Xi)$ by\index{$\cR_\alpha$}
\beq
\cR_\alpha(X)=
X BX-X A_\alpha-A_\alpha^\ast X-C_\alpha,
\label{EQ-RicX}
\eeq
where $A_\alpha$, $B$ and $C_\alpha$ are defined by~\eqref{EQ-BDef} 
and~\eqref{EQ-ACdef}.
The equation $\cR_\alpha(X)=0$ is an algebraic {\sl Riccati equation} for the
unknown {\sl self-adjoint} $X\in L(\Xi)$. We refer the reader to the 
monographs~\cite{LR,AFIJ} for an in depth discussion of such equations.

A solution $X$ of the Riccati equation is called {\sl minimal} ({\sl maximal}) if it 
is such that $X\le X'$ ($X\ge X'$) for any other solution $X'$ of the
equation. We shall investigate the Riccati equation in 
Section~\ref{SEC-Riccati}.
At this point we just mention that, under Assumption~(C), it has a unique 
maximal solution $X_\alpha$ for any $\alpha\in\bar\fI_c$, with the special
values
\beq
X_0=0,\qquad X_1=\theta M^{-1}\theta.
\label{EQ-Xzero_one}
\eeq

\bep\label{PROP-GaussianCocycles}
Suppose that Assumption  (C) is satisfied and
let $\nu$ be the Gaussian measure on $\Xi$ with mean $a$ and
covariance $N\ge0$. Denote by $P_\nu$ the orthogonal projection on $\Ran N$
and by $\widehat{N}$ the inverse of the restriction of $N$ to its range.
Let $F,G\in L(\Xi)$ be self-adjoint
and define $\Phi$, $\Psi$ by~\eqref{EQ-PhiPsiDef}.
\ben
\item For $t>0$ the function
$$
\rr\ni\alpha\mapsto g_t(\alpha)=
\frac1t\log\EE_\nu[\e^{-\alpha (S^t+\Phi(x(t))-\Psi(x(0)))}]
$$
is convex. It is finite and real analytic on some open interval 
$\fI_t=]\alpha_-(t),\alpha_+(t)[\ni0$ and infinite on its complement. 
Moreover, the following alternatives hold:
\begin{itemize}
\item Either $\alpha_-(t)=-\infty$
or $\lim_{\alpha\downarrow\alpha_-(t)}g_t'(\alpha)=-\infty$.
\item Either $\alpha_+(t)=+\infty$ 
or $\lim_{\alpha\uparrow\alpha_+(t)}g_t'(\alpha)=+\infty$.
\end{itemize}
\item Set\index{$\fI_+, \fI_-$}
\begin{align*}
\fI_+&=\{\alpha\in\bar\fI_c\,|\,
\theta X_{1-\alpha}\theta+\alpha(X_1+F)>0\},\\[6pt]
\fI_-&=\{\alpha\in\bar\fI_c\,|\,
\widehat{N}+P_\nu(X_\alpha-\alpha(G+\theta X_1\theta))|_{\Ran N}>0\},
\end{align*}
with the proviso that $\fI_-=\bar\fI_c$ whenever $N=0$.
Then\index{$\fI_\infty$} $\fI_\infty=\fI_-\cap\fI_+$ is a (relatively) open subinterval of 
$\bar\fI_c$ containing $0$. 
\item If $X_1+F>0$ and either $N=0$ or 
$\widehat{N}+P_\nu(X_1-\theta X_1\theta-G)|_{\Ran N}>0$, 
then $[0,1]\subset\fI_\infty$.
\item For $\alpha\in\fI_\infty$ one has
\beq
\lim_{t\to\infty}g_t(\alpha)=e(\alpha).
\label{EQ-etildeDef}
\eeq
\item Set $\alpha_-=\inf\fI_\infty<0$\index{$\alpha_+,\alpha_-$} and $\alpha_+=\sup\fI_\infty>0$. Then,
\beq
\lim_{t\to\infty}\alpha_\pm(t)=\alpha_\pm,
\label{EQ-limalpha+}
\eeq
and for any $\alpha\in\rr\setminus[\alpha_-,\alpha_+]$,
\beq
\lim_{t\to\infty}g_t(\alpha)=+\infty.
\label{EQ-limg}
\eeq
\een
\eep

\ber
The existence and  value of  the limit~\eqref{EQ-etildeDef} for 
$\alpha\in\partial\fI_\infty$ is a delicate problem whose  resolution  requires  
additional information on the two subspaces
$$
\Ker(\theta X_{1-\alpha}\theta+\alpha(X_1+F)),\qquad
\Ker(\widehat{N}+P_\nu(X_\alpha-\alpha(G+\theta X_1\theta))|_{\Ran N})
$$
at the points $\alpha\in\partial\fI_\infty$. Since, as we shall see in the next section, this question 
is irrelevant for the large deviations properties of the functional
$S^t+\Phi(x(t))-\Psi(x(0))$, we shall not discuss it further.
\eer
\ber
We shall see in Section~\ref{SEC-Riccati} that
the maximal solution $X_\alpha$ of the Riccati equation is linked to the 
function $e(\alpha)$ through the identity $e(\alpha)=\lambda_\alpha$, 
where $\lambda_\alpha$ is given by Eq.~\eqref{EQ-ealphaNaive}.
Thus, the large time behavior of the function $\alpha\mapsto g_t(\alpha)$ is 
completely characterized by the maximal solution~$X_\alpha$ through this
formula and the two numbers $\alpha_\pm$.
Riccati equations play an important role in various
areas of engineering mathematics, e.g., control and filtering theory. For 
these reasons, very efficient algorithms are available to numerically compute
their maximal/minimal solutions. Hence, our approach is well designed for
numerical investigation of concrete models.
\eer

\paragraph{Steady state dissipated TDE}
According to Eq.~\eqref{EQ-Heat} and~\eqref{EQ-Xzero_one}, the case 
of TDE dissipation in the stationary process corresponds to the
choice
$$
\widehat{N}=\theta X_1\theta,\qquad
F=-X_1,\qquad
G=-\theta X_1\theta,
$$
and it follows directly from Proposition~\ref{PROP-RicX}~(2) and~(4) below that
$$
\fI_+=\{\alpha\in\bar\fI_c\,|\,X_{1-\alpha}>0\}=[\tfrac12-\kappa_c,1[.
$$
Setting 
$\alpha_-=\inf\{\alpha\in\bar\fI_c\,|\,X_\alpha+\theta X_1\theta>0\}$,
we have either $\alpha_-\in]\tfrac12-\kappa_c,0[$ and
$$
\fI_\infty=]\alpha_-,1[,
$$
or $\alpha_-=\tfrac12-\kappa_c$ and
$$
\fI_\infty=[\alpha_-,1[.
$$

Suppose that $\tfrac12-\kappa_c\le-1$ and let 
$\alpha\in[\tfrac12-\kappa_c,-1]$. From Proposition~\ref{PROP-RicX}~(10)
we deduce that $X_\alpha\le\alpha X_1$.
Since $X_1=\theta M^{-1}\theta>0$, it follows that
\beq
X_\alpha+\theta X_1\theta\le\alpha X_1+\theta X_1\theta
=\alpha(X_1-\theta X_1\theta)+(1+\alpha)\theta X_1\theta
\le\alpha(\theta M^{-1}\theta-M^{-1}).
\label{EQ-IminusBound}
\eeq
Observe that the right-hand side of this inequality is odd under conjugation 
by $\theta$. Moreover, Proposition~\ref{PROP-StrictPositivity2}~(1) implies
that it vanishes iff $\ep=0$. It follows that
$\sp(X_\alpha+\theta X_1\theta)\cap]-\infty,0]\not=\emptyset$. Thus, we can
conclude that one always has $\alpha_+=1$ and $\alpha_-\ge-1$, with
strict inequality whenever $\ep>0$.

By Proposition~\ref{PROP-GaussianCocycles},
\beq
e_{\rm TDE,st}(\alpha)=
\lim_{t\to\infty}\frac1t\log\EE_\mu[
\e^{-\alpha\fS^t}]
=\left\{\begin{array}{ll}
e(\alpha)&\text{for }\alpha\in]\alpha_-,1[\\[6pt]
+\infty&\text{for }\alpha\not\in[\alpha_-,1].
\end{array}
\right.
\label{EQ-RawLimit}
\eeq
An explicit evaluation of the resulting Gaussian integral further shows that
$$
e_{\rm TDE,st}(1)
=\lim_{t\to\infty}\frac1t\log\EE_\mu[\e^{-\fS^t}]
=\12\tr(Q\vartheta^{-1}Q^\ast)>0.
$$
The Gallavotti--Cohen symmetry is broken in the sense that it fails outside
the interval $]0,1[$, in particular 
$e_{\rm TDE,st}(0)=e(0)=0<e_{\rm TDE,st}(1)$. Note also that
$$
\liminf_{\alpha\to1}e_{\rm TDE,st}(\alpha)=e(1)=0<e_{\rm TDE,st}(1)
<\limsup_{\alpha\to1}e_{\rm TDE,st}(\alpha)=+\infty,
$$
i.e., the limiting cumulant generating function for TDE dissipation rate in 
the stationary process is {\sl neither} lower semicontinuous {\sl nor} 
upper semicontinuous.

\ber
We shall see in Section~\ref{SEC-Riccati}
(see Remark~\ref{REM-EquilibriumXalpha}) that in the case of
thermal equilibrium, i.e., $\vartheta=\vartheta_0I$ for some
$\vartheta_0\in]0,\infty[$, one has $X_\alpha=\alpha\vartheta_0I$ and hence
$X_{-1}+\theta X_1\theta=0$. Thus, in this case, $\alpha_-=-1$ and since
$e(\alpha)$ vanishes identically by Proposition~\ref{THM-ealpha}~(3),
$$
e_{\rm TDE,st}(\alpha)=\left\{\begin{array}{ll}
0&\text{for }|\alpha|<1\\[6pt]
+\infty&\text{for }|\alpha|>1.
\end{array}
\right.
$$
\eer
\ber
According to Eq.~\eqref{EQ-qMTDE}, for 
quasi-Markovian networks the steady-state TDE dissipation 
corresponds to
$$
\widehat{N}=\theta X_1\theta,\qquad
F=-X_1+\pi_Q\vartheta^{-1}\pi_Q,\qquad
G=\theta(-X_1+\pi_Q\vartheta^{-1}\pi_Q)\theta.
$$
Since $\theta\pi_Q=\pm\pi_Q=\pi_Q\theta$, one has
$$
[0,1[\subset
\fI_+=\{\alpha\in\bar\fI_c\,|\,
X_{1-\alpha}+\alpha\pi_Q\vartheta^{-1}\pi_Q>0\}
\subset[\tfrac12-\kappa_c,1[,
$$
provided $\partial\Xi\not=\Xi$.
The inequality~\eqref{EQ-IminusBound} yields
$$
(I-\pi_Q)(X_\alpha+\theta X_1\theta
-\alpha\pi_Q\vartheta^{-1}\pi_Q)(I-\pi_Q)
\le\alpha(I-\pi_Q)(\theta M^{-1}\theta-M^{-1})(I-\pi_Q),
$$
for $\tfrac12-\kappa_c\le\alpha\le-1$. From the Lyapunov 
equation~\eqref{EQ-LyapM} one easily deduces that
$$
(I-\pi_Q)(\theta M^{-1}\theta-M^{-1})(I-\pi_Q)=0
$$
iff $\theta M\theta=M$ so that the above argument still applies 
and~\eqref{EQ-RawLimit} holds with $\fS^t$ replaced 
by $\fS_{\rm qM}^t$ and $\alpha_-\ge-1$ with strict inequality
whenever $\ep>0$.
\eer

\paragraph{Transient dissipated TDE}\label{PAR-TrTDE}
 Consider now the functional $\fS^t$ for
the process started with the Dirac measure $\nu=\delta_{x_0}$ for some 
$x_0\in\Xi$. This corresponds to
$$
N=0,\qquad
F=-X_1,\qquad
G=-\theta X_1\theta,
$$
and in this case
$$
\fI_+=[\tfrac12-\kappa_c,1[,\qquad\fI_-=\bar\fI_c,
$$
and hence $\fI_\infty=[\tfrac12-\kappa_c,1[$. 
Proposition~\ref{PROP-GaussianCocycles} yields a cumulant generating
function
\beq
e_{\rm TDE,tr}(\alpha)=
\lim_{t\to\infty}\frac1t\log\EE_{x_0}[
\e^{-\alpha\fS^t}]
=\left\{\begin{array}{ll}
e(\alpha)&\text{for }\alpha\in]\tfrac12-\kappa_c,1[\\[6pt]
+\infty&\text{for }\alpha\not\in[\tfrac12-\kappa_c,1],
\end{array}
\right.
\label{EQ-RawTrLimit}
\eeq
which does not depend on the initial condition $x_0$.

\ber\label{REM-qMTDEtr}
For quasi-Markovian networks it may happen
that $\fI_\infty=]\alpha_-,1[$ with $\alpha_->\tfrac12-\kappa_c$.
For later reference, let us consider the case\footnote{We shall see 
in Section~\ref{SEC-Chain} that this is the case for a large
class of linear chains.} $\kappa_c=\kappa_0$ (recall Remark~\ref{REM-kappa0}).
We deduce from Proposition~\ref{PROP-RicX}~(12) that
$$
X_{1-\alpha}+\alpha\pi_Q\vartheta^{-1}\pi_Q
\ge\frac{1-\alpha}{\vartheta_{\rm max}}(I-\pi_Q)
+\frac{\Delta}{\vartheta_{\rm min}\vartheta_{\rm max}}
(\alpha-\tfrac12+\kappa_0)\pi_Q>0,
$$
for $\alpha\in[\tfrac12-\kappa_0,0]$. Thus, in this case we have
$\fI_\infty=[\tfrac12-\kappa_c,1[$ as in the Markovian case.
\eer

\paragraph{Steady state entropy production rate}
Motivated by~\cite{MNV}, where the functional $\Ep(\mu,t)$ plays a central 
role, we shall also investigate the large time asymptotics of its cumulant
generating function
$$
e_{{\rm ep},t}(\alpha)
=\frac1t\log\EE_\mu\left[\e^{-\alpha\Ep(\mu,t)}\right],
$$
in the stationary process. We observe that this function coincides
with a R\'enyi relative entropy, namely
$$
e_{{\rm ep},t}(\alpha)
=\Ent_\alpha(\widetilde{\PP}_{\mu\Theta}^t|\PP_\mu^t),
$$
so that the symmetry~\eqref{EQ-RenyiSymmetry} yields
$$
e_{{\rm ep},t}(1-\alpha)
=\Ent_\alpha(\PP_\mu^t|\widetilde{\PP}_{\mu\Theta}^t)
=\Ent_\alpha(\widetilde{\PP}_{\mu}^t|\PP_{\mu\Theta}^t)
=\frac1t\log\EE_{\mu\Theta}\left[\e^{-\alpha\Ep(\mu\Theta,t)}\right].
$$
The large time behavior of $e_{{\rm ep},t}(\alpha)$ follows from 
Proposition~\ref{PROP-GaussianCocycles} with the choice
$$
\widehat{N}=\theta X_1\theta,\qquad
F=\theta X_1\theta-X_1,\qquad
G=0.
$$
Thus,
$$
\fI_+=\{\alpha\in\bar\fI_c\,|\,X_{1-\alpha}+\alpha X_1>0\},\qquad
\fI_-=\{\alpha\in\bar\fI_c\,|\,X_{\alpha}+(1-\alpha)\theta X_1\theta>0\},
$$
and since we can write
$X_\alpha+(1-\alpha)\theta X_1\theta=\theta(Y_{1-\alpha}+W_{1-\alpha})\theta$
with $Y_{1-\alpha}=X_{1-\alpha}+\theta X_\alpha\theta$ and 
$W_{1-\alpha}=(1-\alpha)X_1-X_{1-\alpha}$, it follows from
Proposition~\ref{PROP-RicX}~(10) that
$$
\fI_\infty=\{\alpha\in\bar\fI_c\,|\,X_{1-\alpha}+\alpha X_1>0\}.
$$
In particular the limit
$$
e_{\rm ep}(\alpha)=\lim_{t\to\infty}e_{{\rm ep},t}(\alpha),
$$
coincides with $e(\alpha)$ for all $\alpha\in\rr$ iff
the following condition holds:

\begin{quote}
{\bf Condition (R)} $X_{1-\alpha}+\alpha X_1>0$ for all $\alpha\in\bar\fI_c$.
\end{quote}

This condition involves maximal solutions of two algebraic Riccati
equations. Except in some special cases (see Proposition~\ref{PROP-RicX}~(12)),
its validity is not ensured by general principles
(the known comparison theorems for Riccati equations do not apply)
and we shall leave it as an open question. We will come back to it in 
Section~\ref{SEC-Examples} in context of concrete examples.

\paragraph{Transient canonical entropic functional} Assuming for
simplicity that the covariance $N$ of the initial condition $\nu\in\cP(\Xi)$
is positive definite, Proposition~\ref{PROP-GaussianCocycles} applies to
the cumulant generating function of $S_\nu^t$ with
$$
\widehat{N}=N^{-1},\qquad
F=\theta G\theta=\theta N^{-1}\theta-X_1.
$$
It follows that
$$
\fI_\infty=\{\alpha\in\bar\fI_c\,|\,X_\alpha+(1-\alpha)N^{-1}>0
\mbox{ and }X_{1-\alpha}+\alpha N^{-1}>0\},
$$
so that $\alpha_-=1-\alpha_+=\tfrac12-\kappa_\nu$ for some
$\kappa_\nu>\tfrac12$ and
$$
e_\nu(\alpha)=\lim_{t\to\infty}\frac1t\EE_\nu\left[\e^{-\alpha S_\nu^t}\right]
=\left\{\begin{array}{ll}
e(\alpha)&\text{for }|\alpha-\frac12|<\kappa_\nu;\\[6pt]
+\infty&\text{for }|\alpha-\frac12|>\kappa_\nu.
\end{array}
\right.
$$
Note that by the construction of $S_\nu^t$ the Gallavotti--Cohen symmetry holds for  all times. One has
$\kappa_\nu=\kappa_c$ and hence $e_\nu(\alpha)=e(\alpha)$ for all
$\alpha\in\rr$, provided
$$
(\kappa_c-\tfrac12)X_{\frac12+\kappa_c}^{-1}<N
<-(\kappa_c+\tfrac12)X_{\frac12-\kappa_c}^{-1}.
$$

\subsection{Extended fluctuation relations} 
\label{SEC-ExtendedFT}

We finally deal with the 4$^{\rm th}$ and last step of our scheme: we derive 
an LDP for the the entropic functionals considered in the previous section
and illustrate its use in obtaining extended fluctuation relations for
various physical quantities of interest. We start with a complement to
the discussion of Section~\ref{SEC-LDP}.

In most cases relevant to entropic functionals of harmonic networks,  the 
generating function~$\Lambda$ is real analytic and strictly convex on a 
finite interval $D_0=]\alpha_-,\alpha_+[$, is infinite on 
$\rr\setminus[\alpha_-,\alpha_+]$, and  the interval $E=]x_-,x_+[$ is finite. 
In such cases $\Lambda_\pm$ are both finite and~\eqref{EQ-supporting} implies 
that the Legendre transform of $\Lambda$ is given by
$$
\Lambda^\ast(x)=\sup_{\alpha\in\rr}(\alpha x-\Lambda(\alpha))=\left\{
\begin{array}{ll}
x\alpha_--\Lambda_-&\mbox{for }x\le x_-;\\[10pt]
x\ell(x)-\Lambda(\ell(x))
&\mbox{for }x\in ]x_-,x_+[;\\[10pt]
x\alpha_+-\Lambda_+&\mbox{for }x\ge x_+;
\end{array}
\right.
$$
where $\ell:E\to D_0$ is the reciprocal function to $\Lambda'$. Thus, 
$\Lambda^\ast$ is real analytic on $E$, affine on $\rr\setminus E$ and
$C^1$ on $\rr$. The G\"artner-Ellis theorem only provides a local LDP on 
$E$ for which the affine branches of $\Lambda^\ast$ are irrelevant. However, 
exploiting the Gaussian nature of the underlying measure $\PP$, it is 
sometimes possible to extend this local LDP to a global one, with the rate 
function $\Lambda^\ast$. Inspired by the earlier work of Bryc and 
Dembo~\cite{BD}, we have recently obtained such an extension
for entropic functionals of a large class of Gaussian dynamical 
systems~\cite{JPS}. The next result is an adaptation of the
arguments in~\cite{BD,JPS} and applies to the functional
$$
\xi_t=S^t+\Phi(x(t))-\Psi(x(0)),
$$
under the law $\PP_\nu$, with the hypothesis and notations of 
Proposition~\ref{PROP-GaussianCocycles}. We set (recall~\eqref{EQ-spmDef})
$$
\eta_-=\left\{\begin{array}{ll}
-\infty&\mbox{if }\alpha_+=\frac12+\kappa_c;\\[4pt]
-e'(\alpha_+)&\mbox{if }\alpha_+<\frac12+\kappa_c;
\end{array}
\right.\qquad
\eta_+=\left\{\begin{array}{ll}
+\infty&\mbox{if }\alpha_-=\frac12-\kappa_c;\\[4pt]
-e'(\alpha_-)&\mbox{if }\alpha_->\frac12-\kappa_c.
\end{array}
\right.
$$
\bet\label{THM-GaussianLDP}
\ben
\item If Assumption~(C) holds then, under the law $\PP_\nu$, the family 
$(\xi _t)_{t\ge0}$ satisfies a global LDP  with 
the rate function
\beq
J(s)=\left\{
\begin{array}{ll}
I(\eta_-)-(s-\eta_-)\alpha_+=
-s\alpha_+-e(\alpha_+)&\mbox{for }s\le\eta_-;\\[4pt]
I(s)&\mbox{for }s\in]\eta_-,\eta_+[;\\[4pt]
I(\eta_+)-(s-\eta_+)\alpha_-=
-s\alpha_--e(\alpha_-)&\mbox{for }s\ge\eta_+;
\end{array}
\right.
\label{EQ-Jdef}
\eeq
where $I(s)$ is given by~\eqref{EQ-TheRate}. In particular, if $\ep>0$, then
it follows from the strict convexity of $I(s)$ that
$$
J(-s)-J(s)<I(-s)-I(s)=s,
$$
for $s>\max(-\eta_-,\eta_+)$.
\item Under the same assumptions, the family $(\xi_t)_{t\ge0}$ satisfies the 
Central Limit Theorem: For any Borel set $\cE\subset\rr$,
$$
\lim_{t\to\infty}\PP_\nu\left[
\frac{\xi_t-\EE_\nu[\xi_t]}{\sqrt{ta}}\in\cE\right]={\mathrm n}_1(\cE),
$$
where $a=e''(0)$ and $\mathrm{n}_1$ denotes the centered Gaussian measure on $\rr$
with variance $1$.
\een
\eet

\begin{figure}
\centering
\includegraphics[scale=0.4]{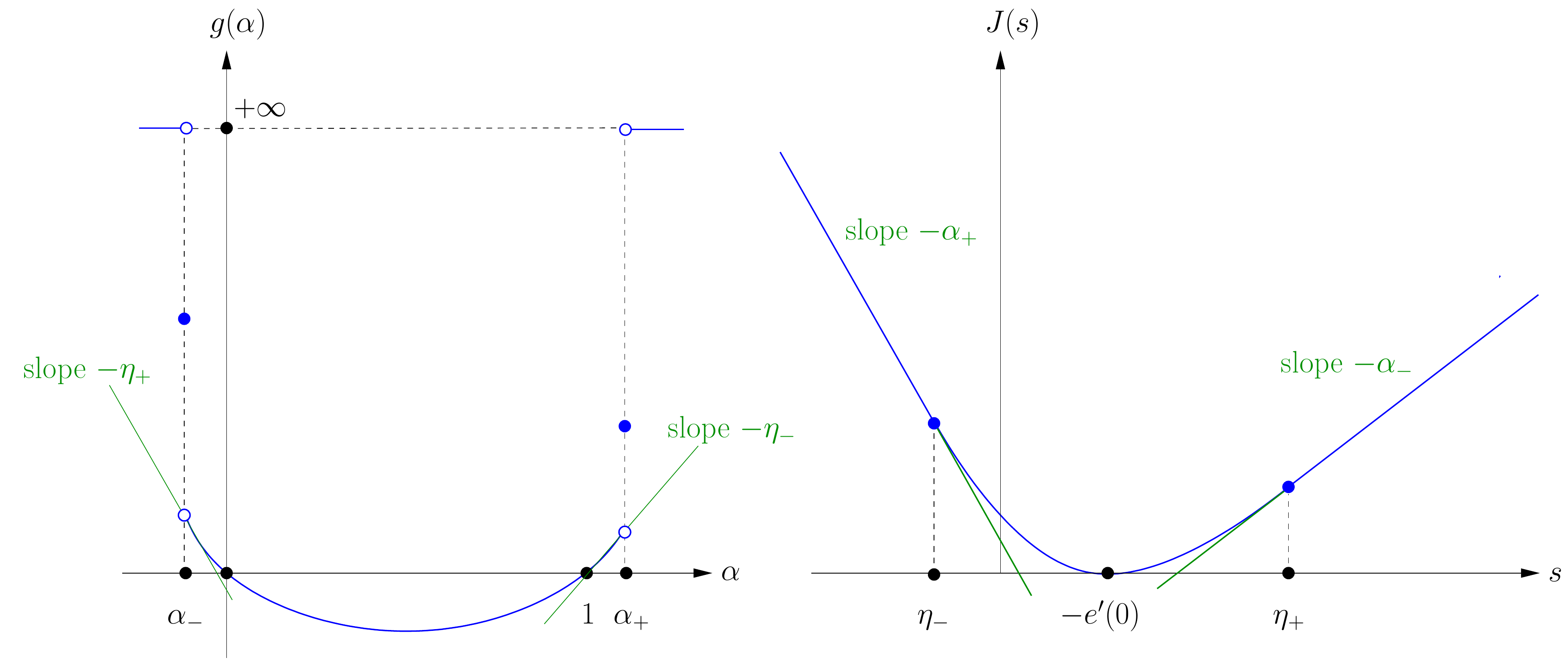}
\caption{The cumulant generating function 
$g(\alpha)=\limsup_{t\to\infty}g_t(\alpha)$ and the rate
function $J(s)$ for the functionals $(\xi_t)_{t\ge0}$ of
Theorem~\ref{THM-GaussianLDP}.}
\label{Fig3}
\end{figure}

If $\fI_\infty=\bar\fI_c$, then we are in the same situation as in 
Section~\ref{SEC-LDP} and $\xi_t$ has the  same large fluctuations as the 
canonical entropic functional $S^t$. In particular it also satisfies the 
Gallavotti--Cohen fluctuation theorem. However, in the more likely event that 
$\fI_\infty$ is strictly smaller than $\bar\fI_c$, then (see Figure~\ref{Fig3}) the function
$g(\alpha)=\limsup_{t\to\infty}g_t(\alpha)$
only coincides with $e(\alpha)$ on $]\alpha_-,\alpha_+[$ and the rate function 
$J(s)$ differs from $I(s)$ outside the closure of the interval 
$]\eta_-,\eta_+[$. Unless $\alpha_-=1-\alpha_+$ (in which case
$\eta_-=-\eta_+$ and $J(-s)-J(s)=s$ for all $s\in\rr$)
the Gallavoti-Cohen symmetry is broken and the universal fluctuation 
relation~\eqref{EQ-UFT} fails. The symmetry function 
$\frak{s}(s)=J(-s)-J(s)$ then satisfies an {\sl ``extended fluctuation relation''}.

\bigskip
Combining Theorem~\ref{THM-GaussianLDP} with the results of 
Section~\ref{SEC-MoreEF} we obtain global LDPs for steady state and
transient dissipated TDE. Let us discuss their features in more detail.

\paragraph{Steady state dissipated TDE}

Assuming $\ep>0$, we have $-1<\alpha_{\mathrm{TDE,st}-}<0$ and
$\alpha_{\mathrm{TDE,st}+}=1$, hence
$\eta_{\mathrm{TDE,st}-}=-e'(1)=-\ep$ and 
$\eta_{\mathrm{TDE,st}+}=-e'(\alpha_{\mathrm{TDE,st}-})>\ep$. In this case,
the symmetry function is
$$
\frak{s}_{\rm TDE,st}(s)=\left\{\begin{array}{ll}
s&\mbox{for }0\le s\le\ep;\\[4pt]
s-I(s)&\mbox{for }\ep\le s\le\eta_{\mathrm{TDE,st}+};\\[4pt]
e(\alpha_{\mathrm{TDE,st}-})+(1+\alpha_{\mathrm{TDE,st}-})s&
\mbox{for } s\ge\eta_{\mathrm{TDE,st}+};
\end{array}
\right.
$$
and in particular $\frak{s}_{\rm TDE,st}(s)<s$ for $s>\ep$. The slope of the 
affine branch of $\frak{s}_{\rm TDE,st}$ satisfies
$$
\frak{s}_{\rm TDE,st}'(s)=1+\alpha_{\mathrm{TDE,st}-}
\in]0,1[,\quad(s\ge\eta_{\mathrm{TDE,st}+}),
$$
so that $s\mapsto\frak{s}_{\rm TDE,st}(s)$ is strictly increasing.

In the equilibrium case ($\vartheta_{\rm min}=\vartheta_{\rm max}$) one has
$\alpha_{\mathrm{TDE,st}\mp}=\mp1$ and $e(\alpha)$ vanishes identically. Hence the rate function for steady state dissipated TDE is the universal function
$$
J_{\rm TDE,st}(s)=|s|,
$$
and $\frak{s}_{\rm TDE,st}(s)=0$ for all $s\in\rr$.

\paragraph{Transient dissipated TDE}

Assuming again $\ep>0$, we have $\alpha_{\mathrm{TDE,tr}-}=\frac12-\kappa_c$ 
and $\alpha_{\mathrm{TDE,tr}+}=1$,
so that $\eta_{\mathrm{TDE,tr}-}=-e'(1)=-\ep$ and
$\eta_{\mathrm{TDE,tr}+}=-e'(\tfrac12-\kappa_c)=+\infty$.
The symmetry function reads
$$
\frak{s}_{\rm TDE,tr}(s)=\left\{\begin{array}{ll}
s&\mbox{for }0\le s\le\ep;\\[4pt]
s-I(s)&\mbox{for }s\ge\ep;
\end{array}
\right.
$$
which coincides with the steady state heat dissipation for 
$0\le s\le\eta_{\mathrm{TDE,st}+}$.
However, the strict concavity of the function $s-I(s)$ implies
$$
\frak{s}_{\rm TDE,tr}(s)<\frak{s}_{\rm TDE,st}(s)
$$
for all $s>\eta_{\mathrm{TDE,st}+}$. By Remark~\ref{REM-ellSymmetry},
$$
\frac{\d\ }{\d s}(s-I(s))=1-\ell(s)=0
$$
iff $s=-e'(-1)>-e'(0)=\ep$. Thus, whenever $\tfrac12-\kappa_c<-1$\footnote{This corresponds to the near equilibrium 
regime.} the function $[0,\infty[\ni s\mapsto\frak{s}_{\rm TDE,tr}(s)$ has a 
unique maximum at $s=-e'(-1)$, and the concavity of $s-I(s)$ implies that
$\frak{s}_{\rm TDE,tr}$ becomes negative for large enough $s$.
In the opposite case where $\tfrac12-\kappa_c>-1$ the symmetry function
$\frak{s}_{\rm TDE,tr}$ is strictly monotone increasing (see Figure~\ref{Fig6} 
in Section~\ref{SEC_Triangle} for an explicit example of this 
somewhat surprising fact.)

\section{Examples}
\label{SEC-Examples}

In this section we turn back to harmonic networks in the setup of 
Section~\ref{SEC-Model}. We denote by $\{\delta_i\}_{i\in\cI}$ the canonical 
basis of the configuration space $\rr^\cI$. 

We start with two general  facts which reduce the phase space 
controllability condition (C) and the non-vanishing of $\ep$
to configuration space controllability (see 
Section~\ref{SSECT_Proof_of_LEM-NetControl} for a proof).

\bel\label{LEM-NetControl}
\ben
\item If $\Ker\omega=\{0\}$, then $(A,Q)$ is controllable iff 
$(\omega^\ast\omega,\iota)$ is controllable.
\item Denote by $\pi_i$, $i\in\partial\cI$, the orthogonal projection on 
$\Ker(\vartheta-\vartheta_i)$. Let $\cC_i=\cC(\omega^\ast\omega,\iota\pi_i)$.
If there exist $i,j\in\partial_\cI$ such that $\vartheta_i\not=\vartheta_j$ 
and $\cC_i\cap\cC_j\not=\{0\}$, then $\mathrm{ep}(\mu)>0$.
\een
\eel
\begin{figure}
\centering
\includegraphics[scale=0.5]{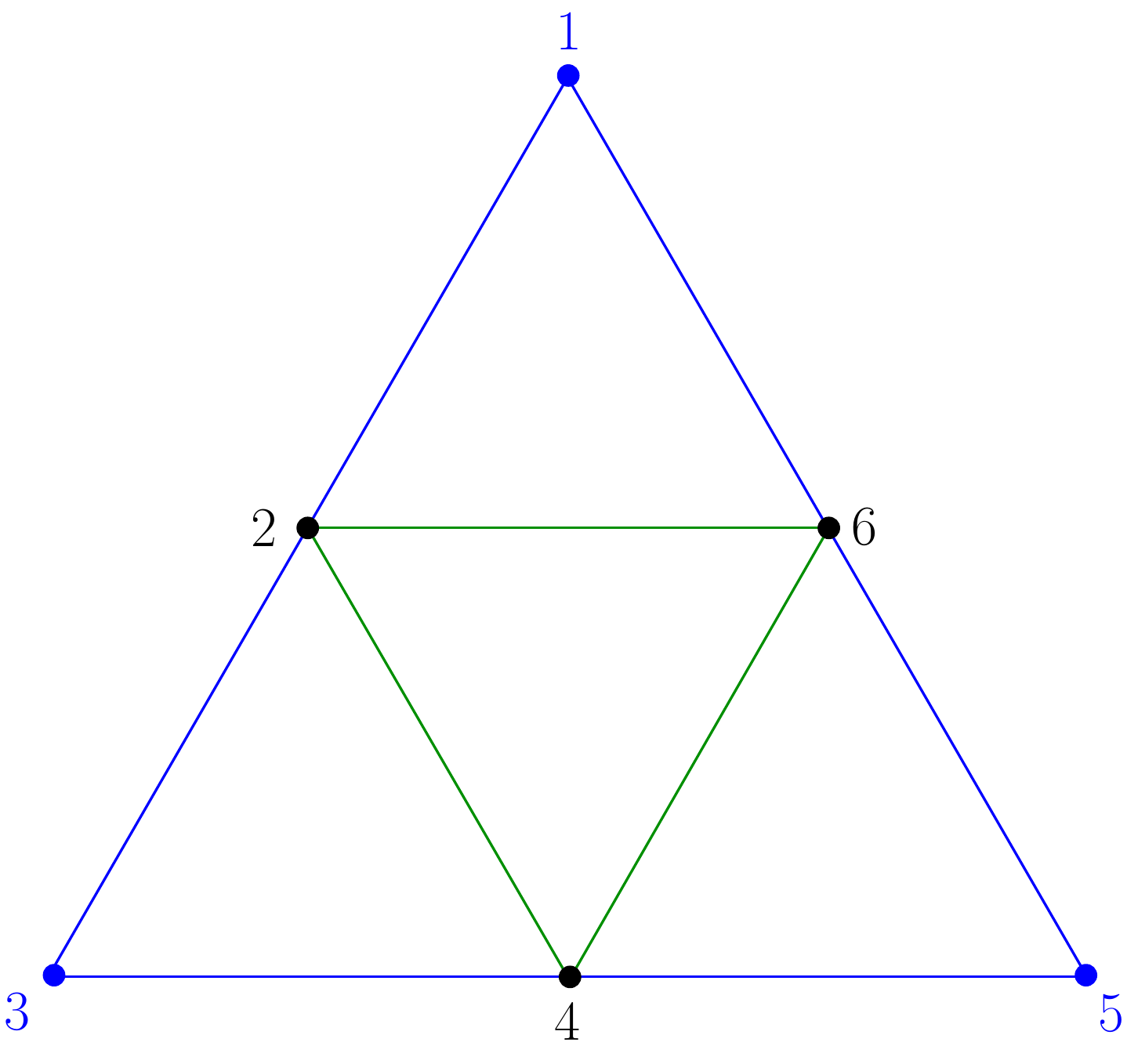}\hskip 0.4cm
\includegraphics[scale=0.55]{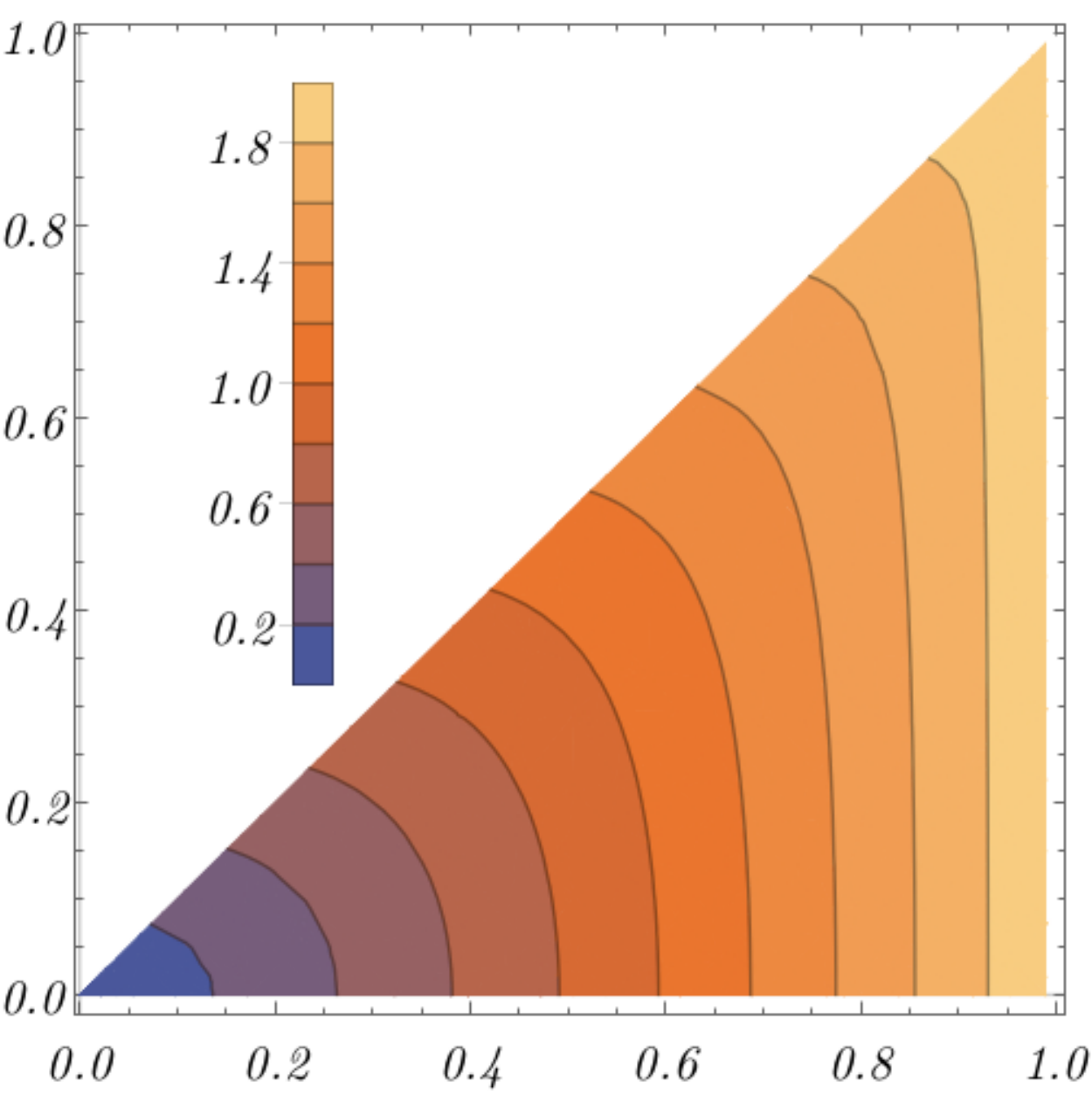}
\caption{A triangular network and a contour plot of $1/\kappa_c$ as function
of the parameters $(u,v)$. See the text for details.}
\label{Fig4}
\end{figure}

\subsection{A triangular network}
\label{SEC_Triangle} 

Consider the triangular network of 
Figure~\ref{Fig4} where $\cI=\zz_6$ and $\partial\cI=\zz_6\setminus2\zz_6$ 
(the  indices arithmetic is modulo 6). The potential
$$
\tfrac12 q\cdot\omega^2q=\tfrac12\sum_{i\in\cI}q_i^2+a\sum_{i\in\cI}q_iq_{i+1}
+b\sum_{i\in\partial\cI}q_{i}q_{i+2},
$$
is positive definite provided $|a|<\12$ and $2a^2-\12<b<1-4a^2$.
One easily checks that $a\not=0$ implies
$\Ran\iota\vee\Ran\omega^2\iota=\rr^\cI$.
Thus Assumption (C) is verified under these conditions.
Noting that $\delta_2\in\cC_1\cap\cC_3$, we conclude that $\ep>0$
if $\vartheta_1\not=\vartheta_3$. By symmetry, $\ep>0$ iff
$$
\Delta=\vartheta_{\rm max}-\vartheta_{\rm min}>0.
$$
We shall fix the parameters of the model to the following values
$$
a=\frac1{2\sqrt2},\qquad b=\frac14,\qquad\gamma_1=\gamma_3=\gamma_5=1,
\qquad\underline{\vartheta}=\frac1{|\partial\cI|}\sum_{i\in\partial\cI}\vartheta_i,
$$
the ``relative temperatures'' being parametrized by
$$
\vartheta_1=\underline{\vartheta}(1-u),\quad
\vartheta_3=\underline{\vartheta}(1+\tfrac12(u+3v)),\quad
\vartheta_5=\underline{\vartheta}(1+\tfrac12(u-3v)).
$$
Under these constraints, the simplex $\{(u,v)\,|\,0\le u\le 1,0\le v\le u\}$ 
is a fundamental domain for the action of the symmetry group $S_3$ of the 
network which corresponds to $\vartheta_{\rm min}=\vartheta_1$,
$\vartheta_{\rm max}=\vartheta_3$.
Factoring $\vartheta=\underline{\vartheta}\hat{\vartheta}$, one easily deduces
from~\eqref{EQ-EomegaDef} that the matrix $E(\omega)$ and hence the cumulant
generating function $e(\alpha)$ do not depend on $\underline{\vartheta}$. We have 
performed our numerical calculations with $\underline{\vartheta}=1$. The thermodynamic drive of 
the system is the ratio $\varrho=\Delta/\underline{\vartheta}=\frac32(u+v)\in[0,3]$. 

Figure~\ref{Fig4} shows the reciprocal of $\kappa_c$ as a function of
$(u,v)$. It was obtained by numerical calculation of the eigenvalues of
the  Hamiltonian matrix $K_\alpha$. The lower-left and upper-right corners
of the plot correspond to $\varrho=0$ and $\varrho=3$ respectively.
Its right edge is the singular limit $\vartheta_{\rm min}=0$.
Our results are compatible with the two limiting behaviors
$$
\lim_{\varrho\downarrow0}\kappa_c=\infty,\qquad
\lim_{\vartheta_{\rm min}\downarrow0}\kappa_c=\tfrac12.
$$
The first limit, which corresponds to thermal equilibrium 
$\vartheta_{\rm min}=\vartheta_{\rm max}=\underline{\vartheta}$,
follows from the lower bound~\eqref{EQ-kappaclow}.
\begin{figure}
\centering
\includegraphics[scale=0.6]{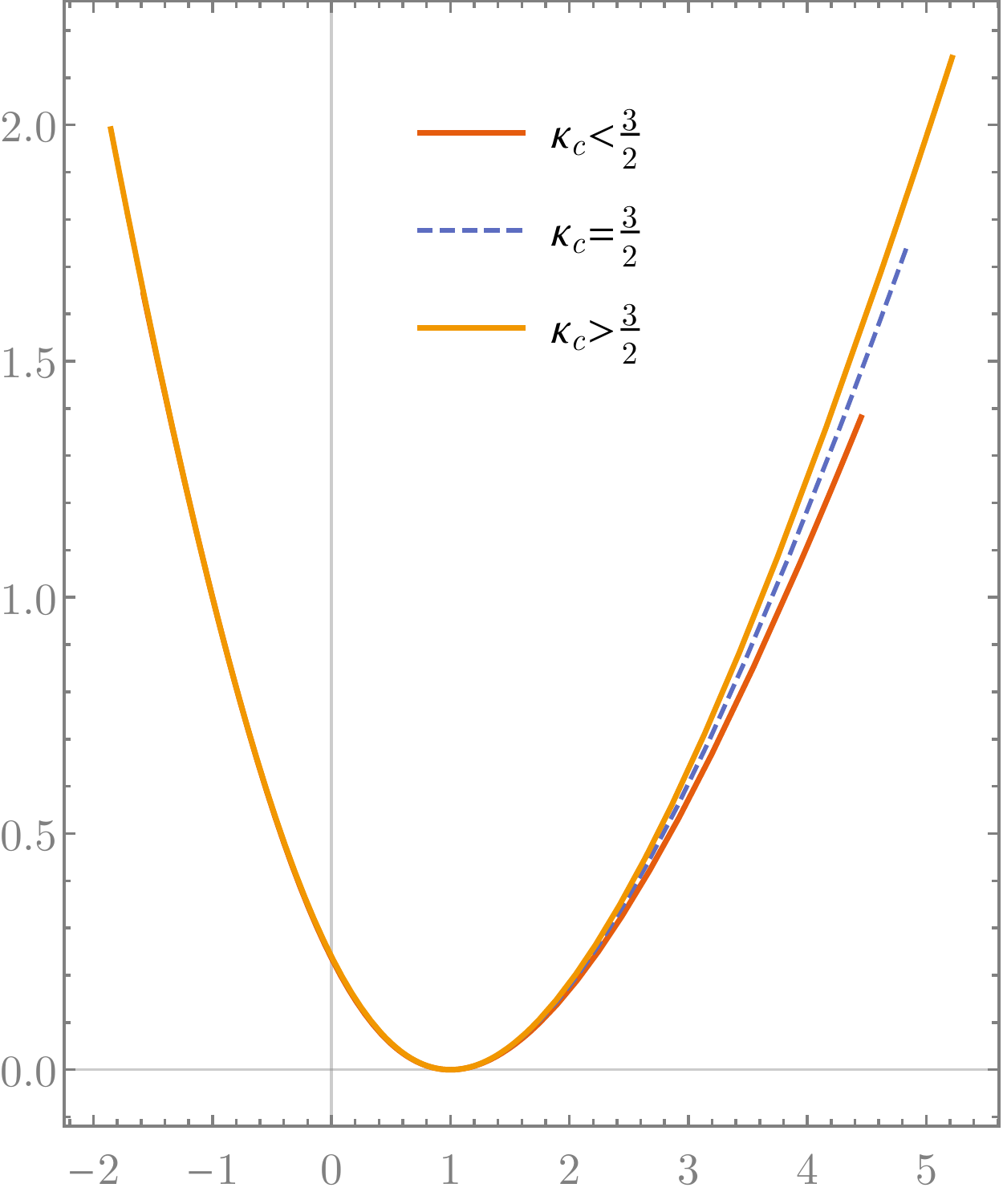}\hskip 0.2cm
\includegraphics[scale=0.6]{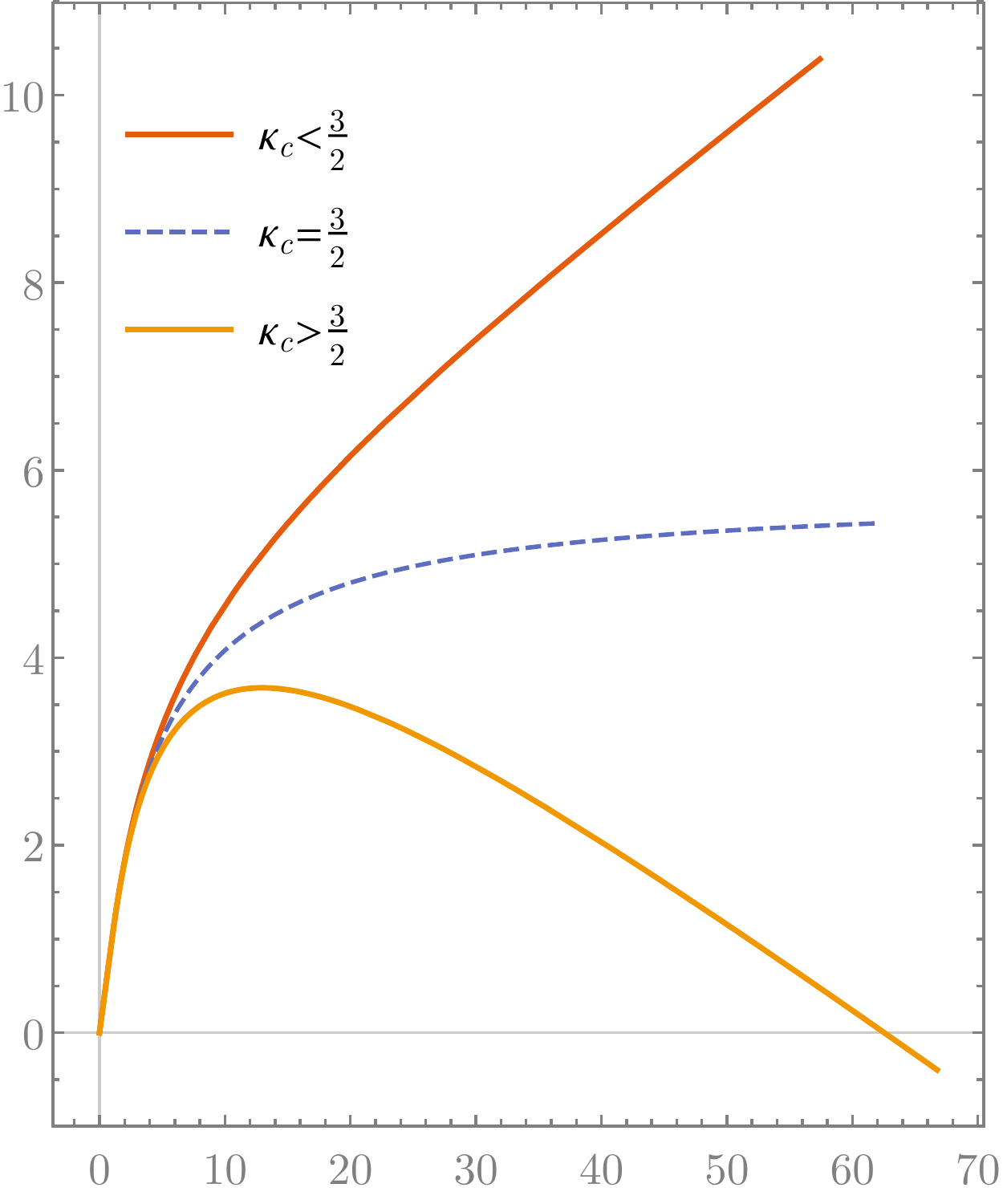}
\caption{The numerically computed rate function $J_{\rm TDE,tr}(s)$ 
and the corresponding symmetry function $\mathfrak{s}_{\rm TDE,tr}(s)$ 
for the transient TDE dissipation of the triangular network
(both the argument $s$ and the value of these functions are in the units of the
corresponding steady state entropy production rate $\ep$).
}
\label{Fig6}
\end{figure}
Computing the generating function $e(\alpha)$ from 
Eq.~\eqref{EQ-ealphaSpecForm}, and its Legendre transform, we have
obtained the symmetry function $\frak{s}_{\rm TDE,tr}(s)$ for transient
TDE dissipation at three points
on the line $v=0.3(1-u)$ where $\kappa_c=1.4, 1.5$ and $1.6$ respectively.
The result, displayed in Figure~\ref{Fig6} confirm our discussion
in Section~\ref{SEC-ExtendedFT}.

\begin{figure}
\centering
\includegraphics[scale=0.5]{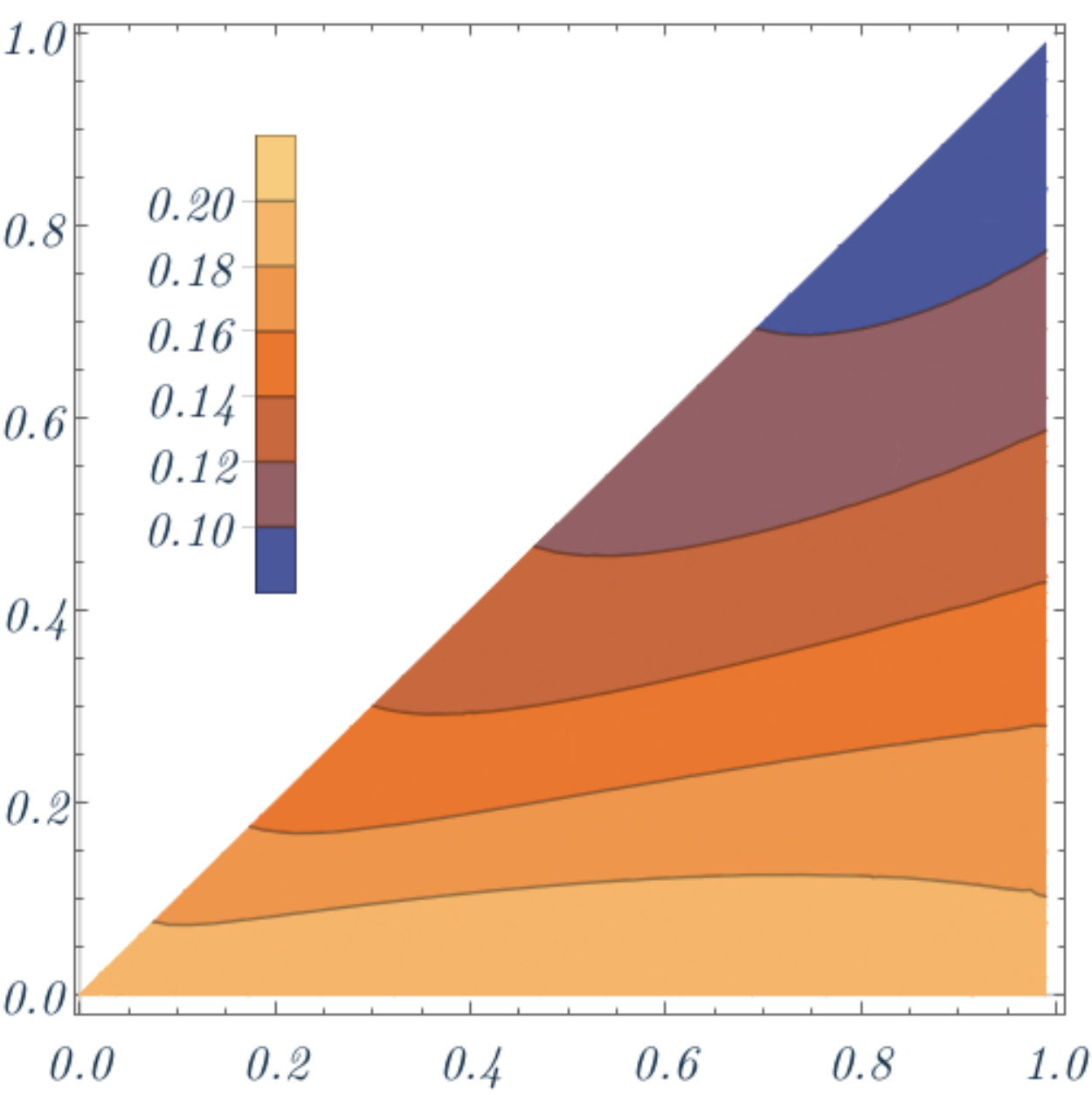}\hskip 0.2cm
\includegraphics[scale=0.45]{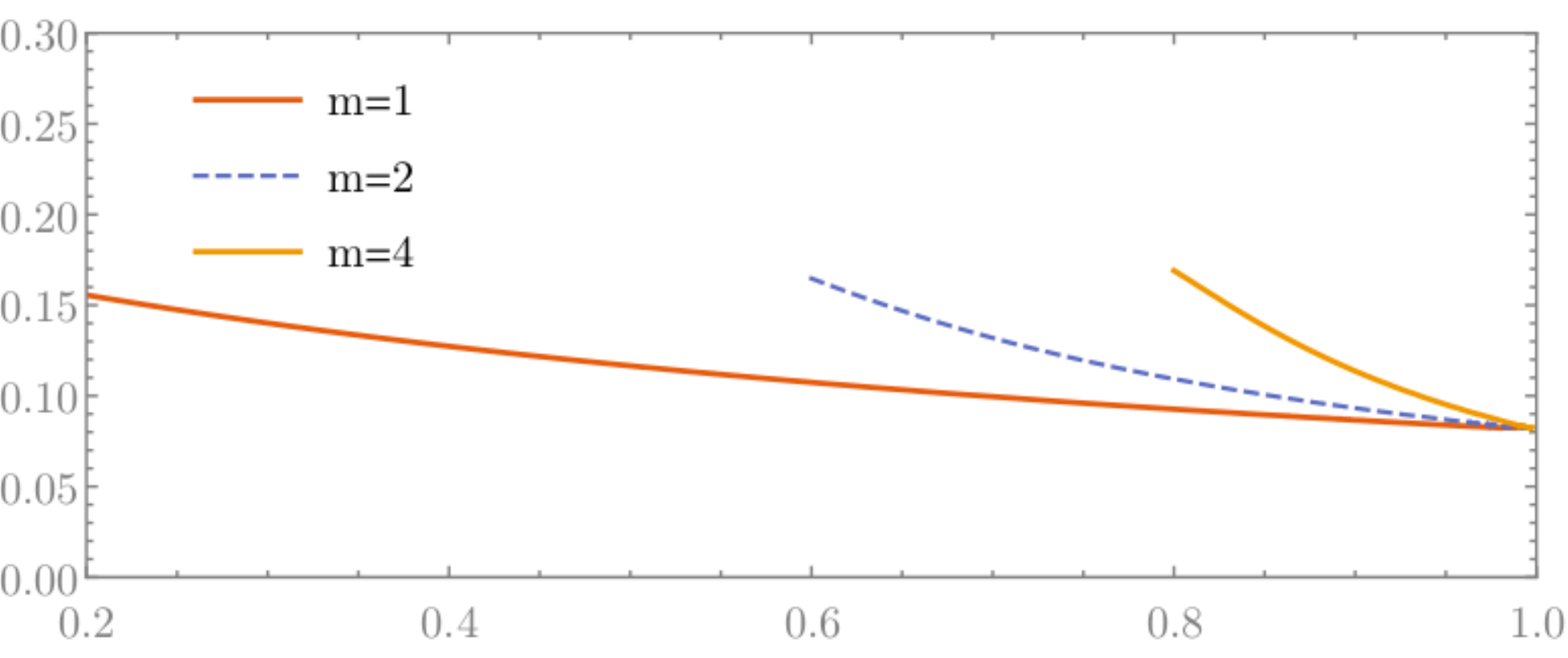}
\caption{Contour plot of $\min\sp(X_{1-\alpha}+\alpha X_1)$ as function
of $(u,v)$ and some sections along the lines $v=1+m(u-1)$
for the triangular network.
}
\label{FigCondR}
\end{figure}

Solving the Riccati equation~\eqref{EQ-RicX} one  can investigate the 
validity of Condition~(R). Figure~\ref{FigCondR} shows a plot of
$\min\sp(X_{1-\alpha}+\alpha X_1)$  as function of $(u,v)$ and a few sections
along the lines $v=1+m(u-1)$. It appears that Condition~(R) is clearly
satisfied for all temperatures.

\subsection{Jacobi chains}
\label{SEC-Chain}

\begin{figure}
\centering
\includegraphics[scale=0.6]{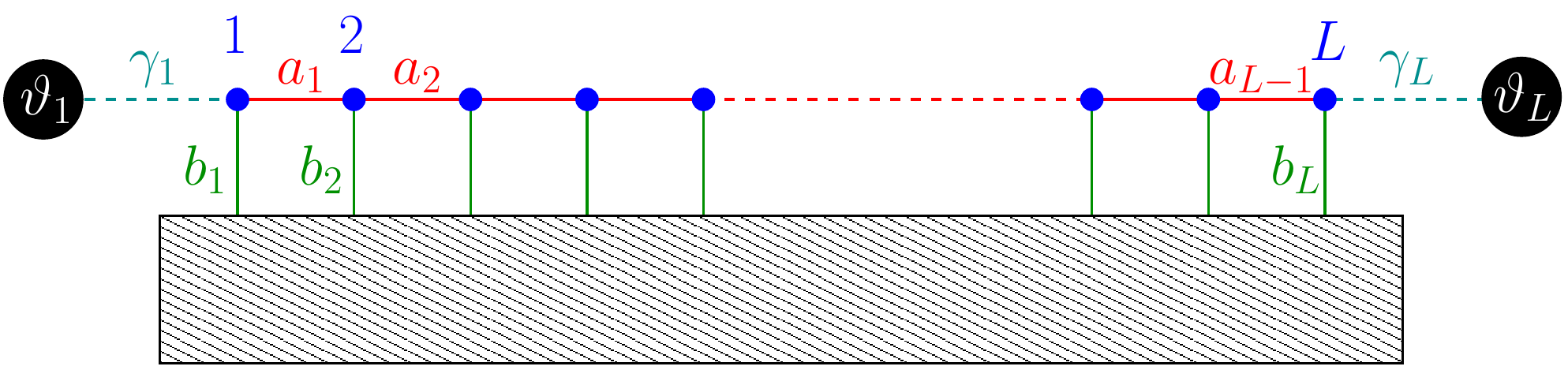}
\caption{A linear chain coupled to two heat baths.}
\label{Fig7}
\end{figure}

In our framework, a chain of $L$ oscillators with nearest neighbour 
interactions coupled to heat baths at its two ends (see Figure~\ref{Fig7}) is 
described by $\cI=\{1,\ldots,L\}$, $\partial\cI=\{1,L\}$, and the potential 
energy
\beq
\tfrac12|\omega q|^2=\tfrac12\sum_{i=1}^Lb_iq_i^2
+\sum_{i=1}^{L-1}a_i q_i q_{i+1},
\label{EQ-JacobiPotential}
\eeq
where, without loss of generality, we may assume $\omega$ to be self-adjoint.
We parametrize the temperature and relaxation rates of the baths by
$$
\bar\vartheta=\tfrac12(\vartheta_1+\vartheta_L),\qquad
\Delta=|\vartheta_L-\vartheta_1|,\qquad
\bar\gamma=\sqrt{\gamma_1\gamma_L},\qquad
\delta=\log\frac{\gamma_1}{\gamma_L},\qquad
\kappa_0=\frac{\bar\vartheta}{\Delta},
$$
and introduce the parity operator
$$
\begin{array}{lccc}
\cS:&\rr^\cI&\to&\rr^\cI\\
&(q_i)_{i\in\cI}&\mapsto&(q_{L+1-i})_{i\in\cI}.
\end{array}
$$
To formulate our main result  (see 
Section~\ref{SSECT_Proof_of_THM-Jacobi} for its proof) we state
\begin{quote}
{\bf Assumption (J)\,} $\omega>0$ and $\hat a=a_1a_2\cdots a_{L-1}\not=0.$
\end{quote}

\begin{quote}
{\bf Assumption (S)\,} The chain is symmetric, i.e., $[\cS,\omega^2]=0$
and $\delta=0$.
\end{quote}

\bet\label{THM-Jacobi}
Under Assumption (J), the following hold for the harmonic chain with 
potential~\eqref{EQ-JacobiPotential}:
\ben
\item Assumption (C) is satisfied.
\item If $\Delta\not=0$, then the covariance of the steady state $\mu$
satisfies
$$
\vartheta_{\mathrm{min}}<M<\vartheta_{\mathrm{max}},
$$
and $\ep>0$.
\item If Assumption (S) also holds, then 
$\ds\kappa_c=\kappa_0$ and Condition (R) is satisfied.
\een
\eet

\begin{figure}
\centering
\includegraphics[scale=0.7]{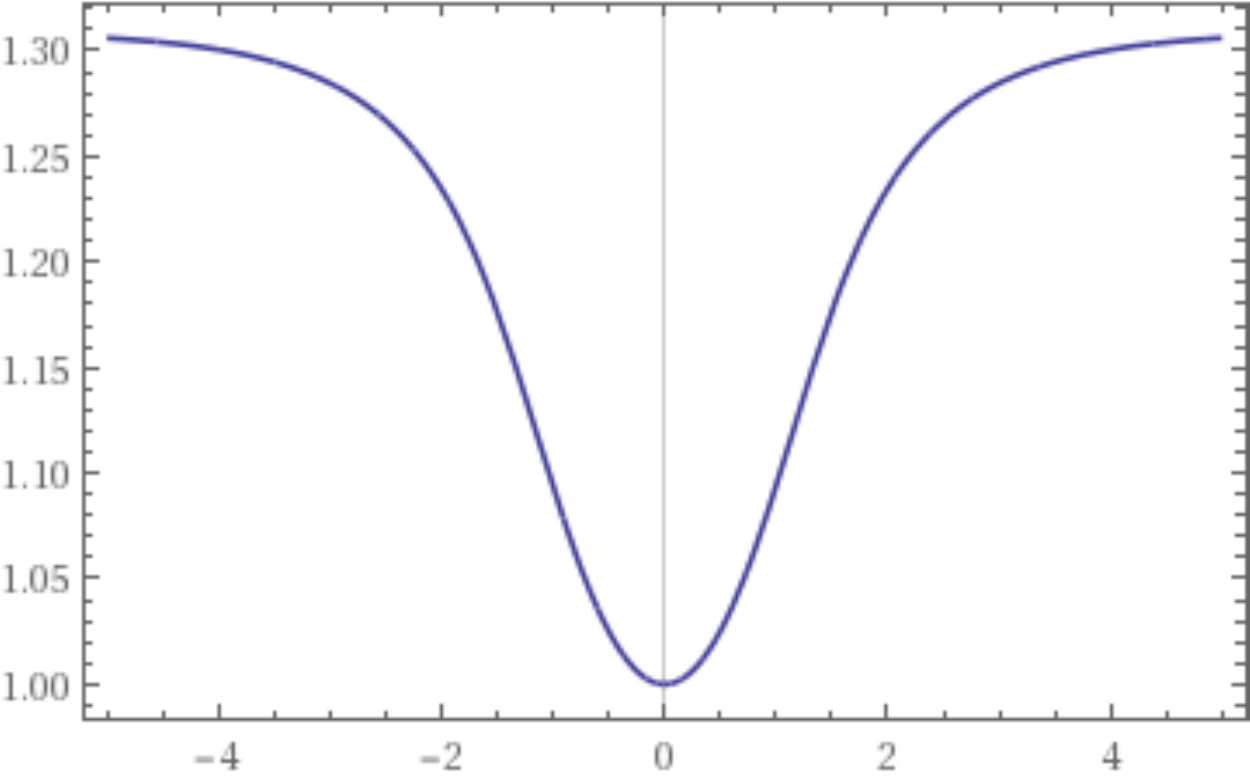}
\caption{The critical value $\kappa_c/\kappa_0$ as a function of $\delta$ for 
an homogeneous chain.}
\label{Fig8}
\end{figure}

\ber
For a class of symmetric quasi-Markovian anharmonic 
chains, Rey-Bellet and Thomas have obtained in~\cite{RT3} a local LDP for 
various entropic functionals of the form $S^t+\Psi(x(t))-\Psi(x(0))$ under the 
law $\PP_{x_0}$, $x_0\in\Xi$. In view of their Hypothesis (H1) (more precisely, the
condition $k_2\ge k_1\ge2$), their results should apply in particular to 
harmonic chains satisfying Assumptions~(J) and~(S). They proved that the 
cumulant generating 
function of these functionals are finite and satisfy the Gallavotti--Cohen 
symmetry on the interval $]\12-\kappa_0,\12+\kappa_0[$. The lower bound
of this interval is consistent with  Part~(4) of 
Theorem~\ref{THM-Jacobi} and Remark~\ref{REM-qMTDEtr}, whereas the upper bound is different from our conclusions in Section~\ref{SEC-ExtendedFT} on the transient TDE.  
There, we found that the cumulant  generating function diverges for $\alpha>1$. In view of this, it appears  that   the analysis of~\cite{RT3} does not apply 
to the harmonic case.
\eer
\ber We believe that Condition (S) is essential for Part~(4) since the proof  indicates that for
non-symmetric chains $\kappa_c>\kappa_0$ is generic. Figure~\ref{Fig8} shows
a plot of $\kappa_c$ vs $\delta$ for a homogeneous chain with $L=4$, $b_i=1$, 
$a_i=\12$, $\bar\gamma=2$, $\bar\vartheta=4$ and $\Delta=2$.
\eer

\section{Proofs}
\label{SEC-Proofs}

Even though the processes induced by Eq.~\eqref{EQ-The_SDE} take values in a 
real vector space, it will  be sometimes  more convenient to work with complex 
vector spaces. With this in mind, we start with some general remarks and notational conventions 
concerning complexifications.

Let $E$ be a real Hilbert space with inner product $\langle\,\cdot\,,\,\cdot\,\rangle$. We denote by 
$\cc E=\{x+\i y\,|\,x,y\in E\}$ the 
complexification of $E$. This complex vector space inherits a natural 
Hilbertian structure with inner product
$$
(x+\i y,u+\i v)
=\langle x,u\rangle+\langle y,v\rangle
+\i\langle x,v\rangle-\i\langle y,u\rangle.
$$
We denote by $|\cdot|$ the induced norm. Any $A\in L(E,F)$ 
extends to an element of $L(\cc E,\cc F)$ which we denote by the 
same symbol: $A(x+\i y)=Ax+\i Ay$. If $A$ is a 
self-adjoint/non-negative/positive 
element of $L(E)$, then this extension is a self-adjoint/non-negative/positive
element of $L(\cc E)$. The conjugation $\cC_E:x+\i y\mapsto x-\i y$ is
a norm-preserving involution of~$\cc E$. For $z\in\cc E$ and 
$A\in L(\cc F,\cc E)$ we set $\bar z=\cC_E z$ and $\bar A=\cC_E A\cC_F$.
We identify $E$ with the set $\{z\in\cc E\,|\,\bar z=z\}$ of real elements of $\cc E$. Likewise, $L(F,E)$ is identified with the set 
$\{A\in L(\cc F,\cc E)\,|\,\bar A=A\}$ of real elements of $L(\cc F,\cc E)$.
A subspace $V\subset\cc E$ is real if it is invariant under $\cC_E$.
$V$ is real iff there exists a subspace $V_0\subset E$ such that $V=\cc V_0$.
If $A\in L(\cc F,\cc E)$ is real, then $\Ran A$ and $\Ker A$ are real 
subspaces of $\cc E$ and $\cc F$.
Finally, we note that if $(A,Q)\in L(E)\times L(F,E)$, then the 
controllability subspace of the corresponding pair in 
$L(\cc E)\times L(\cc F,\cc E)$ is the real subspace $\cc\cC(A,Q)\subset\cc E$.
In particular $(A,Q)$ is controllable as a pair of $\rr$-linear maps iff it is 
controllable as a pair of $\cc$-linear maps.

\bigskip
Note that
\begin{equation} \label{process-xi}
\xi(t)=\int_0^t\e^{(t-s)A}Q\d w(s)
\end{equation}
is a centered Gaussian random variable with covariance
\beq
M_t=\int_0^t\e^{sA}QQ^\ast\e^{sA^\ast}\d s.
\label{EQ-MtDef}
\eeq
The next lemma  concerns some elementary properties of this operator.

\bel\label{LEM-Control_M}
Assume that 
$(A,Q,\vartheta,\theta)\in L(\Xi)\times L(\partial\Xi,\Xi)\times L(\partial\Xi)\times L(\Xi)$
satisfies the structural relations~\eqref{EQ-Structure} and let  $M_t$ be given by 
Eq.~\eqref{EQ-MtDef}.
\ben
\item$\Ran M_t=\cC(A,Q)$ for all $t>0$.
\item The subspace $\cC(A,Q)$ is invariant for both~$A$ and~$A^*$, and $\sp(A|_{\cC(A,Q)}),\sp(A^*|_{\cC(A,Q)})\subset\cc_-$. In particular, there exist constants
$C\ge1$ and $\delta'\ge\delta>0$ such that 
$$
C^{-1}\e^{-\delta' t}|x|\le|\e^{tA}x|\le C\e^{-\delta t}|x|\quad\mbox{for $x\in \cC(A,Q)$}, 
$$ 
and the function $t\mapsto M_t$ converges to a limit~$M$ as $t\to+\infty$. 

\item 
$\Ran M=\cC(A,Q)=\cC(A^\ast,Q)$.
\item 
$A|_{\cC(A,Q)^\bot}=-A^\ast|_{\cC(A,Q)^\bot}$
and $\e^{tA}|_{\cC(A,Q)^\bot}$ is unitary.
\item 
The following inequality holds for all $t\ge0:$
\begin{equation} \label{Mt-estimate}
\vartheta_{\rm min}(I-\e^{tA}\e^{tA^\ast})\le M_t
\le\vartheta_{\rm max}(I-\e^{tA}\e^{tA^\ast})\le\vartheta_{\rm max}.
\end{equation}
In particular,
$$
\vartheta_{\rm min}\le M|_{\Ran M}\le\vartheta_{\rm max},
$$
and if all the reservoirs are at the same temperature $\vartheta_0$, 
then $M|_{\Ran M}=\vartheta_0$.
\item $M-M_t=\e^{tA}M\e^{tA^\ast}\ge0$ and $(M-M_t)|_{\Ran M}>0$.
\item $M$ satisfies the Lyapunov equation
\beq
AM+MA^\ast+QQ^\ast=0.
\label{EQ-LyapM}
\eeq
\item If $(A,Q)$ is controllable, then $\Ran M=\Xi$ and $M$ is the only 
solution of~\eqref{EQ-LyapM}. Moreover, for any $\tau>0$ there exists a 
constant $C_\tau$ such that
$$
0<M_t^{-1}-M^{-1}\le C_\tau\e^{-2\delta t}
\quad\mbox{for all $t\ge\tau$}.
$$
\een
\eel

\noindent{\bf Proof.} (1) Fix $t>0$. From the relation
$$
x\cdot M_tx=\int_0^t|Q^\ast\e^{sA^\ast}x|^2\d s
$$
we deduce that $\Ker M_t=\cap_{s\in[0,t]}\Ker Q^\ast\e^{sA^\ast}$. This
relation is easily seen to be equivalent to
\beq
\Ker M_t=\bigcap_{n\ge0}\Ker Q^\ast A^{\ast n},
\label{EQ-Ker_Mt}
\eeq
and hence to
\beq
\Ran M_t=\bigvee_{n\ge0}\Ran A^nQ.
\label{EQ-Ran_Mt}
\eeq
The right-hand side of the last relation is included in any $A$-invariant 
subspace containing $\Ran Q$, and therefore coincides with the controllability
subspace $\cC(A,Q)$.

(2) 
The invariance of the subspace~$\cC(A,Q)$ under~$A$ follows from the definition. To prove its invariance under~$A^*$, it suffices to recall the relation
\beq
A+A^\ast=-Q\vartheta^{-1}Q^\ast.
\label{EQ-ReA}
\eeq
We now prove that the spectra of the restrictions of~$A$ and~$A^*$ to~$\cC(A,Q)$ are subsets of~$\cc_-$. It suffices to consider the case of~$A$. 

Pick $\alpha\in\sp(A)$ and let $z\in\cc\Xi\setminus\{0\}$ be a 
corresponding eigenvector. It follows from~\eqref{EQ-ReA} that
$$
2\Re\alpha |z|^2=(z,(A+A^\ast)z)=-|\vartheta^{-1/2}Q^\ast z|^2,
$$
which implies $\Re\alpha\le0$. If $\Re\alpha=0$, then $Q^\ast z=0$ 
and~\eqref{EQ-ReA} yields $A^\ast z=-\alpha z$ which
further implies $Q^\ast A^{\ast n}z=(-\alpha)^nQ^\ast z=0$ for all
$n\ge0$. Eq.~\eqref{EQ-Ker_Mt} then gives $z\in\Ker M_t$  and so 
$\sp(A|_{\Ran M_t})\subset\cc_-$. The remaining statements are elementary
consequences of this fact and the observation that $M_t$ vanishes on~$\cC(A,Q)^\bot$.

(3) The proof of the relation $\Ran M=\cC(A,Q)$ is exactly the same as that of~(1). The relation $\cC(A,Q)=\cC(A^*,Q)$ is a simple consequence of~\eqref{EQ-ReA}. 

(4) Combining~\eqref{EQ-Ker_Mt} with~\eqref{EQ-ReA}, we
deduce $\Ker(A+A^\ast)=\Ker Q^\ast\supset \cC(A,Q)^\bot$. Thus $A$ and $-A^\ast$
coincide on $\cC(A,Q)^\bot$.

(5) From Eq.~\eqref{EQ-ReA} we deduce
$$
\int_0^t\e^{sA}Q\vartheta^{-1}Q^\ast\e^{sA^\ast}\d s
=-\int_0^t\frac{\d\ }{\d s}\e^{sA}\e^{sA^\ast}\d s=I-\e^{tA}\e^{tA^\ast}, 
$$
from which we infer
$$
\vartheta_{\rm max}^{-1}M_t\le I-\e^{tA}\e^{tA^\ast}
\le\vartheta_{\rm min}^{-1}M_t.
$$
This is equivalent to~\eqref{Mt-estimate}. 
Restricting these inequalities to $\cC(A,Q)$ and taking the limit
$t\to\infty$ yields the desired result.

(6) The first assertion follows directly from the definition of $M$ 
and the group property of $\e^{tA}$. The second assertion is a consequence
of Parts~(3) and~(5) which imply
$$
(M-M_t)|_{\Ran M}=\e^{tA}M\e^{tA^\ast}|_{\Ran M}
\ge \vartheta_{\mathrm{min}}\e^{tA}\e^{tA^\ast}|_{\Ran M}>0.
$$

(7) Follows from Part~(6) and Eq.~\eqref{EQ-MtDef} by differentiation.

(8) Any solution $N$
of~\eqref{EQ-LyapM} is easily seen to satisfy
$$
N-M_t=\e^{tA}N\e^{tA^\ast}\quad\mbox{for all $t\ge0$}.
$$
Letting $t\to+\infty$ and using the exponential decay of~$\e^{tA}$ and~$\e^{tA^*}$ (see~(2) in the case $\cC(A,Q)=\Xi$), we see that $N=M$. 
The second assertion follows from the 
identity 
$$M_t^{-1}-M^{-1}=M_t^{-}(M-M_t)M^{-1}
$$ 
and the inequalities $M_t\ge c_\tau>0$ for $t\ge\tau$ and $\|M_t-M\|\le Ce^{-2\delta t}$ for $t\ge0$. 
\hfill$\square$

\subsection{Sketch of the proof of Theorem~\ref{THM-InvMeas}}
\label{SSECT_Proof_of_THM_InvMeas}

\paragraph{(1)}
The fact that~$M$ is well defined and satisfies~\eqref{EQ-Mbounds} was established in Lemma~\ref{LEM-Control_M}. Let us prove the invariance of~$\mu$. 

We fix a random variable~$x_0$ that is independent of~$w$ and is distributed by the law~$\mu$. We wish to show that the law of the process
\begin{equation} \label{process-x}
x(t)=\e^{t A}x_0+\xi(t),
\end{equation}
where~$\xi$ is given by~\eqref{process-xi}, coincides with~$\mu$ for all $t\ge0$. To this end, we note that both terms in~\eqref{process-x} are centred Gaussian random variables with covariances~$\e^{tA}M\e^{tA^*}$ and~$M_t$, respectively. Since they are independent, $x(t)$ is also a centred Gaussian random variable with covariance $\e^{tA}M\e^{tA^*}+M_t$. This operator coincides with~$M$ in view of Lemma~\ref{LEM-Control_M}~(6). Hence, the law of~$x(t)$ coincides with~$\mu$. 

\paragraph{(2)}
If the pair~$(A,Q)$ is controllable, then for any initial condition~$x_0$ independent of~$w$ the corresponding solution~\eqref{process-x} converges in law to~$\mu$. It follows that~$\mu$ is the only invariant measure. On the other hand, if the pair~$(A,Q)$ is not controllable, then, by Lemma~\ref{LEM-Control_M}, the subspace $\Ker M=\cC(A,Q)^\bot\not=\{0\}$ is invariant for the group~$\{\e^{t A}\}$, whose restriction to it is a unitary. The latter has infinitely many invariant measures (e.g., the normalized Lebesgue measure on any sphere $\{x\in \cC(A,Q)^\bot\,|\,|x|=R\}$ is invariant). 

To prove the mixing property, we write
$$
P^tf(x)=\EE f\bigr(\e^{t A}x+\xi(t)\bigr)=\int_{\Xi} f(y){\rm n}_t(x,y)\,\d y,
$$
where ${\rm n}_t(x,y)$ denotes the density of the Gaussian measure with mean value~$\e^{t A}x$ and covariance~$M_t$:
$$
{\rm n}_t(x,y)=\det(2\pi M_t)^{-1/2}
\exp\Bigl\{-\frac12\,\bigl(y-\e^{tA}x,M_t^{-1}(y-\e^{tA}x\bigr)\Bigr\}.
$$
The required convergence follows now from assertions~(6) and~(8) of Lemma~\ref{LEM-Control_M} and the Lebesgue theorem on dominated convergence. 

\paragraph{(3)} 
The fact that process~\eqref{EQ-The_Process} is centred and Gaussian follows from linearity of the equation. Let us calculate its covariance operator~$K(t,s)$. It is a straightforward to check that a stationary solution of~\eqref{EQ-The_SDE} defined on  the 
whole real line can be written as 
$$
\hat \xi(t)=\int_{-\infty}^t\e^{(t-r)A}Q\d w(r),
$$
where $w(t)$ stands for a two-sided $\rr^{\partial\cI}$-valued Brownian motion. Assuming without loss of generality that $t>s$, for any $\eta_1,\eta_2\in\Xi$ we write
\begin{align*}
(\eta_1,K(t,s)\eta_2)&=\EE\bigl\{(\hat\xi(t),\eta_1\bigr)\bigl(\hat\xi(s),\eta_2\bigr)\bigr\}\\
&=\EE\biggl\{\int_{-\infty}^t\bigl(\e^{(t-r)A}Q\d w(r),\eta_1\bigr)\,
\int_{-\infty}^s\bigl(\e^{(s-r)A}Q\d w(r),\eta_2\bigr)\biggr\}\\
&=\int_{-\infty}^s\bigl(Q^*\e^{(t-r)A^*}\eta_1,Q^*\e^{(s-r)A^*}\eta_2\bigr)\d r\\
&=\int_0^{+\infty}\bigl(\eta_1,\e^{(t-s+u)A}QQ^*\e^{uA^*}\eta_2\bigr)\d u=(\eta_1,\e^{tA}M\eta_2). 
\end{align*}
This implies the required relation~\eqref{EQ-Mts} and completes the proof of Theorem~\ref{THM-InvMeas}. 
\hfill\qed

\bigskip
For later use, we now formulate and prove two other auxiliary results. 
We start with a few technical facts. Consider the scale of spaces\index{$\fH,\fH_+,\fH_-$}
$$
\fH_+\subset\fH\subset\fH_-,
$$
where $\fH=L^2(\rr)\otimes\cc\Xi$, $\fH_+$ is the Sobolev 
space $H^1(\rr)\otimes\cc\Xi$, and $\fH_-=H^{-1}(\rr)\otimes\cc\Xi$ is  its dual w.r.t.\;the duality 
induced by the inner product of~$\fH$. To simplify notations, we shall also
use the symbols $\fH$, $\fH_\pm$ to denote the corresponding real Hilbert 
spaces (the meaning should remain clear from the context). For $x\in\fH$, we 
denote by
$$
\hat x(\omega)=\int x(s)\e^{-\i\omega s}\d s
$$
its Fourier transform. Since, under Assumption~(C), $A$ is stable, we can use 
$$
\|x\|_\pm=\left(\int|(A-\i\omega)^{\pm1}
\hat x(\omega)|^2\frac{\d\omega}{2\pi}\right)^{\frac12}
$$
as norms on $\fH_\pm$. For $\tau>0$, we denote by $\Pi_\tau$\index{$\Pi_\tau$} the operator of 
multiplication with the characteristic function of the interval $[0,\tau]$. 
Thus, $\Pi_\tau$ is an orthogonal projection in $\mathfrak H$ whose range 
$\fH_\tau$ will be identified with the Hilbert space 
$L^2([0,\tau])\otimes\cc\Xi$. 

\bel\label{LEM-WH}Under Assumption~(C) the following hold.
\ben
\item The Volterra integral operator
$$
(Rx)(s)=\int_{-\infty}^s\e^{(s-s')A}x(s')\d s'
$$
maps isometrically $\fH_-$ onto $\fH$ and $\fH$ onto $\fH_+$.
By duality, its adjoint
$$
(R^\ast x)(s)=\int_s^\infty\e^{(s'-s)A^\ast}x(s')\d s',
$$
has the same properties.
\item $\Pi_\tau R$ is Hilbert-Schmidt, with norm
$$
\|\Pi_\tau R\|_2=\left(\tau
\int_0^\infty\tr(\e^{tA^\ast}\e^{tA})\d t\right)^{\frac12}.
$$
\item For $t_0\in[0,\tau]$, the Hilbert--Schmidt norm of the map 
$R_{t_0}:\fH\to\Xi$ defined by $R_{t_0}x=(Rx)(t_0)$ is given by
$$
\|R_{t_0}\|_2=\left(\int_0^{t_0}\tr(\e^{tA^\ast}\e^{tA})\d t\right)^{\frac12}.
$$
\een
\eel

\noindent{\bf Proof.} (1) Follows from our choice of the norms on $\fH_\pm$
and the fact that 
$(Rx)^{\,\widehat{}}(\omega)=(\i\omega-A)^{-1}\hat x(\omega)$.

\noindent(2) $\Pi_\tau R$ is an integral operator with kernel
$1_{[0,\tau]}(s)\theta(s-s')\e^{(s-s')A}$, where $1_{[0,\tau]}$ denotes
the characteristic function of the interval $[0,\tau]$ and $\theta$
the Heaviside step function. Its Hilbert-Schmidt norm is given by
$$
\|\Pi_\tau R\|_2^2=
\int_0^\tau\d s\int_{-\infty}^s\d s'\, \tr(\e^{(s-s')A^\ast}\e^{(s-s')A})
=\tau\int_0^\infty\d t\, \tr(\e^{tA^\ast}\e^{tA}).
$$

\noindent(3) Follows from a simple calculation.
\hfill\qed

\bigskip
Given $\tau>0$, consider the process 
$\{x(t)\}_{t\in[0,\tau]}$ started with a Gaussian measure 
$\nu\in\cP(\Xi)$. Let $a\in\Xi$ be the mean of~$\nu$ and $0\le N\in L(\Xi)$ 
its covariance. Denote by $(\,\cdot\,|\,\cdot\,)$
the inner product of $\fH_\tau$.

\bel\label{LEM-Cov} Let $T_\tau:\Xi\ni v\mapsto\e^{sA}v\in\fH_\tau$ and define\index{$\cD_\tau$}
$$
\cD_\tau=\left[
\begin{array}{cc}
T_\tau N^{\frac12}&\Pi_\tau RQ
\end{array}\right]:
\Xi\oplus\partial\fH\to\fH_\tau,
$$
where $\partial\fH=L^2(\rr)\otimes\partial\Xi$\index{$\partial\fH$}, and the operator~$Q$ acts on~$\partial\fH$ by the relation $(Qy)(t) =Qy(t)$ for $t\in\rr$.
Then, under Assumption~(C), the following properties hold for any $\tau>0$:
\ben
\item $\cD_\tau$ is Hilbert-Schmidt and has a unique continuous extension to 
$\Xi\oplus\fH_-$.
\item $\cK_\tau=\cD_\tau\cD_\tau^\ast$\index{$\cK_\tau$} is a non-negative trace class operator
on $\fH_\tau$ with integral kernel
\beq
\cK_\tau(s,s')=\e^{(s-s')_+A}(\e^{(s\wedge s')A}N\e^{(s\wedge s')A^\ast}
+M_{s\wedge s'})\e^{(s-s')_-A^\ast},
\label{EQ-CDef}
\eeq
and there exists a constant $C_\nu$, depending
on $A$, $B$ and $N$ but not on $\tau$, and such that 
$$
\cK_\tau\le C_\nu,\qquad\|\cK_\tau\|_1\le C_\nu\tau,
$$
where $\|\cdot\|_1$ denotes the trace norm.
\item The process $\{x(t)\}_{t\in[0,\tau]}$ is Gaussian with mean 
$T_\tau a$ and covariance $\cK_\tau$, i.e.,
\beq
\EE_\nu[\e^{\i(x|u)}]=\e^{\i(T_\tau a|u)-\frac12(u|\cK_\tau u)}
\label{EQ-CovForm}
\eeq
for all $u\in\fH_\tau$.
\een
\eel

\bigskip
{\noindent\bf Proof.} (1) $T_\tau$ is clearly finite 
rank and it follows from Lemma~\ref{LEM-WH}~(2) that the operator $\cD_\tau$ 
is Hilbert-Schmidt. Lemma~\ref{LEM-WH}~(1) further implies that it extends by
continuity to $\Xi\oplus\fH_-$.

(2) It follows immediately that
\beq
\cK_\tau=\cD_\tau\cD_\tau^\ast
=T_\tau NT_\tau^\ast+\Pi_\tau RQQ^\ast R^\ast\Pi_\tau|_{\fH_\tau}
\label{EQ-Cfactor}
\eeq
is non-negative and trace class. Formula~\eqref{EQ-CDef} can be checked by
an explicit  calculation.


Defining  the function $u\in\fH_\tau$ to be  zero 
outside $[0,\tau]$, we can invoke Plancherel's theorem to 
translate~\eqref{EQ-Cfactor} into
$$
(u|\cK_\tau u)=\left|\int_{-\infty}^{\infty}N^\12(A^\ast+\i\omega)^{-1}
\hat u(\omega)\frac{\d\omega}{2\pi}\right|^2
+\int_{-\infty}^{\infty}|Q^\ast(A^\ast+\i\omega)^{-1}
\hat u(\omega)|^2\frac{\d\omega}{2\pi}.
$$
By Lemma~\ref{LEM-Control_M}, Assumption~(C) implies
$\sp(A)\cap\i\rr=\emptyset$ and we conclude that 
$$
\cK_\tau\le\int_{-\infty}^{\infty}\|N^\12(A^\ast+\i\omega)^{-1}\|^2
\frac{\d\omega}{2\pi}+\sup_{\omega\in\rr}\|Q^\ast (A^\ast+\i\omega)^{-1}\|^2
<\infty.
$$
Finally, it is well known \cite[Theorem~3.9]{Si1} that the trace norm
of a non-negative trace class integral operator with continuous kernel 
$\cK_\tau(s,s')$ is given by
$$
\|\cK_\tau\|_1=\tr(\cK_1)=\int_0^\tau\tr(\cK_\tau(s,s))\d s
=\int_0^\tau\tr(\e^{sA}N\e^{sA^\ast}+M_s)\d s\le 
\tau\left(C\,\tr(N)+\tr(M)\right),
$$
where $C$ depends only on $A$.

(3) By Eq.~\eqref{EQ-The_Process} we have, for $u\in\fH_\tau$,
$$
(x|u)=(T_\tau x(0)|u)+\int_0^\tau\left[
\int_0^t\e^{(t-s)A}Q\,\d w(s)\right]\cdot u(t)\d t
=x(0)\cdot T_\tau^\ast u+\int_0^\tau Q^\ast(R^\ast u)(s)\cdot\d w(s)
$$
so that
$$
\EE_\nu[\e^{\i(x|u)}]
=\WW[\e^{\i\int_0^\tau Q^\ast(R^\ast u)(s)\cdot\d w(s)}]
\int\e^{\i x\cdot T_\tau^\ast u}\nu(\d x).
$$
Evaluating Gaussian integrals we get
$$
\int\e^{\i x\cdot T_\tau^\ast u}\nu(\d x)
=\e^{\i a\cdot T_\tau^\ast u-\frac12 T_\tau^\ast u\cdot NT_\tau^\ast u}
=\e^{\i(T_\tau a|u)-\frac12 (u|T_\tau NT_\tau^\ast u)},
$$
and
$$
\WW[\e^{\i\int_0^\tau Q^\ast(R^\ast u)(s)\cdot\d w(s)}]=
\e^{-\frac12(u|RQQ^\ast R^\ast u)},
$$
which provide the desired identity.\hfill\qed

\subsection{Proof of Proposition~\ref{PROP-Strict_Positivity}}
\label{SSECT_Proof_of_PROP-Strict_Positivity}

We start with some results on the Markov semigroup 
\beq
(P^tf)(x)=\int f(\e^{tA}x+M_t^{\frac12}y){\rm n}(\d y).
\label{EQ-PtForm2}
\eeq
For a multi-index $\alpha=(\alpha_1,\alpha_2,\ldots)\in\nn^{\dim\Xi}$
and $p\in[1,\infty]$ set
$$
|\alpha|=\sum_i\alpha_i,\qquad
\partial^\alpha=\prod_i\partial_{x_i}^{\alpha_i},
$$
and define\index{$\cA^p$}
$$
\cA^p=\left\{
\psi\in C^\infty(\Xi)\,\bigg|\,
|\partial^\alpha\psi|
\in L^p(\Xi,\d\mu)\text{ for all }\alpha\in\nn^{\dim\Xi}
\right\}.
$$

\bel\label{LEM-LStar}
Suppose that  Assumption  (C) holds.
\ben
\item For any $\nu\in\cP(\Xi)$ and $t>0$, $\nu_t$ is absolutely continuous
w.r.t. Lebesgue measure. Its Radon-Nikodym derivative
\beq
\frac{\d\nu_t}{\d x}(x)
=\det(2\pi M_t)^{-\12}\int\e^{-\12|M_t^{-\12}(x-\e^{tA}y)|^2}\nu(\d y)
\label{EQ-PtStar}
\eeq
is strictly positive and $S_\mathrm{GS}(\nu_t)>-\infty$. Moreover, if
$\nu(|x|^2)<\infty$, then $S_\mathrm{GS}(\nu_t)<\infty$.
\item For any $\nu\in\cP(\Xi)$, any $t>0$, and any multi-index $\alpha$, 
$$
\partial^\alpha\frac{\d\nu_t}{\d x}\in L^1(\Xi,\d x)\cap L^\infty(\Xi,\d x).
$$
\item For $t>0$, 
$\widetilde{M}_t=M-\e^{t\widetilde{A}}M\e^{t\widetilde{A}^\ast}>0$, and
\beq
\widetilde{M}_t^{-1}=M^{-1}+\e^{tA^\ast}M_t^{-1}\e^{tA}.
\label{EQ-TildeM}
\eeq
\item $P^t$ is a contraction semigroup on $L^p(\Xi,\d\mu)$ for any 
$p\in[1,\infty]$. Its adjoint w.r.t.\;the duality 
$\langle f|g\rangle_\mu=\mu(fg)$ is given by
\beq
(P^{t\ast}\psi)(x)=\int\psi(\e^{t\widetilde{A}}x+\widetilde{M}_t^\12y)
{\rm n}(\d y).
\label{EQ-PstarForm}
\eeq
In particular, $P^{t\ast}$ is positivity improving.
\item For all $t>0$,
$P^{t\ast}L^\infty(\Xi,\d\mu)\subset\cA^\infty$.
\item For $p\in[1,\infty[$, $\cA^p$ is a core of the generator of $P^{t\ast}$ 
on $L^p(\Xi,\d\mu)$ and this generator acts on $\psi\in\cA^p$ as
\beq
L^\ast\psi=\12\nabla\cdot B\nabla\psi+\widetilde{A}x\cdot\nabla\psi.
\label{EQ-LstarForm}
\eeq
\item For $\nu\in\cP_+(\Xi)$ and $p\in[1,\infty[$ there exists 
$t_{\nu,p}>0$ such that $\frac{\d\nu_t}{\d\mu}\in\cA^p$ for all $t>t_{\nu,p}$.
\item For $\nu\in\cP_+(\Xi)$ there exist $t_{\nu,\infty}>0$, $C_\nu$
and $\delta_\nu>0$ such that
$$
\left|\log\frac{\d\nu_t}{\d\mu}(x)\right|
\le C_\nu\e^{-\delta_\nu t}(1+|x|^2)
$$
for $t\ge t_{\nu,\infty}$.
\een
\eel

\noindent{\bf Proof.} (1) We deduce from Eq.~\eqref{EQ-PtForm2} that for any 
bounded measurable function $f$ on $\Xi$ one has
\[
\begin{split}
\nu_t(f)&=\nu(P^t f)=\int f(\e^{tA}x+M_t^\12y)\nu(\d x){\rm n}(\d y)\\[2mm]
&=\det(2\pi M_t)^{-\12}\int f(y)\e^{-\12|M_t^{-\12}(y-\e^{tA}x)|^2}\nu(\d x)\d y,
\end{split}
\]
from which we conclude that $\nu_t$ is absolutely continuous 
w.r.t.  Lebesgue measure with  Radon-Nikodym derivative given by 
Eq.~\eqref{EQ-PtStar}. It follows immediately that
$$
\frac{\d\nu_t}{\d x}(x)\le\det(2\pi M_t)^{-\12},
$$
which implies the lower bound 
$$
S_{\rm GS}(\nu_t)\ge \12\log\det(2\pi M_t)>-\infty.
$$
To derive an upper bound, let $r$ be such that $B_r=\{x\in\Xi\,|\,|x|<r\}$ satisfies 
$\nu(B_r)>\12$. Then one has
\begin{align*}
\frac{\d\nu_t}{\d x}(x)
&\ge\12\det(2\pi M_t)^{-\12}
\inf_{z\in B_r}\e^{-\12|M_t^{-\12}(x-\e^{tA}z)|^2}\\
&\ge\12\det(2\pi M_t)^{-\12}
\e^{-\12\|M_t^{-1}\|\sup_{z\in B_r}|(x-\e^{tA}z)|^2}\\
&\ge\12\det(2\pi M_t)^{-\12}
\e^{-\12\|M_t^{-1}\|(|x|+R\|\e^{tA}\|)^2},
\end{align*}
from which we conclude that
$$
\log\frac{\d\nu_t}{\d x}(x)\ge -C_t(1+|x|^2)
$$
for some constant $C_t>0$, and hence
$$
S_{\rm GS}(\nu_t)\le C_t(1+\nu(|x|^2)).
$$

(2) From Eq.~\eqref{EQ-PtStar} we deduce that
$$
\partial^\alpha\frac{\d\nu_t}{\d x}(x)=
\int p_{\alpha,t}(x-\e^{tA}y)\e^{-\12|M_t^{-\12}(x-\e^{tA}y)|^2}
\nu(\d y),
$$
where $p_{\alpha,t}$ denotes a polynomial whose coefficients are continuous
functions of $t\in]0,\infty[$. It follows that
$$
\sup_{x\in\Xi}\left|\partial^\alpha\frac{\d\nu_t}{\d x}(x)\right|
\le\sup_{z\in\Xi}|p_{\alpha,t}(z)|\e^{-\12|M_t^{-\12}z|^2}<\infty,
$$
and
$$
\int\left|\partial^\alpha\frac{\d\nu_t}{\d x}(x)\right|\d x
\le\int|p_{\alpha,t}(z)|\e^{-\12|M_t^{-\12}z|^2}\d z<\infty.
$$

(3) From Lemma~\ref{LEM-Control_M}~(5) we get 
$$
\e^{tA^\ast}M^{-1}\e^{tA}=(M+\e^{-tA}M_t\e^{-tA^\ast})^{-1}<M^{-1}.
$$
The strict positivity of $\widetilde{M}_t$ follows from
$$
\widetilde{M}_t=M-M(\e^{tA^\ast}M^{-1}\e^{tA})M>M-MM^{-1}M=0.
$$
Using again Lemma~\ref{LEM-Control_M}~(5), it is straightforward to
check the last statement of Part~(3).

(4) For $f\in L^1(\Xi,\d\mu)$ we have
$$
\|P^tf\|_{L^1(\Xi,\d\mu)}=
\mu(|P^tf|)\le\mu(P^t|f|)=\mu(|f|)=\|f\|_{L^1(\Xi,\d\mu)}.
$$ 
The representation~\eqref{EQ-PtForm2} shows that $P^t$ is a contraction on 
$L^\infty(\Xi,\d\mu)$. The Riesz-Thorin interpolation theorem yields that
$P^t$ is a contraction on $L^p(\Xi,\d\mu)$ for all $p\in[1,\infty]$.
To get a representation of the adjoint semigroup $P^{t\ast}$, we start again 
with Eq.~\eqref{EQ-PtForm2},
\begin{align*}
\langle\psi|P^tf\rangle_\mu
&=\int\psi(y)f(\e^{tA}y+M_t^\12 x){\rm n}(\d x)\mu(\d y)\\
&=\int\psi(y)f(\e^{tA}y+ x)
\frac{\e^{-\12|M_t^{-\12}x|^2}}{\det(2\pi M_t)^\12}
\frac{\e^{-\12|M^{-\12}y|^2}}{\det(2\pi M)^\12}\d x\d y\\
&=\int\psi(y)f(x)
\frac{\e^{-\12|M_t^{-\12}(x-\e^{tA}y)|^2}}{\det(2\pi M_t)^\12}
\frac{\e^{-\12|M^{-\12}y|^2}}{\det(2\pi M)^\12}\d x\d y\\
&=\int\psi(y)f(x)
\frac{\e^{-\12|M_t^{-\12}(x-\e^{tA}y)|^2}}{\det(2\pi M_t)^\12}
\e^{\12(|M^{-\12}x|^2-|M^{-\12}y|^2)}\mu(\d x)\d y,
\end{align*}
to conclude that
$$
(P^{t\ast}\psi)(x)=\det(2\pi M_t)^{-\12}\int\e^{-\phi_t(x,y)}\psi(y)\d y,
$$
where, taking~\eqref{EQ-TildeM} into account, 
$$
\phi_t(x,y)=\12 x\cdot(M_t^{-1}-M^{-1})x+\12y\cdot\widetilde{M}_t^{-1}y
-\e^{tA^\ast}M_t^{-1}x\cdot y.
$$
Using Lemma~\ref{LEM-Control_M}~(5) and~\eqref{EQ-TildeM} one shows that
\beq
\phi_t(x,\e^{t\widetilde{A}}x+z)
=\12 z\cdot\widetilde{M}_t^{-1}z,
\label{EQ-phiForm}
\eeq
which leads to
$$
(P^{t\ast}\psi)(x)=\det(2\pi M_t)^{-\12}
\int\e^{-\12|\widetilde{M}_t^{-\12}z|^2}
\psi(\e^{t\widetilde{A}}x+z)\d z.
$$
Noticing that $M_t=(I-\e^{tA}\e^{t\widetilde{A}})M$ and
$\widetilde{M}_t=(I-\e^{t\widetilde{A}}\e^{tA})M$ we conclude that
$\det(M_t)=\det(\widetilde{M}_t)$ and Eq.~\eqref{EQ-PstarForm} follows.

(5) Rewriting Eq.~\eqref{EQ-PstarForm} as
\beq
(P^{t\ast}\psi)(x)=\det(2\pi M_t)^{-\12}
\int\e^{-\12|\widetilde{M}_t^{-\12}(z-\e^{t\widetilde{A}}x)|^2}
\psi(z)\d z,
\label{EQ-PstarAltForm}
\eeq
we derive  that for any multi-index $\alpha$,
$$
(\partial^\alpha P^{t\ast}\psi)(x)=
\int p_{\alpha,t}(z-\e^{t\widetilde{A}}x)
\e^{-\12|\widetilde{M}_t^{-\12}(z-\e^{t\widetilde{A}}x)|^2}
\psi(z)\d z,
$$
where $p_{\alpha,t}$ is  a polynomial whose coefficients are
continuous functions of $t\in]0,\infty[$. For $\psi\in L^\infty(\Xi,\d\mu)$ 
this yields
$$
\left\|\partial^\alpha P^{t\ast}\psi\right\|_{L^\infty(\Xi,\d\mu)}
\le\|\psi\|_{L^\infty(\Xi,\d\mu)}\int|p_{\alpha,t}(z)|
\e^{-\12|\widetilde{M}_t^{-\12}z|^2}\d z,
$$
where the integral on the right-hand side is finite for all $t>0$.

(6) $\cA^p$ is dense in $L^p(\Xi,\d\mu)$ for $p\in[1,\infty[$.
For $\psi\in\cA^p$, Eq.~\eqref{EQ-PstarForm} yields
$$
(\partial^\alpha P^{t\ast}\psi)(x)
=\sum_{|\alpha'|=|\alpha|}C_{\alpha,\alpha'}(t)
\int(\partial^{\alpha'}\psi)(\e^{t\widetilde{A}}x+\widetilde{M}_t^\12y)
{\rm n}(\d y)
=\sum_{|\alpha'|=|\alpha|}C_{\alpha,\alpha'}(t)
(P^{t\ast}\partial^{\alpha'}\psi)(x),
$$
where the $C_{\alpha,\alpha'}$ are continuous functions of $t$.
As a consequence of Part~(4), $\cA^p$ invariant under the semigroup 
$P^{t\ast}$ and Part~(6) follows from the core theorem 
(Theorem X.49 in \cite{RS2}) and a simple calculation.

(7) Assuming $\nu(\e^{m|x-a|^2/2})<\infty$, we deduce from
Eq.~\eqref{EQ-PtStar} that for any $m'<m$
$$
\frac{\d\nu_t}{\d\mu}(x)=\det(M^{-1}M_t)^{-\12}
\int\e^{-\phi_t(x,y)}\nu'(\d y),
$$
where 
$$
\phi_t(x,y)=\12(|M_t^{-\12}(x-\e^{tA}y)|^2+m'|y-a|^2-|M^{-\12}x|^2),
$$
and $\nu'$ is such that $\nu'(\e^{\epsilon|x-a|^2})<\infty$ for
$\epsilon>0$ small enough. It follows that
$$
\partial^\alpha\frac{\d\nu_t}{\d\mu}(x)
=\int p_{\alpha,t}(x,y)\e^{-\phi_t(x,y)}\nu'(\d y),
$$
where $p_{\alpha,t}$ is a polynomial of degree $|\alpha|$ whose coefficients 
are continuous functions of $t\in]0,\infty[$. An elementary calculation
shows that
$$
\phi_t(x)=
\inf_{y\in\Xi}\phi_t(x,y)=|M_t^{-\12}x|^2-|M^{-\12}x|^2+m'|a|^2
-|(m'+\e^{tA^\ast}M_t^{-1}\e^{tA})^{-\12}(m'a+\e^{tA^\ast}M_t^{-1}x)|^2,
$$
and since
$\int|p_{\alpha,t}(x,y)|\nu'(\d y)\le C_{\alpha,t}(1+|x|^{2|\alpha|})$
for some constant $C_{\alpha,t}$ we have
$$
\left|\partial^\alpha\frac{\d\nu_t}{\d\mu}(x)\right|
\le C_{\alpha,t}(1+|x|^{2|\alpha|})\e^{-\phi_t(x)}.
$$
This gives the estimate
$$
\left\|\partial^\alpha\frac{\d\nu_t}{\d\mu}\right\|_{L^p(\Xi,\d\mu)}^p
\le C_{\alpha,t}^p
\int(1+|x|^{2|\alpha|})^p\e^{-p(\phi_t(x)+\frac1{2p}|M^{-\12}x|^2)}\d x,
$$
where the  last integral is finite provided the quadratic form
$$
|M_t^{-\12}x|^2-(1-p^{-1})|M^{-\12}x|^2
-|(m'+\e^{tA^\ast}M_t^{-1}\e^{tA})^{-\12}\e^{tA^\ast}M_t^{-1}x|^2
$$
is positive definite. Since $M_t^{-1}-M^{-1}>0$,  this holds if 
$$
M_t^{-1}\e^{tA}(m'+\e^{tA^\ast}M_t^{-1}\e^{tA})^{-1}\e^{tA^\ast}M_t^{-1}
\le \frac1p M^{-1}.
$$
Finally, the last inequality holds  for  large $t$ since the left-hand side is exponentially small
as $t\to\infty$.

(8) By Lemma~\ref{LEM-Control_M}~(1), $\|\e^{tA}\|=\cO(\e^{-\delta t})$ as 
$t\to\infty$. Repeating the previous analysis with $m'=\e^{-\delta t}$ we 
get, for large enough $t>0$,
$$
\log\frac{\d\nu_t}{\d\mu}(x)\le\12\tr(\log M-\log M_t)
+\log\int\e^{\12m'|x-a|^2}\nu(\d x)-\phi_t(x).
$$
One easily shows that $\tr(\log M-\log M_t)=\cO(\e^{-2\delta t})$ and
$|\phi_t(x)|=\cO(\e^{-\delta t})(1+|x|^2)$. Finally, since
$$
\int\e^{\12m'|x-a|^2}\nu(\d x)=1+\cO(m')
$$
as $m'\to0$, we derive the upper bound
$$
\log\frac{\d\nu_t}{\d\mu}(x)\le\cO(\e^{-\delta t})(1+|x|^2).
$$
To get a lower bound we set $m'=0$ and note that the ball 
$B_{t}=\{x\in\Xi\,|\,m|x-a|^2\le\delta t\}$ satisfies
$$
1-\nu(B_t)=\int_{\Xi\setminus B_t}\nu(\d x)
\le\int_{\Xi\setminus B_t}\e^{-m|x-a|^2} \e^{m|x-a|^2}\nu(\d x)
\le\e^{-\delta t}\int\e^{m|x-a|^2}\nu(\d x)=\cO(\e^{-\delta t}).
$$
Since $\log M>\log M_t$ we get
$$
\log\frac{\d\nu_t}{\d\mu}(x)\ge
-\sup_{y\in B_t}\phi_t(x,y)+\log(\nu(B_t)).
$$
It is straightforward to check that
$$
\sup_{y\in B_t}\phi_t(x,y)=\cO(\e^{-\delta t})(1+\cO(t^\12))(1+|x|^2),
$$
and therefore
$$
-\log\frac{\d\nu_t}{\d\mu}(x)\le\cO(\e^{-\epsilon t})(1+|x|^2)
$$
for any $\epsilon<\delta$.
\hfill$\square$

\bigskip
We are now ready to prove Proposition~\ref{PROP-Strict_Positivity}.
Writing the polar decomposition $Q=V(Q^\ast Q)^{\frac12}$, the 
existence of $\beta\in L(\Xi)$ satisfying~\eqref{EQ-betadef} easily
follows from the structural relations $[\vartheta,Q^\ast Q]=0$ and 
$\theta Q=\pm Q$.

\paragraph{(1)} Follows from Condition~\eqref{EQ-betadef} and
Eq.~\eqref{EQ-OmegaGammaCommute}.

\paragraph{(2)} From Eq.~\eqref{EQ-LDef} we deduce that the formal adjoint
of $L$ w.r.t.\;the inner product of $L^2(\Xi,\d x)$ is
$$
L^T=\12\nabla\cdot B\nabla-\nabla\cdot Ax.
$$
It follows from the structural relations~\eqref{EQ-Structure} and
Condition~\eqref{EQ-betadef} that
\begin{align*}
L^\beta&=\e^{\12|\beta^\12x|^2}L^T\e^{-\12|\beta^\12x|^2}
=\12(\nabla-\beta x)\cdot B(\nabla-\beta x)-(\nabla-\beta x)\cdot Ax\\
&=\12\nabla\cdot B\nabla-(A+Q\vartheta^{-1}Q^\ast)x\cdot\nabla
-\12\tr(Q\vartheta^{-1}Q^\ast+A+A^\ast)+\12x\cdot(Q\vartheta^{-2}Q^\ast+\beta A+A^\ast\beta)x\\
&=\12\nabla\cdot B\nabla+A^\ast x\cdot\nabla-\sigma_\beta(x).
\end{align*}
The desired identity thus follows from~\eqref{EQ-Structure} and Part~(1).

\paragraph{(3)} The It\^o formula gives
\begin{align*}
\d\left(\tfrac12x(t)\cdot Cx(t)\right)&=x(t)\cdot C\d x(t)
+\tfrac12\tr(CB)\d t\\
&=\tfrac12x(t)\cdot(CA+A^\ast C)x(t)\d t
+\tfrac12\tr(CB)\d t+x(t)\cdot CQ\d w(t).
\end{align*}
Therefore, since $\log\tfrac{\d\mu_\beta}{\d x}(x)=-\tfrac12x\cdot\beta x$, we
have
$$
\d\log\frac{\d\mu_\beta}{\d x}(x_t)
=-\tfrac12x(t)\cdot(\beta A+A^\ast\beta)x(t)\d t
-\tfrac12\tr(\beta QQ^\ast)\d t-x(t)\cdot\beta Q\d w(t).
$$
Using~\eqref{EQ-betadef} and the decomposition 
$A=\Omega-\12Q^\ast\vartheta^{-1}Q$, we deduce
$$
\d\log\frac{\d\mu_\beta}{\d x}(x_t)
=\tfrac12x(t)\cdot(\Omega\beta-\beta\Omega)x(t)\d t
+\tfrac12|Q^\ast\beta x(t)|^2\d t
-\tfrac12\tr(Q\vartheta^{-1}Q^\ast)\d t-Q^\ast\beta x(t)\cdot\d w(t),
$$
and, observing that $\nabla\log\frac{\d\mu_\beta}{\d x}(x)=-\beta x$, the result 
follows from Eq.~\eqref{EQ-SAbstractDef} and Condition~\eqref{EQ-betadef}.

\paragraph{(4)}
Let $\nu\in\cP_+(\Xi)$ and denote by $\psi_t$ the density of $\nu_t$
w.r.t.\;$\mu$. By Lemma~\ref{LEM-LStar}, $\psi_t$ is a strictly positive
element of $\cA^2$ for large enough $t$. For $\epsilon>0$ we have
$\log\epsilon\le\log(\psi_t+\epsilon)\le\psi_t+\epsilon-1$, 
and hence $\log(\psi_t+\epsilon)\in L^2(\Xi,\d\mu)$. Thus,
$s_\epsilon(\psi_t)=-\psi_t\log(\psi_t+\epsilon)\in L^1(\Xi,\d\mu)$, 
and the monotone convergence theorem yields
$$
\Ent(\nu_t|\mu)=\lim_{\epsilon\downarrow0}\mu(s_\epsilon(\psi_t)).
$$
From
$$
s_\epsilon(\psi_t(x))-s_\epsilon(\psi_s(x))
=\int_s^ts_\epsilon'(\psi_u(x))(L^\ast\psi_u)(x)\d u
$$
we infer
$$
\mu(s_\epsilon(\psi_t))-\mu(s_\epsilon(\psi_s))
=\int_s^t\langle s_\epsilon'(\psi_u)|L^\ast\psi_u\rangle_\mu\d u.
$$
Since $\psi_u$ and
$s_\epsilon'(\psi_u)=-1-\log(\psi_u+\epsilon)+\epsilon(\psi_u+\epsilon)^{-1}$
are elements of $\cA^2$ we can integrate by parts, using
Eq.~\eqref{EQ-LstarForm}, to get
$$
\langle s_\epsilon'(\psi_u)|L^\ast\psi_u\rangle_\mu
=\langle f_\epsilon(\psi_u)||Q^\ast\nabla\psi_u|^2\rangle_\mu
+\langle g_\epsilon(\psi_u)|(\widetilde{A}-A)x\cdot\nabla\psi_u\rangle_\mu,
$$
where
$$
f_\epsilon(\psi)=\frac12\frac{\psi+2\epsilon}{(\psi+\epsilon)^2}, \qquad
g_\epsilon(\psi)=\frac12\frac{\epsilon^2}{(\psi+\epsilon)^2}.
$$
Since $f_\epsilon\ge0$ and decreases as a function of $\epsilon$, the monotone
convergence theorem yields
$$
\lim_{\epsilon\downarrow0}
\int_s^t\langle f_\epsilon(\psi_u)||Q^\ast\nabla\psi_u|^2\rangle_\mu\d u
=\tfrac12\int_s^t\langle\psi_u^{-1}||Q^\ast\nabla\psi_u|^2\rangle_\mu\d u
=\tfrac12\int_s^t\nu_u(|Q^\ast\nabla\log\psi_u|^2)\d u.
$$
Since $0<g_\epsilon\le\12$, the dominated convergence 
theorem gives
$$
\lim_{\epsilon\downarrow0}
\int_s^t\langle g_\epsilon(\psi_u)|
(\widetilde{A}-A)x\cdot\nabla\psi_u\rangle_\mu\d u=0.
$$
We conclude that for $s$ sufficiently large and $t>s$
$$
\Ent(\nu_t|\mu)-\Ent(\nu_s|\mu)
=\lim_{\epsilon\downarrow0}(\mu(s_\epsilon(\psi_t))-\mu(s_\epsilon(\psi_s)))
=\tfrac12\int_s^t\nu_u(|Q^\ast\nabla\log\psi_u|^2)\d u,
$$
and Eq.~\eqref{EQ-DeBruijn} follows.

\paragraph{(5)} Eq.~\eqref{EQ-SAbstractDef}  gives
$$
\EE_\nu[\fS^t]
=\tfrac12\int_0^t\nu_s\left(\left|Q^\ast\nabla\log
\frac{\d\mu_\beta}{\d x}\right|^2
\right)\d s-\tfrac12 t\,\tr(Q\vartheta^{-1}Q^\ast).
$$
Since $S_{\rm GS}(\nu_t)=\Ent(\nu_t|\mu)+\nu_t(\varphi)$, where 
$$
\varphi(x)=-\log\frac{\d\mu}{\d x}(x)
=\tfrac12|M^{-\12}x|^2+\tfrac12\log\det(2\pi M),
$$
Eq.~\eqref{EQ-DeBruijn} implies
$$
\frac{\d\ }{\d t}(S_{\rm GS}(\nu_t)+\EE_\nu[\fS^t])
=\nu_t\left(\tfrac12\left|Q^\ast\nabla\log\frac{\d\nu_t}{\d\mu}\right|^2
+L\varphi
+\tfrac12\left|Q^\ast\nabla\log\frac{\d\mu_\beta}{\d x}\right|^2
-\tfrac12\tr(Q\vartheta^{-1}Q^\ast)\right).
$$
A simple calculation yields 
$L\varphi=-\12|Q^\ast\nabla\log\frac{\d\mu}{\d x}|^2
+\12\tr(Q\vartheta^{-1}Q^\ast)$ and hence
\begin{align*}
\frac{\d\ }{\d t}(S_{\rm GS}(\nu_t)+\EE_\nu[\fS^t])
&=\tfrac12\nu_t\left(
\left|Q^\ast\nabla\log\frac{\d\nu_t}{\d\mu}\right|^2
+\left|Q^\ast\nabla\log\frac{\d\mu_\beta}{\d x}\right|^2
-\left|Q^\ast\nabla\log\frac{\d\mu}{\d x}\right|^2
\right)\\
&=\tfrac12\nu_t\left(\left|Q^\ast\nabla\log
\frac{\d\nu_t}{\d\mu_\beta}\right|^2\right)
+\nu_t\left(\nabla\log\frac{\d\nu_t}{\d x}\cdot B\nabla\log\frac{\d\mu}{\d\mu_\beta}\right).
\end{align*}
An integration by parts shows that
\begin{align*}
\nu_t\left(\nabla\log\frac{\d\nu_t}{\d x}\cdot B\nabla\log\frac{\d\mu}{\d\mu_\beta}\right)
&=-\nu_t\left(\nabla\cdot B\nabla\log\frac{\d\mu}{\d\mu_\beta}\right)
=\tr\left(B(M^{-1}-\beta)\right),
\end{align*}
and, since $BM^{-1}-B\beta=-A-MAM^{-1}+A+A^\ast$, we have
$\tr(B(M^{-1}-\beta))=0$. The result follows.

\subsection{Proof of Proposition~\ref{PROP-StrictPositivity2}}
\label{SSECT_Proof_of_PROP-StrictPositivity2}

\paragraph{(1)} Since the first equivalence is provided 
by~\eqref{EQ-NESSEp}, it suffices to show the sequence of implications
\beq
MQ=Q\vartheta\Rightarrow[\Omega,M]=0\Rightarrow\mu\Theta=\mu\Rightarrow
\ep=0.
\label{EQ-LogiChain}
\eeq
Writing $\Omega=A+\12 Q\vartheta^{-1}Q^\ast$ and
invoking Lemma~\ref{LEM-Control_M}~(6) (the 
covariance of the steady state satisfies the Lyapunov equation 
$B+AM+MA^\ast=0$) one easily derives
$$
[\Omega,M]=\tfrac12\left((MQ-Q\vartheta)\vartheta^{-1}Q^\ast
+Q\vartheta^{-1}(MQ-Q\vartheta)^\ast\right),
$$
which proves the first implication in~\eqref{EQ-LogiChain}.
The last identity, rewritten as $[A-A^\ast,M]=0$, further implies that
$$
0=AM+MA^\ast+B=A^\ast M+MA+B
=\theta A\theta M+M\theta A^\ast\theta+\theta B\theta
=\theta(A\theta M\theta+\theta M\theta A^\ast+B)\theta,
$$
from which we deduce that $\theta M\theta$ is also solution of
the Lyapunov equation. Lemma~\ref{LEM-Control_M}~(7) allows us to conclude
that $\theta M\theta=M$ which is clearly equivalent to
$\mu\Theta=\mu$ and proves the second implication in~\eqref{EQ-LogiChain}.
Finally, from~\eqref{EQ-musigma} we deduce that if $\mu\Theta=\mu$, 
then 
\begin{align*}
\ep
&=-\mu(\sigma_\beta)=-\tfrac12\tr(M[\Omega,\beta])
=\tfrac12\tr(\beta[\Omega,M])\\
&=\tfrac12\tr(\theta\beta[\Omega,M]\theta)
=\tfrac12\tr(\beta[\theta\Omega\theta,\theta M\theta])
=-\tfrac12\tr(\beta[\Omega,\theta M\theta])=-\ep,
\end{align*}
which gives  the last implication.

\paragraph{(2)} Let $\vartheta_1,\vartheta_2\in\sp(\vartheta)$ be such that $\vartheta_1\not=\vartheta_2$ and 
$\cC_{\vartheta_1}\cap\cC_{\vartheta_2}\ni u\not=0$. Assume that 
$\ep=0$.
By Part~(1) this implies $MQ=Q\vartheta$ and $[\Omega,M]=0$. 
By construction, there exist polynomials $f_1$, $f_2$ and vectors 
$v_1,v_2\in\Xi$ such that
$$
f_1(\Omega)Q\pi_{\vartheta_1}v_1=u=f_2(\Omega)Q\pi_{\vartheta_2}v_2.
$$
The first equality in the above formula yields
$$
Mu=Mf_1(\Omega)Q\pi_{\vartheta_1}v_1=f_1(\Omega)MQ\pi_{\vartheta_1}v_1
=f_1(\Omega)Q\vartheta\pi_{\vartheta_1}v_1
=\vartheta_1f_1(\Omega)Q\pi_{\vartheta_1}v_1=\vartheta_1 u.
$$
Similarly,  the second one yields  $Mu=\vartheta_2 u$. Since $u\not=0$,
this  contradicts  the assumption
$\vartheta_1\not=\vartheta_2$.

\subsection{Proof of Proposition~\ref{PROP-Time_Reversal}}
\label{SSECT_Proof_of_PROP-Time_Reversal}

Let $\tau>0$, $\nu\in\cP^1_{\mathrm{loc}}(\Xi)$, set
$$
\psi_t=\frac{\d\nu_{\tau-t}}{\d x},
$$
and note that since
$\psi_\tau+|\nabla\psi_\tau|\in L^2_{\mathrm{loc}}(\Xi,\d x)$, it follows
from Lemma~\ref{LEM-LStar} that
\beq
\int_0^\tau\left(\|f\psi_t\|_2^2+\|f\nabla\psi_t\|_2^2\right)\d t
<\infty
\label{EQ-psibound}
\eeq
for all $f\in C_0^\infty(\Xi)$.
We consider the process $\bx=\{x(t)\}_{t\in[0,\tau]}$ which is the 
solution of the SDE~\eqref{EQ-The_SDE} with initial law $\nu$.
By Theorem~2.1 in~\cite{PH}, the estimate~\eqref{EQ-psibound} implies that
the process $\bar\bx=\{\bar x_t\}_{t\in[0,\tau]}$ with 
$\bar x_t=x_{\tau-t}$ is a diffusion satisfying the SDE
$$
\d \bar x(t)=\bar b(\bar x(t),t)\d t+Q\d\bar w(t)
$$
with initial law $\nu P^\tau$\!, drift $\bar b(x,t)=-Ax+B\nabla\log\psi_t(x)$,
and a standard $\partial\,\Xi$-valued Wiener process $\bar w(t)$. 
Since $\theta Q=\mp Q$, the time-reversed process  $\boldsymbol{\widetilde{x}}=\Theta^\tau(\bx)
=\{\theta\bar x(t)\}_{t\in[0,\tau]}$ satisfies
$$
\d\widetilde{x}(t)
=\widetilde{b}(\widetilde{x}(t),t)\d t+Q\d\widetilde{w}(t)
$$
with initial law $\nu P^\tau\Theta$, drift
$\widetilde{b}(x,t)=\theta\bar b(\theta x,t)$, and standard Wiener
process $\widetilde{w}(t)=\mp\bar w(t)$. Using the structural 
relations~\eqref{EQ-Structure} and $A+A^\ast=-QQ^\ast\beta$ we derive
$$
\widetilde{b}(x,t)=Ax+QQ^\ast\nabla\log\phi_t(x),\qquad
\phi_t=\Theta\left(\frac{\d\mu_\beta}{\d x}\right)^{-1}\psi_t,
$$
and conclude that we can rewrite the original SDE~\eqref{EQ-The_SDE} as
\beq
\d x(t)=\widetilde{b}(x(t),t)\d t+Q(\d w(t)-Q^\ast\nabla\log\phi_t(x(t))\d t). 
\label{EQ-SDEbar}
\eeq
Set
$$
\eta(t)=\int_0^tQ^\ast\nabla\log\phi_s(x(s))\cdot\d w(s),
$$
and let $Z(t)=\cE(\eta)(t)$ denote its stochastic exponential. We claim that 
\beq 
\EE_{\nu P^\tau\Theta}^\tau[Z(t)]=1
\label{cl-cergy}
\eeq
for all $t\in[0,\tau]$. Delaying the 
proof of this claim and applying Girsanov theorem we conclude that
$$
w(t)-\int_0^tQ^\ast\nabla\log\phi_s(x(s))\d s
$$
is a standard Wiener process under the law $\EE_{\nu P^\tau\Theta}^\tau[Z(\tau)\,\cdot\,]$, 
so that Eq.~\eqref{EQ-SDEbar} implies
\beq
\frac{\d\widetilde{\PP}_\nu^\tau}{\d\PP_{\nu P^\tau\Theta}^\tau}=Z(\tau).
\label{EQ-RNForm}
\eeq
Using It\^o calculus, one  derives from Eq.~\eqref{EQ-The_SDE} that
\begin{align*}
Q^\ast\nabla\log\phi_t(x(t))\cdot\d w(t)
&=\d\log\phi_t(x(t))-((\partial_t+L)\log\phi_t)(x(t))\d t\\
&=\d\log\phi_t(x(t))
-\left(\frac{(\partial_t+L)\phi_t}{\phi_t}(x(t))
-\tfrac12|Q^\ast\nabla\log\phi_t(x(t))|^2\right)\d t, 
\end{align*}
from which we obtain
$$
\eta(t)-\tfrac12[\eta](t)
=\log\phi_t(x(t))-\log\phi_0(x(0))
-\int_0^t\left(
\frac{(\partial_s+L)\phi_s}{\phi_s}\right)
(x(s))\d s.
$$
The generalized detailed balance condition~\eqref{EQ-GenDBC} further yields
$$
(\partial_s+L)\phi_s=-\sigma_\beta\phi_s,
$$
so that
$$
\eta(t)-\tfrac12[\eta](t)
=\log\frac{\d\nu_{\tau-t}}{\d\mu_\beta}(\theta x(t))
-\log\frac{\d\nu_{\tau}}{\d\mu_\beta}(\theta x(0))
+\int_0^t\sigma_\beta(x(s))\d s,
$$
from which we conclude that
\beq
Z(t)=\exp\left[\eta(t)-\tfrac12[\eta](t)\right]
=\left(\frac{\d\nu_{\tau}}{\d\mu_\beta}(\theta x(0))\right)^{-1}
\frac{\d\nu_{\tau-t}}{\d\mu_\beta}(\theta x(t))
\exp\left(\int_{0}^t\sigma_\beta(x(s))\d s
\right),
\label{EQ-ZtForm}
\eeq
and in particular that  $Z(\tau)=\exp(\Ep(\nu,\tau))\circ\Theta^\tau$.  
From~\eqref{EQ-RNForm} we finally get
$$
\frac{\d\widetilde{\PP}_\nu^\tau}{\d\PP_{\nu P^\tau\Theta}^\tau}
=\exp[\Ep(\nu,\tau)]\circ\Theta^\tau.
$$

It remains to prove the claim  (\ref{cl-cergy}). 
Set $\zeta=\nu P^\tau\Theta$ and observe
that it suffices to show that $\EE_\zeta[Z(t)]\ge1$ for $t\in[0,\tau]$ since 
$\EE_\zeta[Z(t)]\le1$ is a well known property of the stochastic exponential. 
The proof of this fact relies  on a sequence of approximations. 

The inequality $\EE_\zeta[Z(t)]\le1$ gives that for $s,s',t\in[0,\tau]$ and bounded 
measurable $f,g$ one has 
\beq
|\EE_\zeta[Z(t)f(x(s))g(x(s'))]|\le\|f\|_\infty\|g\|_\infty.
\label{EQ-EzetaBound}
\eeq
Here and in the 
following we denote by $\|\cdot\|_p$ the norm of $L^p(\Xi,\d x)$. The 
duality between $L^p(\Xi,\d x)$ and $L^q(\Xi,\d x)$ will be written
$\langle\,\cdot\,|\,\cdot\,\rangle$.
Next, we note that Eq.~\eqref{EQ-ZtForm} implies
\begin{align*}
\EE_\zeta[Z(t)&g(x(0))f(x(t))]\\
&=\EE_\zeta\left[\left(\frac{\d\nu_{\tau}}{\d\mu_\beta}(\theta x(0))\right)^{-1}
g(x(0))\frac{\d\nu_{\tau-t}}{\d\mu_\beta}(\theta x(t))f(x(t))
\exp\left(\int_{0}^t\sigma_\beta(x(s))\d s
\right)
\right]\\
&=\int g(x)\chi(x)
\EE_x\left[
\chi(x(t))^{-1}\psi_t(\theta x(t))f(x(t))\e^{V(t)}
\right]\d x=\langle g|\chi P_\sigma^t\chi^{-1}\widetilde{\psi}_t f\rangle,
\end{align*}
where we have set
$$
V(t)=\int_{0}^t\sigma_\beta(x(s))\d s,\qquad \chi=\frac{\d\mu_\beta}{\d x},
\qquad \widetilde{\psi}_t=\Theta\psi_t,
$$
and
$$
(P_\sigma^tf)(x)=\EE_x[\e^{V(t)}f(x(t))].
$$
It follows from the estimate~\eqref{EQ-EzetaBound} that 
$\|\chi P_\sigma^t\chi^{-1}\widetilde{\psi}_t f\|_1\le\|f\|_\infty$.
For $n,m>0$ we define
$$
\sigma_{n,m}(x)=\left\{
\begin{array}{ll}
-n&\text{ if }\sigma_\beta(x)\le-n;\\[4pt]
\sigma_\beta(x)&\text{ if } -n\le\sigma_\beta(x)\le m;\\[4pt]
m&\text{ if }\sigma_\beta(x)\ge m;
\end{array}\right.\qquad
\sigma_{m}(x)=\left\{
\begin{array}{ll}
\sigma_\beta(x)&\text{ if } \sigma_\beta(x)\le m;\\[4pt]
m&\text{ if }\sigma_\beta(x)\ge m;
\end{array}\right.
$$
and  set
$$
V_{n,m}(t)=\int_0^t\sigma_{n,m}(x(s))\d s,\qquad
V_{m}(t)=\int_0^t\sigma_{m}(x(s))\d s.
$$
Since 
\begin{align*}
\lim_{n\to\infty}\sigma_{n,m}(x)&=\sigma_m(x),&
\sigma_{n,m}(x)&\le m\\
\lim_{m\to\infty}\sigma_m(x)&=\sigma_\beta(x),&
\sigma_m(x)&\le\sigma_\beta(x), 
\end{align*}
for all $x\in\Xi$, we have
\begin{align*}
\lim_{n\to\infty}\e^{V_{n,m}(t)}&=\e^{V_m(t)},& 
\e^{V_{n,m}(t)}&\le\e^{mt},\\
\lim_{m\to\infty}\e^{V_{m}(t)}&=\e^{V(t)},&
\e^{V_{m}(t)}&\le\e^{V(t)},
\end{align*}
$\PP_{\zeta}$-almost surely. Hence, 
the dominated convergence theorem yields
\begin{align*}
\langle g|\chi & P_\sigma^t\chi^{-1}\widetilde{\psi}_t f\rangle
=\EE_{\zeta}\left[\chi(x(0))\widetilde{\psi}_{0}(x(0))^{-1}g(x(0))
\chi(x(t))^{-1}\widetilde{\psi}_t(x(t))f(x(t))\e^{V(t)}\right]\\
&=\lim_{m\to\infty}\lim_{n\to\infty}
\EE_{\nu_{\tau-t}}\left[\chi(x(0))\widetilde{\psi}_{0}(x(0))^{-1}g(x(0))
\chi(x(t))^{-1}\widetilde{\psi}_t(x(t))f(x(t))\e^{V_{n,m}(t)}\right]\\
&=\lim_{m\to\infty}\lim_{n\to\infty}
\langle g|\chi P_{\sigma_{n,m}}^t\chi^{-1}\widetilde{\psi}_t f\rangle,
\end{align*}
where, by the Feynman-Kac formula,
$$
(P_{\sigma_{n,m}}^tf)(x)=\EE_x[\e^{V_{n,m}(t)}f(x(t))]
=(\e^{t(L+\sigma_{n,m})}f)(x)
$$
defines a quasi-bounded semigroup on $L^2(\Xi,\d x)$. In the following, we
assume that $f\in C_0^\infty(\Xi)$ is non-negative. It follows from
Eq.~\eqref{EQ-PtStar} that
$\chi^{-1}\widetilde{\psi}_t f\in C_0^\infty(\Xi)\subset\Dom(L)=\Dom(L+\sigma_{n,m})$ 
and we can write
$$
\langle g|\chi P_{\sigma_{n,m}}^t\chi^{-1}\widetilde{\psi}_t f\rangle
=\langle g|\widetilde{\psi}_t f\rangle
+\int_0^t\langle g |
\chi(L+\sigma_{n,m})P_{\sigma_{n,m}}^s\chi^{-1}\widetilde{\psi}_t f\rangle\d s.
$$
Denote by $L^T$ the adjoint of $L$ on $L^2(\Xi,\d x)$ which acts on
$C_0^\infty(\Xi)$ as $L^T=\12\nabla\cdot B\nabla-\nabla\cdot Ax$. Assuming
$g\in C_0^\infty$, we get
$$
\langle g|\chi P_{\sigma_{n,m}}^t\chi^{-1}\widetilde{\psi}_t f\rangle
=\nu_{\tau-t}\Theta(gf)+\int_0^t\langle
\chi^{-1}(L^T+\sigma_{n,m})\chi g
|\chi P_{\sigma_{n,m}}^s\chi^{-1}\widetilde{\psi}_t f\rangle\d s.
$$
The generalized detailed balance condition~\eqref{EQ-GenDBC} yields
$$
\chi^{-1}L^T\chi g=\Theta(L+\sigma_\beta)\Theta g
=(\Theta L\Theta-\sigma_\beta)g,
$$ 
and it follows that
$$
\langle
\chi^{-1}(L^T+\sigma_{n,m})\chi g
|\chi P_{\sigma_{n,m}}^s\chi^{-1}\widetilde{\psi}_t f\rangle
=\langle
(\Theta L\Theta+\sigma_{n,m}-\sigma_\beta) g
|\chi P_{\sigma_{n,m}}^s\chi^{-1}\widetilde{\psi}_t f\rangle.
$$
Since $g$ is compactly supported, if $n$ and $m$ are sufficiently large
we have $(\sigma_{n,m}-\sigma_\beta) g=0$ and so
$$
\langle g|\chi P_{\sigma_{n,m}}^t\chi^{-1}\widetilde{\psi}_t f\rangle
=\nu_{\tau-t}\Theta(gf)+\int_0^t\langle \Theta L\Theta g
|\chi P_{\sigma_{n,m}}^s\chi^{-1}\widetilde{\psi}_t f\rangle\d s.
$$
Taking the limits $n\to\infty$ and $m\to\infty$ we get that
$$
\langle g|\chi P_{\sigma}^t\chi^{-1}\widetilde{\psi}_t f\rangle
=\nu_{\tau-t}\Theta(fg)+\int_0^t\langle \Theta L\Theta g
|\chi P_{\sigma}^s\chi^{-1}\widetilde{\psi}_t f\rangle\d s
$$
holds for all $f,g\in C_0^\infty(\Xi)$. For $k>0$ set 
$$
g_k(x)=(1+\e^{|x|^2/2k})^{-1},
$$
and let $\rho\in C_0^\infty(\rr)$ be such that $0\le\rho\le 1$,
$\rho'\le0$, $\rho(x)=1$ for $x\le0$ and $\rho(x)=0$ for $x\ge1$.
Define $g_{k,r}\in C_0^\infty(\Xi)$ by 
$g_{k,r}(x)=g_k(x)\rho(\langle x\rangle-r)$.
One easily checks that
$$
\lim_{r\to\infty}\|g_{k,r}-g_k\|_\infty+\|L(g_{k,r}-g_k)\|_\infty=0,
$$
and noticing that $g_k$ and $g_{k,r}$ are $\Theta$-invariant, it 
follows that
$$
\langle g_k|\chi P_{\sigma}^t\chi^{-1}\widetilde{\psi}_t f\rangle
=\nu_{\tau-t}\Theta(g_kf)+\int_0^t\langle\Theta L g_k
|\chi P_{\sigma}^s\chi^{-1}\widetilde{\psi}_t f\rangle\d s.
$$
Using the fact that
$$
(Lg_k)(x)\ge-\frac{1}{8k}\tr(B),
$$
and the monotone convergence theorem we conclude that
\begin{align*}
\langle 1|\chi P_{\sigma}^t\chi^{-1}\widetilde{\psi}_t f\rangle
&=\lim_{k\to\infty}\langle g_k|\chi P_{\sigma}^t\chi^{-1}
\widetilde{\psi}_t f\rangle\\
&\ge\lim_{k\to\infty}\nu_{\tau-t}\Theta(g_kf)
-\frac{1}{8k}\tr(B)\int_0^t\langle 1
|\chi P_{\sigma}^s\chi^{-1}\widetilde{\psi}_t f\rangle\d s
=\nu_{\tau-t}(f).
\end{align*}
Finally, letting $f$ converge to $1$ monotonically, we deduce
$$
\EE_\zeta[Z(t)]=\lim_{f\nearrow 1}\EE_\zeta[Z(t)f(x(t))]
=\lim_{f\nearrow 1}\langle 1|\chi P_{\sigma}^t\chi^{-1}
\widetilde{\psi}_t f\rangle
\ge\lim_{f\nearrow 1}\nu_{\tau-t}(f)=\nu_{\tau-t}(1)=1.
$$
This completes the proof of the claim (\ref{cl-cergy}).

\subsection{Proof of Theorem~\ref{THM-ealpha}}
\label{SSECT_Proof_of_THM-ealpha}

\paragraph{(1)}
We start with some algebraic preliminaries. For $\omega\in\rr$,
set
$$
R(\omega)=\vartheta^{-1}Q^\ast(A+\i\omega)^{-1}Q,\qquad
U(\omega)=I+R(\omega),
$$
and note that since the matrices $A$, $Q$ and $\vartheta$ are real
one has
\beq
\cC R(\omega)\cC=R(-\omega),\qquad \cC U(\omega)\cC=U(-\omega),
\label{EQ-ConjugaForm}
\eeq
where $\cC$ denotes complex conjugation on $\cc\partial\Xi$.
Further note that
$$
\det(U(\omega))=\det(I+(A+\i\omega)^{-1}Q\vartheta^{-1}Q^\ast)
=\frac{\det(A^\ast-\i\omega)}{\det(A+\i\omega)},
$$
from which we deduce that
\beq
|\det(U(\omega))|=1.
\label{EQ-inner}
\eeq
From the relations
\begin{align*}
\left(I+\vartheta^{-1}Q^\ast(A-\i\omega)^{-1}Q\right)^{-1}
&=I-\left(I+\vartheta^{-1}Q^\ast(A-\i\omega)^{-1}Q\right)^{-1}
\vartheta^{-1}Q^\ast(A-\i\omega)^{-1}Q\\
&=I-\vartheta^{-1}Q^\ast
\left(I+(A-\i\omega)^{-1}Q\vartheta^{-1}Q^\ast\right)^{-1}(A-\i\omega)^{-1}Q\\
&=I-\vartheta^{-1}Q^\ast
\left(A-\i\omega+Q\vartheta^{-1}Q^\ast\right)^{-1}Q\\
&=I+\vartheta^{-1}Q^\ast(A^\ast+\i\omega)^{-1}Q\\
&=I+\vartheta^{-1}Q^\ast\theta(A^\ast+\i\omega)^{-1}\theta Q\\
&=I+\vartheta^{-1}Q^\ast(A+\i\omega)^{-1}Q=I+R(\omega)
\end{align*}
we also get
\beq
U(-\omega)^{-1}=U(\omega).
\label{EQ-Uinverse}
\eeq

Writing
\begin{align*}
E(\omega)
&=Q^\ast(A^\ast-\i\omega)^{-1}
\left((-A^\ast-\12 Q\vartheta^{-1}Q^\ast+\i\omega)\beta
+\beta(-A-\12 Q\vartheta^{-1}Q^\ast-\i\omega)\right)
(A+\i\omega)^{-1}Q\\
&=Q^\ast\left(
-(A^\ast-\i\omega)^{-1}\beta-\beta(A+\i\omega)^{-1}
-(A^\ast-\i\omega)^{-1}Q\vartheta^{-2}Q^\ast(A+\i\omega)^{-1}
\right)Q\\
&=-R(\omega)-R(\omega)^\ast-R(\omega)^\ast R(\omega)
=I-(I+R(\omega))^\ast(I+R(\omega))=I-U(\omega)^\ast U(\omega),
\end{align*}
shows that $E(\omega)$ is indeed independent of the choice of $\beta$.
The continuity of $\omega\mapsto E(\omega)$ follows from Assumption (C)
and Lemma~\ref{LEM-Control_M}~(1) which ensures that 
$\i\rr\cap\sp(A)=\emptyset$.

\paragraph{(2)} 
Invoking Relation~\eqref{EQ-Uinverse} we infer
$$
E(\omega)=I-U(\omega)^\ast U(\omega)
=U(\omega)^\ast(U(-\omega)^\ast U(-\omega)-I)U(\omega)
=-U(\omega)^\ast E(-\omega)U(\omega),
$$
and
$$
I-\alpha E(\omega)=U(\omega)^\ast
\left(U(-\omega)^\ast U(-\omega)+\alpha E(-\omega)\right)
U(\omega)
=U(\omega)^\ast(I-(1-\alpha)E(-\omega))U(\omega).
$$
Combining the last identity with Eq.~\eqref{EQ-ConjugaForm} 
and~\eqref{EQ-inner} yields
\beq
\det(I-\alpha E(\omega))=\overline{\det(I-(1-\bar\alpha)E(\omega))}.
\label{EQ-DetSymmetry}
\eeq

The simple estimate 
$\|(A+\i\omega)^{-1}\|_2\le c(1+\omega^2)^{-\12}$ implies
\beq
\|E(\omega)\|_1\in L^1(\rr,\d\omega),\qquad
\lim_{\omega\to\pm\infty}\|E(\omega)\|=0.
\label{EQ-EomegaDecay}
\eeq
Thus, the eigenvalues of $E(\omega)$, which are continuous functions of
$\omega$, tend to zero as $\omega\to\pm\infty$. Since~\eqref{EQ-DetSymmetry}
implies that $I-E(\omega)$ is unimodular, $1\not\in\sp(E(\omega))$ for any 
$\omega\in\rr$ and we conclude that $E(\omega)<1$ for all $\omega\in\rr$.
From~\eqref{EQ-DetSymmetry} we further deduce that the elements of 
$\sp(E(\omega))\setminus\{0\}$ can be paired as $(\varepsilon,\varepsilon')$
with $0<\varepsilon<1$ and $\varepsilon'=-\varepsilon/(1-\varepsilon)<0$.
Moreover,  since the function
$]0,1[\ni\varepsilon\mapsto-\varepsilon/(1-\varepsilon)$
is monotone decreasing, one has
$$
\varepsilon_-(\omega)=\min\sp(E(\omega))
=-\frac{\varepsilon_+(\omega)}{1-\varepsilon_+(\omega)},\qquad
\varepsilon_+(\omega)=\max\sp(E(\omega)).
$$
Thus, the following alternative holds: either 
$$
\varepsilon_-=\min_{\omega\in\rr}\varepsilon_-(\omega)=0
=\max_{\omega\in\rr}\varepsilon_+(\omega)=\varepsilon_+,
$$
and hence $E(\omega)=0$ for all $\omega\in\rr$, or
$$
\varepsilon_+\in]0,1[,\qquad
\varepsilon_-=-\frac{\varepsilon_+}{1-\varepsilon_+}\in]0,-\infty[,
$$
and hence
$$
\frac1{\varepsilon_-}+\frac1{\varepsilon_+}=1.
$$
This proves Part~(2).

\paragraph{(3)} By Part~(2), $\det(I-\alpha E(\omega))\not=0$ for
$\alpha\in\mathfrak{C}_c$ and hence the function 
\[\mathfrak{C}_c\ni\alpha\mapsto\log\det(I-\alpha E(\omega))\] is analytic.
Moreover, an elementary analysis shows that for any compact subset 
$K\subset\mathfrak{C}_c$ there is a constant $C_K$ such that
$$
\sup_{\alpha\in K}\|E(\omega)(I-\alpha E(\omega))^{-1}\|_1
\le C_K \|E(\omega)\|_1.
$$
For any $\alpha\in\mathfrak{C}_c$ one has
$$
\log\det(I-\alpha E(\omega))
=-\int_0^\alpha\tr(E(\omega)(I-\gamma E(\omega))^{-1})\d\gamma,
$$
and since the integration path from $0$ to $\alpha$ lies in $\mathfrak{C}_c$
there is a constant $C_\alpha<\infty$ such that
$$
|\log\det(I-\alpha E(\omega))|\le C_\alpha\,\|E(\omega)\|_1.
$$
By~\eqref{EQ-EomegaDecay} and Fubini's theorem
$$
e(\alpha)=-\int_{-\infty}^{\infty}
\log\det(I-\alpha E(\omega))\frac{\d\omega}{4\pi}
=\int_0^\alpha\left(\int_{-\infty}^{\infty}\tr(E(\omega)(I-\gamma E(\omega))^{-1})\frac{\d\omega}{4\pi}\right)\d\gamma.
$$
It follows that $\mathfrak{C}_c\ni\alpha\mapsto e(\alpha)$ is analytic and that 
$$
e'(\alpha)
=\int_{-\infty}^{\infty}
\tr(E(\omega)(I-\alpha E(\omega))^{-1})\frac{\d\omega}{4\pi},
$$
$$
e''(\alpha)
=\int_{-\infty}^{\infty}
\tr(E(\omega)(I-\alpha E(\omega))^{-1}E(\omega)(I-\alpha E(\omega))^{-1}
)\frac{\d\omega}{4\pi}.
$$
Since $I-\alpha E(\omega)>0$ for $\alpha\in\fI_c$,
the last formula shows in particular that $e''(\alpha)\ge0$ for
$\alpha\in\fI_c$, and so the function $\fI_c\ni\alpha\mapsto e(\alpha)$ is convex.
Going back to the alternative of Part~(2), we conclude that either
$e(\alpha)$ vanishes identically, or  is strictly convex on $\fI_c$.
The symmetry $e(1-\alpha)=e(\alpha)$ follows from Eq.~\eqref{EQ-DetSymmetry}
and, since $e(0)=e(1)=0$, convexity implies that $e(\alpha)\le0$
for $\alpha\in[0,1]$ and $e(\alpha)\ge0$ for $\alpha\in\fI_c\setminus[0,1]$.
By Plancherel's theorem
$$
\int_{-\infty}^{\infty}(A+\i\omega)^{-1}QQ^\ast(A^\ast-\i\omega)^{-1}
\frac{\d\omega}{2\pi}=\int_0^\infty\e^{tA}QQ^\ast\e^{tA^\ast}\d t=M,
$$
and so 
$$
e'(0)=-e'(1)=\int_{-\infty}^{\infty}
\tr(\Sigma_\beta(A+\i\omega)^{-1}QQ^\ast(A^\ast-\i\omega)^{-1})
\frac{\d\omega}{4\pi}
=\tfrac12\tr(\Sigma_\beta M)=\mu(\sigma_\beta)=-\ep.
$$
Assume that $\varepsilon_+>0$. By Lemma~\ref{LEM-Control_M}~(1), $A$ is stable 
and hence $E(\omega)$ is an analytic function of $\omega$ in a strip
$|\Im\omega|<\delta$. By~\eqref{EQ-EomegaDecay} there is a compact subset
$K$ of this strip such that $\varepsilon_+(\omega)<\varepsilon_+$ for all
$\omega\in\rr\setminus K$. By regular perturbation theory the eigenvalues
of $E(\omega)$ are analytic in $K$, except for possibly  finitely many exceptional
points where some of these eigenvalues cross. Thus, there is a strip
$\mathcal S=\{\omega\,|\,|\Im(\omega)|<\delta'\}$ such that all exceptional 
points of $E(\omega)$ in $\mathcal S\cap K$ are real. Since $E(\omega)$ is 
self-adjoint for $\omega\in\rr$, its eigenvalues are analytic at these 
exceptional points (see, e.g., \cite[Theorem 1.10]{Ka}). We conclude  that the 
eigenvalues of $E(\omega)$ are analytic in $\mathcal S\cap K$. It follows that
the function $\rr\ni\omega\mapsto\varepsilon_+(\omega)$ reaches its maximum 
$\varepsilon_+$ on a finite subset $\cM\subset K\cap\rr$. To each $\fm\in\cM$
let us associate $\delta_\fm>0$, to be chosen later, in such a way that
the intervals $O_\fm=]\fm-\delta_\fm,\fm+\delta_\fm[$ are pairwise disjoint.
Setting
\beq
e_\fm(\alpha)
=-\int\limits_{O_\fm}\log\det(I-\alpha E(\omega))\frac{\d\omega}{4\pi}
=-\sum_j\int\limits_{O_\fm}\log(1-\alpha\varepsilon_j(\omega))
\frac{\d\omega}{4\pi},
\label{EQ-emForm}
\eeq
where the sum runs over all repeated eigenvalues of $E(\omega)$, we can 
decompose
$$
e(\alpha)=\sum_{\fm\in\cM}e_\fm(\alpha)+e_{\rm reg}(\alpha),
$$
where the function $\alpha\mapsto e_{\rm reg}(\alpha)$ is analytic at
$\alpha=\frac12+\kappa_c$. Since
$\fI_c\ni\alpha\mapsto e(\alpha)$ is convex, to prove that it has a continuous
extension to $\alpha=\tfrac12+\kappa_c$ and that its derivative diverges
to $+\infty$ as  $\alpha\uparrow\tfrac12+\kappa_c$, it suffices to
show that for all $\fm\in\cM$ the function $e_\fm(\alpha)$
remains bounded and its derivative diverges to $+\infty$
in this limit. The same argument links the behavior 
of $e(\alpha)$ and $e'(\alpha)$ as $\alpha\downarrow\tfrac12-\kappa_c$ to the 
minima of $\varepsilon_-(\omega)$, and we shall only consider the case 
$\alpha\uparrow\tfrac12+\kappa_c$.

Let $\fm\in\cM$ and consider an eigenvalue $\varepsilon(\omega)$
of $E(\omega)$ which takes the maximal value $\varepsilon_+$ at $\omega=\fm$. 
There is an integer $n\ge1$ and a function $f$, analytic at $\fm$, such that
$f(\fm)>0$ and
$$
\varepsilon(\omega)=\varepsilon_+-(\omega-\fm)^{2n}f(\omega).
$$
Moreover, we can chose $\delta_\fm>0$ such that $f$ is analytic in 
$O_\fm$ and
$$
\inf_{\omega\in O_\fm}f(\omega)>0,\qquad
\sup_{\omega\in O_\fm}f(\omega)<\infty,\qquad
\inf_{\omega\in O_\fm}\varepsilon(\omega)>0.
$$
Setting
$$
\eta=\left(\frac1\alpha-\varepsilon_+\right)^{\tfrac1{2n}}
=\left(\frac{\12+\kappa_c-\alpha}%
{(\frac12+\kappa_c)\alpha}\right)^{\tfrac1{2n}},
$$
so that $\eta\downarrow 0 \Leftrightarrow\alpha\uparrow\tfrac12+\kappa_c$,
we can write
$$
1-\alpha\varepsilon(\omega)
=\alpha\eta^{2n}
\left(1+\left(\frac{\omega-\fm}{\eta}\right)^{2n}f(\omega)\right)
=\alpha(\omega-\fm)^{2n}
\left(\left(\frac{\eta}{\omega-\fm}\right)^{2n}+f(\omega)\right)
$$
and since
\begin{align*}
\int\limits_{|\omega-\fm|\le\eta}&\log\left[
\alpha\eta^{2n}\left(1+\left(\frac{\omega-\fm}{\eta}\right)^{2n}f(\omega)
\right)\right]\d\omega
=\mathcal{O}(\eta\log\eta),\\
\int\limits_{\eta\le|\omega-\fm|\le\delta_\fm}&\log\left[
\alpha(\omega-\fm)^{2n}\left(\left(\frac{\eta}{\omega-\fm}\right)^{2n}
+f(\omega)\right)\right]\d\omega
=\mathcal{O}(1),
\end{align*}
as $\eta\downarrow0$, it follows that
$$
\int\limits_{O_\fm}\log(1-\alpha\varepsilon(\omega))\d\omega
=\mathcal{O}(1)
$$
as $\alpha\uparrow\tfrac12+\kappa_c$. Since the contributions to the sum
on the right-hand side of Eq.~\eqref{EQ-emForm} arising from eigenvalues of 
$E(\omega)$ that do not reach the maximal value $\varepsilon_+$ at $\fm$ are 
analytic at $\alpha=\tfrac12+\kappa_c$, it follows that $e_\fm(\alpha)$
remains bounded as $\alpha\uparrow\tfrac12+\kappa_c$. 

Let us now consider the derivative $e_\fm'(\alpha)$. Setting 
$\eta=\tfrac12+\kappa_c-\alpha$, we can write
$$
\int\limits_{O_\fm}
\frac{\varepsilon(\omega)}{1-\alpha\varepsilon(\omega)}\d\omega
=\int\limits_{O_\fm}
\left(\eta+\frac{f(\omega)}%
{\varepsilon(\omega)\varepsilon_+}(\omega-\fm)^{2n}\right)^{-1}\d\omega.
$$
Since
$$
D=\sup_{\omega\in O_\fm}
\frac{f(\omega)}{\varepsilon(\omega)\varepsilon_+}>0,
$$
we get
$$
\int\limits_{O_\fm}
\frac{\varepsilon(\omega)}{1-\alpha\varepsilon(\omega)}\d\omega
\ge2\int_0^{\delta_\fm}\frac{\d\omega}{\eta+\omega^{2n}D}
\ge C\eta^{-1+\tfrac1{2n}}\to\infty,
$$
as $\eta\downarrow0$. Since again the contributions 
of the eigenvalues of $E(\omega)$ which do not reach the maximal value 
$\varepsilon_+$ at $\fm$ are analytic at
$\alpha=\tfrac12+\kappa_c$, it follows that $e_\fm'(\alpha)\to\infty$
as $\alpha\uparrow\tfrac12+\kappa_c$.

\paragraph{(4)} For any bounded continuous function 
$f:[\varepsilon_-,\varepsilon_+]\to\cc$ one has
$$
\left|\int_{-\infty}^{\infty}
\tr(E(\omega)f(E(\omega)))\frac{\d\omega}{4\pi}\right|
\le \|f\|_\infty\int_{-\infty}^{\infty}\|E(\omega)\|_1\frac{\d\omega}{4\pi}.
$$
Hence, by the Riesz-Markov representation theorem there is a regular signed 
Borel measure $\varrho$ on $[\varepsilon_-,\varepsilon_+]$ such that
$$
\int_{-\infty}^{\infty}
\tr(E(\omega)f(E(\omega)))\frac{\d\omega}{4\pi}
=\int f(\varepsilon)\varrho(\d\varepsilon),
$$
and
$$
\int |\varrho|(\d\varepsilon)
\le\int_{-\infty}^{\infty}\|E(\omega)\|_1\frac{\d\omega}{4\pi}<\infty.
$$
For $\alpha\in\fC_c$ the function
$$
f_\alpha:[\varepsilon_-,\varepsilon_+]\ni\varepsilon\mapsto
-\frac1\varepsilon\log(1-\alpha\varepsilon)
$$ 
is continuous and we can write
\beq
e(\alpha)=-\int_{-\infty}^{\infty}
\tr(E(\omega)f_\alpha(E(\omega)))\frac{\d\omega}{4\pi}
=\int f_\alpha(\varepsilon)\d\varrho(\varepsilon).
\label{EQ-ealphapotential}
\eeq
We can now proceeds as the proof of Theorem~2.4~(2) in~\cite{JPS}.


\paragraph{(5)} We start with some simple consequences of Assumption (C). 
The reader is referred to Section~4 of~\cite{LR} for a short
introduction to the necessary background material.
Since $A_\alpha=A+\alpha Q\vartheta^{-1}Q^\ast$, the
pair $(A_\alpha,Q)$ is controllable for all $\alpha$. The
relation $A_\alpha^\ast=-A_{1-\alpha}$ shows that the same is true 
for the pair $(A_\alpha^\ast,Q)$. Thus, one has
\beq
\bigcap_{n\ge0}\Ker(Q^\ast A_\alpha^n)
=\bigcap_{n\ge0}\Ker(Q^\ast A_\alpha^{\ast n})=\{0\}
\label{EQ-Control}
\eeq
for all $\alpha$. This implies that if $Q^\ast u=0$ and
$(A_\alpha-z)u=0$ or $(A_\alpha^\ast-z)u=0$, then $u=0$,
i.e., no eigenvector of $A_\alpha$ or $A_\alpha^\ast$ is contained  in
$\Ker Q^\ast$.

Assume that $z\in\sp(A_\alpha)$ and let $u\not=0$ be a corresponding 
eigenvector. Since
$$
A_\alpha+A_\alpha^\ast=2(\alpha-\tfrac12)Q\vartheta^{-1}Q^\ast,
$$
taking the real part of $(u,(A_\alpha-z)u)=0$ yields
$$
(\alpha-\tfrac12)|\vartheta^{-\12}Q^\ast u|^2=\Re z|u|^2.
$$
Thus, controllability of $(A_\alpha,Q)$ implies 
$\sp(A_\alpha)\subset\cc_\pm$ for $\pm(\alpha-\12)>0$. 

For $\alpha\in\rr\setminus\{\frac12\}$ and $\omega\in\rr$, Schur's complement 
formula yields
$$
\det(K_\alpha-\i\omega)
=\frac{\det\left(I+\alpha(1-\alpha)Q^\ast(A_\alpha^\ast-\i\omega)^{-1}
Q\vartheta^{-2}Q^\ast(A_\alpha+\i\omega)^{-1}Q\right)}%
{\det\left((A_\alpha+\i\omega)^{-1}\right)\det\left((A_\alpha^\ast-\i\omega)^{-1}\right)},
$$
and using the relations
\begin{align*}
(A_\alpha+\i\omega)^{-1}
&=(A+\i\omega)^{-1}(I+\alpha Q\vartheta^{-1}Q^\ast(A+\i\omega)^{-1})^{-1},\\
(A_\alpha^\ast-\i\omega)^{-1}
&=(I+\alpha(A^\ast-\i\omega)^{-1}Q\vartheta^{-1}Q^\ast)^{-1}
(A^\ast-\i\omega)^{-1},
\end{align*}
one easily derives
\beq
\det(K_\alpha-\i\omega)=|\det(A+\i\omega)|^2\det(I-\alpha E(\omega)).
\label{EQ-detKalpha}
\eeq
Writing Eq.~\eqref{EQ-KalphaDef} as
$$
K_\alpha=\left[\begin{array}{cc}
-A&QQ^\ast\\
0&A^\ast
\end{array}\right]
+\left[\begin{array}{cc}
-\alpha Q\vartheta^{-1}Q^\ast&0\\
\alpha(1-\alpha) Q\vartheta^{-2}Q^\ast&\alpha Q\vartheta^{-1}Q^\ast
\end{array}\right],
$$
one derives that the identity~\eqref{EQ-detKalpha}, as the equality between two polynomials,  extends to 
all $\alpha\in\cc$.

By Part~(2), we conclude that $\sp(K_\alpha)\cap\i\rr=\emptyset$ for 
$\alpha\in\fC_c$. It follows from the regular perturbation theory that the 
spectral projection $P_\alpha$ of $K_\alpha$ for the part of its spectrum in 
the open right half-plane is an analytic function of $\alpha$ in the cut plane 
$\fC_c$ (see, e.g., \cite[Section II.1]{Ka}). For $\alpha\in\rr$, $K_\alpha$ 
is $\rr$-linear on the real vector space $\Xi\oplus\Xi$. Thus, its spectrum is 
symmetric w.r.t.\;the real axis. Observing that $JK_\alpha+K_\alpha^\ast J=0$,
where $J$ is the unitary operator
$$
J=\left[\begin{array}{cc}
0&I\\
-I&0
\end{array}\right],
$$
we conclude that the spectrum of $K_\alpha$ is also symmetric w.r.t.\;the 
imaginary axis. It follows that for $\alpha\in\fI_c$
\beq
\tfrac12\sum_{\lambda\in\sp(K_\alpha)}|\Re\lambda|\,m_\lambda
=\tr(P_\alpha K_\alpha).
\label{EQ-trplusForm}
\eeq
Denoting the resolvent of $K_\alpha$ by $T_\alpha(z)=(z-K_\alpha)^{-1}$,
we have
$$
P_\alpha=\oint_{\Gamma_+}T_\alpha(z)\frac{\d z}{2\pi\i},
$$
where $\Gamma_+\subset\cc_+$ is a Jordan contour enclosing 
$\sp(K_\alpha)\cap\cc_+$ which can be 
chosen so that it also encloses $\sp(-A)=\sp(K_0)\cap\cc_+$. 
Thus, we can rewrite~\eqref{EQ-trplusForm} 
as
$$
\tfrac12\sum_{\lambda\in\sp(K_\alpha)}|\Re\lambda|\,m_\lambda
=\oint_{\Gamma_+}z\tau_\alpha(z)\frac{\d z}{2\pi\i},
$$
with $\tau_\alpha(z)=\tr(T_\alpha(z))$.

An elementary calculation yields the following resolvent 
formula 
\begin{align*}
T_\alpha(z)=T_0(z)+\left[\begin{array}{cc}
-\alpha r(z)QD(z)(I+R^\circledast(z))\vartheta^{-1}Q^\ast r(z)
&r(z)Q(I-D(z))Q^\ast r^\circledast(z)\\[6pt]
-\lambda r^\circledast(z)Q\vartheta^{-1}(I+R(z))D(z)\vartheta^{-1}Q^\ast r^\circledast(z)
&\alpha r^\circledast(z)Q\vartheta^{-1}(I+R(z))D(z)Q^\ast
r^\circledast(z)
\end{array}\right],
\end{align*}
where
$$
\begin{array}{rclrcl}
r(z)&=&(A+z)^{-1},
&r^\circledast(z)&=&(A^\ast-z)^{-1},\\[8pt]
R(z)&=&\vartheta^{-1}Q^\ast r(z)Q,
&R^\circledast(z)&=&Q^\ast r^\circledast(z)Q\vartheta^{-1},
\end{array}
$$
and
$$
D(z)=\left(I+\alpha(R(z)+R^\circledast(z)+R^\circledast(z)R(z))\right)^{-1}.
$$
It follows that
$$
\tau_\alpha(z)=\tau_0(z)+\tr\left(D(z)\alpha
\partial_z((I+R^\circledast(z))(I+R(z))\right).
$$
Thus, for small enough $\alpha\in\cc$ and $z\in\Gamma_+$ we have
$$
\tau_\alpha(z)=\tau_0(z)+\partial_z\log\det
\left(I+\alpha(R(z)+R^\circledast(z)+R^\circledast(z)R(z))\right).
$$
Since
$$
T_0(z)=\left[\begin{array}{cc}
r(z)&-r(z)QQ^\ast r^\circledast(z)\\[6pt]
0&-r^\circledast(z)
\end{array}\right],
$$
the fact that $\Gamma_+$ encloses $\sp(-A)\subset\cc_+$ but no point of
$\sp(A^\ast)\subset\cc_-$ implies
$$
\oint_{\Gamma_+}z\,\tau_0(z)\frac{\d z}{2\pi\i}
=\oint_{\Gamma_+}z\,\tr\left((z+A)^{-1}+(z-A^\ast)^{-1}\right)
\frac{\d z}{2\pi\i}=-\tr(A)=\frac12\tr(Q\vartheta^{-1}Q^\ast),
$$
and hence
$$
\oint_{\Gamma_+}z\tau_\alpha(z)\frac{\d z}{2\pi\i}
=\frac12\tr(Q\vartheta^{-1}Q^\ast)
-\oint_{\Gamma_+}\log\det(I+\alpha(R(z)+R^\circledast(z)
+R^\circledast(z)R(z)))\frac{\d z}{2\pi\i}.
$$
Noting that
$$
R(z)+R^\circledast(z)+R^\circledast(z)R(z)
=-Q^\ast(A^\ast-z)^{-1}\Sigma_\beta(A+z)^{-1}Q,
$$
and deforming the contour $\Gamma_+$ to the imaginary axis
(which is allowed due to the decay of the above expression as $|z|\to\infty$)
yields
$$
\tr(K_\alpha P_\alpha)
=\tfrac12\tr(Q\vartheta^{-1}Q^\ast)
+\int_{-\infty}^{\infty}\log\det(I-\alpha E(\omega))\frac{\d\omega}{2\pi}.
$$
Since both sides of the last identity are analytic functions of $\alpha$,
this identity extends to all $\alpha\in\fC_c$ and the proof of
Theorem~\ref{THM-ealpha} is complete.

\subsection{The algebraic Riccati equation}
\label{SEC-Riccati}

This section is devoted to  the study the algebraic Riccati equation
$$
\cR_\alpha(X)=
X BX-X A_\alpha-A_\alpha^\ast X-C_\alpha=0
$$
which plays a central role in the proof of Proposition~\ref{PROP-Renyi}.
We summarize our results in  the following proposition.

\bep\label{PROP-RicX}
Under Assumption (C) the following hold:
\ben
\item For $\alpha\in\fI_c$ the Riccati equation $\cR_\alpha(X)=0$
has a unique maximal solution which we denote by $X_\alpha$. It also
has a unique minimal solution, which is given by 
$-\theta X_{1-\alpha}\theta$. Moreover, 
$$
D_\alpha=A_\alpha-BX_\alpha
$$
is stable and
$$
Y_\alpha=X_\alpha+\theta X_{1-\alpha}\theta>0.
$$
\item The function 
$\fI_c\ni\alpha\mapsto X_\alpha\in L(\Xi)$ is real 
analytic, concave, and satisfies
\beq
\left\{\begin{array}{lll}
X_\alpha<0&\text{for}&\alpha\in]\12-\kappa_c,0[;\\[4pt]
X_\alpha>0&\text{for}&\alpha\in]0,\12+\kappa_c[.
\end{array}
\right.
\label{EQ-SolBounds}
\eeq
$Moreover, X_0=0$ and $X_1=\theta M^{-1}\theta$.
\item If, for some $\alpha\in\bar\fI_c$, $X\in L(\Xi)$ is a self-adjoint solution of 
$\cR_\alpha(X)=0$ and $\sp(A_\alpha-BX)\subset\bar{\cc}_-$, then 
$X$ is the unique maximal solution of $\cR_\alpha(X)=0$.
\item If $\kappa_c<\infty$, then the limits
$$
X_{\12-\kappa_c}=\lim_{\alpha\downarrow\12-\kappa_c}X_\alpha,\qquad
X_{\12+\kappa_c}=\lim_{\alpha\uparrow\12+\kappa_c}X_\alpha,
$$
exist and are non-singular. They are the maximal solutions of the corresponding limiting Riccati 
equations $\cR_{\12\pm\kappa_c}(X_{\12\pm\kappa_c})=0$.
\item If $X\in L(\Xi)$ is self-adjoint and satisfies $\cR_\alpha(X)\le0$ for 
some $\alpha\in\bar\fI_c$, then $X\le X_\alpha$.
\item For all $\alpha\in\bar\fI_c$ the pair $(D_\alpha,Q)$ 
is controllable and $\sp(D_\alpha)=\sp(K_\alpha)\cap\bar{\cc}_-$. Moreover, 
for any $\beta\in L(\Xi)$ satisfying Conditions~\eqref{EQ-betadef}
one has
\beq
e(\alpha)=\tfrac12\tr(D_\alpha+\tfrac12 Q\vartheta^{-1}Q^\ast)
=-\tfrac12\tr(Q^\ast(X_\alpha-\alpha\beta)Q).
\label{EQ-ealphaRicForm}
\eeq
\item For  $t>0$ set 
$$
M_{\alpha,t}=\int_0^t\e^{sD_\alpha}B\e^{sD_\alpha^\ast}\d s>0.
$$
Then for all $\alpha\in\bar\fI_c$
$$
\lim_{t\to\infty}M_{\alpha,t}^{-1}
=\inf_{t>0}M_{\alpha,t}^{-1}
=Y_\alpha\ge0,
$$
and $\Ker(Y_\alpha)$ is the spectral subspace of $D_\alpha$ corresponding to 
its imaginary eigenvalues. 
\item Set $\Delta_{\alpha,t}=M_{\alpha,t}^{-1}-Y_\alpha$.
For all $\alpha\in\bar\fI_c$, one has
\beq
\e^{tD_\alpha^\ast}M_{\alpha,t}^{-1}\e^{tD_\alpha}
=\theta\Delta_{1-\alpha,t}\theta,
\label{EQ-MagicDual}
\eeq
and
$$
\lim_{t\to\infty}\frac1t\log\det(\Delta_{\alpha,t})
=4e(\alpha)-\tr(Q\vartheta^{-1}Q^\ast).
$$
In particular, for
$\alpha\in\fI_c$, $\Delta_{\alpha,t}\to0$ exponentially fast as $t\to\infty$.
\item Let $\widetilde{D}_\alpha=\theta D_{1-\alpha}\theta$. Then
$$
Y_\alpha\e^{t\widetilde{D}_\alpha}=\e^{tD_\alpha^\ast} Y_\alpha
$$
for all $\alpha\in\bar\fI_c$ and $t\in\rr$.
\item Let $W_\alpha=\alpha X_1-X_\alpha$. Then 
$$
\left\{\begin{array}{lll}
W_\alpha\le0&\text{for}&|\alpha-\12|\le\12;\\[4pt]
W_\alpha\ge0&\text{for}&\12\le|\alpha-\12|\le\kappa_c;
\end{array}
\right.
$$
 and
$Y_\alpha+W_\alpha>0$ for all $\alpha\in\bar\fI_c$.
\item Set
$\bar\vartheta=\tfrac12(\vartheta_{\rm max}+\vartheta_{\rm min})$
and $\Delta=\vartheta_{\rm max}-\vartheta_{\rm min}$. Then the
following lower bound holds
$$
\kappa_c\ge\kappa_0=\frac{\bar\vartheta}{\Delta}>\frac12.
$$
Moreover, the maximal solution satisfies
\beq
X_\alpha\ge\left\{
\begin{array}{ll}
\alpha\vartheta_{\rm min}^{-1}&\mbox{for }
\alpha\in[\tfrac12-\kappa_0,0];\\[6pt]
\alpha\vartheta_{\rm max}^{-1}&\mbox{for }
\alpha\in[0,\frac12+\kappa_0].
\end{array}
\right.
\label{EQ-XalphaLowerBound}
\eeq
\item Assume that $\kappa_c=\kappa_0$ and that the steady state covariance 
satisfies the strict inequalities (recall~\eqref{EQ-Mbounds})
$$
\vartheta_{\rm min}<M<\vartheta_{\rm max}.
$$
Then Condition~(R) is satisfied.
\een
\eep

\ber\label{REM-EquilibriumXalpha}
In the equilibrium case 
$\vartheta_{\rm min}=\vartheta_{\rm max}=\vartheta_0$
it follows from Part~(11) that  $\kappa_c=\infty$.
One easily checks that in this case
$$
X_\alpha=\alpha\vartheta_0^{-1}I,\qquad 
\theta X_{1-\alpha}\theta=(1-\alpha)\vartheta_0^{-1}I,\qquad
Y_\alpha=\vartheta_0^{-1}I,\qquad
W_\alpha=0,\qquad
D_\alpha=A.
$$
\eer

\smallskip
{\noindent\bf Proof.} For the reader convenience, we have collected the well 
known results on algebraic Riccati equations needed for the proof in the
\hyperref[SEC-Appendix]{Appendix}.

We denote by $\cH$ the complex Hilbert space
$\cc\Xi\oplus\cc\Xi$ on which the Hamiltonian matrix $K_\alpha$ acts and 
introduce the unitary operators 
$$
\Theta=\left[\begin{array}{cc}
0&\theta\\
\theta&0
\end{array}\right],
$$
acting on the same Hilbert space.
We have already observed in the proof of Theorem~\ref{THM-ealpha}
that for $\alpha\in\rr$ the spectrum of $K_\alpha$ is symmetric w.r.t.\;the 
real axis and the imaginary axis. The time-reversal covariance relations
\beq
\theta A_\alpha\theta=A_\alpha^\ast=-A_{1-\alpha},\qquad
\theta B\theta=B^\ast=B,\qquad
\theta C_\alpha\theta=C_\alpha^\ast=C_\alpha=C_{1-\alpha},
\label{EQ-thetaCovariance}
\eeq
which follow easily from the definitions of the operators $A_\alpha$, $B$, 
$C_\alpha$ (recall Eq.~\eqref{EQ-AQdef}, \eqref{EQ-BDef} 
and~\eqref{EQ-ACdef}), further yield 
$\Theta K_\alpha-K_{1-\alpha}^\ast\Theta=0$ which implies
\beq
\sp(K_\alpha)=\sp(K_{1-\alpha}).
\label{EQ-spKalphasymmetry}
\eeq

(1) By Theorem~\ref{THM-ealpha}~(5), $\sp(K_\alpha)\cap\i\rr=\emptyset$
for $\alpha\in\fI_c$ and the existence and uniqueness of the minimal/maximal 
solution of $\cR_\alpha(X)=0$ follows from Corollary~\ref{COR-RicExist}.
The relation between minimal and maximal solutions follows from the identity 
$$
\cR_\alpha(\theta X\theta)=\theta\cR_{1-\alpha}(-X)\theta,
$$
which is a direct consequence of Eq.~\eqref{EQ-thetaCovariance}.
The maximal solution $X_\alpha$ is related to the 
spectral subspace $\cH_-(K_\alpha)$ of $K_\alpha$ for the part of its 
spectrum in the open left half-plane $\cc_-$ by
\beq
\cH_-(K_\alpha)=\Ran\left[\begin{array}{c}I\\X_{\alpha}\end{array}\right],
\label{EQ-SpectralSubspaces}
\eeq
see Section~\ref{APP-Gap}. In particular $\sp(D_\alpha)=\sp(K)\cap\cc_-$.

The matrix $Y_\alpha=X_\alpha-\theta X_{1-\alpha}\theta$ is called the  {\sl gap} of the equation
$\cR_\alpha(X)=0$. It is obviously non-negative. It  has the remarkable 
property  that for any solution $X$, $\Ker(Y_\alpha)$ is the spectral subspace 
of $A_\alpha-BX$ for the part of its spectrum in $\i\rr$ 
(Theorem~\ref{THM-RicGap}~(1)). Since $\sp(D_\alpha)\subset\cc_-$, we must 
have $Y_\alpha>0$.

(2) One deduces from Eq.~\eqref{EQ-SpectralSubspaces} that the spectral 
projection of $K_\alpha$ for the part of its spectrum in $\cc_+$ is given by
$$
P_\alpha=\left[\begin{array}{c}I\\X_\alpha\end{array}\right]
Y_\alpha^{-1}
\left[\begin{array}{cc}\theta X_{1-\alpha}\theta&I\end{array}\right]
=\left[\begin{array}{cc}I-Y_\alpha^{-1}X_\alpha&Y_\alpha^{-1}\\
X_\alpha(I-Y_\alpha^{-1}X_\alpha)&X_\alpha Y_\alpha^{-1}
\end{array}\right].
$$
As already noticed in the proof of Theorem~\ref{THM-ealpha}, $P_\alpha$ is an 
analytic function of $\alpha$ in the cut plane $\fC_c\supset\fI_c$. 
It follows that $Y_\alpha^{-1}$ and $X_\alpha Y_\alpha^{-1}$ are real 
analytic on $\fI_c$. The same holds for  $Y_\alpha$ and 
$X_\alpha=X_\alpha Y_\alpha^{-1}Y_\alpha$.

To prove concavity we shall invoke the implicit function theorem to compute
the first and second derivatives $X'_\alpha$ and $X''_\alpha$ of the maximal
solution. To this end, we must show that the derivative $D\cR_\alpha$ of the 
map $X\mapsto\cR_\alpha(X)$ at $X=X_\alpha$ is injective. 
A simple calculation shows that
$$
D\cR_\alpha: Z\mapsto -ZD_\alpha-D_\alpha^\ast Z.
$$
By (1) one has $\sp(D_\alpha)\subset\cc_-$ for $\alpha\in\fI_c$.
It follows that for any $L\in L(\Xi)$ the Lyapunov equation
$D\cR_\alpha Z=L$ has the unique solution
$$
Z=\int_0^\infty\e^{tD_\alpha^\ast}L\,\e^{tD_\alpha}\d t
$$
(see, e.g., Section~5.3 in~\cite{LR}). This ensures the applicability of
the implicit function theorem and a straightforward calculation yields
the following expressions valid for all $\alpha\in\fI_c$:
\begin{align}
X_\alpha'&=\int_0^\infty\e^{tD_\alpha^\ast}
\left(X_\alpha B\beta+\beta BX_\alpha
+(1-2\alpha)\beta B\beta\right)\e^{tD_\alpha}\d t,
\label{EQ-Xalphaprime}\\[4pt]
X_\alpha''&=-2\int_0^\infty\e^{tD_\alpha^\ast}(X_\alpha'-\beta)
B(X_\alpha'-\beta)\e^{tD_\alpha}\d t.
\label{EQ-Xalphasecond}
\end{align}
From \eqref{EQ-Xalphasecond} we deduce $X_\alpha''\le0$ which yields concavity.

We shall now prove the inequalites~\eqref{EQ-SolBounds}, using again
the Lyapunov equation. Indeed, one can rewrite the Riccati equation
$\cR_\alpha(X_\alpha)=0$ in the  following two distinct forms:
\begin{align}
X_\alpha A_\alpha+A_\alpha^\ast X_\alpha&=X_\alpha BX_\alpha-C_\alpha,
\label{EQ-Lyap1}\\[4pt]
X_\alpha D_\alpha+D_\alpha^\ast X_\alpha&=-X_\alpha BX_\alpha-C_\alpha.
\label{EQ-Lyap2}
\end{align}
Recall that Condition~(C) implies
$\sp(A_\alpha)\subset\cc_-$ for $\alpha<0$ (as established at the beginning of
the proof of Theorem~\ref{THM-ealpha}~(5)). It follows from 
Eq.~\eqref{EQ-Lyap1} that
\beq
X_\alpha=-\int_0^\infty\e^{tA_\alpha^\ast}
(X_\alpha BX_\alpha-C_\alpha)\e^{tA_\alpha}\d t
\le\alpha(1-\alpha)
\int_0^\infty\e^{tA_\alpha^\ast}Q\vartheta^{-2}Q^\ast\e^{tA_\alpha}\d t.
\label{EQ-Xalpha}
\eeq
Since $(A_\alpha^\ast,Q)$ is controllable, we can conclude that $X_\alpha<0$ 
for $\alpha\in]\12-\kappa_c,0[$.

Similarly, for $\alpha>1$, $\sp(A_\alpha)\subset\cc_+$ and 
Eq.~\eqref{EQ-Lyap1} leads to
\beq
X_\alpha=\int_0^\infty\e^{-tA_\alpha^\ast}
(X_\alpha BX_\alpha-C_\alpha)\e^{-tA_\alpha}\d t
\ge\alpha(\alpha-1)
\int_0^\infty\e^{-tA_\alpha^\ast}Q\vartheta^{-2}Q^\ast\e^{-tA_\alpha}\d t.
\label{EQ-Lyap3}
\eeq
Controllability again yields $X_\alpha>0$ for 
$\alpha\in]1,\12+\kappa_c[$.

Finally, for $\alpha\in]0,1[$ we use Eq.~\eqref{EQ-Lyap2} and the fact
that $D_\alpha$ is stable (established in  Part~(1)) to obtain
$$
X_\alpha=\int_0^\infty\e^{tD_\alpha^\ast}
(X_\alpha BX_\alpha+C_\alpha)\e^{tD_\alpha}\d t
\ge\alpha(1-\alpha)
\int_0^\infty\e^{tD_\alpha^\ast}Q\vartheta^{-2}Q^\ast\e^{tD_\alpha}\d t.
$$
It follows that $X_\alpha\ge0$ for $\alpha\in]0,1[$. To show that $X_\alpha>0$,
let $u\in\Ker X_\alpha$. From~\eqref{EQ-Lyap1} we infer
$(u,C_\alpha u)=0$ and hence $u\in\Ker C_\alpha=\Ker Q^\ast$. 
Using~\eqref{EQ-Lyap1} again, we deduce
$A_\alpha u\in\Ker X_\alpha$. Thus, we conclude that
$u\in\Ker Q^\ast A_\alpha^n$ for all $n\ge0$ and~\eqref{EQ-Control} yields that 
$u=0$.

From $X_0=\lim_{\alpha\uparrow0}X_\alpha\le0$ and
$X_0=\lim_{\alpha\downarrow0}X_\alpha\ge0$, we deduce $X_0=0$.

To prove the last assertion, we deduce from~\eqref{EQ-Lyap1} and identities  $A_1=-A^\ast=-\theta A\theta$,  $C_1=0$, that 
$\widehat{M}=\theta X_1^{-1}\theta$ satisfies 
the Lyapunov equation $A\widehat{M}+\widehat{M}A^\ast+B=0$. Since $A$ is 
stable, this equation has a unique solution and Lemma~\ref{LEM-Control_M} (5)
yields $\widehat{M}=M$.

(3) is a well known property of the Riccati equation (Theorem~\ref{THM-RicMax}~(3)).

(4) Since $X_\alpha$ is concave and vanishes at $\alpha=0$, the function
$\alpha\mapsto X_\alpha-\alpha X_0'$ is monotone decreasing/increasing for
$\alpha$ negative/positive. Thus, to prove the 
existence of the limits $X_{\12\pm\kappa_c}$ it suffices to show that
the set $\{X_\alpha\,|\,\alpha\in\fI_c\}$ is bounded in $L(\Xi)$.  For positive $\alpha$, this
follows directly from Part~(2) which implies $0\le X_\alpha\le\alpha X_0'$.
For negative $\alpha$, taking the trace on
both sides of the first equality in Eq.~\eqref{EQ-Xalpha} and using the
fact that $C_\alpha\le0$, we obtain
$$
\tr(X_\alpha)=-\int_0^\infty\tr((X_\alpha BX_\alpha-C_\alpha)
\e^{tA_\alpha^\ast}\e^{tA_\alpha})\d t
\ge-\tr(X_\alpha BX_\alpha-C_\alpha)\int_0^\infty\|\e^{tA_\alpha}\|^2\d t.
$$
Thus, an upper bound on $\tr(X_\alpha BX_\alpha-C_\alpha)$ will conclude the proof. 
Taking the trace of Riccati's equation yields
$$
\tr(X_\alpha BX_\alpha-C_\alpha)=\tr(X_\alpha(A_\alpha+A_\alpha^\ast))
=(2\alpha-1)\tr(X_\alpha Q\vartheta^{-1}Q^\ast)
\le\frac{2\alpha-1}{\vartheta_\mathrm{min}}\tr(\widehat X_\alpha),
$$
where $\widehat{X}_\alpha=Q^\ast X_\alpha Q$. Combining the last
inequality with the estimate
$$
\tr(\widehat X_\alpha)^2\le|\partial\cI|\tr(\widehat X_\alpha^2)
=|\partial\cI|\tr(Q^\ast X_\alpha QQ^\ast X_\alpha Q)
\le|\partial\cI|\,\|Q\|^2\tr(X_\alpha BX_\alpha)
$$
yields a quadratic inequality for $\tr(\widehat{X}_\alpha)$ which gives
$$
\tr(\widehat X_\alpha)\ge-(1-2\alpha)|\partial\cI|\,\|Q\|^2\vartheta_\mathrm{min}^{-1}.
$$
Summing up, we have obtained the required lower bound
$$
\tr(X_\alpha)\ge-(1-2\alpha)^2|\partial\cI|\,
\|Q\|^2\vartheta_\mathrm{min}^{-2}
\int_0^\infty\|\e^{tA_\alpha}\|^2\d t.
$$
By continuity, we clearly have $\cR_{\12\pm\kappa_c}(X_{\12\pm\kappa_c})=0$.
Continuity also implies that $\sp(D_{\12\pm\kappa_c})\subset\bar{\cc}_-$
and the maximality of $X_{\12\pm\kappa_c}$ follows from Part~(3).

Since $C_{\12\pm\kappa_c}\le0$, the fact that $X_{\12\pm\kappa_c}$ is regular 
follows from the same argument we have used to prove the regularity of
$X_\alpha$ for $\alpha\in]0,1[$.

(5) is another well known property of the Riccati equation (Theorem~\ref{THM-RicGap}~(3)).

(6) Since $D_\alpha=A+Q(\alpha\vartheta^{-1}Q^\ast-Q^\ast X_\alpha)$, the controllability of 
$(D_\alpha,Q)$ follows from that of $(A,Q)$. The relation between $\sp(K_\alpha)$ and
$\sp(D_\alpha)$ is a direct consequence of the relation
$$
-K_\alpha\left[\begin{array}{c}I\\X_{\alpha}\end{array}\right]=
\left[\begin{array}{c}I\\X_{\alpha}\end{array}\right]D_\alpha,
$$
which follows from Eq.~\eqref{EQ-SpectralSubspaces}. Formula~\eqref{EQ-ealphaRicForm}
is obtained by combining this information with Eq.~\eqref{EQ-ealphaSpecForm}. The last
assertion is deduced from  controllability of $(D_\alpha,Q)$ in the same way as in the proof
of Lemma~\ref{LEM-Control_M}~(1).

(7) To prove the existence of the limit, we note that (6) implies that
for any $\alpha\in\bar\fI_c$ and $t_0>0$ the function 
$[t_0,\infty[\ni t\mapsto M_{\alpha,t}^{-1}$ takes strictly positive values 
and is bounded and decreasing. Thus, we have
$$
Z_\alpha=
\lim_{t\to\infty}M_{\alpha,t}^{-1}=\inf_{t>0}M_{\alpha,t}^{-1}\ge0.
$$
Since $M_{\alpha,t}^{-1}$ is easily 
seen to satisfy the differential Riccati equation
\beq
\frac{\d\ }{\d t}M_{\alpha,t}^{-1}=
-\left(M_{\alpha,t}^{-1}BM_{\alpha,t}^{-1}+M_{\alpha,t}^{-1}D_\alpha
+D_\alpha^\ast M_{\alpha,t}^{-1}\right),
\label{EQ-DiffRic}
\eeq
it follows that for any $t>0$ and $\tau\ge0$
$$
M_{\alpha,t}^{-1}-M_{\alpha,t+\tau}^{-1}
=\int_0^\tau\left(M_{\alpha,t+s}^{-1}BM_{\alpha,t+s}^{-1}
+M_{\alpha,t+s}^{-1}D_\alpha
+D_\alpha^\ast M_{\alpha,t+s}^{-1}\right)\d s.
$$
Letting $t\to\infty$, we conclude that $Z_\alpha$ satisfies
\beq
Z_\alpha BZ_\alpha+Z_\alpha D_\alpha+D_\alpha^\ast Z_\alpha=0.
\label{EQ-RicY}
\eeq
Expressing the last equation in terms of 
$V_\alpha=\theta(Z_\alpha-X_\alpha)\theta$ and 
using~\eqref{EQ-thetaCovariance}, we derive
$\cR_{1-\alpha}(V_\alpha)=0$. By a well known property of
Lyapunov equation (see, e.g., Theorem~4.4.2 in~\cite{LR}), one has
$\sp(D_\alpha+BM_{\alpha,t}^{-1})\subset\cc_+$ for all $t>0$, which
implies
$\sp(D_\alpha+BZ_\alpha)\subset\bar {\cc}_+$. Since 
$D_\alpha+BZ_\alpha=-\theta(A_{1-\alpha}-BV_\alpha)\theta$,
we have $\sp(A_{1-\alpha}-BV_\alpha)\subset\bar{\cc}_-$.
From Part~(3) we conclude that $V_\alpha$ is the maximal solution
to the Riccati equation $\cR_{1-\alpha}(X)=0$, i.e., that
$V_\alpha=X_{1-\alpha}$. Thus, 
$$
Z_\alpha=X_\alpha+\theta X_{1-\alpha}\theta=Y_\alpha,
$$
is the gap of the  Riccati equation. It is a well known
property of this gap that
$\Ker(Y_\alpha)$ is the spectral subspace of $D_\alpha$ associated
to its imaginary eigenvalues (Theorem~\ref{THM-RicGap}~(1)).

(8) Combining~\eqref{EQ-DiffRic} and~\eqref{EQ-RicY}, one shows that 
$\Delta_{\alpha,t}=M_{\alpha,t}^{-1}-Y_\alpha$ 
satisfies the differential Riccati equation
\beq
\frac{\d\ }{\d t}\Delta_{\alpha,t}=-\Delta_{\alpha,t}B\Delta_{\alpha,t}
+\Delta_{\alpha,t}\widetilde{D}_\alpha
+\widetilde{D}_\alpha^\ast\Delta_{\alpha,t},
\label{EQ-DeltaRiccati}
\eeq
where $\widetilde{D}_\alpha=-(A_\alpha+B\theta X_{1-\alpha}\theta)
=\theta D_{1-\alpha}\theta$. Since
$$
\Delta_{\alpha,t}^{-1}=(I-M_{\alpha,t}Y_\alpha)^{-1}M_{\alpha,t},
$$
we further have $\lim_{t\to0}\Delta_{\alpha,t}^{-1}=0$. We deduce that
$S_{\alpha,t}=\Delta_{\alpha,t}^{-1}$ satisfies the linear Cauchy problem
$$
\frac{\d\ }{\d t}S_{\alpha,t}
=B-\widetilde{D}_\alpha S_{\alpha,t}
-S_{\alpha,t}\widetilde{D}_\alpha^\ast,\qquad 
S_{\alpha,0}=0,
$$
whose solution is easily seen to be given by
\begin{align*}
S_{\alpha,t}
&=\int_0^t\e^{-s\widetilde{D}_\alpha}B\e^{-s\widetilde{D}_\alpha^\ast}\d s
=\theta\left(\int_0^t\e^{-sD_{1-\alpha}}B\e^{-sD_{1-\alpha}^\ast}\d s
\right)\theta\\
&=\theta\e^{-tD_{1-\alpha}}\left(\int_0^t\e^{sD_{1-\alpha}}
B\e^{sD_{1-\alpha}^\ast}\d s
\right)\e^{-tD_{1-\alpha}^\ast}\theta\\[4pt]
&=\theta\e^{-tD_{1-\alpha}}M_{1-\alpha,t}\e^{-tD_{1-\alpha}^\ast}\theta.
\end{align*}
We thus conclude that
$$
\Delta_{\alpha,t}
=\theta\e^{tD_{1-\alpha}^\ast}M_{1-\alpha,t}^{-1}\e^{tD_{1-\alpha}}\theta,
$$
which immediately yields~\eqref{EQ-MagicDual}.

Since $\Delta_{\alpha,t}$ is strictly positive for $t>0$, we infer from 
Eq.~\eqref{EQ-DeltaRiccati} that
\begin{align*}
\frac{\d\ }{\d t}\log\det(\Delta_{\alpha,t})
&=\tr(\dot\Delta_{\alpha,t}\Delta_{\alpha,t}^{-1})
=-\tr(\Delta_{\alpha,t}B-\tilde D_\alpha-\tilde D_\alpha^\ast)\\
&=-\tr(Q^\ast\Delta_{\alpha,t}Q)+2\tr(D_{1-\alpha}).
\end{align*}
By Part~(3) and Theorem~\ref{THM-ealpha}~(5), we have 
$$
\tr(D_{1-\alpha})=-\12\sum_{\lambda\in\sp(K_{1-\alpha})}|\Re\lambda|m_\lambda
=2e(1-\alpha)-\frac12\tr(Q\vartheta^{-1}Q^\ast)
=2e(\alpha)-\frac12\tr(Q\vartheta^{-1}Q^\ast).
$$
Since $\Delta_{\alpha,t}\to0$ for $t\to\infty$, given $\epsilon>0$ there 
exists $t_0>0$ such that
$$
4e(\alpha)-\tr(Q\vartheta^{-1}Q^\ast)
-\epsilon\le\frac{\d\ }{\d t}\log\det(\Delta_{\alpha,t})
\le4e(\alpha)-\tr(Q\vartheta^{-1}Q^\ast)
$$
for all $t>t_0$. It is straightforward to derive from these estimates that 
$$
\lim_{t\to\infty}\frac1t\log\det(\Delta_{\alpha,t})
=4e(\alpha)-\tr(Q\vartheta^{-1}Q^\ast).
$$

(9) Using~\eqref{EQ-thetaCovariance}, one rewrites the Riccati 
equation~\eqref{EQ-RicY} as
\begin{align*}
D_\alpha^\ast Y_\alpha&=-Y_\alpha(D_\alpha+BY_\alpha)
=-Y_\alpha(A_\alpha+B(Y_\alpha-X_\alpha))\\
&=-Y_\alpha(A_\alpha+B\theta X_{1-\alpha}\theta)
=-Y_\alpha\theta(-A_{1-\alpha}+BX_{1-\alpha})\theta\\
&=Y_\alpha\theta D_{1-\alpha}\theta=Y_\alpha\widetilde{D}_\alpha.
\end{align*}
Thus, the result immediately follows from the fact that
$$
\frac{\d\ }{\d t}\,\e^{tD_\alpha^\ast}Y_\alpha\e^{-t\widetilde{D}_\alpha}
=\e^{tD_\alpha^\ast}
(D_\alpha^\ast Y_\alpha-Y_\alpha\widetilde{D}_\alpha)
\e^{-t\widetilde{D}_\alpha}=0.
$$

(10) For any $u\in\Xi$ we infer from Parts~(2) and~(4) that the function
$\alpha\mapsto(u,W_\alpha u)$ 
is convex, real analytic on the interval $\fI_c$, and 
continuous on its closure. Since it vanishes for $|\alpha-\12|=\12$ 
one has either $(u,W_\alpha u)=0$ for all $\alpha\in\bar\fI_c$ or
$(u,W_\alpha u)<0$ for $|\alpha-\12|<\12$ and
$(u,W_\alpha u)>0$ for $\12<|\alpha-\12|\le\kappa_c$. This proves
the first assertion.

Since $Y_\alpha+W_\alpha=\alpha X_1+\theta X_{1-\alpha}\theta$, we deduce 
from Part~(2) that $Y_\alpha+W_\alpha>0$ for $|\alpha-\12|\le\12$. Consider now
$\12<|\alpha-\12|\le\kappa_c$. If $u\in\Xi$ is such that 
$(u,W_\alpha u)>0$, then Part~(7) yields $(u,(Y_\alpha+W_\alpha)u)>0$. 
Thus, it remains 
to consider the case of $u\in\Xi$ such that $(u,W_\alpha u)=0$ for all
$\alpha\in\bar\fI_c$. Using~\eqref{EQ-Xalphasecond} we get that 
$$
(u,W_\alpha''u)
=-(u,X_\alpha''u)
=2\int_0^\infty|Q^\ast(X_\alpha'-\beta)\e^{tD_\alpha}u|^2\d t=0
$$
for $\alpha\in\fI_c$. Since $QQ^\ast(X_\alpha'-\beta)=-D_\alpha'$, this 
further implies $D_\alpha'\e^{tD_\alpha}u=0$ for all 
$(\alpha,t)\in\fI_c\times\rr$. Duhamel's  formula 
$$
\frac{\d\ }{\d\alpha}\e^{tD_\alpha}u
=\int_0^t\e^{(t-s)D_\alpha}D_\alpha'\e^{sD_\alpha}u\,\d s=0
$$
allows us to conclude that $\e^{tD_\alpha}u=\e^{tD_0}u=\e^{tA}u$, a relation 
which extends by continuity to all $(\alpha,t)\in\bar\fI_c\times\rr$.
Thus,
$$
\lim_{t\to\infty}\e^{tD_\alpha}u=\lim_{t\to\infty}\e^{tA}u=0,
$$
which, using~(7) again, further implies that $u\not\in\Ker(Y_\alpha)$
and hence $(u,(Y_\alpha+W_\alpha)u)=(u,Y_\alpha u)>0$.

\begin{figure}
\centering
\includegraphics[scale=0.6]{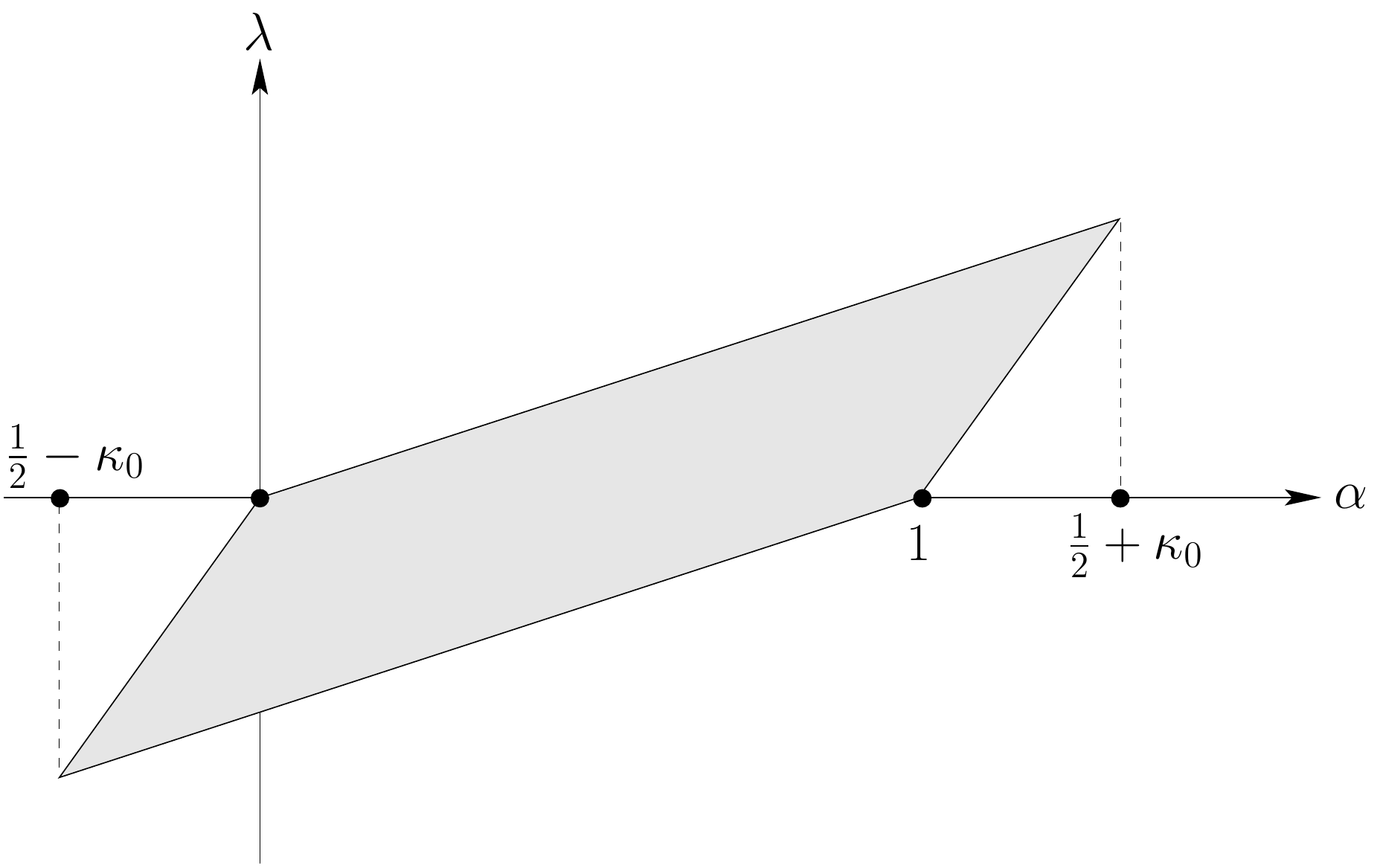}
\caption{The parallelogram $\cP$.}
\label{Fig-Paral}
\end{figure}

(11) For $\lambda\in\rr$, one has
$$
\cR_\alpha(\lambda I)=Q\vartheta^{-1}
\left(\lambda\vartheta-(\alpha-1)\right)
\left(\lambda\vartheta-\alpha\right)
\vartheta^{-1}Q^\ast,
$$
so that $\cR_\alpha(\lambda I)\le0$ iff 
$\alpha-1\le\lambda\vartheta\le\alpha$. It follows that
$\cP=\{(\alpha,\lambda)\in\rr^2\,|\,\cR_\alpha(\lambda I)\le0\}$ is the
closed parallelogram limited by the 4 lines (see Figure~\ref{Fig-Paral})
$$
\lambda=\frac{\alpha}{\vartheta_{\rm max}},\quad
\lambda=\frac{\alpha}{\vartheta_{\rm min}},\quad
\lambda=\frac{\alpha-1}{\vartheta_{\rm max}},\quad
\lambda=\frac{\alpha-1}{\vartheta_{\rm min}}.
$$
The projection of $\cP$ on the $\alpha$-axis is the closed interval
$[\tfrac12-\kappa_0,\tfrac12+\kappa_0]$. Thus, Theorem~\ref{THM-ARI} implies
that the Riccati equation has a self-adjoint solution for all 
$\alpha\in[\tfrac12-\kappa_0,\tfrac12+\kappa_0]$. 
By Theorem~\ref{THM-RicMax}~(2) it also has a maximal solution $X_\alpha$ 
which, by Theorem~\ref{THM-RicGap}~(3), satisfies the lower 
bound~\eqref{EQ-XalphaLowerBound}. From this lower bound we further deduce
that for $\alpha\in[0,\tfrac12+\kappa_0[$, the gap satisfies
$$
Y_\alpha=X_\alpha+\theta X_{1-\alpha}\theta
\ge\frac{\alpha}{\vartheta_{\rm max}}+\frac{1-\alpha}{\vartheta_{\rm min}}
=\frac{\Delta}{\vartheta_{\rm max}
\vartheta_{\rm min}}\left(\tfrac12+\kappa_0-\alpha\right)>0.
$$
Since $\Ker Y_{\frac12+\kappa_c}\not=\{0\}$
by Parts~(6) and~(7), we conclude that $\kappa_c\ge\kappa_0$.

(12) The concavity of $R_\alpha=X_\alpha+(1-\alpha)X_1$ and the fact that
$R_0=R_1=X_1>0$ imply that for $|\alpha-\12|\le\12$ one has $R_\alpha\ge X_1>0$. 
For $\12<\alpha-\12\le\kappa_0$, Part~(11) gives 
$X_\alpha\ge\alpha\vartheta_{\mathrm{max}}^{-1}$. Since $M>\vartheta_{\rm min}$,
Part~(2) yields $X_1=\theta M^{-1}\theta<\vartheta_{\mathrm{min}}^{-1}$ and hence
$$
R_\alpha>\frac{\alpha}{\vartheta_{\mathrm{max}}}
+\frac{1-\alpha}{\vartheta_{\mathrm{min}}}
=\frac{\kappa_0-(\alpha-\12)}{\Delta(\kappa_0^2-\tfrac14)}\ge0.
$$
The case $-\kappa_0\le\alpha-\12<-\12$ is similar.
\hfill\qed

\subsection{Proof of Proposition~\ref{PROP-Renyi}}
\label{SSECT_Proof_of_PROP-Renyi}

\subsubsection{A Girsanov transformation}

By Proposition~\ref{PROP-RicX}, for $\alpha\in\bar\fI_c$ we have
$A=D_\alpha+QQ^\ast(X_\alpha-\alpha\beta)$, and we can
rewrite the equation of motion~\eqref{EQ-The_SDE} as
\beq
\d x(t)=D_\alpha x(t)\d t+Q\d w_\alpha(t),
\label{EQ-SecondSDE}
\eeq
where
$$
w_\alpha(t)=w(t)-\int_0^tQ^\ast(\alpha\beta-X_\alpha)x(s)\d s.
$$
Let $Z_\alpha(t)$ be the stochastic exponential of the local martingale
$$
\eta_\alpha(t)=\int_0^tQ^\ast(\alpha\beta-X_\alpha)x(s)\cdot\d w(s).
$$
Combining the Riccati equation with the
relations $\beta QQ^\ast=QQ^\ast\beta=Q\vartheta^{-1}Q$ and
$\beta QQ^\ast\beta=Q\vartheta^{-2}Q^\ast$, we derive
$$
\12|Q^\ast(\alpha\beta-X_\alpha)x|^2
=-\alpha\sigma_\beta(x)-(\alpha\beta-X_\alpha)x\cdot Ax,
$$
and we can  write the quadratic variation of $\eta_\alpha$ as
$$
\12[\eta_\alpha](t)=-\alpha\int_0^t\sigma_\beta(x(s))\d s
-\int_0^t(\alpha\beta-X_\alpha)x(s)\cdot Ax(s)\d s.
$$
Hence
$$
\eta_\alpha(t)-\12[\eta_\alpha](t)
=\int_0^t(\alpha\beta-X_\alpha)x(s)\cdot\d x(s)
+\alpha\int_0^t\sigma_\beta(x(s))\d s.
$$
The  It\^o calculus and Proposition~\ref{PROP-Strict_Positivity}~(3)
give
$$
\eta_\alpha(t)-\12[\eta_\alpha](t)
=-\left(\lambda_\alpha t
+\alpha\fS^t+\chi_\alpha(x(t))-\chi_\alpha(x(0))\right),
$$
with $\lambda_\alpha=\12\tr(QQ^\ast(\alpha\beta-X_\alpha))$ and
$$
\chi_\alpha(x)=\12 x\cdot X_\alpha x.
$$
Finally, we note that Proposition~\ref{PROP-RicX}~(6) yields
$$
\lambda_\alpha
=-\frac12\tr(Q^\ast(X_\alpha-\alpha\beta)Q)=e(\alpha).
$$

\bel\label{LEM-ZalphaMartingale}
The process
\beq
Z_\alpha(t)=\cE(\eta_\alpha)(t)
=\e^{-\left[e(\alpha) t+\alpha\fS^t
+\chi_\alpha(x(t))-\chi_\alpha(x(0))\right]}
\label{EQ-Zdef}
\eeq
is a $\PP_x$-martingale for all $x\in\Xi$.
\eel
{\noindent\bf Proof.}
We wish to apply the Girsanov theorem; see Section~3.5 in~\cite{KS}. However, it is not clear that  the Novikov condition is satisfied on a given finite interval. To overcome this difficulty, we follow the argument used in the proof of Corollary~5.14 in~\cite[Chapter~3]{KS}. 

Fix $\tau>0$. By Lemma~\ref{LEM-Cov}, $\{x(t)-\e^{tA}x\}_{t\in[0,\tau]}$ 
is a centered 
Gaussian process under the law~$\PP_x$. Since
$$
\int_{s}^{s'}|Q^\ast(\alpha\beta-X_\alpha)x(t)|^2\d t\le C|s-s'|\left(|x|^2+\sup_{t\in[0,\tau]}|x(t)-\e^{tA}x|^2\right)
$$
for some constant $C$, Fernique's theorem implies that there exists 
$\delta>0$ such that
$$
\EE_x\left[\exp\left(\12\int_{s}^{s'}
|Q^\ast(\alpha\beta-X_\alpha)x(t)|^2\d t\right)\right]<\infty,
$$
provided $0\le s\le s'\le \tau$ and $s'-s<\delta$. Novikov criterion 
implies that under the same conditions, 
\begin{align*}
\EE_x&\left[
\frac{\cE(\eta_\alpha)(s')}{\cE(\eta_\alpha)(s)}\bigg|\cW_s
\right]\\
&=\EE_x\left[\exp\left(
\int_s^{s'}Q^\ast(\alpha\beta-X_\alpha)x(t)\cdot\d w(t)
-\12\int_s^{s'}|Q^\ast(\alpha\beta-X_\alpha)x(t)|^2
\d t\right)\bigg|\cW_s\right]
=1.
\end{align*}
For $0\le s\le s'\le s''\le\tau$, $s'-s<\delta$ and $s''-s'<\delta$ we deduce
\begin{align*}
\EE_x\left[\frac{\cE(\eta_\alpha)(s'')}{\cE(\eta_\alpha)(s)}\bigg|\cW_s\right]
&=\EE_x\left[\frac{\cE(\eta_\alpha)(s'')}{\cE(\eta_\alpha)(s')}
\frac{\cE(\eta_\alpha)(s')}{\cE(\eta_\alpha)(s)}
\bigg|\cW_s\right]\\[4pt]
&=\EE_x\left[\EE_x\left[\frac{\cE(\eta_\alpha)(s'')}{\cE(\eta_\alpha)(s')}\bigg|\cW_{s'}\right]
\frac{\cE(\eta_\alpha)(s')}{\cE(\eta_\alpha)(s)}
\bigg|\cW_s\right]\\[4pt]
&=\EE_x\left[\frac{\cE(\eta_\alpha)(s')}{\cE(\eta_\alpha)(s)}
\bigg|\cW_s\right]=1,
\end{align*}
and an induction argument gives
$$
\EE_x\left[\cE(\eta_\alpha)(\tau)\right]
=\EE_x\left[\frac{\cE(\eta_\alpha)(\tau)}{\cE(\eta_\alpha)(0)}\bigg|\cW_0\right]=1.
$$
Since $\tau>0$ is  arbitrary, the proof is complete.\hfill\qed

The previous lemma allows us to apply Girsanov theorem and to conclude that 
$\{w_\alpha(t)\}_{t\in[0,\tau]}$ is  a standard Wiener process under the law
$\QQ_{\alpha,\nu}^\tau[\,\cdot\,]=\EE_\nu[Z_\alpha(\tau)\,\cdot\,]$.
This change of measure will be our main tool in the next section.

\subsubsection{Completion of the proof}

From Eq.~\eqref{EQ-TwistedEp} and the results of the previous section we 
deduce that  for $\alpha\in\bar\fI_c$,
\begin{align*} 
e_t(\alpha)&=\frac1t\log\EE_\mu\left[
\left(\frac{\d\mu}{\d x}(\theta x(t))\right)^\alpha
\left(\frac{\d\mu}{\d x}(x(0))\right)^{-\alpha}
\e^{-\alpha\fS^t}
\right]\\
&=\frac1t\log\EE_\mu\left[Z_\alpha(t)
\left(\frac{\d\mu}{\d x}(\theta x(t))\right)^\alpha
\left(\frac{\d\mu}{\d x}(x(0))\right)^{-\alpha}
\e^{-\chi_\alpha(x(0))+\chi_\alpha(x(t))+e(\alpha)t}
\right]\\
&=e(\alpha)+\frac1t\log\QQ_{\alpha,\mu}^t\left[
\left(\frac{\d\mu}{\d x}(\theta x(t))\right)^\alpha
\left(\frac{\d\mu}{\d x}(x(0))\right)^{-\alpha}
\e^{-\chi_\alpha(x(0))+\chi_\alpha(x(t))}
\right],
\end{align*}
where $\chi_\alpha(x)=\frac12 x\cdot X_\alpha x$. Denoting by  $Q_\alpha^t$ the 
Markov semigroup associated with Eq.~\eqref{EQ-SecondSDE}, we can write
$$
e_t(\alpha)=e(\alpha)
+\frac1t\log\langle\eta_\alpha|Q_\alpha^t\xi_{\alpha}\rangle,
$$
where
\beq
\eta_\alpha(x)=\left(\frac{\d\mu}{\d x}(x)\right)^{1-\alpha}
\e^{-\chi_\alpha(x)},\qquad
\xi_{\alpha}(x)=\left(\frac{\d\mu}{\d x}(\theta x)\right)^{\alpha}
\e^{\chi_\alpha(x)}.
\label{EQ-falphadef}
\eeq
Thus, to prove Eq.~\eqref{EQ-ealphaForm} we must show that the
``prefactor'' $\langle\eta_\alpha|Q_\alpha^t\xi_{\alpha}\rangle$
satisfies
$$
\lim_{t\to\infty}\frac1t\log\langle\eta_\alpha|Q_\alpha^t\xi_{\alpha}\rangle
=0.
$$

To this end, let us note that the Markov semigroup for~\eqref{EQ-The_Process} can be written as 
\beq
(P^tf)(x)=\int_{\cX}f(\e^{tA}x+M_t^\12 y){\rm n}(\d y),
\label{EQ-PtForm}
\eeq
where $\rm n$ denotes the centered Gaussian measure on $\cX$ with covariance~$I$. For $\alpha\in\bar\fI_c$, this yields the representation
\beq
(Q_\alpha^tf)(x)=\det(2\pi M_{\alpha,t})^{-\frac12}\int
\e^{-\frac12|M_{\alpha,t}^{-\frac12}(y-\e^{tD_\alpha}x)|^2}f(y)\d y.
\label{EQ-QalphaForm}
\eeq
Using Eq.~\eqref{EQ-falphadef}, a simple calculation leads to
$$
\langle\eta_\alpha|Q_\alpha^t\xi_{\alpha}\rangle
=\det(2\pi M_{\alpha,t})^{-\frac12}\det(2\pi M)^{-\frac12}
\int\e^{-\frac12z\cdot N_{\alpha,t}z}
\d z
=\det(M_{\alpha,t}^{-1})^{\frac12}\det(M^{-1})^{\frac12}
\det(N_{\alpha,t})^{-\frac12},
$$
provided
$$
N_{\alpha,t}
=\left[\begin{array}{cc}
\theta(Y_{1-\alpha}+W_{1-\alpha}+\Delta_{1-\alpha,t})\theta
&-\e^{tD_\alpha^\ast}M_{\alpha,t}^{-1}\\[4pt]
-M_{\alpha,t}^{-1}\e^{tD_\alpha}&
Y_\alpha+W_\alpha+\Delta_{\alpha,t}
\end{array}\right]
$$
is positive definite. By Schur's complement formula, we have
$$
\det(N_{\alpha,t})=\det(Y_\alpha+W_\alpha+\Delta_{\alpha,t})
\det(Y_{1-\alpha}+W_{1-\alpha}+\Delta_{1-\alpha,t}-T_{\alpha,t}),
$$
where
$$
T_{\alpha,t}=\theta(\e^{tD_\alpha^\ast}M_{\alpha,t}^{-\frac12})
M_{\alpha,t}^{-\frac12}
(Y_\alpha+W_\alpha+\Delta_{\alpha,t})^{-1}
M_{\alpha,t}^{-\frac12}(M_{\alpha,t}^{-\frac12}\e^{tD_\alpha})\theta.
$$
It follows that
$$
\langle\eta_\alpha|Q_\alpha^t\xi_{\alpha}\rangle
=\left(\det(M)\det(M_{\alpha,t})
\det(Y_\alpha+W_\alpha+\Delta_{\alpha,t})
\det(Y_{1-\alpha}+W_{1-\alpha}
+\Delta_{1-\alpha,t}-T_{\alpha,t})\right)^{-\frac12}.
$$

For any $\alpha\in\bar\fI_c$, Proposition~\ref{PROP-RicX} 
implies that $Y_\alpha+W_\alpha>0$ while, as $t\to\infty$,
$M_{\alpha,t}^{-\frac12}\searrow
 Y_\alpha^\frac12$, $\Delta_{\alpha,t}\searrow0$ and 
$\|M_{\alpha,t}^{-\frac12}\e^{tD_\alpha}\|\searrow0$ monotonically 
(and exponentially fast  for $\alpha\in\fI_c$ ). It follows that
\beq
f(\alpha)=
\lim_{t\to\infty}\langle\eta_\alpha|Q_\alpha^t\xi_{\alpha}\rangle
=\left(\frac{\det(Y_\alpha)}{
\det(M)
\det(Y_\alpha+W_\alpha)
\det(Y_{1-\alpha}+W_{1-\alpha})}\right)^{\frac12}.
\label{EQ-falphaDef}
\eeq
For $\alpha\in\fI_c$, $Y_\alpha>0$, and we conclude that
\beq
\lim_{t\to\infty}\frac1t\log\langle\eta_\alpha|Q_\alpha^t\xi_{\alpha}\rangle=0.
\label{EQ-Prefactor}
\eeq
Consider now the limiting cases $\alpha=\12\pm\kappa_c$. We shall
denote by $C$ and $r$ generic positive constants which may vary from
one expression to the other. Since $Y_\alpha$ is singular, one has
$\log\det(M_{\alpha,t}^{-1})\to-\infty$. However, the obvious estimate
$\|\e^{tD_\alpha}\|\le C(1+t)^r$ implies $M_{\alpha,t}\le C(1+t)^r$ and hence
$M_{\alpha,t}^{-1}\ge C(1+t)^{-r}$ from which we conclude that 
\beq
\lim_{t\to\infty}\frac1t\log\det(M_{\alpha,t}^{-1})=0.
\label{EQ-MalphatLimit}
\eeq
It follows  that~\eqref{EQ-Prefactor} also holds in the limiting cases  $\alpha=\12\pm\kappa_c$.

By H\"older's inequality $\rr\ni\alpha\mapsto e_t(\alpha)$ is a convex 
function. The above analysis shows  that it is a proper convex function 
differentiable on $\fI_c$ for any $t>0$, and such that
$\lim_{t\to\infty}e_t(\alpha)=e(\alpha)$ for $\alpha\in\fI_c$. 
Since $\lim_{\alpha\uparrow\frac12+\kappa_c}e'(\alpha)=+\infty$ by
Theorem~\ref{THM-ealpha}~(3), the fact that
$$
\lim_{t\to\infty}e_t(\alpha)=+\infty
$$
for $\alpha\in\rr\setminus\bar\fI_c$ is a consequence of the following lemma and the
symmetry~\eqref{EQ-TheSymmetry2}.

\bel\label{LEM-infty}
Let $(f_t)_{t>0}$ be a family of proper convex functions 
$f_t:\rr\to]-\infty,\infty]$ with the following properties:
\ben
\item For each $t>0$, $f_t$ is differentiable on $]a,b[$.
\item The limit $f(\alpha)=\lim_{t\to\infty}f_t(\alpha)$ exists  for $\alpha\in]a,b[$ and is differentiable on $]a,b[$.
\item $\lim_{\alpha\uparrow b}f'(\alpha)=+\infty$.
\een
Then, for all $\alpha>b$, one has $\lim_{t\to\infty}f_t(\alpha)=+\infty$.
\eel

{\bf Proof.} By convexity, for any $\gamma\in]a,b[$ and any $\alpha\in\rr$
one has
$$
f_t(\alpha)\ge f_t(\gamma)+(\alpha-\gamma)f_t'(\gamma),
$$
and Properties~(1) and~(2) further imply
$$
\lim_{t\to\infty}f_t'(\gamma)=f'(\gamma).
$$
It follows that
\beq
\liminf_{t\to\infty}f_t(\alpha)\ge f(\gamma)+(\alpha-\gamma)f'(\gamma).
\label{EQ-convder}
\eeq
As a limit of a family of convex functions, $f$ is convex on $]a,b[$ and, hence, $\inf_{\gamma\in]a,b[}f(\gamma)>-\infty$. Thus,
Property~(3) and Inequality~\eqref{EQ-convder} yield
$$
\liminf_{t\to\infty}f_t(\alpha)\ge\liminf_{\gamma\uparrow b}
f(\gamma)+(\alpha-\gamma)f'(\gamma)=+\infty\quad
\mbox{for all $\alpha>b$}.
$$
\hfill\qed

\subsection{Proof of Proposition~\ref{PROP-GaussianCocycles}}
\label{SSECT_Proof_of_PROP-GaussianCocycles}

\paragraph{(1)} 
The required properties of the function~$g_t(\alpha)$ are consequences of more general results concerning integrals of exponentials of quadratic forms with respect to a Gaussian measure on an infinite-dimensional space. However, we shall derive here more detailed information about~$g_t(\alpha)$ which will be used later (see the proof of Theorem~\ref{THM-GaussianLDP}). 

We shall invoke Lemmata~\ref{LEM-WH} and~\ref{LEM-Cov},
and use the notations introduced in their proofs.
By Proposition~\ref{PROP-Strict_Positivity}, we can write
$$
g_t(\alpha)=\frac1t\log\int\e^{-\frac{\alpha}{2}(x|\cL_t x)}\gamma_t(\d x),
$$
where $\gamma_t$ is the Gaussian measure on $\fH_t$ with mean $T_t a$ and 
covariance $\cK_t=\cD_t\cD_t^\ast$. The convexity of $g_t$ is a consequence
of H\"older's inequality. The operator $\cL_t$, given by
\beq
(\cL_t x)(s)=-\Sigma_\beta x(s)
+\delta(s-t)(F+X_1-\beta)x(t)-\delta(s)(G+\theta X_1\theta-\beta)x(0),
\label{EQ-LtDef}
\eeq
maps $\fH_+$ to $\fH_-$ in such a way
that $(x|\cL_t y)=(\cL_t x|y)$ for all $x,y\in\Ran\cD_t$. 
It follows that the operator $\cS_t=\cD_t^\ast\cL_t\cD_t$ acting in the space $\Xi\oplus\p\fH$ is self-adjoint, and a simple calculation shows that $\cS_t-\cD_t^\ast[\beta,\Omega]\cD_t$ is finite rank, so that~$\cS_t$ is trace class. Using  explicit formulas for Gaussian measures, we derive
\beq
g_t(\alpha)=-\frac1{2t}\log\det(I+\alpha\cS_t)
-\frac\alpha{2t}(T_ta|\cL_tT_ta)
+\frac{\alpha^2}{2t}(\cD_t^\ast\cL_tT_ta|(I+\alpha\cS_t)^{-1}
\cD_t^\ast\cL_tT_ta)
\label{EQ-gtForm}
\eeq
if $I+\alpha\cS_t>0$, and $g_t(\alpha)=+\infty$ otherwise.
Set $s_-(t)=\min\sp(\cS_t)\le0$, $s_+(t)=\max\sp(\cS_t)\ge0$, and 
$$\alpha_-(t)=\left\{\begin{array}{cl}
-s_+(t)^{-1}&\mbox{if } s_+(t)>0;\\[4pt]
-\infty&\mbox{if } s_+(t)=0;
\end{array}
\right.\qquad
\alpha_+(t)=\left\{\begin{array}{cl}
-s_-(t)^{-1}&\mbox{if } s_-(t)<0;\\[4pt]
+\infty&\mbox{if } s_-(t)=0;
\end{array}
\right.
$$
so that $I+\alpha\cS_t>0$ iff 
$\alpha\in\fI_t=]\alpha_-(t),\alpha_+(t)[$.
Analyticity of~$g_t$ on $\fI_t$ follows from the Fredholm theory (e.g., see~\cite{Si1}), and a simple calculation yields
\beq
\begin{split}
g_t'(\alpha)=&-\frac1{2t}\tr((I+\alpha\cS_t)^{-1}\cS_t)
-\frac1{2t}(T_ta|\cL_tT_ta)\\
&+\frac\alpha{2t}(\cD_t^\ast\cL_tT_ta|
\left((I+\alpha\cS_t)^{-1}+(I+\alpha\cS_t)^{-2}\right)
\cD_t^\ast\cL_tT_ta).
\end{split}
\label{EQ-gtprimeform}
\eeq
Suppose $\alpha_+(t)<\infty$ and denote by $P_-$ the spectral projection
of $\cS_t$ associated to its minimal eigenvalue $s_-(t)<0$. By the 
previous formula, for any $\alpha\in[0,\alpha_+(t)[$ one has
$$
g_t'(\alpha)\ge\frac1{2t}\frac{\tr(P_-)}{\alpha_+(t)-\alpha}
-\frac1{2t}\tr((I+\alpha\cS_t)^{-1}\cS_t(I-P_-))
-\frac1{2t}(T_ta|\cL_tT_ta),
$$
which implies that $g_t'(\alpha)\to+\infty$ 
as $\alpha\to\alpha_+(t)$. The analysis of the lower bound $\alpha_-(t)$ 
is similar.

\paragraph{(2)} Is a simple 
consequence of the continuity and concavity of the maps
$$
\bar\fI_c\ni\alpha\mapsto F_\alpha
=\theta X_{1-\alpha}\theta+\alpha(X_1+F),\qquad
\bar\fI_c\ni\alpha\mapsto G_\alpha
=\widehat{N}+P_\nu(X_\alpha-\alpha(G+\theta X_1\theta))|_{\Ran(N)},
$$
and the fact that $F_0=\theta X_1\theta>0$ and 
$G_0=\widehat{N}+P_\nu X_1|_{\Ran N}>0$.

\paragraph{(3)} If $X_1+F>0$ and 
$\widehat{N}+P_\nu(X_1-G-\theta X_1\theta)|_{\Ran N}>0$, 
then we also have $F_1>0$ and $G_1>0$ and the result is again a consequence of 
the concavity of $F_\alpha$ and $G_\alpha$.

\paragraph{(4)} Proceeding as in the proof of Proposition~\ref{PROP-Renyi}, 
we start from the expression
$$
g_t(\alpha)
=e(\alpha)+\frac1t\log\langle\eta_\alpha|Q_\alpha^t\xi_\alpha\rangle,
$$
where
$$
\eta_\alpha(\d x)
=\e^{-\frac12 x\cdot(X_\alpha-\alpha(G+\theta X_1\theta))x}\nu(\d x),\qquad
\xi_\alpha(x)=\e^{-\frac12 x\cdot(-X_\alpha+\alpha(F+X_1))x}.
$$
Setting
$$
C_{\alpha,t}=\left[\begin{array}{cc}
G_\alpha+P_\nu\theta\Delta_{1-\alpha,t}\theta|_{\Ran N}
&-P_\nu\e^{tD_\alpha^\ast}M_{\alpha,t}^{-1}\\[4pt]
-M_{\alpha,t}^{-1}\e^{tD_\alpha}|_{\Ran N}&
F_\alpha+\Delta_{\alpha,t}
\end{array}\right],
$$
$$
c_{\alpha,t}=\tfrac12 a\cdot(X_\alpha-\alpha(G+\theta X_1\theta)+\theta\Delta_{1-\alpha,t}\theta)a,\quad
b_{\alpha,t}=\left[\begin{array}{c}
P_\nu(X_\alpha-\alpha(G+\theta X_1\theta)+\theta\Delta_{1-\alpha,t}\theta) a\\[4pt]
-M_{\alpha,t}^{-1}\e^{tD_\alpha}a
\end{array}\right],
$$

evaluation of a Gaussian integral leads to
\begin{align}
\langle\eta_\alpha|Q_\alpha^t\xi_\alpha\rangle
&=\det(2\pi\widehat{N}^{-1})^{-\frac12}\det(2\pi M_{\alpha,t})^{-\frac12}
\int_{\Ran(N)\oplus\Xi}
\e^{-\frac12z\cdot C_{\alpha,t}z-z\cdot b_{\alpha,t}-c_{\alpha,t}}
\d z\label{EQ-ztegral}\\
&=\det(\widehat{N}^{\frac12})\det(M_{\alpha,t})^{-\frac12}
\det(C_{\alpha,t})^{-\frac12}
\,\e^{\frac12b_{\alpha,t}\cdot C_{\alpha,t}^{-1}b_{\alpha,t}
-c_{\alpha,t}},\nonumber
\end{align}
provided $C_{\alpha,t}>0$. By Schur's complement formula, the last condition
is equivalent to
$$
F_\alpha+\Delta_{\alpha,t}>0,\qquad
G_\alpha+P_\nu\theta\Delta_{1-\alpha,t}\theta|_{\Ran N}-T_{\alpha,t}>0,
$$
where
$$
T_{\alpha,t}=P_\nu(\e^{tD_\alpha^\ast}M_{\alpha,t}^{-\12})M_{\alpha,t}^{-\12}
(F_\alpha+\Delta_{\alpha,t})^{-1}
M_{\alpha,t}^{-\12}(M_{\alpha,t}^{-\12}\e^{tD_\alpha})|_{\Ran N}.
$$
Moreover, one has
$$
\det(C_{\alpha,t})
=\det(F_\alpha+\Delta_{\alpha,t})
\det(G_\alpha
+P_\nu\theta\Delta_{1-\alpha,t}\theta|_{\Ran N}-T_{\alpha,t}).
$$
For $\alpha\in\fI_\infty$, it follows from Proposition~\ref{PROP-RicX} that
$$
\lim_{t\to\infty}T_{\alpha,t}=0,
$$
and $F_\alpha+\Delta_{\alpha,t}$ and 
$G_\alpha+P_\nu\theta\Delta_{1-\alpha,t}\theta|_{\Ran N}-T_{\alpha,t}$ are 
both positive definite for large $t$. As in the proof of 
Proposition~\ref{PROP-Renyi} we can conclude that
$$
\lim_{t\to\infty}\frac1t\log\langle\eta_\alpha|Q_\alpha^t\xi_{\alpha}\rangle=0.
$$

\paragraph{(5)} Suppose that $\alpha_+<\frac12+\kappa_c$. If
$\alpha\in]\alpha_+,\frac12+\kappa_c]$, then the 
matrix $C_{\alpha,t}$ acquires a negative eigenvalue as $t$ increases. 
Consequently, the integral in~\eqref{EQ-ztegral} diverges and 
$g_t(\alpha)=+\infty$ for large $t$, proving~\eqref{EQ-limg}. The case
$\alpha_->\frac12-\kappa_c$ and $\alpha\in[\frac12-\kappa_c,\alpha_-[$
is similar. Suppose now that $\alpha_+=\frac12+\kappa_c$. Since
$e'(\alpha)\to\infty$ as $\alpha\uparrow\frac12+\kappa_c$ by 
Theorem~\ref{THM-ealpha}~(3), Lemma~\ref{LEM-infty} applies to $g_t$ and 
yields~\eqref{EQ-limg} again. The same argument works in the case 
$\alpha_-=\frac12-\kappa_c$.

Combined with Parts~(1) and~(4), the above analysis shows that for any 
$\alpha<\alpha_+$ one has $\alpha_+(t)\ge\alpha$ for large enough $t$ while 
for any $\alpha>\alpha_+$, $\alpha_+(t)\le\alpha$ for large enough $t$.
We deduce
$$
\alpha_+\le\liminf_{t\to\infty}\alpha_+(t)
\le\limsup_{t\to\infty}\alpha_+(t)\le\alpha_+,
$$
and~\eqref{EQ-limalpha+} follows.\hfill\qed

\subsection{Proof of Theorem~\ref{THM-GaussianLDP}}
\label{SSECT_Proof_of_THM-GaussianLDP}

We use the notation of Proposition~\ref{PROP-GaussianCocycles} and its proof.
We start with a few technical facts that will be used in the proof.

\bel\label{LEM-PrepExtLDP} Assume that Condition~(C) holds and that
$\ep>0$. Then, for some constants $c>0$ and $T>0$, the following hold true.
\ben
\item $\|\cS_t\|\le c$ and $\|\cS_t\|_1\le ct$ for $t\ge T$.
\item 
The function $g_t(\alpha)$ has an analytic continuation from $\fI_t$ to
the cut plane $\cc\setminus(]-\infty,\alpha_-(t)]\cup[\alpha_+(t),\infty])$. 
Moreover, for any compact subset 
$K\subset\cc\setminus(]-\infty,\alpha_-]\cup[\alpha_+,\infty])$ there is
$T_K>0$ such that
$$
\sup_{\alpha\in K\atop t\ge T_K}\left|g_t(\alpha)\right|<\infty.
$$
\item For $t\ge T$
the interval $\fI_t$ is finite and is mapped bijectively to  $\rr$ by the 
function $g_t'$. In the following, we set
$$
\alpha_{s,t}=(g_t^{\prime})^{-1}(s)
$$
for $t\ge T$ and $s\in\rr$.
\item Let
$$
s_\pm=\lim_{\fI_\infty\ni\alpha\to\alpha_\pm}e'(\alpha),
$$
and suppose that $s\in]-\infty,s_-]$ (resp. $s\in[s_+,+\infty[$). Then
we have
\beq
\lim_{t\to\infty}\alpha_{s,t}=\alpha_-\mbox{ (resp. }\alpha_+\mbox{)},
\qquad
\liminf_{t\to\infty}g_t(\alpha_{s,t})\ge e(\alpha_-)
\mbox{ (resp. }e(\alpha_+)\mbox{)}.
\label{EQ-alphastlim}
\eeq
\item For $t\ge T$ and $s\in]-\infty,s_-]\cup[s_+,+\infty[$, let
$$
\fM_{s,t}=\frac1t\cS_t(I+\alpha_{s,t}\cS_t)^{-1},\qquad 
b_{s,t}=\frac1t(I+\alpha_{s,t}\cS_t)^{-\frac34}\cD_t^\ast\cL_t T_ta.
$$
The operator $\fM_{s,t}$ is trace class on $\Xi\oplus\partial\fH$, with trace norm
$$
\|\fM_{s,t}\|_1\le c+|s|,
$$
and $b_{s,t}\in\Xi\oplus\partial\fH$ is such that
$$
\lim_{t\to\infty}\|b_{s,t}\|=0.
$$

\een
\eel

\noindent{\bf Proof.} (1) Writing~\eqref{EQ-LtDef} as
$$
(\cL_tx)(s)=L^{(1)}x(s)+\delta(s-t)L^{(2)}x(t)+\delta(s)L^{(3)}x(0)
$$
with $L^{(j)}\in L(\Xi)$, we decompose 
$\cS_t=\cD_t^\ast\cL_t\cD_t=\cS_t^{(1)}+\cS_t^{(2)}+\cS_t^{(3)}$.
Lemma~\ref{LEM-Cov}~(4) yields
$$
\|\cS_t^{(1)}\|\le\|L^{(1)}\|\,\|\cD_t\|^2=\|L^{(1)}\|\,\|\cK_t\|\le c_1,
$$
$$
\|\cS_t^{(1)}\|_1=\tr(\cD_t^\ast|L^{(1)}|\cD_t)
\le\|L^{(1)}\|\tr(\cD_t\cD_t^\ast)
=\|L^{(1)}\|\|\cK_t\|_1\le c_1 t,
$$
for $t\ge0$. A simple calculation further gives 
$\cS_t^{(2)}=\widetilde{\cD}_t^\ast L^{(2)}\widetilde{\cD}_t$,
$\cS_t^{(3)}=\widetilde{\cD}_0^\ast L^{(3)}\widetilde{\cD}_0$,
where
$$
\widetilde{\cD}_s=\left[
\begin{array}{cc}
\e^{sA}N^{\frac12}&R_sQ
\end{array}
\right].
$$
It follows from Lemma~\ref{LEM-WH}~(3) that
$$
\|\cS_t^{(2)}\|\le
\|\cS_t^{(2)}\|_1=\tr(\widetilde{\cD}_t^\ast|L^{(2)}|\widetilde{\cD}_t)
\le\|L^{(2)}\|\tr(\widetilde{\cD}_t\widetilde{\cD}_t^\ast)
=\|L^{(2)}\|\tr(\e^{tA}N\e^{tA^\ast}+R_tQQ^\ast R_t^\ast)\le c_2,
$$
and
$$
\|\cS_t^{(3)}\|\le
\|\cS_t^{(3)}\|_1=\tr(\widetilde{\cD}_0^\ast|L^{(3)}|\widetilde{\cD}_0)
\le\|L^{(3)}\|\tr(\widetilde{\cD}_0\widetilde{\cD}_0^\ast)
=\|L^{(3)}\|\tr(N)\le c_3,
$$
for $t\ge0$.
We conclude that $\|\cS_t\|\le c_1+c_2+c_3$ and
$\|\cS_t\|_1\le (c_1+c_2+c_3)t$ for $t\ge1$.

(2) Since $g_t(0)=0$ for all $t>0$, it suffices to show that the
function $g_t'$ has the claimed properties. By definition,
\beq
\cc\setminus(]-\infty,\alpha_-(t)]\cup[\alpha_+(t),\infty])
\subset\{\alpha\in\cc\,|\,-\alpha^{-1}\not\in\sp(\cS_t)\},
\label{EQ-SpSt}
\eeq
and the analyticity of $g'_t$ on this set follows directly from 
Eq.~\eqref{EQ-gtprimeform}. 
Let $K\subset\cc\setminus(]-\infty,\alpha_-]\cup[\alpha_+,\infty])$
be compact. By Proposition~\ref{PROP-GaussianCocycles}~(5)~(6)
there exists $T_K\ge T$ such that
\beq
\dist(K,]-\infty,\alpha_-(t)]\cup[\alpha_+(t),\infty])\ge\delta>0
\label{EQ-Kdist}
\eeq
for all $t\ge T_K$. By Part~(1), $\|\alpha\cS_t\|\le\frac12$ so that 
$\|(I+\alpha\cS_t)^{-1}\|\le 2$ for all $t\ge T$ and all $\alpha\in\cc$ satisfying $|\alpha|\le (2c)^{-1}$. By the spectral theorem, it follows 
from~\eqref{EQ-SpSt} and\eqref{EQ-Kdist} that
$$
\|(I+\alpha\cS_t)^{-1}\|\le\frac{2c}{\delta}
$$
for all $t\ge T_K$ and all $\alpha\in K$ such that $|\alpha|\ge (2c)^{-1}$.
Hence $\|(I+\alpha\cS_t)^{-1}\|$ is bounded on $K$ uniformly in $t\ge T_K$.
The boundedness of $g'_t$ now easily follows from Eq.~\eqref{EQ-gtprimeform}
and Part~(1).

(3) By Part~(5) of Proposition~\ref{PROP-GaussianCocycles},
if $T>0$ is large enough then the interval $\fI_t$ is finite for all
$t\ge T$. By Part~(1) of the same Proposition, the function $g_t'$ is 
strictly increasing on $\fI_t$ and maps this interval onto $\rr$.

(4) We consider $s\ge s_+$, the case $s\le s_-$ is similar.
Since $\alpha_{s,t}\in\fI_t$, Part~(5) of 
Proposition~\ref{PROP-GaussianCocycles} gives
$$
\underline{\alpha}=
\liminf_{t\to\infty}\alpha_{s,t}\le
\limsup_{t\to\infty}\alpha_{s,t}\le\lim_{t\to\infty}\alpha_+(t)=\alpha_+.
$$
Suppose that $\underline{\alpha}<\alpha_+$. Invoking convexity, we 
deduce from the definition of $\alpha_{s,t}$ and Part~(4) of  
Proposition~\ref{PROP-GaussianCocycles}
$$
s=\liminf_{t\to\infty}g_t'(\alpha_{s,t})\le
\liminf_{t\to\infty}g_t'(\underline{\alpha})=e'(\underline{\alpha}).
$$
The strict convexity of $e(\alpha)$ leads to $s\le e'(\underline{\alpha})<s_+$
which contradicts our hypothesis and yields the first relation 
in~\eqref{EQ-alphastlim}. 

To prove the second one, notice that
for any $\gamma\in[0,\alpha_+[$ one has $\gamma<\alpha_{s,t}\le\alpha_+(t)$
provided $t$ is large enough. By convexity
$$
g_t(\alpha_{s,t})
\ge g_t(\gamma)+(\alpha_{s,t}-\gamma)g_t'(\gamma)
\ge g_t(\gamma)+(\alpha_{s,t}-\gamma)g_t'(0),
$$
and letting $t\to\infty$ yields
$$
\liminf_{t\to\infty}g_t(\alpha_{s,t})\ge 
e(\gamma)+(\alpha_+-\gamma)e'(0).
$$
Taking $\gamma\to\alpha_+$ gives the desired inequality.

(5) We consider $s\ge s_+$, the case $s\le s_-$ is again similar. By
Part~(3), if $T>0$ is large enough then
$\alpha_{s,t}\in]0,\alpha_+(t)[\subset\fI_t$ for all $t\ge T$.
Since $I+\alpha\cS_t>0$ for $\alpha\in\fI_t$, Part~(1) allows us 
to conclude
$$
\|\fM_{s,t}^+\|_1=\frac1t\tr(\cS_t^+(I+\alpha_{s,t}\cS_t^+)^{-1})
\le\frac1t\|\cS_t^+\|_1\le\frac1t\|\cS_t\|_1\le c.
$$
By Eq.~\eqref{EQ-gtprimeform} and the definition of $\alpha_{s,t}$ we have
$$
s=-\frac12\tr(\fM_{s,t})-\frac1{2t}a\cdot T_t^\ast\cL_t T_ta
+\frac{t\alpha_{s,t}}{2}(b_{s,t}|
((I+\alpha_{s,t}\cS_t)^{\frac12}+(I+\alpha_{s,t}\cS_t)^{-\frac12})b_{s,t}),
$$
from which we deduce
\beq
\|\fM_{s,t}^-\|_1
=s+\|\fM_{s,t}^+\|_1+\frac1ta\cdot T_t^\ast\cL_t T_ta
-t\alpha_{s,t}(b_{s,t}|
((I+\alpha_{s,t}\cS_t)^{\frac12}+(I+\alpha_{s,t}\cS_t)^{-\frac12})b_{s,t}).
\label{EQ-trMform}
\eeq
One easily checks that
$$
\lim_{t\to\infty}\frac1t\|T_t^\ast\cL_t T_t\|=0,
$$
so that $\|\fM_{s,t}^-\|_1\le s+2c$ and hence 
$\|\fM_{s,t}\|_1=\|\fM_{s,t}^-\|_1+\|\fM_{s,t}^+\|_1\le |s|+3c$
for $t$ large enough. Finally, from~\eqref{EQ-trMform} we derive
$$
\|b_{s,t}\|^2\le
\frac12(b_{s,t}|
((I+\alpha_{s,t}\cS_t)^{\frac12}+(I+\alpha_{s,t}\cS_t)^{-\frac12})b_{s,t})
\le\frac1{2t\alpha_{s,t}}
\left(
s+\frac12\tr(\fM_{s,t})+\frac1{2t}a\cdot T_t^\ast\cL_tT_ta
\right),
$$
from which we conclude that $\|b_{s,t}\|\to0$ as $t\to\infty$.
\hfill\qed

\paragraph{(1)} By Proposition~\ref{PROP-GaussianCocycles}~(4) one has
$$
\lim_{t\to\infty}\frac1t\log\EE_\nu[\e^{\alpha\eta_t}]=e(-\alpha),
$$
for $-\alpha\in\fI_\infty$. By the G\"artner-Ellis theorem, the local LDP
holds on the interval $]\eta_-,\eta_+[$ with the rate function
$$
I(s)=\sup_{\alpha\in\rr}(\alpha s-e(-\alpha)).
$$
Note that $I(s)=\sup_{\alpha\in\fI_\infty}(\alpha s-e(-\alpha))$
for $s\in]\eta_-,\eta_+[$. To prove that the global LDP holds
we must show that for all open sets $O\subset\rr$
$$
\liminf_{t\to\infty}\frac1t\log\PP_\nu\left[
\frac{\eta_t}t\in O\right]\ge-\inf_{s\in O}J(s).
$$
By a simple and well known argument (see, e.g., \cite[Section V.2]{dH}), 
it suffices to show that for any $s\in\rr$
$$
\lim_{\epsilon\downarrow0}\liminf_{t\to\infty}\frac1t\log\PP_\nu\left[
|\hat\eta_t|<\epsilon\right]\ge-J(s),
$$
where $\hat\eta_t=\frac{\eta_t}t-s$. The latter holds for any 
$s\in]\eta_-,\eta_+[$ by the G\"artner-Ellis theorem.
Next, we observe that whenever 
$\alpha_\pm=\frac12\pm\kappa_c$, then by 
Proposition~\ref{PROP-GaussianCocycles}~(4) we have 
$\eta_\pm=\pm\infty$. Thus, it suffices to consider the cases where 
$\alpha_->\frac12-\kappa_c$ or/and $\alpha_+<\frac12+\kappa_c$. We shall only 
discuss the second case, the analysis of the first one is similar.

Fix $s\le\eta_-$ and set $\alpha_t=-\alpha_{-s,t}$ so that 
$g'_t(-\alpha_t)=-s$ and, by Lemma~\ref{LEM-PrepExtLDP}~(3),
\beq
\lim_{t\to\infty}\alpha_t=-\alpha_+,\qquad
\liminf_{t\to\infty}g_t(-\alpha_t)\ge e(\alpha_+).
\label{EQ-tiltlim}
\eeq
Defining the tilted probability $\widehat{\PP}_\nu^t$ on $C([0,t],\Xi)$ by
$$
\frac{\d\widehat{\PP}^t_\nu}{\d\PP_\nu^t}
=\e^{\alpha_t\eta_t-tg_t(-\alpha_t)},
$$
we immediately get the estimate
$$
\PP_\nu^t\left[|\hat\eta_t|<\epsilon\right]
\ge\e^{-t(s\alpha_t+\epsilon|\alpha_t|-g_t(-\alpha_t))}
\widehat{\PP}_\nu^t\left[|\hat\eta_t|<\epsilon\right],
$$
and hence,
\beq
\frac1t\log\PP_\nu^t\left[|\hat\eta_t|<\epsilon\right]
\ge g_t(-\alpha_t)-s\alpha_t-\epsilon|\alpha_t|
+\frac1t\log\widehat{\PP}_\nu^t\left[|\hat\eta_t|<\epsilon\right].
\label{EQ-TiltEstimate}
\eeq
We claim that for any sufficiently small $\epsilon>0$,
\beq
p_\epsilon=\liminf_{t\to\infty}\widehat{\PP}_\nu^t\left[
|\hat\eta_t|<\epsilon\right]>0.
\label{EQ-pepsilon}
\eeq
Using~\eqref{EQ-tiltlim} we derive from~\eqref{EQ-TiltEstimate} that 
$$
\liminf_{t\to\infty}\frac1t\log\PP_\nu^t\left[|\hat\eta_t|<\epsilon\right]
\ge e(\alpha_+)+s\alpha_+-\epsilon|\alpha_+|,
$$
provided $\epsilon>0$ is small enough. Letting $\epsilon\downarrow0$,
we finally get
$$
\lim_{\epsilon\downarrow0}\liminf_{t\to\infty}\frac1t\log\PP_\nu^t\left[
|\hat\eta_t|<\epsilon\right]\ge e(\alpha_+)+s\alpha_+,
$$
which, in view of~\eqref{EQ-Jdef}, is the desired relation.

Thus, it remains to prove our claim~\eqref{EQ-pepsilon}. To this end,
note that for $\lambda\in\rr$,
$$
\widehat{\EE}_\nu^t\left[\e^{-\i\lambda\hat\eta_t}\right]
=\EE_\nu^t\left[
\e^{(\alpha_t-\i\lambda/t)\eta_t-tg_t(-\alpha_t)
-\i\lambda g_t'(-\alpha_t)}\right]
=\e^{t(g_t(-\alpha_t+\i\lambda/t)-g_t(-\alpha_t)
-\i g_t'(-\alpha_t)\lambda/t)},
$$
and a simple calculation using Eq.~\eqref{EQ-gtForm}, \eqref{EQ-gtprimeform}
yields 
\beq
\widehat{\EE}_\nu^t\left[\e^{-\i\lambda\hat\eta_t}\right]
=\left(\det(I+\i\lambda\fM_{-s,t})^{-1}
\e^{\i\lambda\tr(\fM_{-s,t})
-\lambda^2(b_{-s,t}|(I+\i\lambda\fM_{-s,t})^{-1}b_{-s,t})}\right)^{\frac12}.
\label{EQ-charform}
\eeq
Let $\cS(\rr,\Xi)$ be the Schwartz space of rapidly decaying 
$\Xi$-valued smooth functions on $\rr$ and $\cS'(\rr,\Xi)$ its dual
w.r.t.\;the inner product of $\fH$.  Denote by $\hat{\gamma}$ the centered Gaussian measure on $\cK_-=\Xi\oplus\cS'(\rr,\Xi)$ with covariance $I$
and let
$$
\widetilde{\eta}_t(k)=-\frac12(k|\fM_{-s,t}k-2b_{-s,t}).
$$
By Lemma~\ref{LEM-PrepExtLDP}~(5),
$|\hat{\eta}_t(k)|<\infty$ for $\hat{\gamma}$-a.e.\;$k\in\cK_-$ and
$$
\bar\eta_t=
\int\widetilde{\eta}_t(k)\hat{\gamma}(\d k)=-\frac12\tr(\fM_{-s,t}).
$$
It follows that for $\lambda\in\rr$,
$$
\int\e^{-\i\lambda(\hat{\eta}_t(k)-\bar\eta_t)}\hat{\gamma}(\d k)
=\left(\det(I+\i\lambda\fM_{-s,t})^{-1}
\e^{\i\lambda\tr(\fM_{-s,t})
-\lambda^2(b_{-s,t}|(I+\i\lambda\fM_{-s,t})^{-1}b_{-s,t})}
\right)^{\frac12},
$$
and comparison with~\eqref{EQ-charform} allows us to conclude that the law 
of $\hat{\eta}_t$ under $\widehat{\PP}_\nu^t$ coincides with the one of 
$\widetilde{\eta}_t-\bar\eta_t$ under $\hat{\gamma}$, so that
$$
p_\epsilon=\liminf_{t\to\infty}\hat{\gamma}
\left[|\widetilde{\eta}_t-\bar\eta_t|
<\epsilon \right].
$$
For $m>0$ let $P_m$ denote the spectral projection of $\fM_{s,t}$ for
the interval $[-m,m]$ and define
\begin{align*}
\zeta_t^<(k)&=-\frac12(P_m k|\fM_{-s,t}k-2b_{-s,t})
+\frac12\tr(P_m\fM_{-s,t}),\\[4pt]
\zeta_t^>(k)&=-\frac12((I-P_m) k|\fM_{-s,t}k-2b_{-s,t})
+\frac12\tr((I-P_m)\fM_{-s,t}),
\end{align*}
so that $\widetilde{\eta}_t-\bar\eta_t=\zeta_t^<+\zeta_t^>$ and
$\hat{\gamma}[\zeta_t^<]=\hat{\gamma}[\zeta_t^>]=0$. Since $\zeta_t^<$
and $\zeta_t^>$ are independent under $\hat{\gamma}$, we have
\beq
p_\epsilon\ge\liminf_{t\to\infty}
\hat{\gamma}\left[|\zeta_t^<|<\epsilon/2\right]
\hat{\gamma}\left[|\zeta_t^>|<\epsilon/2\right].
\label{EQ-pfactor}
\eeq
The Chebyshev inequality gives
$$
\hat{\gamma}\left[|\zeta_t^<|<\epsilon/2\right]
=1-\hat{\gamma}\left[|\zeta_t^<|\ge\epsilon/2\right]
\ge 1-\frac4{\epsilon^2}\hat{\gamma}\left[|\zeta_t^<|^2\right].
$$
Choosing 
$m=\frac13(c+|s|)\epsilon^2$,
the estimate
$$
\hat{\gamma}\left[|\zeta_t^<|^2\right]
=\frac12\tr(P_m\fM_{s,t}^2)+\|P_mb_{s,t}\|^2
\le\frac12\|P_m\fM_{s,t}\|\,\|\fM_{s,t}\|_1+\|b_{s,t}\|^2
\le\frac{m}2(c+|s|)+\|b_{s,t}\|^2,
$$
together with Lemma~\ref{LEM-PrepExtLDP}~(5) shows that
$$
\liminf_{t\to\infty}\hat{\gamma}\left[|\zeta_t^<|<\epsilon/2\right]
\ge 1-\frac{2m}{(c+|s|)\epsilon^2}=\frac13.
$$
To deal with the second factor on the right-hand side of~\eqref{EQ-pfactor}
we first note that $(I-P_m)|\fM_{s,t}|\ge m(I-P_m)$, so that, using again
Lemma~\ref{LEM-PrepExtLDP}~(5),
$$
N_m=\tr(I-P_m)\le\frac1m\tr((I-P_m)|\fM_{s,t}|)\le\frac{\|\fM_{s,t}\|_1}m
\le\frac{c+|s|}m=\frac3{\epsilon^2}.
$$
Setting 
$$
\epsilon_j=\frac{|\mu_j|}{c+|s|}\epsilon,\qquad(j=1,\ldots,N_m),
$$
where the $\mu_j$ denote the repeated eigenvalues of $(I-P_m)\fM_{s,t}$
we have $\sum_j\epsilon_j\le\epsilon$ and hence, passing to an orthonormal
basis of eigenvectors of $(I-P_m)\fM_{s,t}$, we obtain 
$$
\hat{\gamma}\left[|\zeta_t^>|<\epsilon/2\right]
={\rm n}_{N_m}\left[
\left|\sum_{j=1}^{N_m}\mu_jk_j^2-2b_jk_j-\mu_j\right|<\epsilon\right]
\ge\prod_{j=1}^{N_m}{\rm n}_1\left[
\left|k^2-2\frac{b_j}{\mu_j}k-1\right|<\frac{\epsilon_j}{|\mu_j|}\right],
$$
where ${\rm n}_N$ denotes the centered Gaussian measure of unit covariance 
on $\rr^N$ and the $b_j\in\rr$ are such that $|b_j|\le\|b_{s,t}\|$.
An elementary analysis shows that if $|b|\le1$ and $0<\delta\le1$, then
$$
{\rm n}_1\left[|k^2-2bk-1|<\delta\right]
\ge\frac\delta{\e\sqrt{6\pi}}.
$$
Thus, provided $\epsilon<c+|s|$, we can conclude that
$$
\liminf_{t\to\infty}\hat{\gamma}\left[|\zeta_t^>|<\epsilon/2\right]
\ge\left(\frac{\epsilon}{\e(c+|s|)\sqrt{6\pi}}\right)^{3\epsilon^{-2}}>0,
$$
which shows that $p_\epsilon>0$ and concludes the proof of Part~(2).

\paragraph{(2)} According to Bryc's lemma (see~\cite{Br} 
or~\cite[Section 4.8.4]{JOPP})
the Central Limit Theorem for the family $(\eta_t)_{t>0}$ holds,
provided that the generating function~$g_t$ has an analytic 
continuation to the disc $D_\epsilon=\{\alpha\in\cc\,|\,|\alpha|<\epsilon\}$ for some $\epsilon>0$ and satisfies the estimate
$$
\sup_{\alpha\in D_\epsilon\atop t>t_0}|g_t(\alpha)|<\infty,
$$
for some $t_0>0$. These properties clearly follow from 
Lemma~\ref{LEM-PrepExtLDP}~(2).

\subsection{Proof of Lemma~\ref{LEM-NetControl}}
\label{SSECT_Proof_of_LEM-NetControl}

\paragraph{(1)} Let
$$
\cC=\bigvee_{j\ge0}\Ran (\omega^\ast\omega)^{j}\iota
$$
be the controllable subspace of $(\omega^\ast\omega,\iota)$. 
From~\eqref{EQ-AQdef} and~\eqref{EQ-OmegaDef} we derive
$$
\Omega^{2j} Q=(-1)^j\left[\begin{array}{c}
(\omega^\ast\omega)^{j}\iota\\0
\end{array}\right]\vartheta^\12,
$$
and hence
$$
\bigvee_{j\ge0}\Ran(\Omega^{2j} Q)=\cC\oplus\{0\}.
$$
The last relation and
$\Omega(\cC\oplus\{0\})=\{0\}\oplus\omega^\ast\cC$ yield that the 
controllable subspace of $(\Omega,Q)$ is $\cC\oplus\omega^\ast\cC$.
Since $A=\Omega-\12 Q\vartheta^{-1}Q^\ast$, $(A,Q)$ has the same
controllable subspace. Finally, since $\Ker\omega=\{0\}$, we conclude that
$\cC\oplus\omega^\ast\cC=\Xi$ iff $\cC=\rr^\cI$.

\paragraph{(2)} The same argument yields 
$\cC(\Omega,Q\pi_i)=\cC_i\oplus\omega^\ast\cC_i$. Thus if 
$0\not=u\in\cC_i\cap\cC_j$, we have 
$0\not=u\oplus 0\in\cC(\Omega,Q\pi_i)\cap\cC(\Omega,Q\pi_j)$ and the result
follows from Proposition~\ref{PROP-StrictPositivity2}~(2).


\subsection{Proof of Theorem~\ref{THM-Jacobi}}
\label{SSECT_Proof_of_THM-Jacobi}

\paragraph{(1)} By assumption (J), the Jacobi matrix
$$
\omega^2=\left[\begin{array}{ccccccc}
b_1&a_1&0&0&\cdots&0&0\\
a_1&b_2&a_2&0&&0&0\\
0&a_2&b_3&a_3&&0&0\\
\vdots&&\ddots&\ddots&\ddots&&\vdots\\
0&0&0&\ddots&\ddots&a_{L-2}&0\\[-6pt]
0&0&0&0&\ddots&b_{L-1}&a_{L-1}\\
0&0&0&0&\cdots&a_{L-1}&b_L
\end{array}\right],
$$
is positive and $a_i\not=0$ for all $i\in\cI$. Denote by $\{\delta_i\}_{i\in\cI}$ the
canonical basis of $\rr^\cI$. Starting with the obvious fact that
$\Ran(\iota)=\mathrm{span}(\{\delta_i\,|\,i\in\partial\cI\})$,
a simple induction yields
$$
\bigvee_{0\le j\le k}\Ran(\omega^{2j}\iota)
=\mathrm{span}(\{\delta_i\,|\,\dist(i,\partial\cI)\le k\}).
$$
Hence the pair $(\omega^2,\iota)$ is controllable.

\paragraph{(2)} The argument in the proof of Part~(1)
yields $\cC_1=\cC_L=\rr^\cI$ and the first statement follows directly from 
Proposition~\ref{PROP-StrictPositivity2}~(2). To prove the second one, we may assume
that $\vartheta_{\mathrm{min}}=\vartheta_1$ and $\vartheta_{\mathrm{max}}=\vartheta_L$.
From Theorem~\ref{THM-InvMeas}~(3) we already know that $\vartheta_1\le M\le\vartheta_L$
and that 
$$
M-\vartheta_1=\vartheta_1\int_0^\infty\e^{tA}Q
(\vartheta_1^{-1}-\vartheta^{-1})Q^\ast\e^{tA^\ast}\d t.
$$
Since $\vartheta_1^{-1}-\vartheta^{-1}\ge0$ it follows that 
$$
\Ker(M-\vartheta_1)\subset
\bigcap_{n\ge0}\Ker(\vartheta_1^{-1}-\vartheta^{-1})Q^\ast A^{\ast n}
=\left(\bigvee_{n\ge0}A^nQ\delta_1\right)^\perp=\cC_1^\perp=\{0\},
$$
which implies $M-\vartheta_1>0$. A similar argument shows that $\vartheta_2-M>0$.

\paragraph{(3)} Set $\kappa=\alpha-\12$ and
$\kappa_0=\frac{\bar\vartheta}{\Delta}>\12$. Writing
$$
\i\nu-K_\alpha
=\left[\begin{array}{cc}
\Omega+\i\nu&0\\
0&\Omega+\i\nu
\end{array}\right]
+\left[\begin{array}{cc}
Q\vartheta^{-\12}&0\\
0&Q\vartheta^{-\12}
\end{array}\right]
\left[\begin{array}{cc}
\kappa&-\vartheta\\
(\kappa^2-\frac14)\vartheta^{-1}&-\kappa
\end{array}\right]
\left[\begin{array}{cc}
\vartheta^{-\12}Q^\ast&0\\
0&\vartheta^{-\12}Q^\ast
\end{array}\right],
$$
one derives $\det(\i\nu-K_\alpha)=\det(\Omega+\i\nu)^2\det(I+\Sigma(\i\nu))$, where
\beq
\Sigma(z)=\left[\begin{array}{cc}
\kappa R(z)&-R(z)\vartheta\\
(\kappa^2-\frac14)R(z)\vartheta^{-1}&-\kappa R(z)
\end{array}\right],\qquad
R(z)=\vartheta^{-\12}Q^\ast(\Omega+z)^{-1}Q\vartheta^{-\12}.
\label{EQ-Sform}
\eeq
A simple calculation further gives
$$
\det(\Omega+\i\nu)=\det(\omega^2-\nu^2),\qquad
R(\i\nu)=\i\nu\iota^\ast(\omega^2-\nu^2)^{-1}\iota.
$$
Denote by $D(\nu^2)$ the adjugate of $\omega^2-\nu^2$.
Expressing $(\omega^2-\nu^2)^{-1}$ with Cramer's formula and observing that
$D_{1L}(\nu^2)=D_{L1}(\nu^2)=\hat a$, we get
\beq
R(\i\nu)=\frac{2\bar\gamma\i\nu}{d(\nu^2)}\left[\begin{array}{cc}
b(\nu^2)\e^{\12\delta}&\hat a\\
\hat a&c(\nu^2)\e^{-\12\delta}
\end{array}\right],
\label{EQ-Rform}
\eeq
where
\begin{align*}
b(\nu^2)=D_{11}(\nu^2),\qquad
c(\nu^2)=D_{LL}(\nu^2),\qquad
d(\nu^2)=\det(\omega^2-\nu^2),
\end{align*}
are polynomials in $\nu^2$ with real coefficients. Inserting~\eqref{EQ-Rform}
into~\eqref{EQ-Sform}, an explicit calculation of $\det(I+\Sigma(\i\nu))$ 
yields
$$
\det(\i\nu-K_\alpha)=
\left(d(\nu^2)+(\bar\gamma\nu)^2\frac{b(\nu^2)c(\nu^2)-\hat a^2}{d(\nu^2)}\right)^2
+(\bar\gamma\nu)^2\left(\e^{\delta/2}b(\nu^2)-\e^{-\delta/2}c(\nu^2)\right)^2
-4\hat a^2\frac{\kappa^2-\kappa_0^2}{\kappa_0^2-\frac14}(\bar\gamma\nu)^2.
$$
By the Desnanot-Jacobi identity,

$$
\frac{b(\nu^2)c(\nu^2)-\hat a^2}{d(\nu^2)}=\det(\widetilde{\omega^2}-\nu^2)
=\tilde d(\nu^2),
$$
where $\widetilde{\omega^2}$ is the matrix obtained from $\omega^2$ by
deleting its first and last rows and columns. Thus, we finally obtain
$$
\det(\i\nu-K_\alpha)=
\left(d(\nu^2)+(\bar\gamma\nu)^2\tilde d(\nu^2)\right)^2
+(\bar\gamma\nu)^2
\left(\e^{\delta/2}b(\nu^2)-\e^{-\delta/2}c(\nu^2)\right)^2
-4\hat a^2\frac{\kappa^2-\kappa_0^2}{\kappa_0^2-\frac14}(\bar\gamma\nu)^2,
$$
where $b$, $c$, $d$ and $\tilde d$ 
are polynomials with real coefficients. Since $d(0)=\det(\omega^2)>0$, 
$K_\alpha$ is regular for all $\alpha\in\rr$ and we can rewrite the eigenvalue 
equation as 
\beq
g(\nu^2)=\frac{\kappa^2-\kappa_0^2}{\kappa_0^2-\frac14},
\label{EQ-Secular}
\eeq
where the rational function 
$$
g(x)=\frac1{4\hat a^2}\left[
\frac{(d(x)+\bar\gamma^2x\tilde d(x))^2}{\bar\gamma^2x}
+\left(\e^{\delta/2}b(x)-\e^{-\delta/2}c(x)\right)^2\right]
$$
has real coefficients, a simple pole at $0$, a pole of order $2L$ at infinity
and is non-negative on $]0,\infty[$. It follows that
$$
\kappa_c=\sqrt{\kappa_0^2+g_0(\kappa_0^2-\frac14)},
$$
where
$$
g_0=\min_{x\in]0,\infty[}g(x).
$$
Since $\kappa_0>\12$, we conclude that $\kappa_c\ge\kappa_0$, with
equality iff $g_0=0$.

Under Assumption (S) the polynomials $b$ and $c$ coincide
and $\delta=0$. Thus, $g_0=0$ iff the polynomial
\beq
f(x)=d(x)+\bar\gamma^2x\tilde d(x)
\label{EQ-fDef}
\eeq
has a positive zero. If $L$ is odd, then this property follows immediately 
from the fact that
$$
f(0)=\det(\omega^2)>0,\qquad 
f(x)=(-x)^L+\mathcal{O}(x^{L-1})<0\quad (x\to\infty).
$$
A more elaborate argument is needed in the case of even $L$.
We shall invoke the deep connection between spectral analysis of
Jacobi matrices and orthogonal polynomials. We refer the reader
to~\cite{Si2} for a detailed introduction to this vast subject.

Let $\rho$ be the spectral measure of $\omega^2$ for the vector $\delta_1$.
The argument in the proof of Part~(1) shows that $\delta_1$ is cyclic for
$\omega^2$. Thus, $\omega^2$ is unitarily equivalent to multiplication
by $x$ on $L^2(\rr,\rho(\d x))$ and in this Hilbert space $\delta_1$
is represented by the constant polynomial $p_0=1$. Starting with
$\delta_2=a_1^{-1}(\omega^2-b_1)\delta_1=p_1(\omega^2)\delta_1$, a simple 
induction shows that there are real polynomials 
$\{p_k\}_{k\in\{0,\ldots,L-1\}}$ satisfying the recursion
\beq
a_kp_{k-1}(x)+(b_{k+1}-x)p_k(x)+a_{k+1}p_{k+1}(x)=0,
\qquad(k\in\{0,\ldots,L-2\},p_{-1}=0,p_0=1),
\label{EQ-pRecursion}
\eeq
and such that $\delta_k=p_{k-1}(\omega^2)\delta_1$. Thus, these 
polynomials form an orthonormal basis of $L^2(\rr,\rho(\d x))$ such that
\beq
\langle\delta_k|(\omega^2-x)^{-1}\delta_j\rangle
=\int\frac{p_{k-1}(\lambda)p_{j-1}(\lambda)}{\lambda-x}\rho(\d \lambda).
\label{EQ-GreenForm}
\eeq

\begin{figure}
\centering
\includegraphics[scale=0.35]{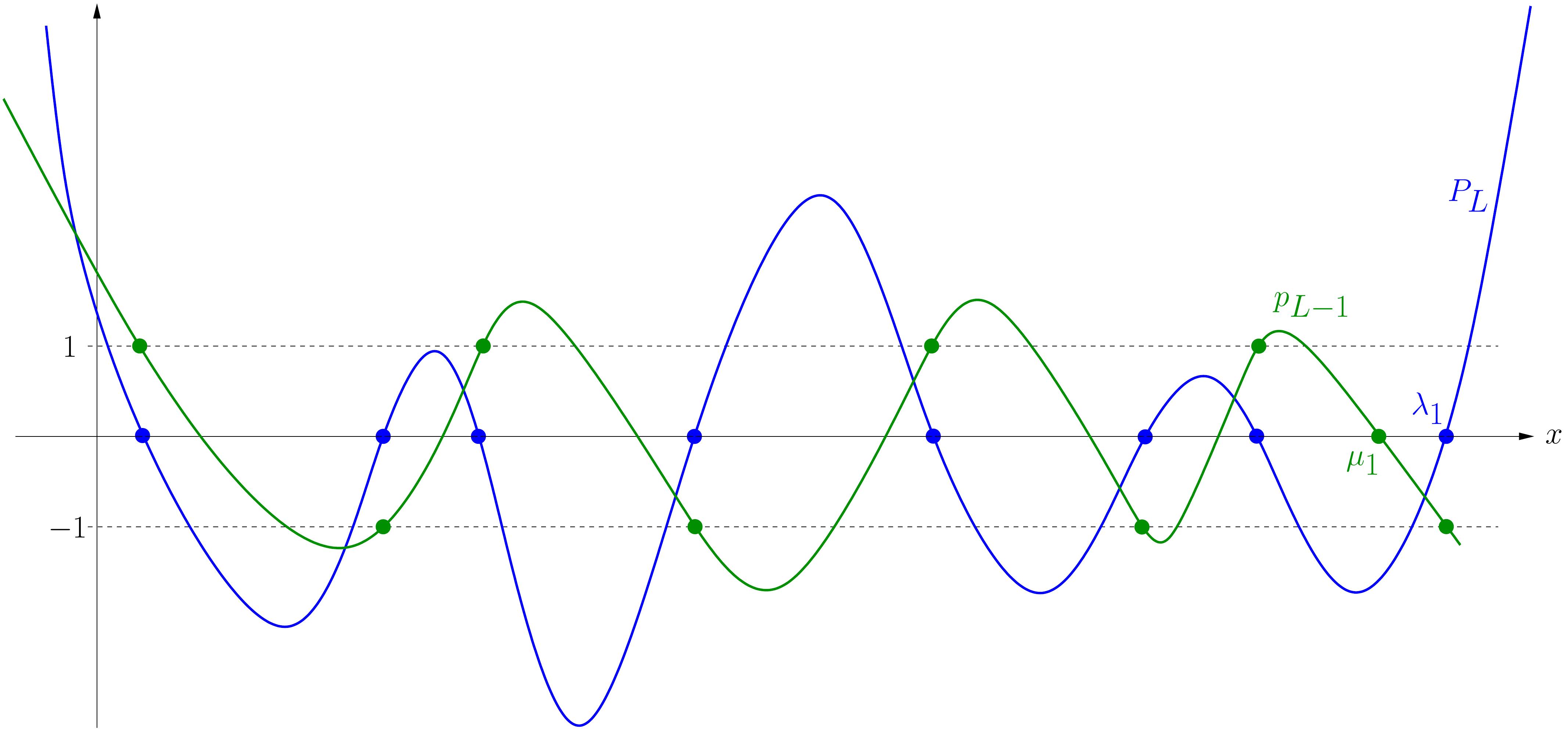}
\caption{The zeros of the polynomials $P_L$ and $p_{L-1}$ interlace.}
\label{Fig9}
\end{figure}

For $1\le j\le k\le L$, define
$$
d_{[j,k]}(x)=\det(x-J_{[j,k]}),\qquad
J_{[j,k]}=\left[\begin{array}{ccccccc}
b_j&a_j&0&0&\cdots&0&0\\
a_j&b_{j+1}&a_{j+1}&0&&0&0\\
0&a_{j+1}&b_{j+2}&a_{j+2}&&0&0\\
\vdots&&\ddots&\ddots&\ddots&&\vdots\\
0&0&0&\ddots&\ddots&a_{k-2}&0\\[-6pt]
0&0&0&0&\ddots&b_{k-1}&a_{k-1}\\
0&0&0&0&\cdots&a_{k-1}&b_k
\end{array}\right].
$$
Laplace expansion of the determinant $P_{k+1}(x)=d_{[1,k+1]}(x)$ on its last 
row yields the recursion
$$
P_{k+1}(x)=(x-b_{k+1})P_k(x)-a_k^2P_{k-1}(x).
$$
Comparing this relation with~\eqref{EQ-pRecursion} one easily deduces
\beq
a_1\cdots a_k\, p_k(x)=P_k(x),\quad (k\in\{1,\ldots,L-1\}),
\qquad d(x)=P_L(x).
\label{EQ-pPrelation}
\eeq
Polynomials of the second kind $\{q_k\}_{k\in\{0,\ldots L-1\}}$
associated to the measure $\rho$ are defined by
\beq
q_k(x)=\int\frac{p_k(\lambda)-p_k(x)}{\lambda-x}\rho(\d x).
\label{EQ-qnDef}
\eeq
Note in particular that $q_0(x)=0$ and $q_1(x)=a_1^{-1}$.
Applying the recursion relation~\eqref{EQ-pRecursion} to both sides of
this definition, we obtain
$$
a_kq_{k-1}(x)+(b_{k+1}-x)q_k(x)+a_{k+1}q_{k+1}(x)
=\int p_k(\lambda)\rho(\d\lambda)=0,\qquad(k\in\{1,\ldots,L-2\}).
$$
Set $\tilde q_k(x)=a_1q_{k+1}(x)$ and observe that these polynomials
satisfy the recursion
$$
a_{k+1}\tilde q_{k-1}(x)+(b_{k+2}-x)\tilde q_k(x)+a_{k+2}\tilde q_{k+1}(x)=0,
\qquad(k\in\{0,\ldots,L-3\},\tilde q_{-1}=0,\tilde q_0=1).
$$
Comparing this Cauchy problem with~\eqref{EQ-pRecursion} and 
repeating the argument leading to~\eqref{EQ-pPrelation} we deduce that
$a_2\cdots a_{k+1}\,\tilde q_k(x)=d_{[2,k+1]}(x)$, so that
$$
a_1\cdots a_k\,q_k(x)=d_{[2,k]}(x),\qquad(k\in\{2,\ldots,L-1\}).
$$
In particular, we can rewrite Definition~\eqref{EQ-fDef} as
\beq
f(x)=P_L(x)+\bar\gamma^2xq_{L-1}(x).
\label{EQ-ffinaly}
\eeq
Taking now Assumption (S) into account we derive from~\eqref{EQ-GreenForm}
that for any $z\in\cc\setminus\sp(\omega^2)$,
\begin{align*}
\int\frac{|p_{L-1}(\lambda)|^2}{\lambda-z}\rho(\d\lambda)
&=\langle\delta_L|(\omega^2-z)^{-1}\delta_L\rangle\\
&=\langle\cS\delta_1|(\omega^2-z)^{-1}\cS\delta_1\rangle
=\langle\delta_1|(\omega^2-z)^{-1}\delta_1\rangle
=\int\frac{\rho(\d\lambda)}{\lambda-z},
\end{align*}
from which we conclude that $|p_{L-1}(\lambda)|=1$ for all 
$\lambda\in\sp(\omega^2)$. 
Denote by $\lambda_L\ge\lambda_{L-1}\ge\cdots\ge\lambda_1$ the eigenvalues
of $\omega^2=J_{[1,L]}$ and by $\mu_{L-1}\ge\mu_{L-2}\ge\cdots\ge\mu_1$ 
that of $J_{[1,L-1]}$. It is a well known property of Jacobi matrices
(or equivalently of orthogonal polynomials) that
$$
\lambda_L<\mu_{L-1}<\lambda_{L-1}<\cdots<\mu_{1}<\lambda_1
$$
(see Figure~\ref{Fig9}). These interlacing inequalities and the previously 
established property allow us to conclude that
$$
p_{L-1}(\lambda_j)=(-1)^j,\qquad
p_{L-1}'(\lambda_1)<0.
$$

From Eq.~\eqref{EQ-ffinaly} and Definition~\eqref{EQ-qnDef}, we deduce
\begin{align*}
f(\lambda_1)
&=\bar\gamma^2\lambda_1\left(p_{L-1}'(\lambda_1)\rho(\{\lambda_1\})
+\sum_{j=2}^{L-1}
\frac{p_{L-1}(\lambda_1)-p_{L-1}(\lambda_j)}{\lambda_1-\lambda_j}
\rho(\{\lambda_j\})
\right)\\
&=\bar\gamma^2\lambda_1\left(p_{L-1}'(\lambda_1)\rho(\{\lambda_1\})
-2\sum_{j=1}^{\12 L-1}
\frac{\rho(\{\lambda_{2j}\})}{\lambda_1-\lambda_{2j}}
\right)<0,
\end{align*}
which, together with $f(0)>0$, shows that $f$ has a positive root.

By Proposition~\ref{PROP-RicX}~(12), the validity of Condition~(R) follows from 
Part~(2) and the fact that $\kappa_c=\kappa_0$.

\addcontentsline{toc}{section}{Appendix}
\refstepcounter{section}
\section*{Appendix: Basic theory of the algebraic Riccati equation}
\label{SEC-Appendix}
\renewcommand{\thesection}{A}
\init
\setcounter{subsection}{0}
\setcounter{theoreme}{0}

In this appendix, for the reader convenience we briefly expose the basic 
results on algebraic Riccati equation used in this work. We refer the reader 
to~\cite{LR,AFIJ,Sc} for detailed expositions and proofs.

 Let $\fh$ be
a  $d$-dimensional complex Hilbert space. We denote by $(\,\cdot\,,\,\cdot\,)$ 
the inner product of $\fh$. We equip the vector space $\cH=\fh\oplus\fh$ with 
the Hilbertian structure induced by $\fh$ and the symplectic form 
$$
\omega(x\oplus y,x'\oplus y')=(x\oplus y,J(x'\oplus y'))=(x,y')-(y,x').
$$
The symplectic complement of $\cV\subset\cH$ is
the subspace $\cV^\omega=\{v\,|\,\omega(u,v)=0\mbox{ for all } u\in\cV\}$.
A subspace $\cV\subset\cH$ is isotropic if $\cV\subset\cV^\omega$ and
Lagrangian if $\cV=\cV^\omega$. $\cV$ is Lagrangian iff it is isotropic and 
$d$-dimensional. For $Y,Z\in L(\fh)$, we denote by $Y\oplus Z$ the element of 
$L(\fh,\cH)$ defined by $(Y\oplus Z)x=Yx\oplus Zx$. In the block-matrix
notation,
$$
Y\oplus Z=\left[\begin{array}{cc}
Y\\Z
\end{array}\right],\qquad
(Y\oplus Z)^\ast=\left[\begin{array}{cc}
Y^\ast&Z^\ast
\end{array}\right].
$$
The graph of $X\in L(\fh)$ is the $d$-dimensional subspace of~$\cH$ defined by 
$$
\cG(X)=\Ran G_X,\quad G_X=I\oplus X.
$$
A subspace $\cV\subset\cH$ is a graph iff $\cV\cap(\{0\}\oplus\fh)=\{0\oplus0\}$.

The algebraic Riccati equation associated to the 
triple $(A,B,C)$ of elements of $L(\fh)$  is the following quadratic equation 
for the unknown self-adjoint $X\in L(\fh)$:
\beq
\cR(X)=XBX-XA-A^\ast X-C=0.
\label{EQ-RicGeneric}
\eeq
In the following, we shall assume that $C$ is self-adjoint, that $B\ge0$
and that the pair $(A,B)$ is controllable. 
We denote by $\fR(A,B,C)$ the set of self-adjoint elements of $L(\fh)$ 
satisfying Eq.~\eqref{EQ-RicGeneric}, which we can also write
as
$$
\cR(X)=
G_X^\ast L G_X =0,\qquad
L=\left[\begin{array}{cc}
C&A^\ast\\
A&-B^\ast
\end{array}\right].
$$

\subsection{Existence of self-adjoint solutions}

The Hamiltonian associated
to the Riccati equation~\eqref{EQ-RicGeneric} is the unique element of 
$L(\cH)$ such that $(u,Lv)=\omega(u,Kv)$ for all 
$u,v\in\cH$. One easily checks that
$$
K=\left[\begin{array}{cc}
-A&B\\C&A^\ast
\end{array}\right].
$$
Note that since $L=L^\ast$, $K$ is $\omega$-skew adjoint:
\beq
\omega(u,Kv)+\omega(Ku,v)=\omega(u,Kv)-\omega(v,Ku)=(u,Lv)-(v,Lu)=0.
\label{EQ-omegaskew}
\eeq

The first result we recall  is a characterization of the set $\fR(A,B,C)$.

\bet[Theorem~7.2.4 in~\cite{LR}]
\label{THM-RicChar}
The map $X\mapsto\cG(X)$ is a bijection from $\fR(A,B,C)$ onto the set of 
$K$-invariant Lagrangian subspaces of $\cH$.
\eet

The following are elementary symplectic geometric properties of projections:

\bel\label{LEM-Kisotropy}
\ben
\item The range of a projection $P\in L(\cH)$ is isotropic iff
$P^\ast JP=0$ and Lagrangian iff $I-P=J^\ast P^\ast J$.
\item Denote by $P_\kappa$ the spectral projection of $K$ for 
$\kappa\in\sp(K)$. Then $JP_\kappa J^\ast=P_{-\bar\kappa}^\ast$ and in
particular $\Ran P_\kappa$ is isotropic iff $\kappa\not\in\i\rr$.
\item Let $\Sigma\subset\sp(K)$ be such that 
$\Sigma\cap(-\bar\Sigma)=\emptyset$. Then the spectral subspace of $K$  for 
$\Sigma$ is isotropic.
\een
\eel

Note that $JK+K^\ast J=0$, which implies that the spectrum of $K$, including
multiplicities, is symmetric w.r.t.\;the imaginary axis.
If $\sp(K)\cap\i\rr=\emptyset$, then the spectral subspace of $K$ for
$\Sigma=\sp(K)\cap\cc_+$ is $d$-dimensional and hence, by 
Lemma~\ref{LEM-Kisotropy}~(3), Lagrangian. Thus, Theorem~\ref{THM-RicChar}
yields (see Theorems~7.2.4 and~7.5.1 in~\cite{LR})
\bec
\label{COR-RicExist}
If $\sp(K)\cap\i\rr=\emptyset$, then $\fR(A,B,C)\not=\emptyset$.
\eec

\ber
In cases where $\sp(K)\cap\i\rr\not=\emptyset$,
and under our controllability assumption, a necessary and sufficient
condition for the existence of self-adjoint solution is that all
Jordan blocks of $K$ corresponding to eigenvalues in $\i\rr$ are
even-dimensional. For the Riccati equations arising in our analysis
of harmonic networks, the singular case $\sp(K_\alpha)\cap\i\rr\not=\emptyset$ 
only occurs at the boundary points $\alpha=\frac12\pm\kappa_c$. There, the 
existence of solutions follows by continuity (Part~(4) of 
Theorem~\ref{PROP-RicX}).
\eer

\smallskip
Another powerful criterion for the existence of self-adjoint solutions is
the following

\bet[Theorem~9.1.1 in~\cite{LR}]\label{THM-ARI}
If there exists a self-adjoint $X\in L(\fh)$ such that $\cR(X)\le0$,
then $\fR(A,B,C)\not=\emptyset$.
\eet

\subsection{Extremal solutions}

The set $\fR(A,B,C)$ inherits the partial order of $L(\fh)$.
A minimal/maximal solution of~\eqref{EQ-RicGeneric} is a minimal/maximal
element of $\fR(A,B,C)$. Clearly, a minimal/maximal solution, if it exists, 
is unique.

\bet
\label{THM-RicMax}
Assume that $\fR(A,B,C)\not=\emptyset$. 
\ben
\item $\fR(A,B,C)$ is compact.
\item $\fR(A,B,C)$ contains a minimal element $X_-$ and a maximal element 
$X_+$. In the following, we set 
$$
D_\mp=A-BX_\mp.
$$
\item $X\in\fR(A,B,C)$ is minimal/maximal iff 
$\sp(A-BX)\subset\overline{\cc}_\pm$.
\item $\fR(A,B,C)=X_-+\fR(D_-,B,0)=X_+-\fR(-D_+,B,0)$.
\een
\eet

Parts~(2) and~(3) are stated as Theorems~7.5.1 in~\cite{LR}. Part~(4) follows
from simple algebra. Since $X\mapsto\cR(X)$ is continuous, $\fR(A,B,C)$ is 
closed. Its boundedness follows from from Part~(4) and the fact that
$$
\|X-X_-\|_1=\tr(X-X_-)\le\tr(X_+-X_-),
$$
for all $X\in\fR(A,B,C)$. The Heine-Borel theorem thus yields Part~(1).

\subsection{The gap}
\label{APP-Gap}

In this section, we assume that $\fR(A,B,C)\not=\emptyset$ and use the
notations introduced in Theorem~\ref{THM-RicMax}. 

The gap of the Riccati equation~\eqref{EQ-RicGeneric} is the non-negative 
element of $L(\fh)$ defined by
$$
Y=X_+-X_-.
$$
We set $\cK=\Ker Y$, so that $\cK^\perp=\Ran Y$. For $X\in L(\fh)$, we define
$$
D_X=A-BX.
$$

\bet\label{THM-RicGap}
\ben
\item For any $X\in\fR(A,B,C)$, $\cK$ is the spectral subspace of
$D_X$ for $\sp(D_X)\cap\i\rr$ and $\cK^\perp$ is the spectral subspace of $D_X^\ast$ for $\sp(D_X^\ast)\setminus\i\rr$. Moreover,
$D_X|_\cK$ is independent of $X\in\fR(A,B,C)$.
\item The map $X\mapsto\Ker X$ is a bijection from $\fR(D_-,B,0)$
onto the set of all $D_-$-invariant subspaces containing the spectral
subspace of $D_-$ to the part of its spectrum in $\i\rr$.
Moreover, $X\le X'$ iff $\Ker X'\subset\Ker X$.
\item If $\cR(X)\le0$ for some self-adjoint $X\in L(\fh)$, then
$X_-\le X\le X_+$.
\item If $\cR(X)<0$ for some self-adjoint $X\in L(\fh)$, then
$\sp(K)\cap\i\rr=\emptyset$.
\een
\eet

The first and last Assertions of Part~(1) is Theorem~7.5.3 in~\cite{LR}.
The second Assertion is dual to the first one. Part~(2) is a special
case of Theorem~1 and Part~(3) is Theorem~14(b) in~\cite{Sc}.
Part~(4) is the first assertion of Theorem~9.1.3 in~\cite{LR}.

Note that Theorem~\ref{THM-RicChar} implies that for $X\in\fR(A,B,C)$ one has
$$
-KG_X=G_XD_X,
$$
so that $\sp(D_X)=\sp(-K|_{\cG(X)})$. Whenever $\sp(K)\cap\i\rr=\emptyset$,
it follows that $\sp(D_X)\cap\i\rr=\emptyset$ and hence $\cK=\{0\}$ and $Y>0$.
By Part~(3) of Theorem~\ref{THM-RicMax}, we further have 
$\sp(D_+)\subset\cc_-$ so that $G_{X_+}$ is the spectral subspace of $K$
to the part of its spectrum in $\cc_-$.

\subsection{Real Riccati equations and real solutions}

In this section, we assume that $\cE$ is a $d$-dimensional real Hilbert space
and $(A,B,C)$ a triple of elements of $L(\cE)$ such that $(A,B)$ is 
controllable, $B\ge0$, and $C$ self-adjoint.

Denote by $\fh=\cc\cE$ the complexification of $\cE$ equipped with its natural 
Hilbertian structure and conjugation $\cC$. The $\cc$-linear extensions of 
$A$, $B$ and $C$ to $\fh$ (which we denote by the same symbols) are such that 
$(A,B)$ is controllable, $B\ge0$, and $C$ is self-adjoint on $\fh$. Let 
$\fR(A,B,C)$ be the set of self-adjoint solutions of~\eqref{EQ-RicGeneric}, 
interpreted as a Riccati equation in $L(\fh)$, and define
$$
\fR_\rr(A,B,C)=\{X\in\fR(A,B,C)\,|\,X=\bar X\}.
$$
Clearly, $\fR_\rr(A,B,C)$ is the set of real self-adjoint solutions 
of~\eqref{EQ-RicGeneric} viewed as a Riccati equation on $L(\cE)$.

\bet\label{THM-RRic}
\ben
\item If $\fR(A,B,C)\not=\emptyset$, then its minimal/maximal element
is real and hence coincides with the minimal/maximal element
of $\fR_\rr(A,B,C)$.
\item Under the same assumption, the gap $Y=X_+-X_-$ is real and so is
$\cK=\Ker Y$.
\item For any $X\in\fR_\rr(A,B,C)$, $\cK$ is the spectral subspace of
$D_X$ for $\sp(D_X)\cap\i\rr$ and $\cK^\perp$ is the spectral subspace of $D_X^\ast$ for $\sp(D_X^\ast)\setminus\i\rr$. Moreover,
$D_X|_\cK$ is independent of $X\in\fR(A,B,C)$.
\een
\eet

To prove Part~(1), note that $\overline{X}\in\fR(A,B,C)$ whenever 
$X\in\fR(A,B,C)$. In particular, one has $\overline{X}_+\in\fR(A,B,C)$ and 
hence $X_+-\overline{X}_+\ge0$. It follows that
$$
\|X_+-\overline{X}_+\|_1=\tr(X_+-\overline{X}_+)=\tr(X_+-X_+^\ast)=0.
$$
The remaining statements are simple consequences of the reality of $X_\pm$.

\newpage
\printindex

\end{document}